\documentclass[a4,oneside,12pt]{Latex/Classes/PhDthesisPSnPDF}

\newcommand{\figuremacroW}[4]{
	\begin{figure}[!htbp]
		\centering
		\includegraphics[width=#4\textwidth]{#1}
		\caption{\footnotesize{#2}}
		\label{#3}
	\end{figure}
}

\newcommand{\ChangeFigFolder}[1]{
\ifpdf
    \graphicspath{{#1/figures/PNG/}{#1/figures/PDF/}{#1/figures/}}
\else
    \graphicspath{{#1/figures/EPS/}{#1/figures/}}
\fi
}

\DeclareMathSymbol{\minus}{\mathord}{operators}{"2D}

\newcommand{\der}[2]{
	\frac{\text{d}{#1}}{\text{d}{#2}}
}

\newcommand{\parder}[2]{
	\frac{\partial{#1}}{\partial{#2}}
}

\newcommand{\vect}[1]{\bm{#1}}

\newcommand{\transpose}[1]{{#1}^\mathsf{T}}
\newcommand{\ctranspose}[1]{{#1}^*}

\newcommand{\tabref}[1]{Table\;\ref{#1}}
\newcommand{\figref}[1]{Figure\;\ref{#1}}
\newcommand{\sectref}[1]{Section\;\ref{#1}}
\newcommand{\chapref}[1]{Chapter\;\ref{#1}}

\DeclareMathSymbol{\minus}{\mathord}{operators}{"2D}
\providecommand{\oforder}{\raise.17ex\hbox{$$ \mbox{\scriptsize $\sim$ }$$}}
\providecommand{\e}[1]{\ensuremath{\!\times\!10^{#1}}}
\providecommand{\cov}{\ensuremath{\text{Cov}}}
\providecommand{\var}{\ensuremath{\text{Var}}}

\ifpdf
    \hypersetup{
        pdfinfo={
                Keywords={LISA Pathfinder, spacetime metrology, geodesic motion, Doppler link, residual acceleration noise, system identification},
                }
    }
    \pdfcatalog { /PageMode (/UseOutlines)
                  /OpenAction (fitbh)  }
\fi

\title{Spacetime Metrology with LISA Pathfinder}

\ChangeFigFolder{0_frontmatter}
\ifpdf
  \author{\href{mailto:congedo@science.unitn.it}{Giuseppe Congedo}}
  \advisor{\href{mailto:vitale@science.unitn.it}{Prof.\;Stefano Vitale}}
  \coadvisor{\href{mailto:hueller@science.unitn.it}{Dr.\;Mauro Hueller}}
  \collegeordept{\href{http://www.unitn.it/en/dphys}{Department of Physics}}
  \university{\href{http://www.unitn.it/en}{University of Trento}}

  \crest{\includegraphics[width=4cm]{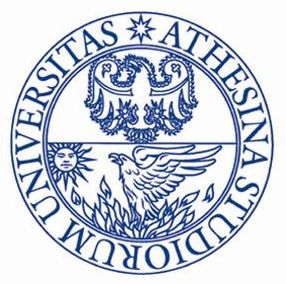}}

\else
  \author{Giuseppe Congedo}
  \advisor{Prof.\;Stefano Vitale}
  \coadvisor{Dr.\;Mauro Hueller}
  \collegeordept{Department of Physics}
  \university{University of Trento}
  \crest{\includegraphics[width=4cm]{logoUNITN}}
\fi

\degree{Philosophi\ae Doctor (PhD)}
\degreedate{26th March 2012}

\begin{document}


\renewcommand\baselinestretch{1.2}
\baselineskip=18pt plus1pt


\maketitle  


%
%
%




\thispagestyle{empty}

\begin{alwayssingle}

\pagestyle{empty}

\vspace*{\fill}
\begin{center}
\copyright~2012 \\
Giuseppe Congedo \\
All Rights Reserved
\end{center}

\end{alwayssingle}





\thispagestyle{empty}
\cleardoublepage

\begin{abstractseparate}

LISA is the proposed ESA-NASA space-based gravitational wave detector in the $\unit[0.1]{mHz}\text{--}\unit[0.1]{Hz}$ band. LISA Pathfinder is the down-scaled version of a single LISA arm. In this thesis it is shown that the arm -- named Doppler link -- can be treated as a differential accelerometer, measuring the relative acceleration between test masses. LISA Pathfinder -- the in-flight test of the LISA instrumentation -- is currently in the final implementation and planned to be launched in 2014. It will set stringent constraints, with unprecedented pureness, on the ability to put test masses in geodesic motion to within the required differential acceleration of $\unit[3\e{-14}]{m\,s^{-2}\,Hz^{-\nicefrac{1}{2}}}$ and track their relative motion to within the required differential displacement measurement noise of $\unit[9\e{-12}]{m\,Hz^{-\nicefrac{1}{2}}}$, at frequencies relevant for the detection of gravitational waves. Given the scientific objectives, it will carry out -- for the first time with such high accuracy required for gravitational wave detection -- the science of spacetime metrology, in which the Doppler link between two free-falling test masses measures the spacetime curvature. This thesis contains a novel approach to the calculation of the Doppler response to gravitational waves. It shows that the parallel transport of 4-vectors records the history of gravitational wave signals passing through photons exchanged between an emitter and a receiver. In practice, the Doppler link is implemented with 4 bodies (two test masses and two spacecrafts) in LISA and 3 bodies (two test masses within a spacecraft) in LISA Pathfinder. Different non-idealities may originate in the measurement process and noise sources couple the motion of the test masses with that of the spacecraft. To compensate for such disturbances and stabilize the system a control logic is implemented during the measurement. The complex closed-loop dynamics of LISA Pathfinder can be condensed into operators acting on the physical coordinates describing the relative motion. The formalism can handle the couplings between the test masses and the spacecraft, the sensing noise, as well as the cross-talk, and allows for the system calibration. It suppresses the transients in the estimated residual acceleration noise between the test masses. The scope of system identification is indeed the calibration of the instrument and the compensation of different effects. After introducing a model for LISA Pathfinder along the optical axis and an example of cross-talk from other degrees of freedom to the optical axis, this thesis describes some data analysis procedures applied to synthetic experiments and tested on a realistic simulator provided by ESA. The same procedures will also be adopted during the mission. Those identification experiments can also be optimized to get an improvement in precision of the noise parameters that the performances of the mission depend on. This thesis demonstrates the fundamental relevance of system identification for the success of LISA Pathfinder in demonstrating the principles of spacetime metrology needed for all future space-based missions.

\end{abstractseparate}




\frontmatter


\begin{dedication} 

To my wife Laura, \\
source of inspiration and happiness.

\end{dedication}



\thispagestyle{empty}
\cleardoublepage

\begin{acknowledgements}      

Firstly, I would like to acknowledge my research advisor Prof.\,Stefano Vitale which offered me to work on the LISA Pathfinder project, gave me the chance to know the many people around the Data Analysis team and, most of all, taught me how to proficiently work as a researcher in physics. The co-advisor Mauro Hueller for his guidance through the open problems in data analysis and the organization of the research program.

Luigi Ferraioli (now in APC, Universit\`e Paris Diderot) for the scientific rigor he taught me, the many times he helped me during these years with both research and traveling issues, without whom I would never have realized a big part of this work. Fabrizio De Marchi (now in University of Rome, Tor Vergata) for the many interesting discussions in astronomy, science (and lots more), the scientific collaboration and the great company during these years. Bill Weber who bore the reading of this thesis and helped me with English. Prof.\;Giovanni Prodi for the initial involvement in research just after my arrival in Trento. Renato Mezzena for the interesting discussions during lunch time. Karine Frisinghelli for her strong support and friendship.

Abroad, Martin Hewitson (AEI, Max-Planck-Institut f\"ur Gravitationsphysik und Universit\"at Hannover) head of the LISA Pathfinder Data Analysis team; Michele Armano (ESAC, ESA, Madrid) and Miquel Nofrarias (IEEC, Universitat de Barcelona) for useful suggestions and discussions; Michele Vallisneri (JPL NASA and California Institute of Technology, Pasadena) for clarifications on the response of the LISA detector.

Again, Prof.\;Stefano Vitale for the teaching opportunity that increased my self-control and my ability as a speaker, much more than I expected. For the teaching adventure, Rita Dolesi, Barbara Rossi, David Tombolato and Fabrizio De Marchi with all whom I collaborated successfully.

Finally, I would like to thank my family in Lecce for the support during the academic program in physics started years ago -- this thesis is its natural conclusion -- and my wife for her role in significantly contributing to both my physical and mental wellness during the PhD.

\end{acknowledgements}

\thispagestyle{empty}
\cleardoublepage


\setcounter{page}{1}
\setcounter{secnumdepth}{3} 
\setcounter{tocdepth}{3}    
\tableofcontents            








\nomenclature{LTP}{LISA Technology Package}
\nomenclature{LPF}{LISA Pathfinder}
\nomenclature{LISA}{Laser Interferometer Space Antenna}
\nomenclature{TM}{Test Mass}
\nomenclature{SC}{SpaceCraft}
\nomenclature{GW}{Gravitational Wave}
\nomenclature{FW}{Fermi-Walker}
\nomenclature{MIMO}{Multi-Input-Multi-Output}
\nomenclature{PSD}{Power Spectral Density}
\nomenclature{SNR}{Signal-to-Noise Ratio}
\nomenclature{FFT}{Fast Fourier Transform}
\nomenclature{IFFT}{Inverse Fast Fourier Transform}
\nomenclature{ESA}{European Space Agency}
\nomenclature{DFACS}{Drag-Free and Attitude Control System}
\nomenclature{PDF}{Probability Density Function}
\nomenclature{GRS}{Gravity Reference Sensor}
\nomenclature{OMS}{Optical Metrology Subsystem}
\nomenclature{TDI}{Time-Delay Interferometry}
\nomenclature{LTPDA}{LTP Data Analysis}
\nomenclature{ST}{Star-Tracker}
\nomenclature{FEEP}{Field Emission Electric Propulsion}
\nomenclature{IFO}{InterFerOmeter}
\nomenclature{TT}{Traceless-Transverse}
\nomenclature{OB}{Optical Bench}
\nomenclature{OSE}{Offline Simulation Environment}
\nomenclature{GR}{General Relativity} 

\begin{multicols}{2} 
\begin{small} 

\printnomenclature[1.5cm] 
\label{nom} 

\end{small}
\end{multicols}

\markboth{\MakeUppercase{\nomname}}{\MakeUppercase{\nomname}}


\mainmatter

\renewcommand{\chaptername}{} 






\ChangeFigFolder{1_introduction}


\chapter{Introduction} \label{chap:introduction}



40 years ago the binary pulsar 1913+16 \cite{hulse1975} opened up a long series of observations aimed at determining various relativistic effects -- like the periastron shift -- that were confirmed to be in very good agreement with General Relativity (GR). The discovery of the pulsar gave the first strong indication of the existence of Gravitational Waves (GWs). Yet to date no direct detection has been made, in spite of many efforts of disparate experiments still in progress. The detection of GW signals requires the development of sophisticated devices capable in accurately measuring very small accelerations between nominally free-falling test particles subjected to different noise sources. The same measurement principle, with slight modifications, is shared among the 1st, the 2nd and the 3rd generation of ground-based detectors, as well as the planned spaced-based detectors.

\section{LISA, a space-borne gravitational wave detector}

A passing GW would cause a change in the relative velocities between test particles in nominal free fall. As a Michelson interferometer, a GW detector measures such a physical quantity. Ground-based GW detectors have currently reached almost their design sensitivities, and the 2nd generation, Adv.\,LIGO \cite{harry2010}, Adv.\,Virgo \cite{accadia2011} and GEO-HF \cite{willke2006}, promises an improvement in detection rates and a wider horizon to be explored in the $\unit[10]{Hz}\text{--}\unit[10]{kHz}$ band. The 3rd generation with the Einstein Telescope \cite{punturo2010,hild2011} will provide further enhancements in both sensitivity and frequency band, especially toward the low-frequency end that, at $\unit[1]{Hz}$, is limited by the Earth gravity noise. It's not just by chance that the proposed design for the Einstein Telescope is an underground $\unit[100]{km}$-wide equilateral triangular scheme of Michelson interferometers as the triangle can be considered the optimal configuration in resolving both source polarization and position with extremely high confidence. Years ago, the Laser Interferometer Space Antenna (LISA) \cite{bender1998,danzmann2011} -- a joint ESA \cite{esa} - NASA \cite{nasa} mission -- was discovered to offer the possibility of exploring a much lower frequency band, $\unit[0.1]{mHz}\text{--}\unit[0.1]{Hz}$, expected to be saturated by the huge population of GW binaries.

\begin{figure}[!htbp]
\centering
\begin{tabular}{*{2}{@{}c@{}}} 
\includegraphics[width=0.5\columnwidth]{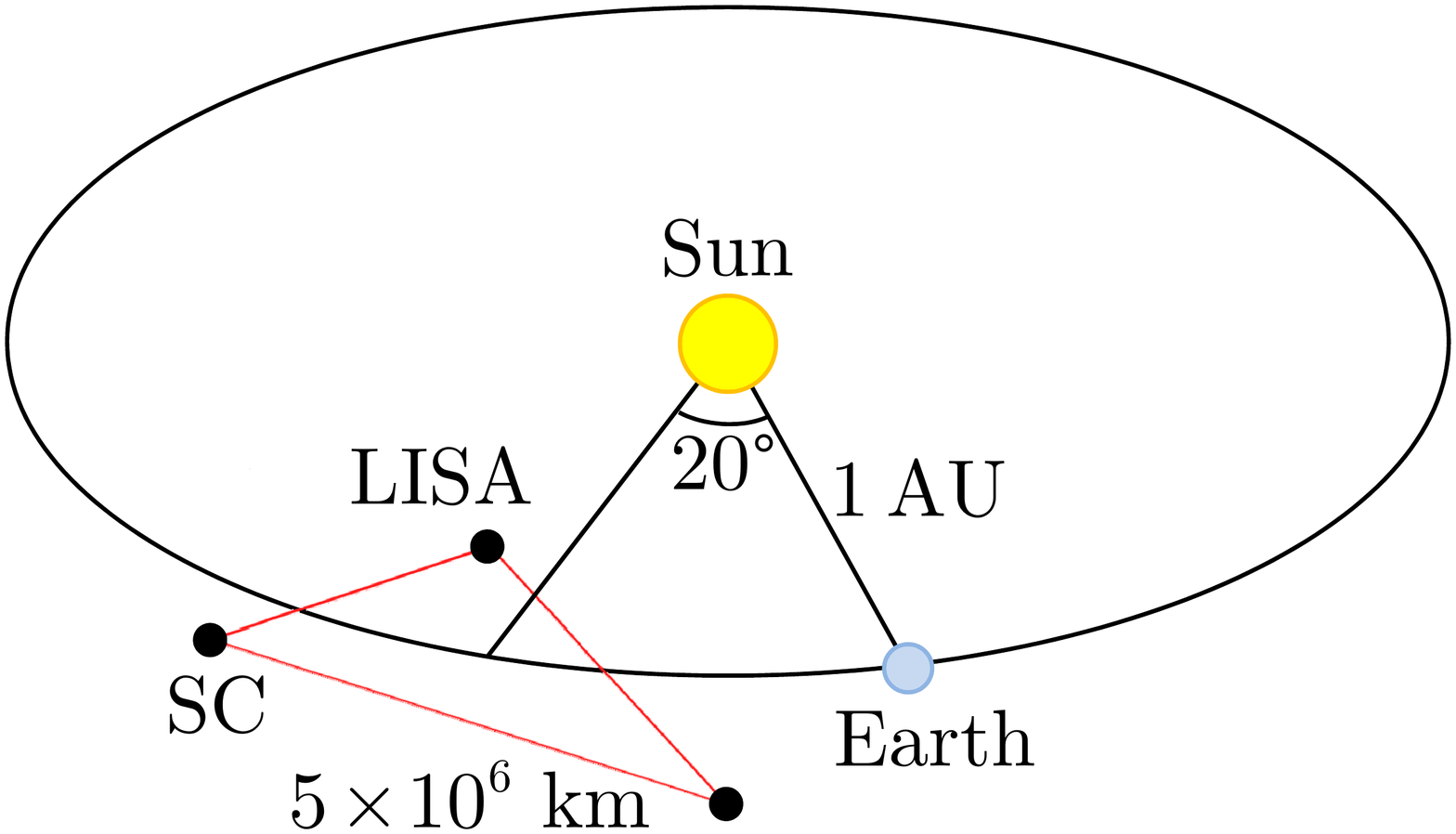} &
\includegraphics[width=0.5\columnwidth]{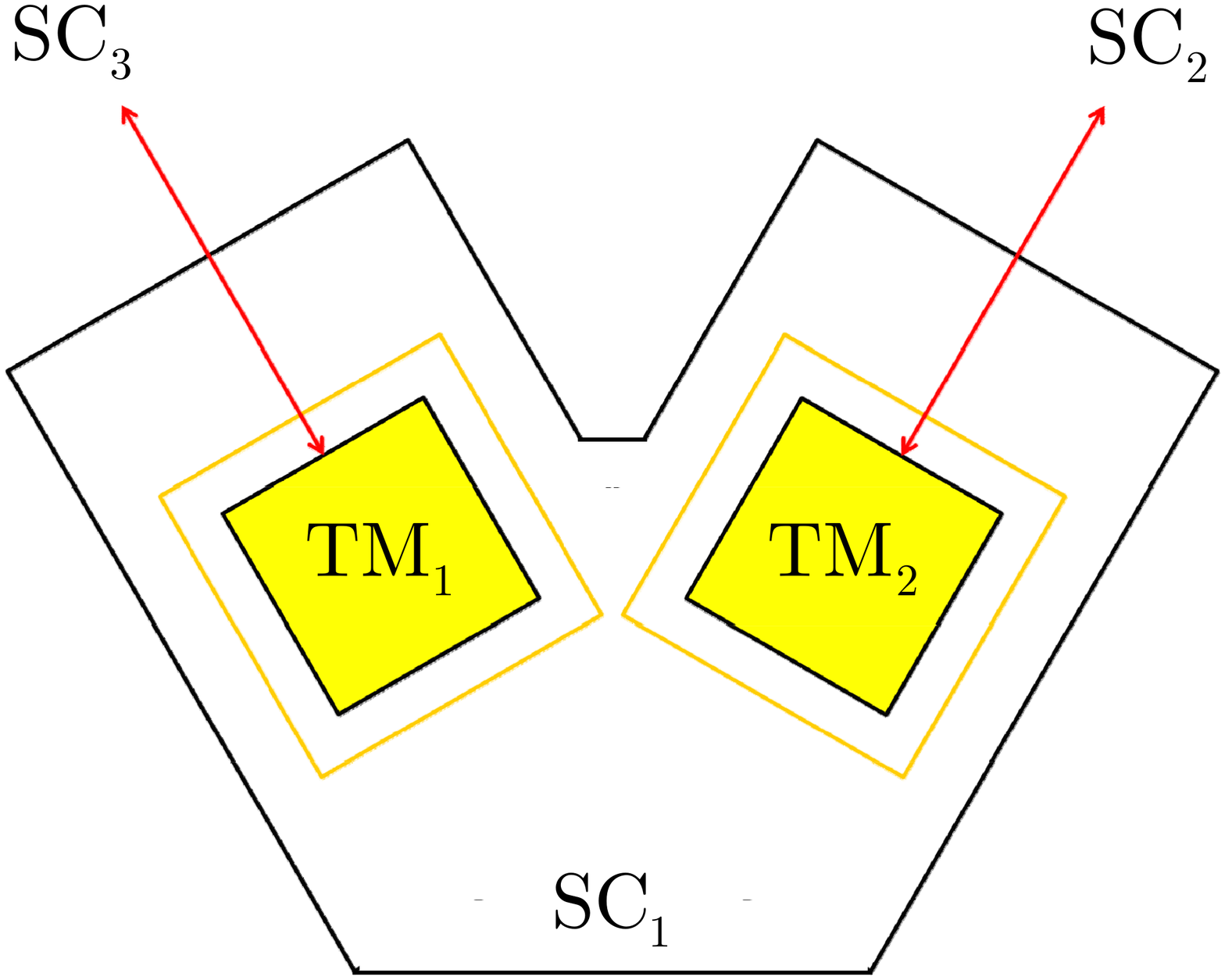} \\
\footnotesize{(a)} & \footnotesize{(b)}
\end{tabular}
\caption{\footnotesize{Scheme of the LISA orbit (not in scale) around the Sun and details of a single SC. (a) the triangular formation follows the Earth and maintains its arm length approximately constant within few percents. (b) each SC contains two TMs and the relative displacements to the faraway counterparts are detected by a laser-interferometric technique. 6 TMs constitute 6 Doppler links, two per LISA arm, tracking the local curvature variations around the Sun and are sensitive to GW signals in the $\unit[0.1]{mHz}\text{--}\unit[0.1]{Hz}$ band.}}
\label{fig:introduction:lisa_scheme}
\end{figure}

The key concept of LISA is the constellation flight of three SpaceCrafts (SCs) -- each hosting and protecting two Test Masses (TMs) in nominal free fall -- in a $\unit[5\e{6}]{km}$ sided equilateral triangle around the Sun at $\unit[1]{AU}$ as shown in \figref{fig:introduction:lisa_scheme}. The arm length is approximately constant within a fractional tolerance of few percent. The angles are allowed to vary over the year within $\oforder1^\text{o}$ at most. No frequent orbit corrections are actually needed and the formation follows the Earth with a trailing angle of $\oforder20^\text{o}$, a compromise solution between gravitational perturbations and communication/fuel constraints \footnote{Recently, due to funding cuts, the US side has withdrawn its participation in a GW mission in the 2020-2025 timeframe. Meanwhile, the European has started a feasibility study of a descoped version named eLISA/NGO \cite{jennrich2011b,elisa} fitting the cost of an ESA L-class mission and at the same time maintaining most of the scientific objectives. Some of the modifications include a shorter lifetime, shorter arm lengths ($\unit[1\e{6}]{km}$), a smaller trailing angle and the possible suppression of two Doppler links. The adopted ``mother-daughter'' configuration would be the first Michelson interferometer in space allowing for the detection of many continuous sources with revolutionary scientific returns \cite{amaro2012}.
This mission is being evaluated by ESA at the time of writing down this thesis. However, this thesis refers to LISA without any loss of generality, while keeping in mind that all discussions and results are still valid for \textit{any} variant of LISA based on the same detection principle.}.

In LISA the relative velocities between the TMs change as a GW passes through the constellation. LISA is a combination of 3 quasi-independent Michelson interferometers and, as such, detects oscillating signals. Given the very low frequency band compared to the ground-based, LISA will be sensitive to continuous signals arising from inspiral, merger and ringdown of binaries. Among many astrophysical targets, the detection and characterization of the following objects will be of fundamental importance during the nominal 5-year mission:
\begin{enumerate}
  \item Super-Massive Black Holes (SMBHs) with very high Signal-to-Noise Ratio (SNR), out to redshift $z\oforder15$, from the merging of galactic nuclei;
  \item a dozen of galactic verification binaries for each of which an electromagnetic counterpart is available;
  \item hundreds (or even thousands) of galactic binaries, continuous or chirping, that can be distinctively resolved;
  \item unresolved galactic binaries appearing as noise foreground at low frequency;
  \item Extreme Mass Ratio Inspirals (EMRIs) to study GR in highly curved spacetimes;
  \item stochastic cosmic background.
\end{enumerate}
These scientific objectives make LISA a GW telescope with a potentially huge impact in whole physics. Contrary to the ground-based detectors, LISA can be considered a signal-dominated detector where the interferometric outputs are three correlated time-series containing the superposition of many signals in whole sky: its conceptual and practical complexities make the extraction of such signals sophisticated. A typical feature of LISA is its ability in resolving sources with very high position accuracy. This is due to a double Doppler modulations induced by the revolution around the Sun and the intrinsical rotation of the normal to the constellation plane (Appendix \ref{sect:appendix:introduction_binary} shows an example of the LISA response to a single galactic binary).

The LISA objectives in astrophysics requires that the TMs must be kept in free fall with a residual acceleration noise as low as $\unit[3\e{-15}]{m\,s^{-2}\,Hz^{-\nicefrac{1}{2}}}$ around $\unit[1]{mHz}$ -- a goal achievable thanks to the sophisticated design and technology employed onboard.


\section{LISA Pathfinder: spacetime metrology and verification of the detection principle}

In the last decade LISA Pathfinder (LPF) \cite{vitale2002} was proposed to fly as a targeted ESA mission \cite{lpf} to verify the detection principle of LISA. LPF is a down-scaled version of a single LISA arm to the size of about $\unit[40]{cm}$. The main scope of LPF is to give an in-flight test of the LISA instrumentation and demonstrate that parasitic forces are constrained such that the measured differential acceleration between two TMs is below the level of $\unit[3\e{-14}]{m\,s^{-2}\,Hz^{-\nicefrac{1}{2}}}$ around $\unit[1]{mHz}$.

\begin{figure}[!htbp]
\centering
\begin{tabular}{*{2}{@{}c@{}}} 
\includegraphics[width=0.5\columnwidth]{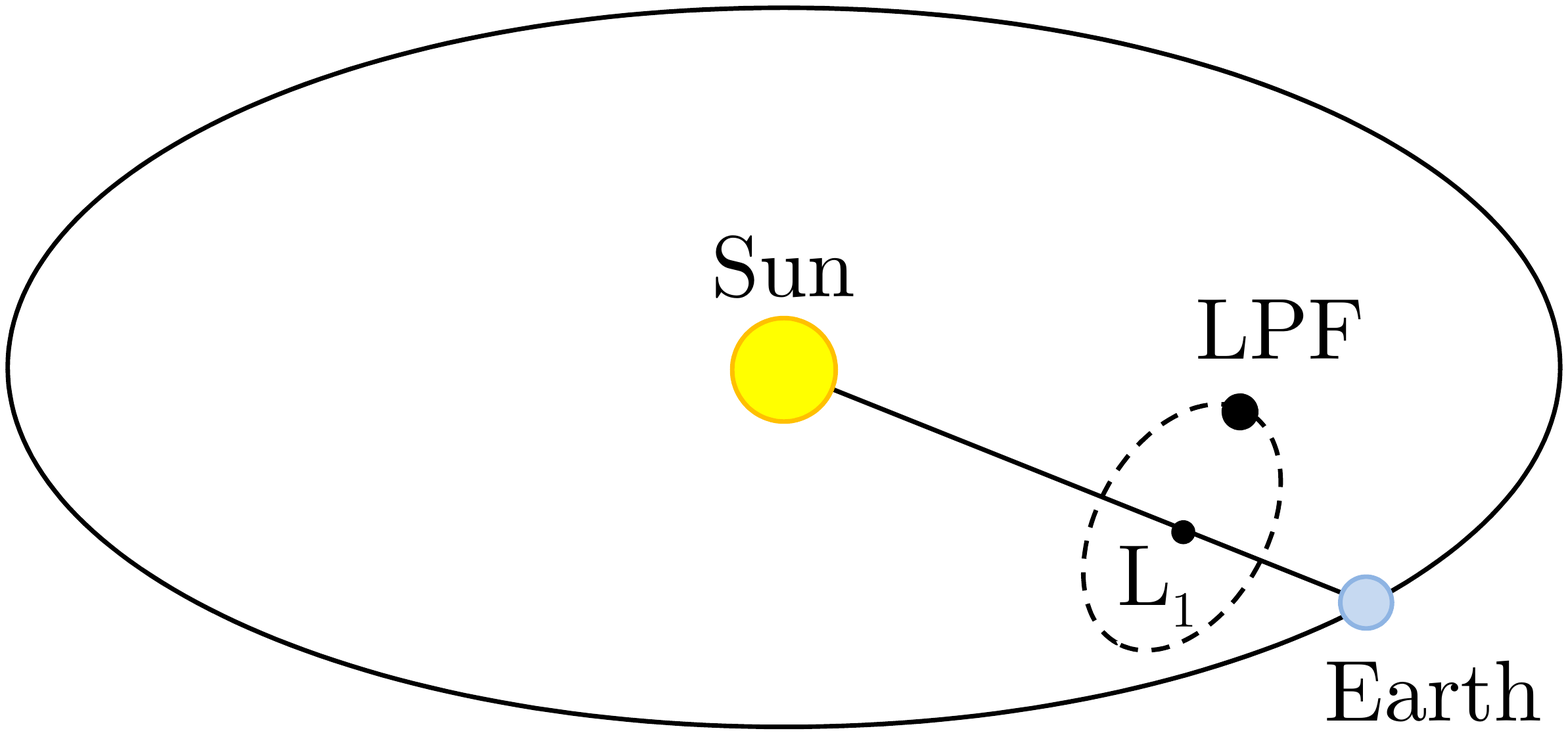} &
\includegraphics[width=0.5\columnwidth]{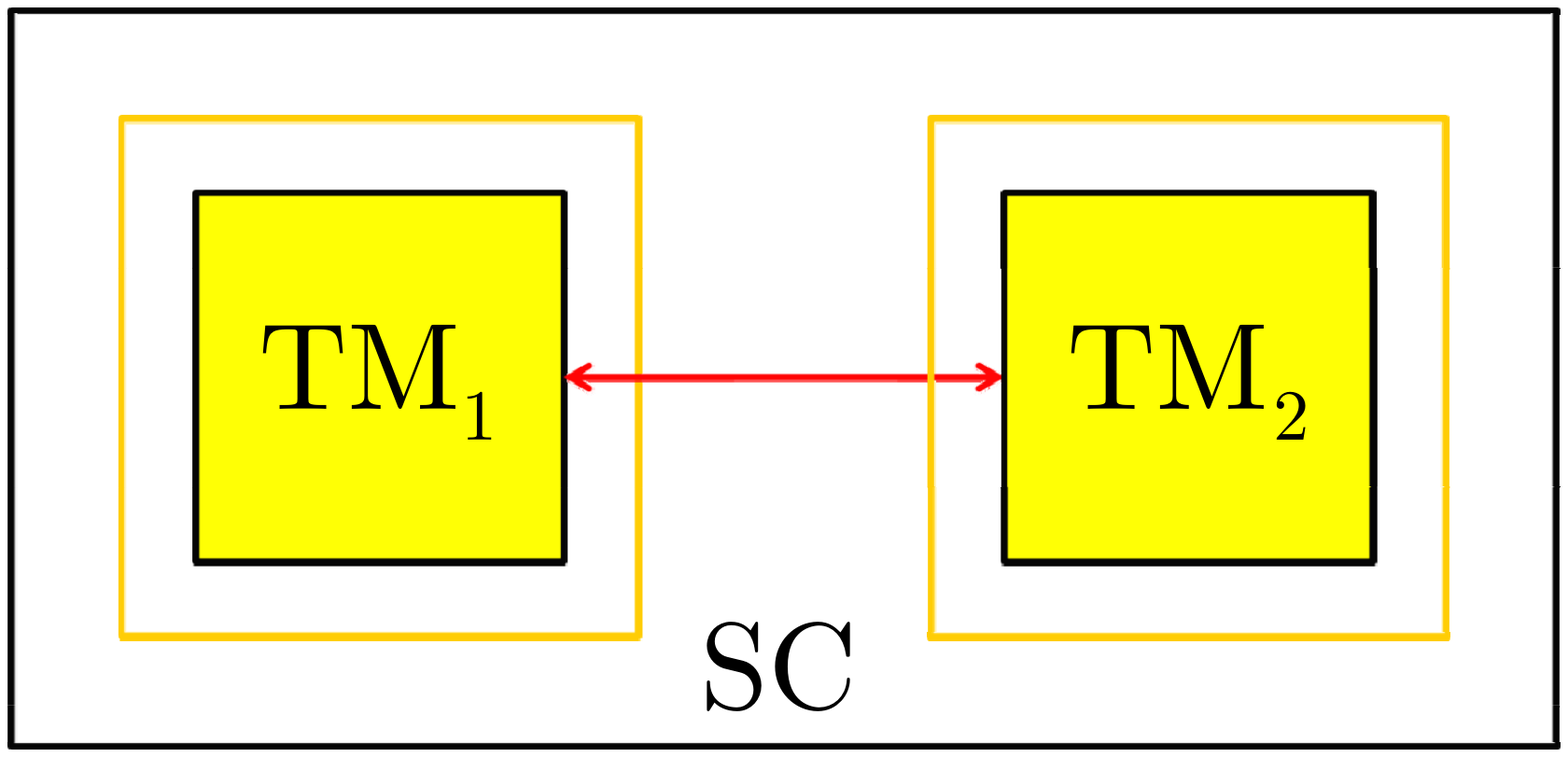} \vspace{-20pt} \\
\footnotesize{(a)} & \footnotesize{(b)}
\end{tabular}
\caption{\footnotesize{Scheme of the LPF orbit (not in scale) around L$_1$ and details of the SC. (a) the SC is in a halo orbit and station-keeping forces must be applied orthogonally for its stabilization. (b) the SC contains two TMs whose relative displacements are detected by a laser-interferometric technique.}}
\label{fig:introduction:lpf_scheme}
\end{figure}

Currently in the final implementation and planned to be launched in 2014 \cite{antonucci2011c}, LPF will fly in a Lissajous orbit around the L$_1$ Lagrange point ($\unit[1.5\e{6}]{km}$ away from the Earth toward the Sun). See \figref{fig:introduction:lpf_scheme} for reference. Even though such orbits are periodic, they are unstable and station-keeping forces must be applied orthogonally to the orbit plane (and parallel to the axis joining the two celestial bodies). The solar array, also working as a shield to the SC underneath, will point the Sun to within a few degrees. A residual spin around the same axis is kept lower than 3$^\text{o}$ per day for scientific requirements. An alternate possibility has been also considered as backup option in case the propulsion module may fail in transferring the payload from the low Earth orbit to the target. The SC may be injected in a highly eccentric orbit around the Earth with a period of 27 days. Even though this solution does not allow for continuous measurements at the optimal sensitivity close to the perigee for 2--3 days, it is an interesting test-bench for utilizing the Moon as a calibrator of the instrument \cite{S2-UTN-TN-3069}.

LPF is expected to provide an accurate noise model for LISA and put stringent constraints, with unprecedented results, on \cite{danzmann2007}:
\begin{enumerate}
  \item the ability to keep TMs in free fall -- the so-called \textit{differential acceleration noise requirement} -- to within the level of
  \begin{equation}\label{eq:introduction:acc_noise}
    S_{\text{n},\delta a}^{\nicefrac{1}{2}}\lesssim3\e{-14}\left[1+\left(\frac{f}{f_0}\right)^2\right]\,\unit{m\,s^{-2}\,Hz^{-\nicefrac{1}{2}}}~;
  \end{equation}
  \item the ability to track relative displacements between the TMs with a laser interferometer -- the so-called \textit{differential displacement noise requirement} -- to within the level of
  \begin{equation}\label{eq:introduction:ifo_noise}
    S_{\text{n},\delta x}^{\nicefrac{1}{2}}\lesssim9\e{-12}\left[1+\left(\frac{f_0}{f}\right)^2\right]\,\unit{m\,Hz^{-\nicefrac{1}{2}}}~;
  \end{equation}
\end{enumerate}
where $f_0=\unit[3]{mHz}$ and over the $\unit[1\text{--}30]{mHz}$ band. The LPF requirements are relaxed by almost an order of magnitude to LISA. The high frequency regime is dominated by the displacement requirement of $\unit[9]{pm\,Hz^{-\nicefrac{1}{2}}}$, whereas the acceleration requirement of $\unit[30]{fN\,Hz^{-\nicefrac{1}{2}}}$ has much more importance to the low frequency assessment of the LISA noise. \figref{fig:introduction:lisa_lpf_reqs} compares the requirements in Power Spectral Density (PSD) of the residual acceleration noise for LISA and LPF. Even though LPF shares the same hardware design with LISA, a relaxation in both acceleration noise level and frequency band is allowed for the first.

\figuremacroW{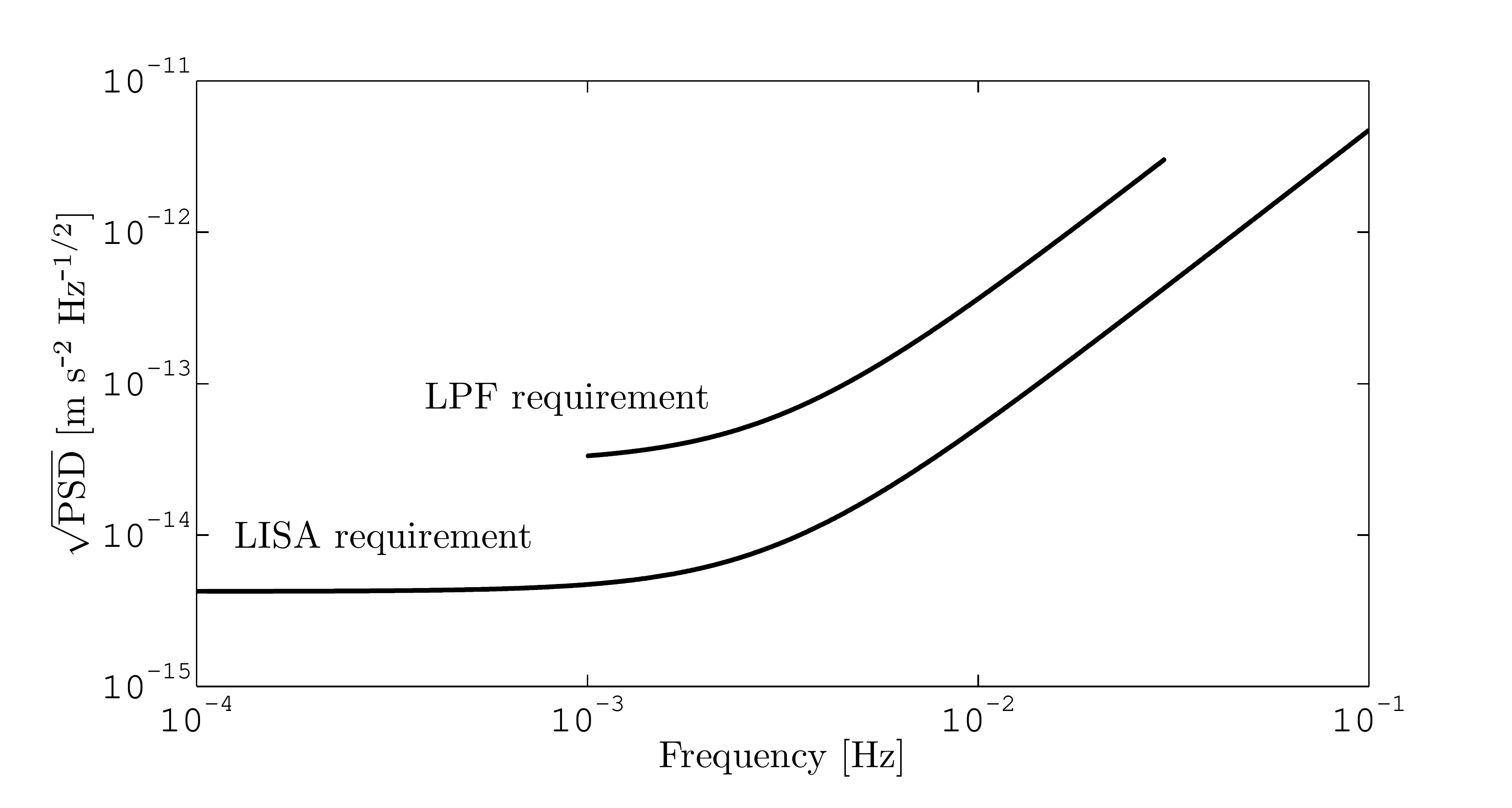}{Comparison between the residual acceleration noise requirement of LPF and LISA. LPF is relaxed with respect to LISA by a factor $\oforder7$ in amplitude. The required LISA band ($\unit[0.1]{mHz}\text{--}\unit[0.1]{Hz}$) is extended toward the low frequency compared to the LPF band ($\unit[1\text{--}30]{mHz}$). Obviously, during the mission a lower acceleration and a wider frequency band will be easily reached.}{fig:introduction:lisa_lpf_reqs}{0.8}

\section{LISA Technology Package}

LPF and its main scientific payload, the LISA Technology Package (LTP) \cite{antonucci2011a}, will give an in-flight test of the LISA hardware and effectively measure the differential acceleration noise that pollutes the sensitivity of LISA below $\unit[3\e{-14}]{m\,s^{-2}\,Hz^{-\nicefrac{1}{2}}}$ around $\unit[1]{mHz}$ -- the minimum performance level for LISA to carry on its science program in astrophysics. As said, the observational horizon of LISA will include thousands of GW sources. Among all, those with the highest SNR will be surely the SMBHs. However, there are sources that are expected to lay at the limit of the LISA sensitivity for which an accurate assessment of the instrumental noise is mandatory. The population of EMRIs \cite{amaro2007} is the most important example: they are a valuable instrument to test GR and curvature in the strong gravity regime. Different EMRI search methods have been developed. After having subtracted the highest signals (SMBHs and calibration binaries), in order to extract the EMRI signatures, all methods strictly have to deal with the instrumental noise level, for which the LPF mission has a crucial role. In fact, a systematic error in the reconstructed noise shape would dramatically affect the identification of such sources. This thesis shows the importance of LPF and system identification for the correct assessment of the noise parameters and the noise shape. A numerical example will be provided by \chapref{chap:sys_identification}.

During the 3 months of operations, the LTP experiment on board LPF will be used in an extensive characterization campaign to measure all force disturbances and systematics, like the TMs couplings, the various cross-talks, the TM charging due to cosmic particles and its interaction with the electrostatic environment, the thermal and magnetic effects, etc. The impact of the effects on the differential acceleration noise can be inferred by simulations and through on-ground measurements. In fact, two facilities (single-mass and 4-mass torsion pendulum) have been employed during the last years to investigate the one-degree-of-freedom behavior of a replica of the Au-Pt TM of $\unit[1.96]{kg}$ and its electrostatic housing, including all sensing and actuation capacitive electrodes, entirely named Gravitational Reference Sensor (GRS) \cite{dolesi2003}. A comprehensive review of the current status of the on-going measurement activities and their extrapolations to LISA are given in \cite{antonucci2011a} and references therein.

The LTP experiment comprises the following key subsystems shown in \figref{fig:introduction:lpf_scheme_subsystems}: two GRSs, the Optical Metrology System (OMS) (InterFerOmeters (IFOs) and the optical bench), Star-Trackers (STs), an on-board computer, the Drag-Free and Attitude Control System (DFACS) and the Field Emission Electric Propulsion (FEEP) thrusters. The experiment is also equipped with magnetometers, thermometers and a cosmic charge counter. The sensors with the relative sensed motions are reported in \tabref{tab:introduction:subsystems_sensors}. The noise requirements are reported in \tabref{tab:introduction:subsystems_requirements}.

\figuremacroW{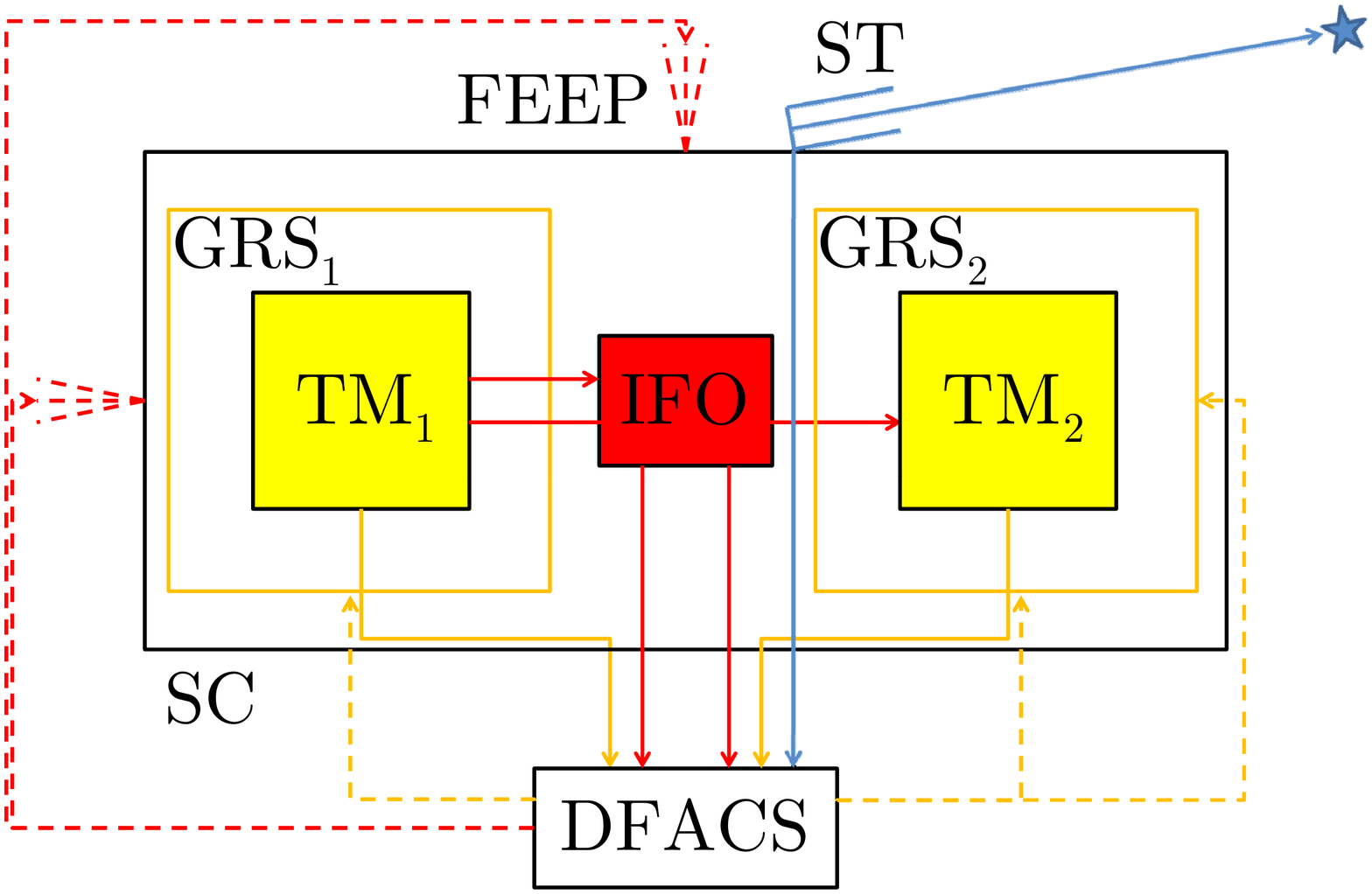}{Scheme of the key subsystems of the LPF mission. The SC contains two GRSs and an optical bench with four interferometers. The relative displacements and attitudes between the TMs and the optical bench are read out by the interferometers and the capacitive sensors. The interferometric, capacitive and star-tracker readouts (solid lines) are fed into the DFACS that computes the forces that shall be actuated by the FEEP thrusters and the capacitive actuators (dashed lines). In the main science mode the reference TM is not actuated along the optical axis.}{fig:introduction:lpf_scheme_subsystems}{0.8}

\begin{table}[!htbp]
\caption{\label{tab:introduction:subsystems_sensors}\footnotesize{LTP sensors and the relative sensed motions.}}
\centering
\begin{tabular}{l p{9cm}}
\hline
\hline
Sensor & Motion \\
\hline
GRS & linear and angular motion of the TMs relative to their housings  \\
OMS & linear and angular motion of the reference TM relative to the optical bench \\
    & linear and angular motion of the second TM relative to the reference TM \\
ST & absolute attitude of the SC \\
\hline
\hline
\end{tabular}
\end{table}

\begin{table}[!htbp]
\caption{\label{tab:introduction:subsystems_requirements}\footnotesize{LTP key subsystems and the main noise requirements around $\unit[1]{mHz}$.}}
\centering
\begin{tabular}{l l l}
\hline
\hline
Subsystem & Requirement & Note \\
\hline
\multirow{2}{*}{GRS} & $\unit[1.8]{nm\,Hz^{-\nicefrac{1}{2}}}$ & displacement sensing \\
& $\unit[20]{fN\,Hz^{-\nicefrac{1}{2}}}$ & actuation \\
\multirow{2}{*}{OMS} & $\unit[9]{pm\,Hz^{-\nicefrac{1}{2}}}$ & displacement sensing \\
& $\unit[20]{nrad\,Hz^{-\nicefrac{1}{2}}}$ & attitude sensing \\
ST & $\unit[32]{''\,Hz^{-\nicefrac{1}{2}}}$ & -\\
\multirow{4}{*}{DFACS} & \multirow{2}{*}{$\unit[5\text{--}6]{nm\,Hz^{-\nicefrac{1}{2}}}$} & displacement control \\
& & (main science mode) \\
& \multirow{2}{*}{$\unit[0.4\text{--}0.5]{\mu rad\,Hz^{-\nicefrac{1}{2}}}$} & attitude control \\
& & (main science mode) \\
FEEP & $\unit[0.1]{\mu N\,Hz^{-\nicefrac{1}{2}}}$ & - \\
\hline
\hline
\end{tabular}
\end{table}

\subsection{Gravitational reference sensor}

Each GRS comprises an Au-Pt cubic TM of size $\unit[46]{mm}$ and a surrounding electrostatic housing containing capacitive sensors and actuators in all 6 degrees of freedom. Each GRS senses the relative displacement and attitude of the TM to its housing and provides actuation along the same degrees of freedom. Gaps between the TM and its housing are $\unit[3\text{--}4]{mm}$, a compromise between noise minimization and efficient sensing/actuation. The GRS vacuum chamber allows for a residual gas pressure at the level of $\unit[10]{\mu Pa}$. 
UV light illumination is utilized to control the accumulated charge with a discharging threshold of $\oforder\unit[10^7]{e}$ -- the accumulated charge in one day for an expected charging rate of $\oforder\unit[10^2]{e\,s^{-1}}$.
The sensing requirements of each GRS are $\unit[1.8]{nm\,Hz^{-\nicefrac{1}{2}}}$ in displacement and $\unit[200]{nrad\,Hz^{-\nicefrac{1}{2}}}$ in attitude. The actuation requirement is $\unit[20]{fN\,Hz^{-\nicefrac{1}{2}}}$ with a maximum range of $\unit[2.5]{nN}$.

\subsection{Optical metrology system}


The OMS \cite{steier2009} comprises: a Zerodur$^\circledR$ monolithic optical bench, 4 Mach-Zehnder heterodyne $\unit[1.024]{\mu m}$ interferometers and redundant quadrant photodiodes. 
The first IFO, $X_1$, senses the relative displacement and attitude of one reference TM to the optical bench itself. The differential IFO, $X_{12}$, senses the relative displacement and attitude between the two TMs.
Relative displacements are measured by averaging among the four quadrants, whereas relative angles are measured by taking the difference between opposite quadrants (differential wavefront sensing). The ``reference'' IFO is subtracted from the previous ones for compensating spurious fiber optical path length variations before the first beam splitter. The ``frequency'' IFO is utilized for laser frequency stabilization. The sensing requirements are $\unit[9]{pm\,Hz^{-\nicefrac{1}{2}}}$ in displacement, as in \eqref{eq:introduction:ifo_noise}, and $\unit[20]{nrad\,Hz^{-\nicefrac{1}{2}}}$ in attitude with a maximum range of $\unit[100]{\mu m}$. A rotation around the optical axis is not sensed, but can be provided by the GRS.

\subsection{Star-trackers}

The STs are small telescopes reading out the inertial attitude of the SC with respect to the star field. The sensing requirement is $\unit[32]{''\,Hz^{-\nicefrac{1}{2}}}$ ($\unit[160]{\mu rad\,Hz^{-\nicefrac{1}{2}}}$).

\subsection{Drag-free and attitude control system}

The outputs of all sensors, GRSs, OMS and STs, are elaborated by the on-board computer and fed into the DFACS \cite{fichter2005}. The DFACS has the responsibility of computing the control forces that shall be passed to capacitive and thruster actuators in order to stabilize the system and meet the acceleration requirement in \eqref{eq:introduction:acc_noise}.

There are different operational control modes for the LPF mission. To avoid large transients in the data, the transition between two modes is implemented with overlapping sub-modes. In the \textit{accelerometer mode} LPF acts as a standard accelerometer in which the TMs are both electrostatically actuated along the optical axis and controlled to follow the SC motion. The resulting noise is much higher than the requirement. In the \textit{main science mode}, the DFACS is responsible in maintaining a reference TM in free fall along the optical axis and forcing both the second TM and the SC to follow it by capacitive and thruster actuation.

The need for the DFACS is explained not only by the scientific requirements, but also by the fact that noise sources can destabilize the system on a time scale of few minutes and the gaps between the TM and its housing are just $\unit[3\text{--}4]{mm}$. One of the proposed activities, the free flight experiment \cite{grynagier2009}, is aimed at obtaining an improvement in differential acceleration noise at low frequency by turning off the capacitive actuation also on the second TM which is left in ``parabolic'' free fall and impulsively kicked every $\unit[200]{s}$.

In the main science mode the DFACS is conceptually divided into three control loops \cite{S2-ASD-TN-2001} with the following priority:
\begin{enumerate}
  \item \textit{drag-free control loop}, controlling the relative displacement and attitude of the SC with respect to the reference TM through thruster actuation;
  \item \textit{electrostatic suspension control loop}, controlling the relative displacement and attitude between the TMs through capacitive actuation on the second TM;
  \item \textit{attitude control loop}, controlling the inertial (absolute) attitude of the TMs through capacitive actuation.
\end{enumerate}
The drag-free requirement are $\unit[5\text{--}6]{nm\,Hz^{-\nicefrac{1}{2}}}$ in displacement and $\unit[0.4\text{--}0.5]{\mu rad\,Hz^{-\nicefrac{1}{2}}}$ in attitude.

\subsection{Thrusters}

The FEEP is attained by an ensemble of 3 clusters, of 4 thrusters each, attached to the SC. An electron flux keeps the SC neutral. The force requirement is $\unit[0.1]{\mu N\,Hz^{-\nicefrac{1}{2}}}$ with a maximum range of $\unit[100]{\mu N}$. The FEEP thruster authority is the only means by which the reference TM can be maintained in free-fall along the optical axis, hence mitigating the SC jitter at low frequency. The SC is also equipped with colloid thrusters provided by NASA for complementary experiments.

Recently, ESA has considered the possibility to employ cold gas thrusters in place of the FEEP. The new design is expected to perform to within the requirements as well. However, the considerations and the results of this thesis are still valid and are not appreciably affected by the possible change in design.

\section{Outline of the work}

In LISA a total of 6 TMs, whose relative displacements \footnote{Throughout this thesis an extensive use (and abuse) of terms like ``relative displacement'', ``frequency shift'', ``phase difference'', etc. will be made without any relevant distinction. The explanation is that a relative displacement is proportional to a phase difference, $\delta r\simeq\lambda\,\delta\phi$ (with $\lambda$ the light wavelength), and a relative velocity is proportional to a frequency shift, $\delta v\simeq\lambda\,\delta\omega$. The two are obviously related by a time derivative. The fractional frequency shift is also useful, as in the next chapter, and its relation to phase difference is $\delta\omega/\omega=\dot{\delta\phi}/\omega$. The following table shows the equivalence between the mentioned quantities:
\begin{center}
\begin{tabular}{l l l}
\hline
\hline
Relative displacement & Phase shift \\
Relative velocity & Frequency shift \\
Relative acceleration & Frequency shift rate \\
\hline
\hline
\end{tabular}
\end{center}} are tracked by a laser-interferometric technique, constitute 6 Doppler links, two per LISA arm, tracking the local curvature variations around the Sun and sensing the small fluctuations induced by GW signals in the $\unit[0.1]{mHz}\text{--}\unit[0.1]{Hz}$ band. LISA can be viewed as a combination of three quasi-independent nominally equal-arm Michelson interferometers with vertices at each SC. In the ground-based detectors the laser frequency noise is common-mode between the two arms and can be subtracted with very high accuracy. In LISA a relatively small difference between two arms of order of a few percent makes such a subtraction impossible and a laser frequency fluctuation noise as large as $\oforder\unit[10^{-13}]{Hz^{-\nicefrac{1}{2}}}$ around $\unit[1]{mHz}$ corrupts the GW detection. The Time-Delay Interferometry (TDI) \cite{tinto2005a} provides for a solution of the problem: the Doppler measurements are properly time-shifted, to take into account on the photon flight times, and linearly combined, to get the suppression of the laser frequency fluctuation noise by 7 orders of magnitudes. Scope of the entire LPF mission is the accurate modeling of the unsuppressed part of the noise (except for the relative motion between the SCs), the residual acceleration noise affecting the geodesic motion of the TMs after the TDI compensation.

In LISA 6 TMs, whose frequency shifts are optically sensed along each arm of the triangle, build up 6 Doppler links, two per single arm in both directions, forth and back. The fundamental Doppler link can be described as a \textit{four-body TM-SC-SC-TM sequence of measurements}. Referring to \figref{fig:introduction:lisa_links}, the relative velocity of one TM to the optical bench of its hosting SC is measured by a local interferometer; at the same time, the laser signal is sent toward the second SC; finally, a new local measurement is performed between the second TM and the optical bench. Therefore, three measurements, TM to SC, SC to SC and TM to SC, are combined to form the TM-to-TM Doppler link that carries the GW signal. It is easy to recognize that the two local signals carry no GW information, but they are affected by noise, mostly due to parasitic forces that couple the TMs to the SC motion and interferometric sensing, which enter into the noise budget of the Doppler link. The single LISA arm is efficiently reformulated in \chapref{chap:metrology} as a \textit{time-delayed differential accelerometer} whose input signals and noise sources are effectively described as equivalent differential accelerations between the TMs. The most important disturbances affecting the GW detection are due to:
\begin{enumerate}
  \item real forces, relevant at low frequency, say below few $\unit{mHz}$, with red spectrum;
  \item readout sensing coming from all noise sources in the interferometric readout, except for the frequency fluctuation subtracted by TDI;
  \item mixing of motion from degrees of freedom other than the axis joining the TMs, named cross-talk from other degrees of freedom into the main optical axis.
\end{enumerate}

As the main aim is the measuring of the total equivalent differential accelerations, for the rest, all disturbances above will be treated as equivalent accelerations, inputs to a time-delayed differential accelerometer.


\figuremacroW{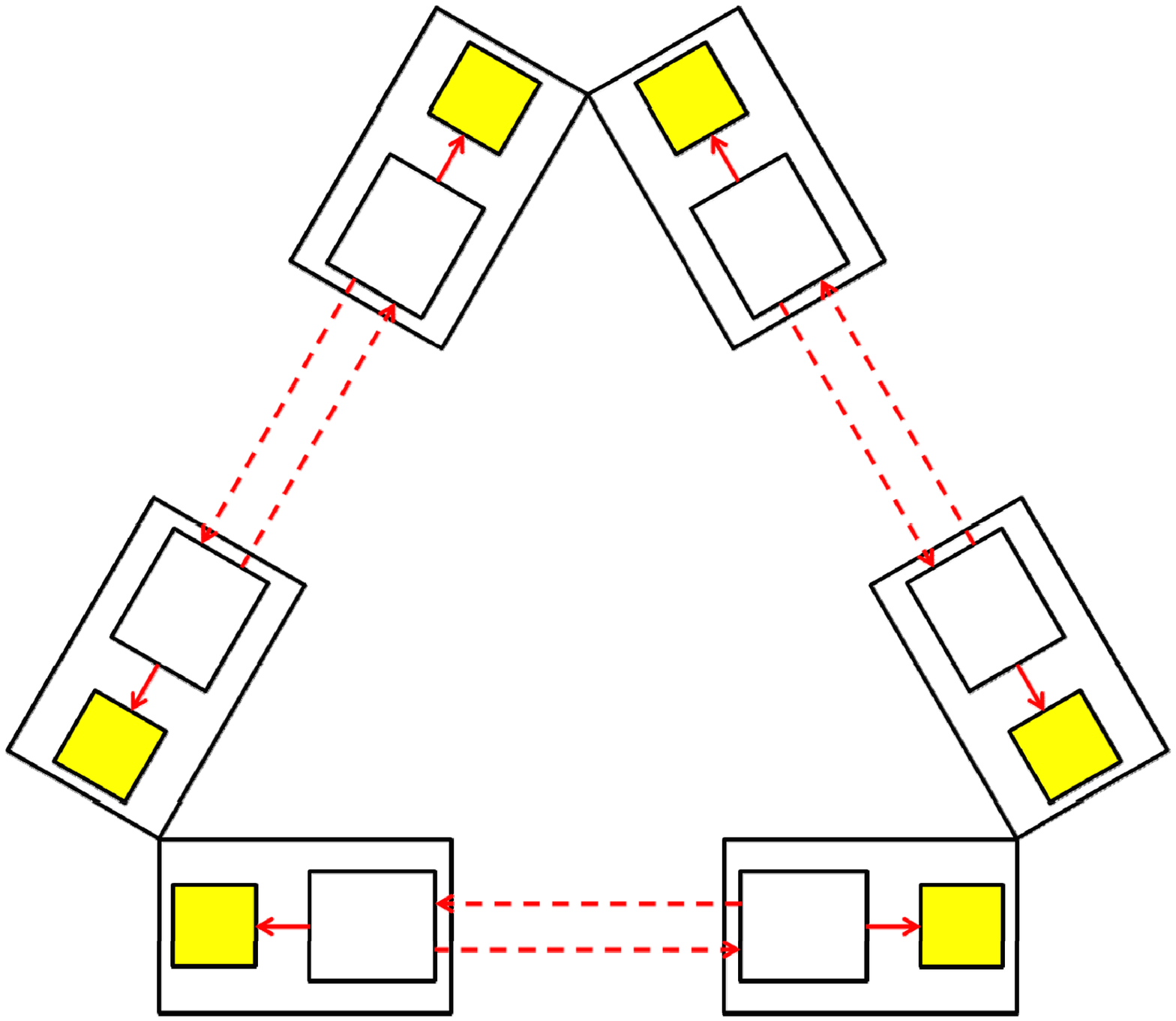}{LISA measurement scheme. The solid arrows show the local links measuring the relative motion of the TMs to their hosting SCs. The dashed arrows show the links measuring the relative motion between the SCs.}{fig:introduction:lisa_links}{0.8}

LPF aims at estimating the residual noise affecting the LISA link through measurements performed in closed loop. One (any) arm of LISA is virtually shrunk \cite{vitale2009} to $\unit[38]{cm}$ and implemented in the LPF mission with some differences. LPF is essentially a SC carrying two TMs in nominal free fall and employs a \textit{three-body TM-SC-TM sequence of measurements}. It measures the relative motion of a TM with respect to the SC and the relative motion between the TMs. All TMs in LISA are controlled along the degrees of freedom orthogonal to the measurement axes and the control is said \textit{off}-axis. Instead, as the measurement axis for LPF is within the SC, a TM must be controlled along the same degree of freedom and the control is said \textit{on}-axis. In this way it is not yet possible to maintain both TMs in free fall along the optical axis: while a reference TM is nominally in free fall, the second must be actuated in order for the differential force disturbances can be compensated. As the control has a fundamental importance in the system stabilization, applied forces must be taken into account as inputs to the differential accelerometer and subtracted from the data.

The LISA arm viewed as a time-delayed differential accelerometer is practically implemented in LPF in a closed-loop differential measurement based on three main concepts: dynamics, sensing and control. \chapref{chap:dynamics} will give an extensive description of the equations governing the link, showing how known couplings, cross-talks and control forces can be taken into account. In the approximation of small TM motion and weak force couplings, the system is linear and the dynamical equations can be rewritten as linear operators acting on the relevant coordinates. As will be demonstrated, the construction of a differential operator then allows:
\begin{enumerate}
  \item the conversion of the sensed motion into total equivalent acceleration;
  \item the subtraction of the couplings, the control forces and the cross-talk from the data;
  \item the suppression of the system transients, at least to within the accuracy to which the system parameters have been measured.
\end{enumerate}

The assessment of the final level of the total equivalent differential acceleration noise -- the key scientific target of LPF -- is literally an iterative process, since the quality of free fall achieved at a given stage of the mission depends on the results of the previous experiments and the accuracy and precision to which the noise parameters have been estimated. Examples of the adopted data analysis procedures will be given in \chapref{chap:sys_identification}, showing the relevance of system identification to achieve the free-fall level needed for LISA. A whole data analysis pipeline will be described and applied to data generated with the model described in \chapref{chap:dynamics} and a realistic simulator provided by industry, hence putting constraints on the accuracy to which the noise parameters can be estimated. The precision of those extracted parameters can also be inferred and optimized as shown in \chapref{chap:optimal_design}. All analysis has been performed under the framework of the LTP Data Analysis (LTPDA) Toolbox \cite{ltpda}, an objected-oriented extension of MATLAB$^\circledR$ \cite{matlab} that will be extensively employed during the mission.

\begin{description}
  \item[\chapref{chap:metrology}.] The chapter discusses on the Doppler link between two TMs in free fall and the GW perturbation of the link through the parallel transport of the emitter 4-velocity. The chapter shows that the parallel transport induces a time delay in the physical quantities. It presents a novel derivation of the response of the Doppler link to the GW, an analogous result already found in literature. The Doppler link can be reformulated as a time-delayed differential accelerometer where all inputs (signals and noise) are equivalent differential accelerations. In the end, it introduces the concept of cross-talk from other degrees of freedom to the optical axis.
  \item[\chapref{chap:dynamics}.] The mathematical description achieved so far is translated into equations governing dynamics, sensing, and control for LPF, i.e. the implementation of a single down-scaled LISA arm. The chapter introduces an operator formalism capable of managing the complex and coupled equations in a compact form. The main advantage of such an abstract formalism is that transfer matrices can be easily extracted, in particular the one representing the conversion from the sensed coordinates to the total equivalent acceleration. The extent to which the suppression of system transients can be achieved is also a novel result of this thesis. The cross-talk from other degrees of freedom can be viewed as a first-order perturbation of the nominal dynamics and all relevant transfer matrices are derived for this case. A model of LPF along the optical axis and an example of cross-talk are given in the end of the chapter.
  \item[\chapref{chap:sys_identification}.] System identification is the key method for the calibration of the system modeled by transfer matrices, allowing for confident noise projections and, most of all, the unbiased estimation of the total equivalent acceleration noise. The chapter discusses examples of the data analysis pipelines adopted for the LPF mission. The relevance of system identification for non-standard scenarios, its impact to the estimation of the total equivalent acceleration noise and the suppression of system transients are given in the end of the chapter.
  \item[\chapref{chap:optimal_design}.] Parameter accuracy is the main target of system identification, whereas precision is the main target of the design of optimal experiments. The chapter focuses on the search of optimal experiments for the LPF mission allowing for a more precise identification of the system parameters that are crucial for the estimation of the total equivalent acceleration noise.
\end{description}





\ChangeFigFolder{2_metrology}


\chapter{Spacetime metrology} \label{chap:metrology}

This chapter is devoted to discussing on the significance of the Doppler link as a detector to track the spacetime curvature and show the road toward the real detection of GWs. The Doppler link comprises two free-falling particles exchanging photons. As a GW passes through that region, the relative velocity between the particles changes as well and produces a frequency shift in the detected photon. The calculation of the natural physical observable discussed here -- the fractional frequency shift -- is formally equivalent to the well-known integration of null geodesics found in literature. This thesis presents a novel derivation by employing the fact that the underlying mathematical operation producing the shift is the parallel transport of 4-vectors.

Subsequently, the chapter stresses that many problems may worsen the real extraction of GW signals from Doppler measurements. In fact, (i) the particles are nearly in free fall, which means that noise forces push the masses away from the reference optimal geodesics; (ii) there are sensing inaccuracies; (iii) the TMs are extended bodies; (iv) the SCs are extended body coupling with the motion of the TMs. In realistic conditions like these, a useful concept is to describe the Doppler link as a differential accelerometer whose inputs are equivalent accelerations. Therefore, GW signals, real forces, sensing noise, pointing inaccuracies and extended body dynamics can be all treated as equivalent input accelerations. One more benefit is that performances of different gravitational experiments whose measurement principle is based on free-falling TMs can be compared at the level of equivalent differential acceleration noise.


\section{Metrology without noise} \label{sect:metrology:doppler_link}

The fundamental measurement scheme of LISA and LPF is the Doppler link between two free-falling TMs embedded into a gravitational field. This section introduces the physics of the Doppler link, viewed as the rod to track the spacetime curvature in a purely idealistic viewpoint where noise does not affect the measurement and the TMs are in perfect free fall \footnote{Otherwise, the TMs would have non-zero acceleration and even in this idealistic situation theory needs some care. See Appendix \ref{sect:appendix:metrology_fermi-walker} for a discussion.}. An emitter sends a photon to a faraway counterpart; the receiver measures the photon frequency and compares it to a reference frequency of a locally emitted photon. The comparison requires the emitter and receiver to have their clocks previously synchronized to a common reference. As such possible error is a subject of TDI, the following assumes a perfect synchronization.

Denoting with $k^\mu$ the photon wave 4-vector, the frequency of the photon measured by any observer with 4-velocity $v^\mu$ is the scalar product $\omega = k_\mu v^\mu$ \cite{misner}. The measured frequency shift of a photon produced by an emitter with velocity $v^\mu_\text{e}$ at the event $x^\mu_\text{e}$ and detected by a receiver with velocity $v^\mu_\text{r}$ at the event $x^\mu_\text{r}$, both in free fall, is given by \citep{vitale2009,chauvineau2005}
\begin{equation}\label{eq:metrology:freqshift}
\delta\omega_{\text{e}\rightarrow\text{r}} = k_\mu \Delta v^\mu_{\text{e}\rightarrow\text{r}}~,
\end{equation}
where all quantities are measured by the receiver and the operation $\Delta v^\mu_{\text{e}\rightarrow\text{r}}$ implements the difference between $v^\mu_\text{r}$ and $v^\mu_\text{e}$, parallel-transported from $x^\mu_\text{e}$ to $x^\mu_\text{r}$
\begin{equation}\label{eq:metrology:covdiff_parallel}
\Delta v^\mu_{\text{e}\rightarrow\text{r}} =  v^\mu_{\text{r}}(x^\alpha_{\text{r}}) - v^\mu_{\text{e}}(x^\alpha_{\text{e}}
\xrightarrow{\text{\tiny{parallel}}} x^\alpha_{\text{r}})~,
\end{equation}
where by definition $v^\mu_\text{e}$ is parallel-transported along the photon path if $v^\mu_{\text{e}~;\alpha} k^\alpha = 0$ and the photon path is defined by the null geodesic equation $k^\mu_{~;\alpha}k^\alpha = 0$. As usual in GR, a semicolon is a covariant derivative, whereas a comma is an ordinary derivative. In \eqref{eq:metrology:covdiff_parallel} an $\alpha$-index is used for clearness, but it does not have relevance for all tensor operations. A representative pictorial view of the operation being performed is shown in \figref{fig:metrology:geodesics}.

\figuremacroW{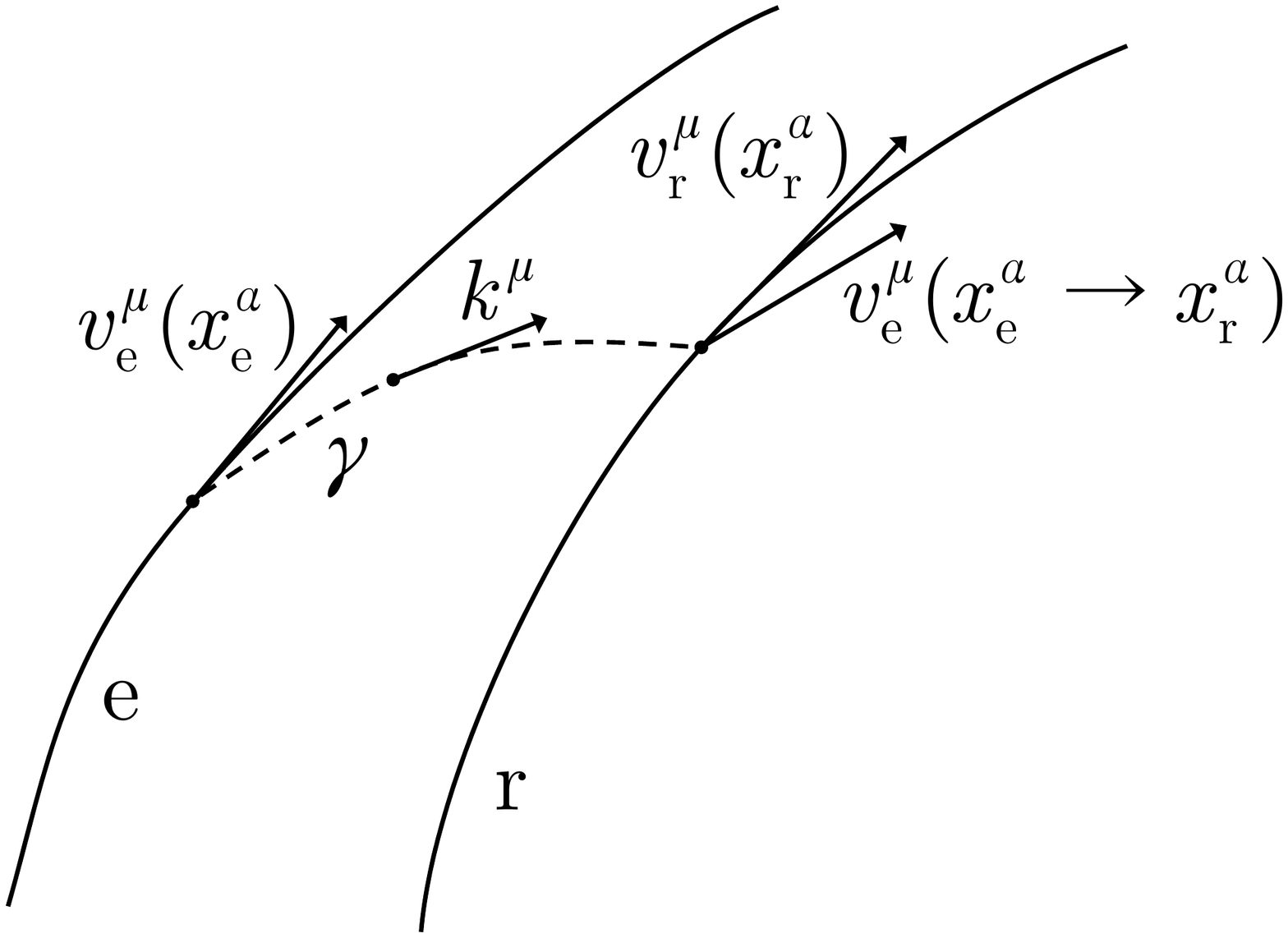}{Pictorial view of the Doppler link. A free-falling emitter with 4-velocity $v^\mu_\text{e}$ sends a photon at the event $x^\mu_\text{e}$. The photon has wave vector $k^\mu$ and is detected by a free-falling receiver with 4-velocity $v^\mu_\text{r}$ at the event $x^\mu_\text{r}$. In order for the Doppler frequency shift to be recorded, $v^\mu_\text{e}$ must be parallel-transported from $x^\mu_\text{e}$ to $x^\mu_\text{r}$, in this way tracking the spacetime curvature along the null geodesic $\gamma$.}{fig:metrology:geodesics}{0.6}

The formula \eqref{eq:metrology:freqshift} can be split into two terms that make the understanding easier. In order to do that, it is necessary to integrate the equation governing the parallel transport of $v^\mu_\text{e}$ in \eqref{eq:metrology:covdiff_parallel}. Firstly, it is worth observing that $k^\mu=\text{d}x^\mu/\text{d}\lambda$, where $\lambda$ is an affine parameter and $x^\mu$ spans the photon geodesic. Therefore, using the definition of the covariant derivative it holds
\begin{equation}
\begin{split}
0 & = v^\mu_{\text{e}\,;\alpha} k^\alpha = \left(v^\mu_{\text{e}\,,\alpha}+\Gamma^\mu_{\alpha\beta} v^\beta_\text{e}\right) \der{x^\alpha}{\lambda} \\
& \implies \parder{v^\mu_\text{e}}{x^\alpha} = - \Gamma^\mu_{\alpha\beta} v^\beta_\text{e} \\
& \implies \text{d}v^\mu_\text{e} = - \Gamma^\mu_{\alpha\beta} v^\alpha_\text{e}\,\text{d}x^\beta~,
\end{split}
\end{equation}
where $\Gamma^\mu_{\alpha\beta}$ are the Christoffel symbols for the underlying curved spacetime. Substituting the preceding in \eqref{eq:metrology:covdiff_parallel} the following expression turns out
\begin{equation}\label{eq:metrology:covdiff_explicit}
\begin{split}
\Delta v^\mu_{\text{e}\rightarrow\text{r}} = \delta v^\mu_{\text{e}\rightarrow\text{r}} +
\int_\gamma \Gamma^\mu_{\alpha\beta} v^\alpha_\text{e}\,\text{d}x^\beta~.
\end{split}
\end{equation}
where $\gamma:\,x^\mu_\text{e}\rightarrow x^\mu_\text{r}$, parameterized by $\lambda$, is the photon geodesic from the emitter to the receiver and $\delta v^\mu_{\text{e}\rightarrow\text{r}} = v^\mu_\text{r}(x^\alpha_\text{r}) - v^\mu_\text{e}(x^\alpha_\text{e})$ is the difference in velocity \textit{without} the parallel-transport of $v^\mu_\text{e}$. Finally, the total frequency shift measured by the receiver reads
\begin{equation}\label{eq:metrology:freqshift_explicit}
\delta\omega_{\text{e}\rightarrow\text{r}} = \delta\omega_v + \delta\omega_\Gamma~,
\end{equation}
where
\begin{subequations}\label{eq:metrology:freqshift_terms}
\begin{align}
\delta\omega_v & = k_\mu\delta v^\mu_{\text{e}\rightarrow\text{r}}~, \label{eq:metrology:freqshift_terms_a} \\
\delta\omega_\Gamma & = k_\mu \int_\gamma \Gamma^\mu_{\alpha\beta} v^\alpha_\text{e}\,\text{d}x^\beta~, \label{eq:metrology:freqshift_terms_b}
\end{align}
\end{subequations}
which correspond to the following two contributions:
\begin{enumerate}
  \item the relativistic Doppler shift just due to the relative velocity between the emitter and the receiver, as if it was in absence of gravity;
  \item the parallel transport term written as a global path integral on the light beam and dominated by the spacetime curvature between the emitter and the observer. 
\end{enumerate}
Inspecting \eqref{eq:metrology:freqshift_terms_b}, since $\Gamma^\mu_{\alpha\beta}$ goes like a space derivative of the metric, it can be found that $\Gamma^\mu_{\alpha\beta}v^\alpha_\text{e}$ goes like a time derivative of the metric itself. The consequence is that the Doppler shift due to curvature can be seen as the space integral of the first time derivative of the metric over the light beam. It is worth noting that such operation of comparing far apart vectors is not local. Indeed, in GR locality implies flatness and, if the operation was local, gravity would have no influence on it: the global behavior of the parallel transport gives gravity a central role in the Doppler link.

\subsection{Weak field limit} \label{sect:metrology:weak_field}

To better understand the meaning of \eqref{eq:metrology:freqshift_explicit} and how curvature affects the Doppler link through \eqref{eq:metrology:freqshift_terms_b}, it is a good practice to take the weak field limit of it. This is also of crucial importance since it shows how GWs can be effectively detected.

The metric $g_{\mu\nu}$ can be expanded to first order like
\begin{equation}
g_{\mu\nu}=\eta_{\mu\nu}+h_{\mu\nu}~,
\end{equation}
with $h_{\mu\nu}$ a perturbation to the flat Minkowski metric $\eta_{\mu\nu}$. The proper expansion of the Christoffel symbols to first order is
\begin{equation}\label{eq:metrology:christoffel}
\Gamma^\mu_{\alpha\beta} = \frac{1}{2} \left(h^\mu_{~\alpha\,,\beta}+h^\mu_{~\beta\,,\alpha}-h_{\alpha\beta}^{\quad,\mu}\right)~,
\end{equation}
and all indices are raised up by means of $\eta_{\mu\nu}$.

The aim is to estimate the contribution of the perturbation $h_{\mu\nu}$ to the Doppler shift $\delta\omega_\Gamma$, now renamed $\delta\omega_h$. When the underlying spacetime metric is flat the photon geodesic connecting emitter and receiver can be considered a straight line: hence, the only effect that parallel transport can cause is a time delay on the emitter 4-vectors. In this case, $k^\mu$ is constant all along the light path with good approximation and \eqref{eq:metrology:freqshift_terms_b} becomes
\begin{equation}\label{eq:metrology:freqshift_nearby}
\begin{split}
\text{d}\omega_h & = k_\mu \Gamma^\mu_{\alpha\beta} v^\alpha_\text{e}\,\text{d}x^\beta \\
& = k_\mu \Gamma^\mu_{\alpha\beta} v^\alpha_\text{e} k^\beta C_\lambda\,\text{d}\tau~,
\end{split}
\end{equation}
where $C_\lambda=\text{d}\lambda/\text{d}\tau$ is a constant for the linear transformation \cite{misner} that connects the photon affine parameter to a reference proper time assumed here to be the one measured by the receiver. Considering that
\begin{equation}
\begin{split}
\Gamma^\mu_{\alpha\beta} k_\mu k^\beta & = \frac{1}{2} \left(h^\mu_{~\alpha\,,\beta}+h^\mu_{~\beta\,,\alpha}-h_{\alpha\beta}^{\quad,\mu}\right) k_\mu k^\beta \\
& = \frac{1}{2} h^\mu_{~\beta\,,\alpha} k_\mu k^\beta~,
\end{split}
\end{equation}
since the first and third terms cancel out \footnote{Indeed, the third term is $h_{\alpha\beta}^{\quad,\mu}k_\mu k^\beta = h_{\alpha\beta\,,\mu}k^\mu k^\beta = h_{\alpha~,\mu}^{~\beta}k^\mu k_\beta$ which is exactly the first term by considering that $\mu$ and $\beta$ are contracted indices and $h_{\alpha\beta}$ is symmetric.}, then \eqref{eq:metrology:freqshift_nearby} can be recast as
\begin{equation}\label{eq:metrology:freqshift_nearby_h}
\text{d}\omega_h = \frac{1}{2} h^\mu_{~\beta\,,\alpha} k_\mu k^\beta v^\alpha_\text{e} C_\lambda\,\text{d}\tau~.
\end{equation}

The GW theory usually assumes the well-known traceless-transverse (TT) gauge
\begin{equation}
h_{\mu\nu} =
\begin{pmatrix}
0 & \multicolumn{2}{c}{\cdots} & 0 \\
\multirow{2}{*}{\vdots} & h_+ & h_\times & \multirow{2}{*}{\vdots} \\
& h_\times & -h_+ \\
0 & \multicolumn{2}{c}{\cdots} & 0
\end{pmatrix}~,
\end{equation}
which further simplifies the computation of \eqref{eq:metrology:freqshift_nearby_h}. Moreover, the so-called \textit{wave coordinate system} can be readily exploited. The $z$ axis is the direction of the incoming GW and $x$ and $y$ define the polarization plane. See \figref{fig:metrology:wave_coordinates} for a graphical definition.

\figuremacroW{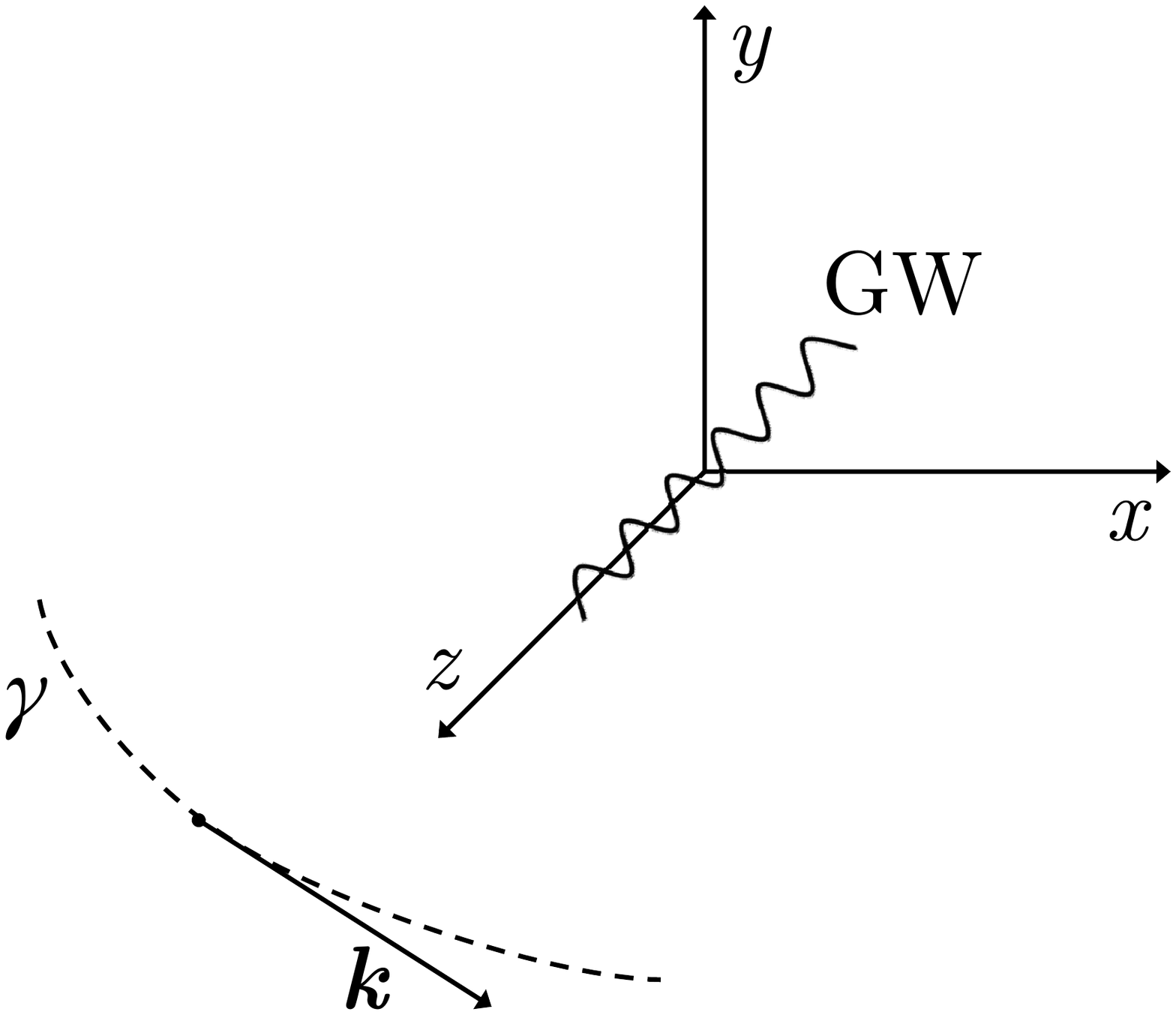}{Definition of the instantaneous wave coordinate system. The GW propagates along the direction $z$. $x$ and $y$ define the polarization plane. The 3-vector $\vect{k}$ is firstly projected onto the polarization plane and then to each of two polarization axes. The concept is better clarified in \eqref{eq:metrology:polarization_states}.}{fig:metrology:wave_coordinates}{0.6}

Therefore, in the TT gauge and in the wave instantaneous coordinate system, it holds (see Appendix \ref{sect:appendix:metrology_calculation} for details)
\begin{equation}\label{eq:metrology:freqshift_nearby_h_term}
h^\mu_{~\beta\,,\alpha} k_\mu k^\beta = H_{,\alpha}~,
\end{equation}
where $H$ is the response to the GW
\begin{equation}\label{eq:metrology:freqshift_h}
H = K_+ h_+ + K_\times h_\times~,
\end{equation}
and the coefficients $K_+$ and $K_\times$ are defined by
\begin{subequations}
\begin{align}
K_+ & = k_x^2-k_y^2~, \\
K_\times & = 2 k_x k_y~.
\end{align}
\end{subequations}
The meaning of \eqref{eq:metrology:freqshift_h} is readily clarified: the photon wave vector is decomposed along the two polarization states of the GW. To look for the response of the Doppler link to the GW signal, \eqref{eq:metrology:freqshift_nearby_h_term} is substituted in \eqref{eq:metrology:freqshift_nearby_h} and the following equation turns out
\begin{subequations}
\begin{align}
\text{d}\omega_h & = \frac{1}{2} H_{,\alpha} v^\alpha_\text{e} C_\lambda\,\text{d}\tau \\
& = \frac{1}{2} \parder{H}{x^\alpha} \der{x^\alpha_\text{e}}{\tau} C_\lambda\,\text{d}\tau \\
& = \frac{1}{2} C_\lambda\,\text{d}H~.
\end{align}
\end{subequations}
The preceding can be easily integrated between the instants at which the photon is emitted and received, $\tau_\text{e}$ and $\tau_\text{r}$. For instance, the right-end side is
\begin{equation}
\delta H = H(\tau_\text{r}) - H(\tau_\text{e})~,
\end{equation}
and the equation finally reads
\begin{equation}\label{eq:metrology:freqshift_nearby_final}
\delta\omega_h = \frac{1}{2} C_\lambda \delta H~.
\end{equation}
The result obtained above shows that an incoming GW induces a Doppler frequency shift on a photon exchanged between two geodesics. The effect is proportional to the difference between the GW signal at the time of the receiver and the one, time-delayed, at the time of the emitter, as a strict consequence of the parallel transport.

The formula in \eqref{eq:metrology:freqshift_nearby_final} can even be put in a more explicit and physically interpretable form. So far, the following facts have been considered: (i) weakness of the gravitational field, such that the underlying metric is flat; (ii) calculation in the TT gauge and in the wave coordinate system. The last reasonable assumption is about the non-relativistic regime of the test particles. As a matter of fact, the emitter and the receiver can be assumed to fall in the gravitational field at low velocities compared to $c$. Hence, all 4-vector equations can be rewritten in terms of 3-vectors. In this approximation, the definition of the photon wave vector implies $C_\lambda = c/k$ \footnote{Indeed, from the definition of $k^\mu$ along the null geodesics it follows $k^\mu\text{d}\lambda=\text{d}x^\mu$ and differentiating with respect to the proper time of the receiver implies $k^\mu\,\text{d}\lambda/\text{d}\tau=\text{d}x^\mu/\text{d}\tau$. In the non-relativistic regime $\text{d}x^\mu/\text{d}\tau\rightarrow c$ and using the definition of $C_\lambda$ the relation is finally demonstrated.\label{foot:metrology:c_over_k}}, where $k$ is the module of $\vect{k}$, the space part of $k^\mu$. The GW polarization responses are symmetric if the wave coordinate system is written in spherical coordinates \footnote{From the definitions, $K_+=k_x^2-k_y^2=k^2\sin^2\theta\left(\cos^2\phi-\sin^2\phi\right)=k^2\sin^2\theta\cos2\phi$ and $K_\times=2k_xk_y=2k^2\sin^2\theta\sin\phi\cos\phi=k^2\sin^2\theta\sin2\phi$.}
\begin{subequations}
\begin{align}
K_+(k,\theta,\phi) & = k^2 \xi_+(\theta,\phi)~, \\
K_\times(k,\theta,\phi) & = k^2 \xi_\times(\theta,\phi)~,
\end{align}
\end{subequations}
and
\begin{subequations}\label{eq:metrology:polarization_states}
\begin{align}
\xi_+(\theta,\phi) & = \sin^2\theta\cos{2\phi}~, \\
\xi_\times(\theta,\phi) & = \sin^2\theta\sin{2\phi}~,
\end{align}
\end{subequations}
are the \textit{directional sensitivities} to each of the two GW polarizations. $\theta$ is the projection angle, named \textit{declination}, of $\vect{k}$ onto the GW polarization plane orthogonal to the $z$ axis defining the GW propagation direction. Notice in \eqref{eq:metrology:polarization_states} that the Doppler response is null, both in $\xi_+$ and $\xi_\times$, for $\theta=0$, i.e., when the photon wave vector is parallel to $z$, whereas is maximum for $\theta=\nicefrac{\pi}{2}$, i.e., when is orthogonal to $z$. $\phi$ is the projection angle, named \textit{polarization}, onto the two polarization states. In fact, when $\phi=0,\nicefrac{\pi}{2}$, then $\xi_+$ is maximum and $\xi_\times=0$; when $\phi=\nicefrac{\pi}{4},\nicefrac{3\pi}{4}$, then $\xi_\times$ is maximum and $\xi_+=0$. See \figref{fig:metrology:polarizations} for a graphical interpretation. Since the degeneracy around $\vect{k}$, the \textit{right ascension} is not measured with a single photon, but it can be inferred from the modulation induced by the rotation of the beam.

\figuremacroW{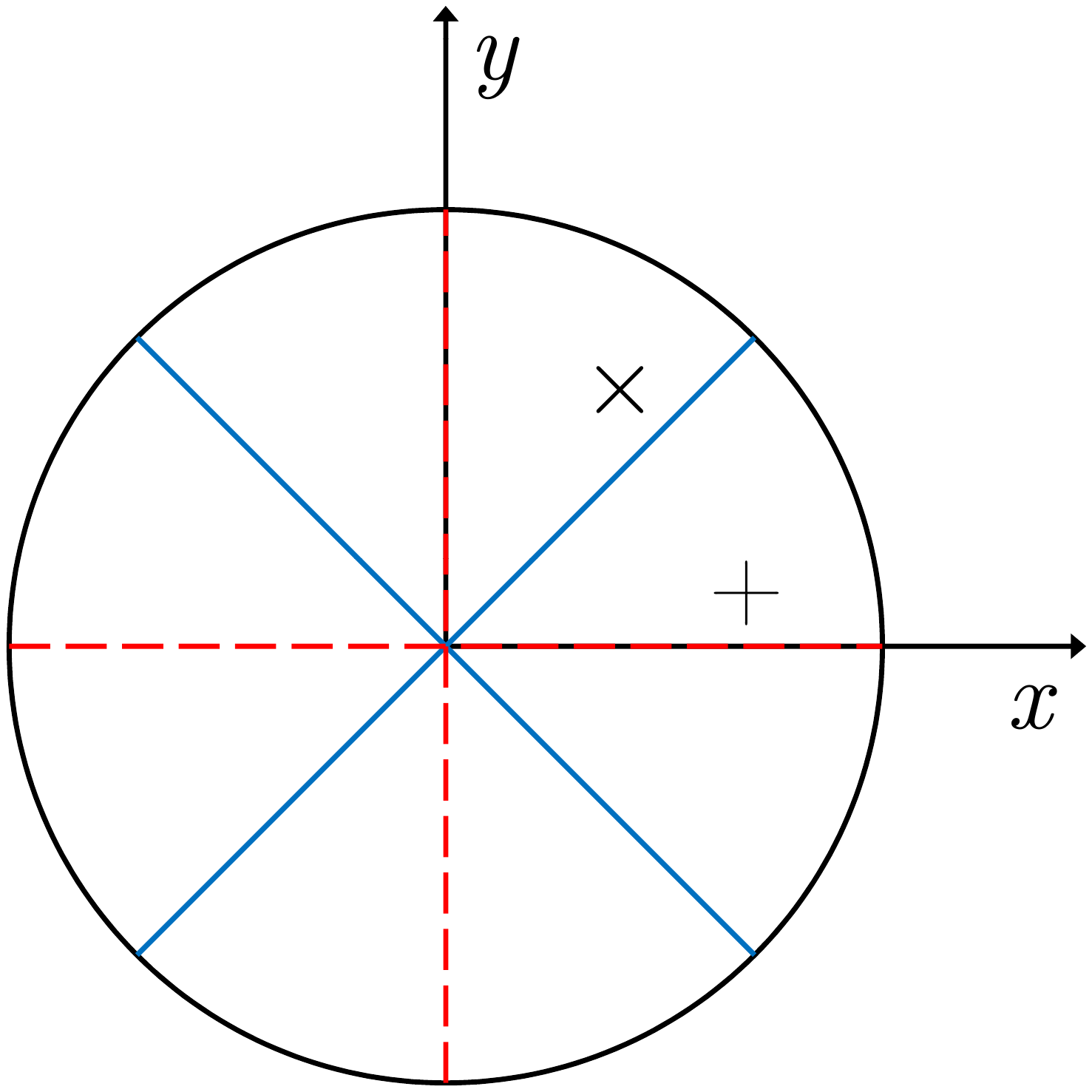}{Graphical interpretation of the $\phi$ polarization angle (measured counterclockwise around $z$). When $\phi=0,\nicefrac{\pi}{2}$, then $\xi_+$ is maximum and $\xi_\times=0$ (dashed lines); when $\phi=\nicefrac{\pi}{4},\nicefrac{3\pi}{4}$, then $\xi_\times$ is maximum and $\xi_+=0$, as predicted by \eqref{eq:metrology:polarization_states}. Hence, any GW signal can be decomposed into the $+$ and $\times$ polarization states in the $xy$ plane.}{fig:metrology:polarizations}{0.6}

The polarization states can be viewed as two independent bases of the fundamental decomposition
\begin{equation}
h(t,\theta,\phi) = \xi_+(\theta,\phi) h_+(t) + \xi_\times(\theta,\phi) h_\times(t)~,
\end{equation}
where $h_+$ and $h_\times$ are the two GW polarization states, $\xi_+$ and $\xi_\times$ the two directional sensitivities of the Doppler link and $h$ the Doppler response. \eqref{eq:metrology:freqshift_nearby_final} can be elaborated as
\begin{equation}
\begin{split}
\delta\omega_h & = \frac{1}{2} C_\lambda \delta H \\
& = \frac{1}{2} \frac{c}{k} k^2 \delta h \\
& = \frac{1}{2} \omega_\text{e} \delta h~.
\end{split}
\end{equation}
where $\omega_\text{e}$ is the frequency of the emitted photon and $\delta h$ denotes the difference between the signal evaluated at detection and emission. The final result is the fractional frequency shift measured by the Doppler link
\begin{equation}\label{eq:metrology:freqshift_nearby_final_explicit}
\frac{\delta\omega_h}{\omega_\text{e}} = \frac{1}{2} \delta h~.
\end{equation}
Therefore, if $\delta x$ is the separation between two geodesics, the fractional frequency shift -- the natural physical observable -- is proportional to the difference between the GW response evaluated at the instant of detection and the one time-delayed to the instant of emission,
\begin{equation}
\delta h(t) = h(t) - h(t-\delta x/c)~.
\end{equation}

A particularly interesting discussion is about the \textit{long-wavelength limit}, for which the GW wavelength $\lambda \gg \delta x$. By taking the limit for infinitely small $\delta x/c$, i.e., assuming that the two geodesics are infinitely close each other or the photon flight time is infinitely small, there is no parallel transport and $\delta h$ becomes a time derivative
\begin{equation}\label{eq:metrology:h_longwave}
\delta h \simeq \frac{\delta x}{c} \dot{h}~.
\end{equation}
Analogously, in Fourier domain for $\omega \ll c/\delta x$ it holds
\begin{equation}
\begin{split}
\delta h & = \left(1-e^{-i\,\frac{\delta x}{c}\,\omega}\right)h \\
& \simeq i\,\frac{\delta x}{c}\,\omega\,h~.
\end{split}
\end{equation}
Therefore, the time delay due to the parallel transport can be effectively ignored at low frequency. For example, in LISA the long-wavelength limit applies below $\unit[60]{mHz}$ for a photon one-way trip. The fractional frequency shift becomes proportional to the time derivative of the GW signal or, equivalently, the phase shift becomes directly proportional to the GW signal, in fact
\begin{equation}
\begin{split}
\frac{\dot{\delta\phi}_h}{\omega_\text{e}} & \simeq \frac{1}{2} \frac{\delta x}{c} \dot{h} \implies \\
\delta\phi_h & \simeq \frac{1}{2} \omega_\text{e} \frac{\delta x}{c} h~.
\end{split}
\end{equation}

This section has shown how the GW signal convolves with the Doppler link and produces a frequency shift measured by the receiver. The results are well-known in literature \cite{vinet2007,dhurandhar2002}, but the difference here is in the derivation. Instead of integrating the null geodesic, the calculations have been performed employing the parallel transport of 4-vectors, a very fundamental concept in GR.

\section{Doppler link as differential accelerometer} \label{sect:metrology:doppler_link_diff_acc}

This section reformulates the Doppler link as a \textit{differential time-delayed accelerometer}. The result is that the Doppler link measures the spacetime curvature between emitter and receiver, corrupted by differential parasitic accelerations and non-inertial forces due to the particular choice of the detector reference frame in which the measurement is performed. 

Consider the frequency shift in \eqref{eq:metrology:freqshift_explicit}, induced by the classical Doppler contribution in \eqref{eq:metrology:freqshift_terms_a} and the contribution due to the parallel transport in \eqref{eq:metrology:freqshift_terms_b}. For LISA, in the weak-field limit the metric can be decomposed as
\begin{equation}
g_{\mu\nu}=\eta_{\mu\nu}+h_{\mu\nu\,\odot}+h_{\mu\nu\,\oplus}+h_{\mu\nu}~,
\end{equation}
where $\eta_{\mu\nu}$ is the flat Minkowski metric, $|h_{\mu\nu\,\odot}|\oforder2\e{-12}$ is the perturbation due to the Sun gravity and $|h_{\mu\nu\,\odot}|\oforder2\e{-17}$ is the perturbation due to the Earth gravity. $h_{\mu\nu}$ is the perturbation due to GWs; since $|h_{\mu\nu}|\oforder|h_{\mu\nu\,\odot}|^2$, it is clearly smaller than the average local gravity of the Solar System. Expanding the Christoffel symbols to second order for the local gravity and to first order for the GW perturbation, it becomes
\begin{equation}
\Gamma^\mu_{\alpha\beta} \rightarrow \Gamma^\mu_{\alpha\beta\,\odot} + \Gamma^\mu_{\alpha\beta\,\oplus} + \Gamma^\mu_{\alpha\beta}~,
\end{equation}
since the mixed products between $h_{\mu\nu}$ and $h_{\mu\nu\,\odot}$ are negligible with respect to $h_{\mu\nu}$. Hence, the effect of the local gravity within the Solar System can be separated from the effect of GWs. Moreover, these effects intervenes at typical frequencies \footnote{Around $\unit[3\e{-8}]{Hz}$ for the revolution about the Sun and around $\unit[4\e{-7}]{Hz}$ for the revolution of the Moon about the Earth.\label{foot:metrology:doppler_dc}} below the LISA measurement band.

In the same way, for low velocities, i.e. small compared to $c$, all mixed products between $h_{\mu\nu}$ and velocity are second order. Analogously, the parallel transport of the acceleration contributes to second order. To first order, $k^\mu$ is constant along the light path and differentiating \eqref{eq:metrology:freqshift_explicit}, with respect to the proper time of the receiver $\tau$, it holds
\begin{equation}\label{eq:metrology:freq_shift_der}
\dot{\delta\omega}_{\text{e}\rightarrow\text{r}} = k_\mu\delta a^\mu_{\text{e}\rightarrow\text{r}}
+ k_\mu \int_\gamma \der{\Gamma^\mu_{\alpha\beta}}{\tau}v^\alpha_\text{e}\,\text{d}x^\beta~,
\end{equation}
where the derivative commutes with the integral as the variation of the extremes of integration contributes to second order. 
Hence, the differential time-delayed accelerometer measures the effect of the parallel transport and differential parasitic accelerations between emitter and receiver. For the rest, $\Gamma^\mu_{\alpha\beta}$ describes only the GW perturbation, bearing in mind that there the gravity of the Solar System falls below the measurement band.

To first approximation, the frequency shift is now evaluated in a reference frame in where emitter and receiver appear at rest. Since a net relative velocity is a Doppler effect, this is eventually included in the first term and, in fact, in LISA it must be considered in the calculation, even though it intervenes at frequencies again lower than the measurement band. In such a reference frame, $v^\alpha_\text{e}=(c,0,0,0)$, the $x$-axis is aligned to $k^\mu$ to have $k^\mu=k\,(1,1,0,0)$ and $\text{d}x^\beta=(c\,\text{d}t,\text{d}x,0,0)$. In addition, $\text{d}/\text{d}\tau=\text{d}t/\text{d}\tau\,\text{d}/\text{d}t$, where $\text{d}t/\text{d}\tau=1$ is the Lorentz factor and $\text{d}/\text{d}t=\partial/\partial t=c\,\partial_0$ for low relative velocities. From the definition, $\text{d}x^\beta=k^\beta\text{d}\lambda$, the second term in \eqref{eq:metrology:freq_shift_der} becomes
\begin{equation}
k_\mu \int_\gamma \der{\Gamma^\mu_{\alpha\beta}}{\tau}v^\alpha_\text{e}\,\text{d}x^\beta =
c^2k^2\int_\gamma \partial_0\left(\Gamma^0_{00}+\Gamma^0_{01}-\Gamma^1_{00}-\Gamma^1_{01}\right)\text{d}\lambda~.
\end{equation}
where a $c$ comes from the velocity and another one from $\partial_0$. Using the expansion of the Christoffel symbols in \eqref{eq:metrology:christoffel} it follows that
\begin{equation}
\Gamma^0_{00}+\Gamma^0_{01}-\Gamma^1_{00}-\Gamma^1_{01} = \frac{1}{2}\left(h_{00,0}+2h_{01,0}+h_{11,0}\right)~.
\end{equation}
Applying the derivative, the integrand becomes
\begin{equation}
\partial_0\left(\Gamma^0_{00}+\Gamma^0_{01}-\Gamma^1_{00}-\Gamma^1_{01}\right) =
\frac{1}{2}\left(h_{00,00}+2h_{01,00}+h_{11,00}\right)~.
\end{equation}
In these approximations, the only independent component of the Riemann tensor that can be observed along the beam is $R_{0110}$ \footnote{The number of independent components of the Riemann tensor are $\nicefrac{1}{12}\,n^2(n^2-1)$, where $n$ is the number of dimensions. The particular choice of the reference frame is equivalent to working within a 2-dimensional space.} that, to first order, is given by
\begin{equation}
R_{0110} = \frac{1}{2}\left(h_{00,11}-2h_{01,01}+h_{11,00}\right)~.
\end{equation}
The integral can be recast as
\begin{equation}
\begin{split}
k_\mu & \int_\gamma \der{\Gamma^\mu_{\alpha\beta}}{\tau}v^\alpha_\text{e}\,\text{d}x^\beta = \\
& c^2k\int_\gamma \left[R_{0110}+\frac{1}{2}\left(h_{00,00}+2h_{01,00}+2h_{01,01}-h_{00,11}\right)\right]\text{d}x~.
\end{split}
\end{equation}
Dividing by $\omega_\text{e}$ the result is the derivative of the fractional frequency shift; multiplying this by $c$ the result is the equivalent input acceleration in terms of curvature
\begin{equation}
\delta a_R = c^2 \int_\gamma R_{0110}\,\text{d}x~,
\end{equation}
The equivalent input acceleration in terms of the additional contribution is
\begin{equation}
\delta a_\text{gauge} = \frac{1}{2}c^2 \int_\gamma \left(h_{00,00}+2h_{01,00}+2h_{01,01}-h_{00,11}\right)\,\text{d}x~.
\end{equation}
Solving the linearized Einstein equations in the $(ct,x)$ coordinates, it follows $h_{00,11}=h_{11,11}$ (see Appendix \ref{sect:appendix:metrology_einstein}); since this does not simplify the above formula, the additional contribution is interpreted merely as a gauge effect depending on the particular choice of the reference frame. The fixing of a proper gauge should be able, in principle, to suppress those terms. A local gauge transformation in $h_{\mu\nu}$ is defined by
\begin{equation}
h'_{\mu\nu} = h_{\mu\nu} - \xi_{\mu,\nu} - \xi_{\nu,\mu}~,
\end{equation}
where $h'_{\mu\nu}$ is the transformed perturbation and $\xi_\mu$ are infinesimal shifts in the coordinates
\begin{equation}
x'^\mu = x^\mu + \xi^\mu~.
\end{equation}
As the above is a tranformation between two reference frames, the gauge terms are interpreted as non-inertial forces.

The conclusion of the section is that the LISA arm can be viewed as a differential time-delayed accelerometer measuring equivalent input acceleration. It measures the spacetime curvature between emitter and receiver along the light beam. The measurement is corrupted by: (i) parasitic differential forces affecting the geodesic motion of the TMs; (ii) the curvature due to the Solar System at frequencies below the measurement band; (iii) non-inertial forces mainly due to the rotation of the arm.

\section{Metrology with noise} \label{sect:metrology:noise}

This section presents a series of issues in the actual measurement of frequency shifts by means of the Doppler link. The results of \sectref{sect:metrology:doppler_link} can be summarized in \eqref{eq:metrology:freqshift_explicit}, \eqref{eq:metrology:freqshift_nearby_final} and subsequently in \eqref{eq:metrology:freqshift_nearby_final_explicit}, but have been obtained in very idealistic conditions.

There are many points where noise, non-idealities, etc., may enter into the measurement. However, taking a look on \eqref{eq:metrology:freq_shift_der}, noise sources and disturbances corrupt the GW detection at the level of differential accelerations.
The emitter and the receiver are faraway of being in free fall because of the presence of many external non-gravitational forces. The environment can be chosen to be as quiet as possible, but in reality many disturbances can take the emitter and the receiver away from the purely gravitational geodesic
\begin{equation}
\frac{\text{d}^2x^\mu}{\text{d}\tau^2}+\Gamma^\mu_{\alpha\beta}\der{x^\alpha}{\tau}\der{x^\beta}{\tau} = \frac{f^\mu}{m}~,
\end{equation}
where $m$ is the particle mass and $f^\mu$ are the external non-gravitational noise forces affecting the exact knowledge of $x^\mu_\text{e}$ and $x^\mu_\text{r}$. In this way the photon geodesic is determined only to within a given uncertainty given by the noise in the coordinates.
Actually, emitter and receiver are not pointlike, but are extended bodies introducing more degrees of freedom in the dynamics and an extra source of indetermination as it is discussed in the next section. In addition, the future position of the receiver can not be determined a priori with absolute precision and there are surely pointing misalignments affecting the measurement.

To defend the TM from the ``polluted environment'' in which it is embedded, an isolating box, the SC, contains and protects it. This prevents the TM from being disturbed by external non-gravitational forces, but introduces a series of parasitic couplings to the SC, mostly electromagnetic and self-gravity, which must be measured and compensated.

The classical Doppler shift is a deterministic signal that does contribute, but at much lower frequencies and can be effectively subtracted from the data. In LISA the Doppler effect is minimized in advance in the experimental design by optimizing the SC orbits, so that the maximum allowed relative fluctuation of the arm lengths is few percents.


\tabref{tab:metrology:disturbances} summarizes some types of disturbances, starting from the most relevant ones, playing the role of imperfections for the detection of GWs through the Doppler link in LISA.
\begin{table}[htb]
\caption{\label{tab:metrology:disturbances}\footnotesize{Sources of indetermination for the GW detection through the Doppler link in LISA.}}
\centering
\begin{tabular}{p{5.5cm}p{7.5cm}}
\hline
\hline
Disturbances & Note \\
\hline
classical Doppler shift & minimized in orbit design, but out of band \\
laser frequency fluctuation & abated in post-processing by 7 orders of magnitude with TDI \\
differential forces & mostly coupling forces between the TM and the SC, estimated and characterized by LPF \\
displacement sensing between the TM and the SC & readout noise, pointing inaccuracies, estimated and characterized by LPF \\
displacement sensing between two SCs & readout noise, pointing inaccuracies, peculiarity of LISA \\
extended body dynamics & dynamical, sensing and actuation cross-talk, estimated and characterized by LPF \\
clock stability & required by TDI \\
\hline
\hline
\end{tabular}
\end{table}

The next subsections introduce in turn the three most relevant noise contributions in LISA: (i) the frequency noise due to laser instability and largely compensated on-ground through TDI; (ii) the acceleration noise due to force couplings between the TM and the SC; (iii) the readout noise due to the interferometric sensing.


\subsection{Laser frequency noise}

The practical implementation of the Doppler link between two faraway TMs in nominal free fall, like in LISA and all spaced-based missions, has a fundamental problem. The laser interferometry on ground is based on equal arms and power recycling: therefore it can not be of any help for the space-based detectors. In fact, there are mostly two reasons for this. On one hand, it is impossible to put two satellites in space with fixed and constant separation without taking into account of the Keplerian evolution. On the other hand, there is a huge light power dispersion among million of kilometers preventing the same signal of being bounced back in order to do the usual interferometry.

\begin{figure}[!htbp]
\centering
\begin{tabular}{*{2}{@{\hspace{10pt}}c@{\hspace{-10pt}}}} %
\includegraphics[width=0.4\columnwidth]{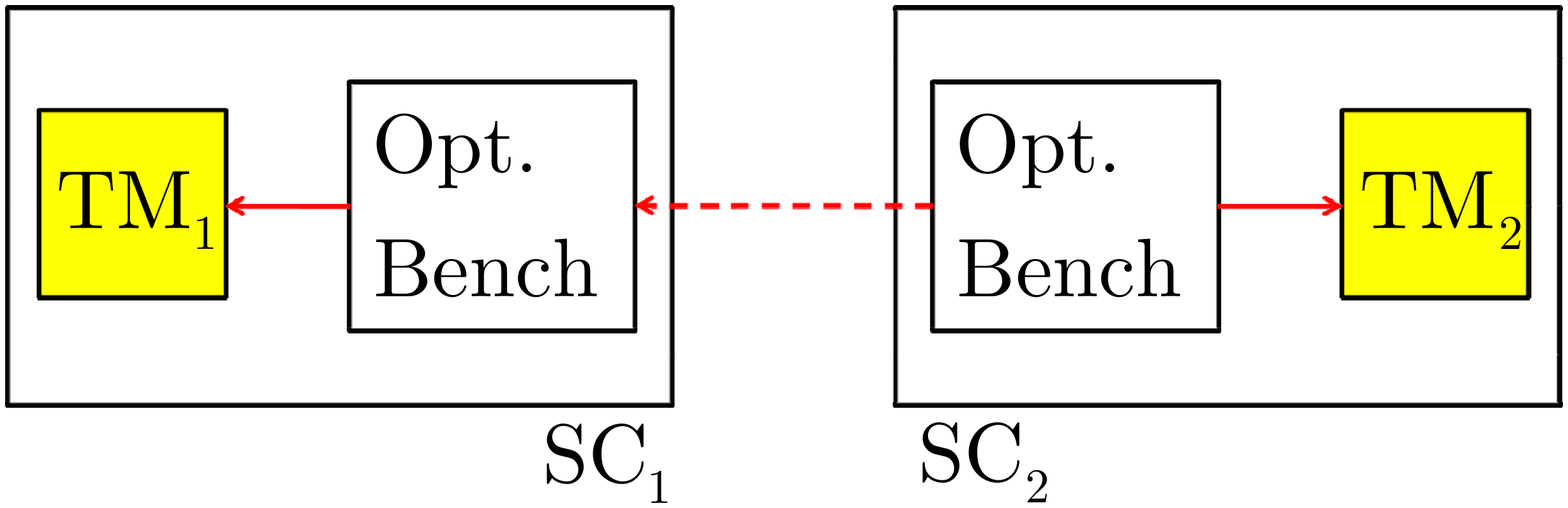} &
\includegraphics[width=0.6\columnwidth]{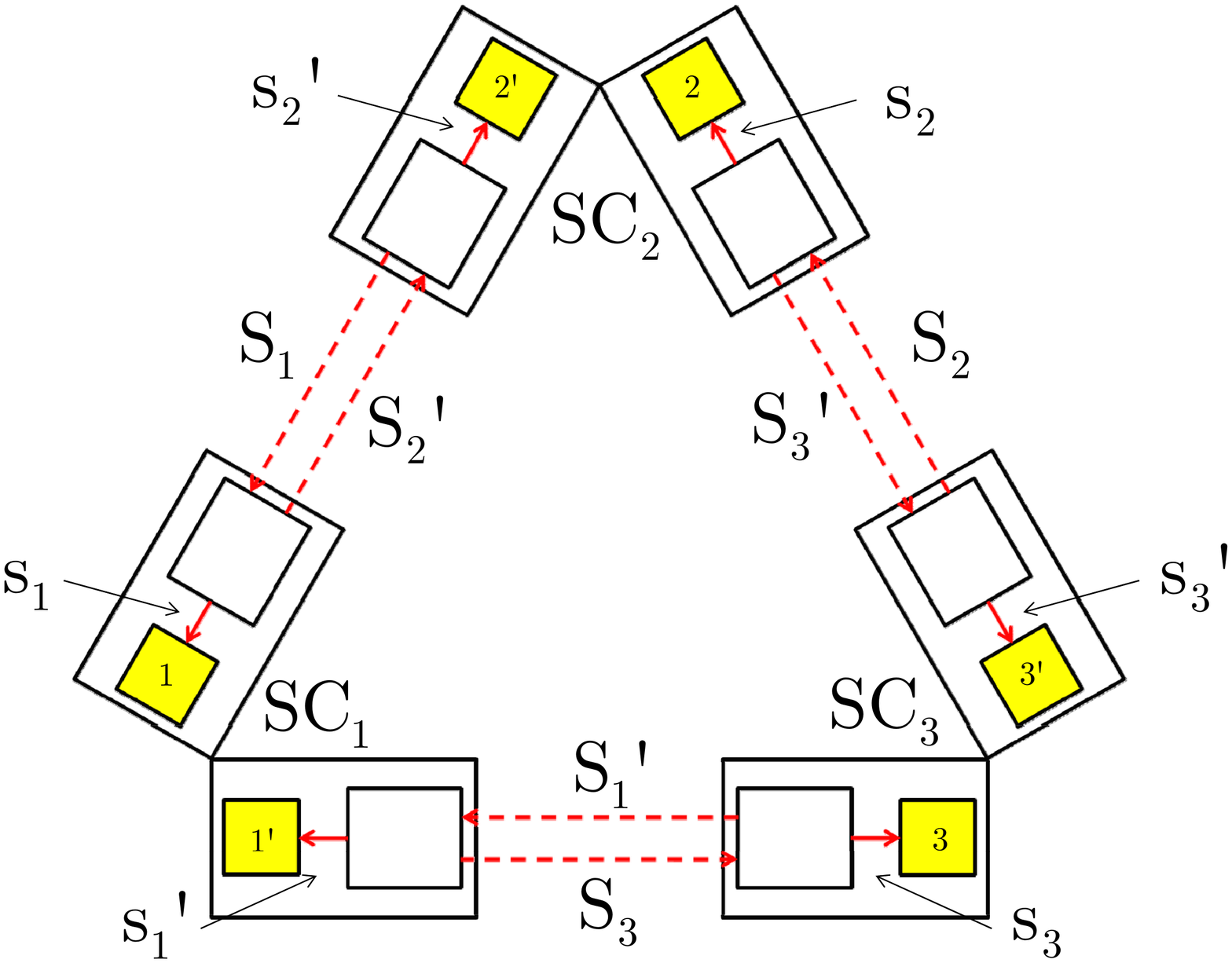} \\
\footnotesize{(a)} & \footnotesize{(b)}
\end{tabular}
\caption{\footnotesize{Scheme of the LISA constellation. (a) the single arm is made of two TMs, each contained in two faraway SCs; the Doppler link is obtained by three independent measurements: two local measurements (TMs to their optical benches) and a faraway measurement (SC to SC). (b) the constellation comprises 6 Doppler links, forth and back for every arm.}}
\label{fig:metrology:lisa_arm_tdi}
\end{figure}

In a LISA arm, the light signal is sent toward the other SC where it is compared to a local reference signal. Therefore, a LISA arm, as shown in \figref{fig:metrology:lisa_arm_tdi}, is obtained by a combination of lower-level measurements between four bodies: two TMs and two SCs. In the language of the preceding section the emitter coincides nominally with $\text{TM}_2$ and the receiver coincides nominally with $\text{TM}_1$. Hence, the TM-to-TM link can be effectively depicted with three interferometric measurements: $\text{TM}_2$ to its hosting $\text{SC}_2$, $\text{SC}_2$ to $\text{SC}_1$, and $\text{TM}_1$ to its hosting $\text{SC}_1$. The frequency shift between two faraway TMs for a light ray from $\text{TM}_2$ to $\text{TM}_1$, can be constructed as follows
\begin{equation}\label{eq:metrology:doppler_link_3masses}
\begin{split}
\frac{\delta\omega_{2\rightarrow1}}{\omega_\text{e}~} & =
\frac{1}{c}\hat{\vect{k}}\cdot\left(\vect{v}_{\text{TM}_2} - \vect{v}_{\text{TM}_1}\right) \\
& = \frac{1}{c}\hat{\vect{k}}\cdot\left(\vect{v}_{\text{TM}_2}^{(\text{SC})} + \vect{v}_{\text{SC}_2}
- \vect{v}_{\text{TM}_1}^{(\text{SC})} - \vect{v}_{\text{SC}_1}\right) \\
& = \frac{1}{c}\hat{\vect{k}}\cdot\left(\delta\vect{v}_\text{SC} + \vect{v}_{\text{TM}_2}^{(\text{SC})}
- \vect{v}_{\text{TM}_1}^{(\text{SC})}\right)~,
\end{split}
\end{equation}
where $\delta\vect{v}_\text{SC}$ is the measurement between the two SCs containing the time delay due to the photon flight time (about $\unit[17]{s}$ for LISA); $\vect{v}_{\text{TM}_1}^{(\text{SC})}$ is the local measurement between $\text{TM}_1$ and its hosting SC; $\vect{v}_{\text{TM}_2}^{(\text{SC})}$ is the local measurement between $\text{TM}_2$ and its hosting SC, but time-delayed by the photon flight time. Obviously, the three measurements contain noise sources at different levels, but the GW signal is masked within the first one.

When extended to whole LISA configuration with 6 TMs, 6 faraway links and 6 local links, as shown in \figref{fig:metrology:lisa_arm_tdi}, the adopted scheme contains an unavoidable large laser frequency fluctuation noise of $\oforder\unit[10^{-13}]{Hz^{-\nicefrac{1}{2}}}$ due to arm length imbalances of a few percent. Such disturbance can be mitigated by TDI \cite{tinto2005a,tinto2005b} in data post-processing allowing for the compensation of arm length imbalances and optical bench vibrations (1st generation TDI), as well as arm flexing (2nd generation TDI).

A more abstract notation can be introduced for describing the problem. Referring to \figref{fig:metrology:lisa_arm_tdi} and assuming the standard naming convention of TDI, the SCs are numbered clockwise with index $k$, each arm is labeled with the number of the opposing SC, each TM is numbered as the hosting SC, but it is primed if it is on the right side of the SC. The photodiode outputs corresponding to the local measurements are named, $s_k$ for the $k$-th TM and $s_k'$ for the $k'$-th TM. The photodiode outputs corresponding to the faraway incoming link between the SCs are named, $S_k$ for the side of the $k$-th TM and $S_k'$ for the side of the $k'$-th TM. The result in \eqref{eq:metrology:doppler_link_3masses} can be generalized to any incoming Doppler links on the left and right sides of the $k$-th SC
\begin{subequations}
\begin{align}
\sigma_k(t) & = s_k(t) + S_k(t) - s_{p[k]}'(t-T_{p^2[k]})~, \\
\sigma_k'(t) & = s_k'(t) + S_k'(t) - s_{p^2[k]}(t-T_{p[k]})~,
\end{align}
\end{subequations}
where $p[k]$ is the cyclic permutation of $(1 2 3)$ and $p^2[k]=p[p[k]]$. $T_k$ is the time delay in the $k$-th arm assumed constant within the 1st generation TDI. Notice the symmetry of the preceding equations: an unprimed index goes to a primed one (and vice-versa) and $p[k]$ goes to $p^2[k]$ (and vice-versa).

The 1st generation TDI solution corresponding to an unequal-arm Michelson interferometer with the $k$-th SC at its vertex is a linear combination of time-shifted photodiode outputs given by
\begin{equation}
\begin{split}
X_k(t) & = \sigma_k'(t) + \sigma_{p^2[k]}(t-T_{p[k]}) + \sigma_k(t-2T_{p[k]}) + \sigma_{p[k]}'(t-T_{p^2[k]}-2T_{p[k]}) \\
& \quad - \left[\sigma_k(t) + \sigma_{p[k]}'(t-T_{p^2[k]}) + \sigma_k'(t-2T_{p^2[k]}) + \sigma_{p^2[k]}(t-T_{p[k]}-2T_{p^2[k]})\right]~,
\end{split}
\end{equation}
which contains the round-trip delay of $\sigma_k$ in the $p[k]$-th arm and $\sigma_k'$ in the $p^2[k]$-th arm. Such combinations are able to cancel out the frequency fluctuation noise of arm length imbalances and optical bench vibrations \footnote{The 2nd generation TDI solution can be derived considering that the photon flight times are not constant and the time delays do not commute anymore. Such combinations can compensate the arm flexing, but introduces much more complexity in the system.} to $\oforder\unit[10^{-20}]{Hz^{-\nicefrac{1}{2}}}$ corresponding to the differential acceleration requirement of $\unit[3\e{-15}]{m\,s^{-2}\,Hz^{-\nicefrac{1}{2}}}$ around $\unit[1]{mHz}$ \footnote{In fact, $\unit[3\e{-15}]{m\,s^{-2}\,Hz^{-\nicefrac{1}{2}}}/[(2\pi\times\unit[1]{mHz})^2\times\unit[5\e{6}]{km}]=\unit[1.5\e{-20}]{Hz^{-\nicefrac{1}{2}}}$.}.

\subsection{Residual acceleration noise}

The first remaining contribution after the TDI compensation of the frequency fluctuation noise is the residual acceleration noise, also frequently named force (per unit mass) noise, whose characterization is one of the main scientific targets of the LPF mission.

Considering the low velocity regime of \eqref{eq:metrology:freqshift_explicit}, whose GW signal is given by \eqref{eq:metrology:freqshift_nearby_final_explicit}, the Doppler link expressed as the time derivative of the fractional frequency shift is
\begin{equation}\label{eq:metrology:freqshift_low_v}
\frac{\dot{\delta\omega}_{\text{e}\rightarrow\text{r}}}{\omega_\text{e}~} = \frac{1}{c}\delta a_\parallel + \frac{1}{c}\delta a_\bot + \frac{1}{2}\dot{\delta h}~,
\end{equation}
where the first two terms are accelerations, parallel and orthogonal to the line of sight $\hat{\vect{k}}$ defined by the light beam
\begin{subequations}
\begin{align}
\delta a_\parallel & = \hat{\vect{k}}\cdot\delta\vect{a}_{\text{e}\rightarrow\text{r}}~, \\
\delta a_\bot & = \dot{\hat{\vect{k}}}\cdot\delta\vect{v}_{\text{e}\rightarrow\text{r}}~.
\end{align}
\end{subequations}
This shows that the Doppler link reads the GW signal, but also accelerations longitudinal and transversal to the line of sight.

The deep meaning of \eqref{eq:metrology:freqshift_low_v} is that signals and all unwanted noise sources effectively enter into the Doppler link as equivalent time-delayed accelerations that can be modeled as
\begin{equation}
\frac{\dot{\delta\omega}_{\text{e}\rightarrow\text{r}}}{\omega_\text{e}~} = \frac{1}{c}\left(\delta a_\text{n} + \delta a_h\right)~,
\end{equation}
analogous to the reformulation of the Doppler link as a differential accelerometer in \eqref{eq:metrology:freq_shift_der}. In fact, the GW signal equivalent acceleration is
\begin{equation}
\delta a_h = \frac{c}{2}\dot{\delta h}~,
\end{equation}
which becomes proportional to the second time-derivative of the GW signal in the long-wavelength limit \eqref{eq:metrology:h_longwave}
\begin{equation}
\delta a_h \simeq \frac{\delta x}{2}\ddot{h}~.
\end{equation}

Spurious sources are overall contained in $\delta a_\text{n}$ and expressed as equivalent accelerations. They can all be categorized in two types of contributions following the idea of \eqref{eq:metrology:freqshift_low_v}: those along the light path -- the most important contribution due to real differential forces (per unit mass) $\nicefrac{\delta f}{m}$ acting between the TMs, with a typical spectral shape $\oforder\omega^{-n}$, $n\simeq1,2,4$ -- and those orthogonal -- the cross-talk from other degrees of freedom to the optical axis introduced with some examples in \sectref{sect:metrology:fiducial_points}.

\subsection{Readout noise}

The second noise contribution after the TDI compensation is the interferometric sensing noise due to various unsuppressed frequency fluctuations. The interferometric sensing is usually expressed in terms of displacement $\delta x$ having the typical spectral shape of \eqref{eq:introduction:ifo_noise}, i.e., flat at high frequency and $\oforder\omega^{-2}$ at low frequency.

As already discussed for the GW signal, even the readout noise can be expressed in terms of equivalent acceleration as input to the differential accelerometer, by multiplying the displacement spectrum by $\omega^2$. In fact, if the noise PSD in displacement is
\begin{equation}
S_{\text{n},\delta x}^{\nicefrac{1}{2}}(\omega) = \delta x_0 \left[1+\left(\frac{\omega_0}{\omega}\right)^2\right]~,
\end{equation}
where $\delta x_0$ and $\omega_0$ are two scaling constants, the corresponding equivalent acceleration is found to be
\begin{equation}
S_{\text{n},\delta a_x}^{\nicefrac{1}{2}}(\omega) = \delta a_0 \left[1+\left(\frac{\omega}{\omega_0}\right)^2\right]~,
\end{equation}
where $\delta a_0=\omega_0^2\,\delta x_0$.

The equivalent acceleration to the readout noise can be summed up to the acceleration noise and assuming the two contributions are uncorrelated, the noise PSD of the \textit{total equivalent acceleration noise} is
\begin{equation}
S_{\text{n},\delta a}(\omega) = S_{\text{n},\nicefrac{\delta f}{m}}(\omega) + \omega^4 S_{\text{n},\delta x}(\omega)~.
\end{equation}
As a matter of fact, the total equivalent acceleration -- the main focus of this thesis, whose results can be easily extrapolated to LISA and all space-based missions -- is dominated by sensing at high frequency due to the $\omega^4$ factor and differential forces at low frequency. The preceding shows again that the sensing can be described as input equivalent acceleration to the LISA arm viewed as differential accelerometer.

\subsection{Summary}

The Doppler link is a de facto differential time-delayed accelerometer: it measures relative time-delayed accelerations between nominal freely-falling particles, where the accelerations come from direct forces at low frequency, sensing at high frequency and the cross-talk from other degrees of freedom that couples with the dynamics along the optical axis. This approach has two very practical and useful consequences:
\begin{enumerate}
  \item it puts signals, force noise, readout noise and whatever noise sources at the same level, treating them as equivalent differential accelerations, and provides for a benchmark to compare them all; even though the aim of this thesis is not to give a comprehensive review of all noise sources and systematics, nor a full noise projection, a general idea is given throughout this work.
  \item it is a means by which very disparate gravitational experiments, ground-based and space-based missions, with different scientific targets and frequency bands, can be really qualified within a unified viewpoint; for example, see \cite{danzmann2007} (Figure 5 and references therein) for a comparison of the experimental performances of few missions on gravitational physics, based at some extent on the ability of putting test particles in geodesics motion.
\end{enumerate}

\section{Dynamics of fiducial points} \label{sect:metrology:fiducial_points}

This section describes in more details two important effects that enter into the Doppler measurement previously introduced: the body finite extension and the pointing inaccuracies due to misalignments in the optical device. Both cases can be traced back to the fact that the \textit{fiducial points} in which light is reflected do not coincide with the centers of mass.


A toy model is now introduced in order to give a first understanding of the problem. Let $x$ be the Doppler measurement axis and $y$ and $z$ the respective orthogonal ones. Consider a single cubic TM of latus $l$, subjected to:
\begin{enumerate}
  \item a small rotation $\delta\phi$ due to an unsuppressed torque along $z$;
  \item a small translation $\delta y$ due to an unsuppressed force along $y$.
\end{enumerate}
Conversely, both cases may correspond to small misalignments of the optical bench performing the measurement along $x$. \figref{fig:metrology:tm_misalignment} gives the proper geometrical representation where the effects are purposely enlarged for the sake of clarity.

\figuremacroW{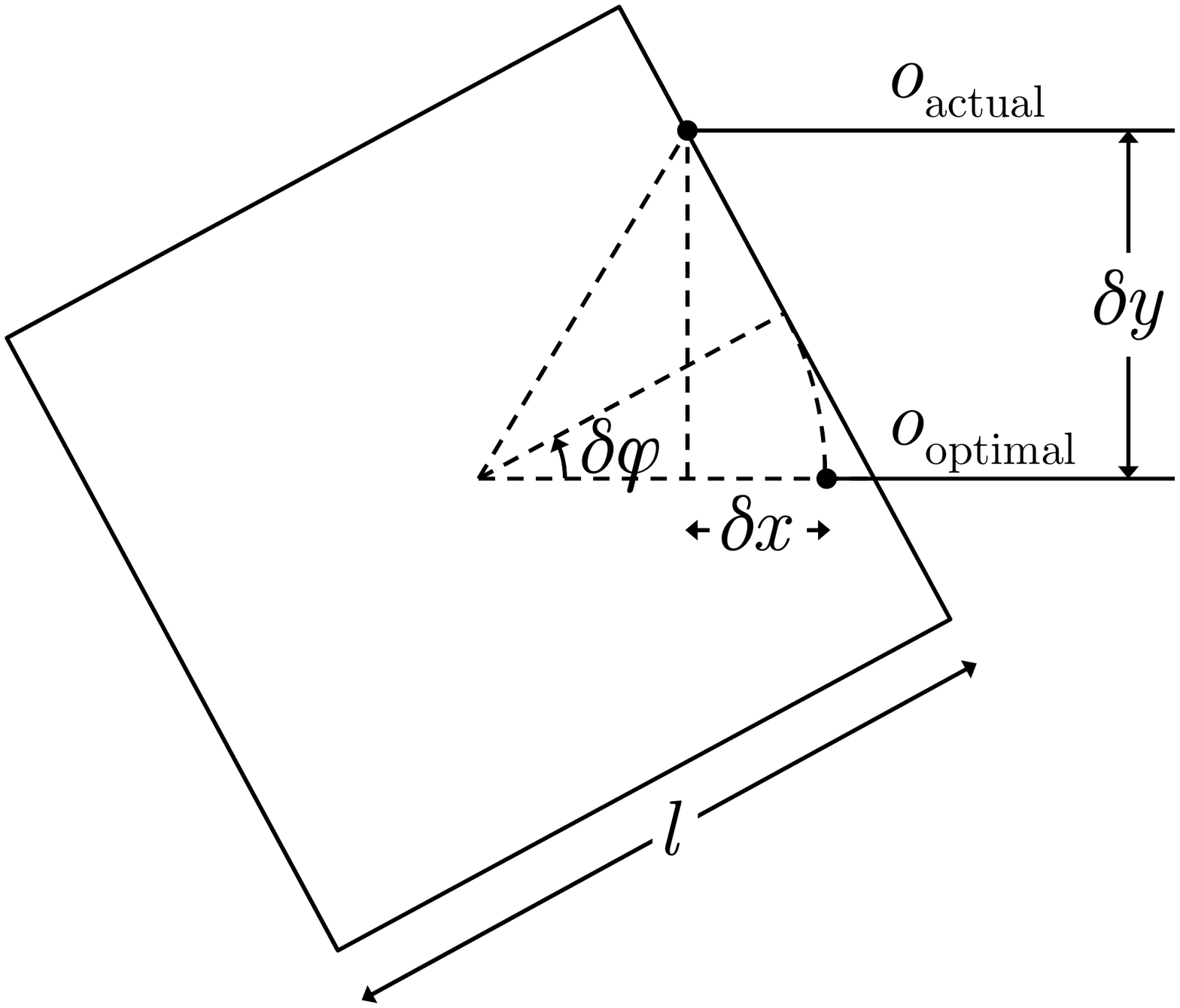}{Geometrical representation of misalignments in the measurement axis. The actual measurement $o_\text{actual}$ contains small imperfections due to unsuppressed TM translations and rotations to the optimal $o_\text{optimal}$ direction pointing the TM center of mass. The fiducial points where light is reflected are highlighted as big dots. $o_\text{actual}$ differs from $o_\text{optimal}$ by $\delta x=\delta y \tan{\delta\phi} - \left(\frac{l}{2}\frac{1}{\cos{\delta\phi}} - \frac{l}{2}\right)$.}{fig:metrology:tm_misalignment}{0.6}

The optimal measurement to the TM center of mass is named $o_\text{optimal}$; the actual misaligned measurement $o_\text{actual}$ differs from this by a small amount $\delta x$,
\begin{equation}
o_\text{actual} = o_\text{optimal} + \delta x~,
\end{equation}
and $\delta x$ contains the cross-talk from both type of imperfections
\begin{equation}
\delta x = \delta x_y + \delta x_\phi~.
\end{equation}
With simple considerations (see \figref{fig:metrology:tm_misalignment}), to second order, it turns out
\begin{subequations}
\begin{align}
\delta x_y & \simeq \delta y \delta\phi~, \\
\delta x_\phi & \simeq - \frac{l}{4}\delta\phi^2~.
\end{align}
\end{subequations}
In fact, when the measurement is performed along the optimal axis, but the TM is rotated, then only $\delta x_\phi$ survives and the contribution is negative since it subtracts displacement to $o_\text{optimal}$. Instead, when the TM is not rotated $\delta x_y$ vanishes. In the general situation when the TM is both translated and rotated $\delta x_y$ is intrinsically coupled with the $\phi$ motion. The above will be referred as dynamical cross-talk.

This simple calculation suggests that any detector measuring the relative motion between two extended bodies reads out a fake signal due to an unavoidable cross-talk from other degrees of freedom to the optical axis.

\figref{fig:metrology:tm_sc_misalignment} shows a scheme of a misalignment between a TM and its hosting SC, affecting the local link within the LISA arm. As said, the local link is corrupted by force noise coupling the SC motion with the TM. However, small misalignments of the optical bench and the (linear and angular) motion of the TM enter into the link. Among the things, the local link will be characterized by the LPF mission.

\figuremacroW{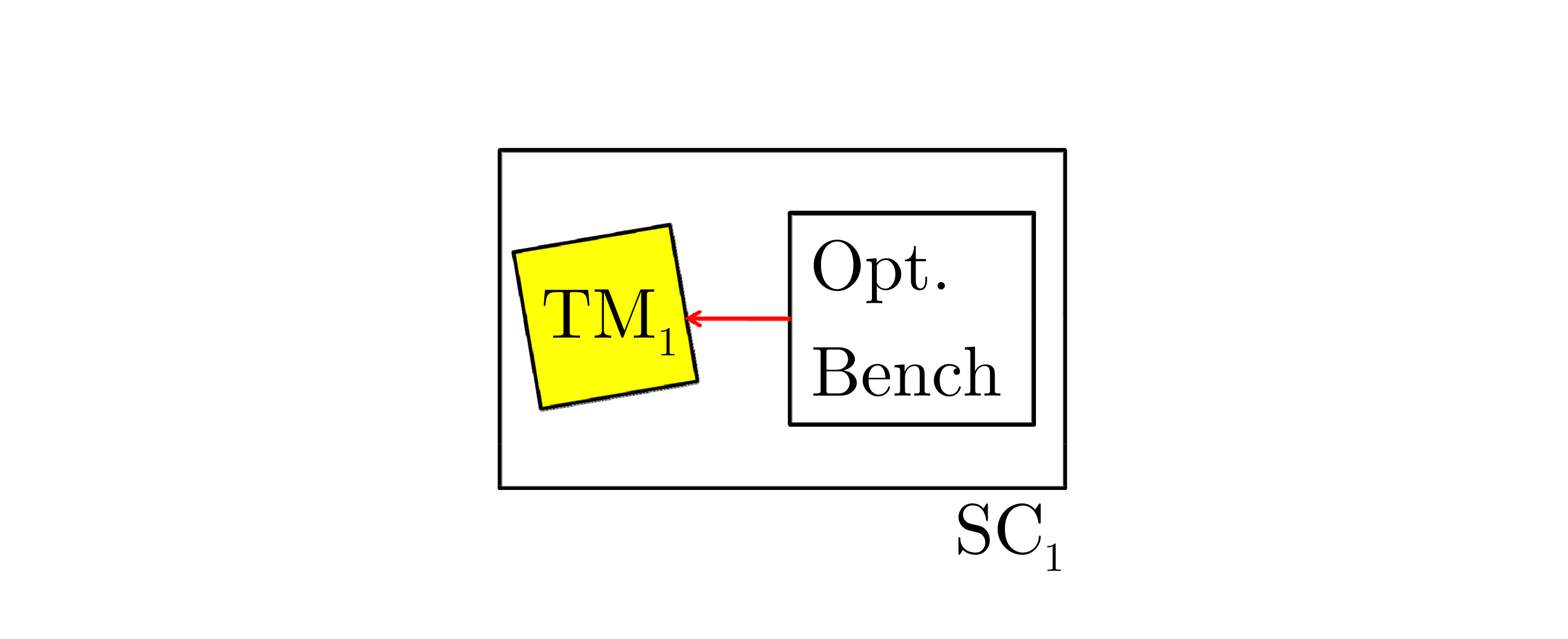}{Scheme of a misalignment between a TM and its hosting SC in the LISA arm. The (linear and angular) motion of the TM relative to its hosting SC couples with the Doppler link.}{fig:metrology:tm_sc_misalignment}{0.6}

Analogously, \figref{fig:metrology:sc_sc_misalignment} shows a scheme of a misalignment between two SCs, affecting the faraway link within the LISA arm. Again, the (linear and angular) motion between the SCs corrupts the link. Despite to LISA, in LPF there is only one SC, so LPF will not characterize the link between the SCs.

\figuremacroW{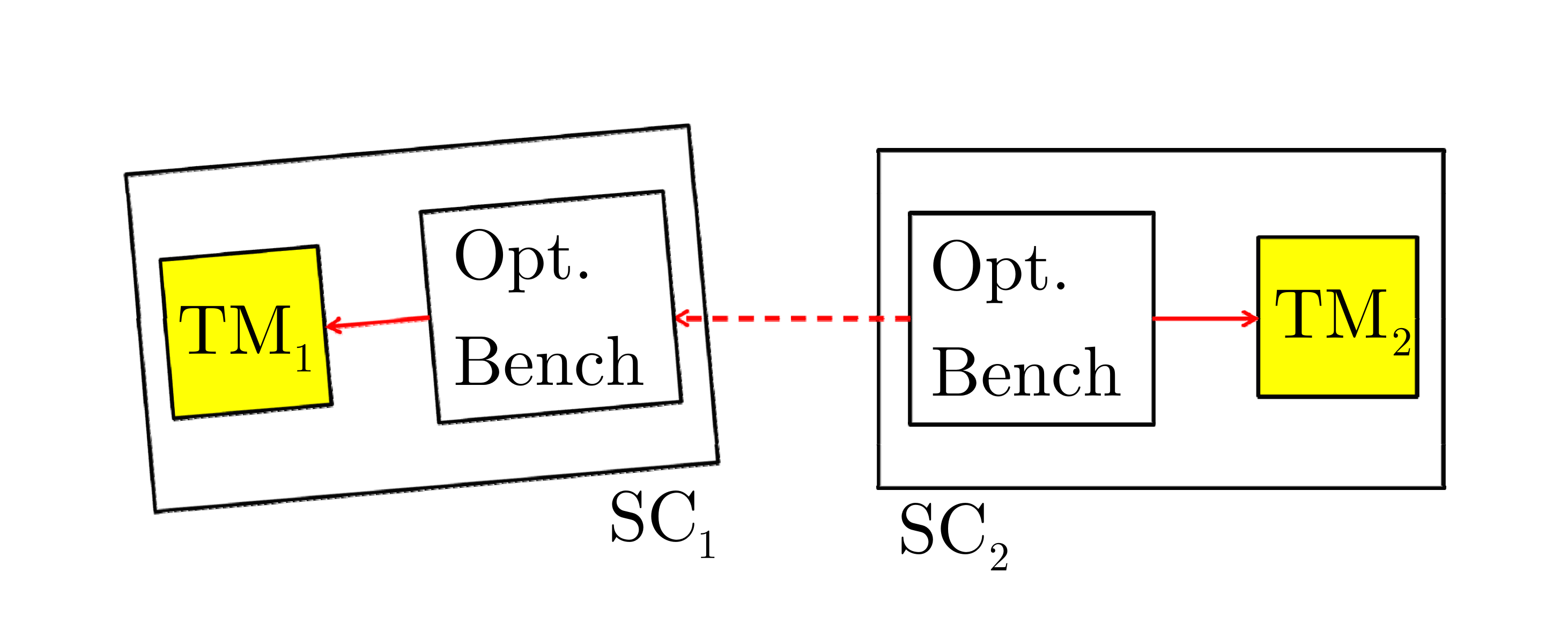}{Scheme of a misalignment between two SCs in the LISA arm. The (linear and angular) motion of a SC relative to the other couples with the Doppler link.}{fig:metrology:sc_sc_misalignment}{0.6}

It is worth to stress that the Doppler link implemented between two faraway extended bodies measures an unavoidable acceleration coming from the \textit{differential (time-delayed) dynamics of fiducial points}.

The LISA arm (4-body system) is a sensor measuring the relative motion of the TMs with respect to the hosting SCs and the relative motion between the two SCs. Instead, its down-scaled version to LPF (3-body system) is a sensor measuring only the relative motion of the TMs with respect to the common hosting SC. Such difference between LISA and LPF implies that the total number of degrees of freedom is 24 for LISA and 18 for LPF. However, the linear motion of the center of mass must be subtracted from this figure as it is common-mode. Since the relative motion between the TMs is the scientific degree of freedom, the spurious degrees of freedom are 20 for LISA and 14 for LPF. \tabref{tab:metrology:dof} shows each contribution affecting the differential measurement. As expected, LPF reproduces the LISA arm up to the two local measurements between the TMs and the SC, but the differential motion between the SCs is a peculiarity of LISA.
\begin{table}[!htbp]
\caption{\label{tab:metrology:dof}\footnotesize{Spurious sources coming from the dynamics of other degrees of freedom and affecting the main sensitive axis of LISA and LPF. The interferometric arm respectively reads 20 and 14 spurious degrees of freedom in LISA and LPF. The main difference is that in LPF the TMs fit a common SC.}}
\centering
\begin{tabular}{lD{.}{.}{4.3}D{.}{.}{2.1}} 
\hline
\hline
\multirow{2}{*}{Extra-contribution} & \multicolumn{2}{c}{Degrees of freedom} \\
& \multicolumn{1}{c}{LISA} & \multicolumn{1}{c}{LPF} \\
\hline
Linear motion between the SCs & 2 & \text{-} \\
Linear motion of the TMs & 6 & 5 \\
Angular motion of the SCs & 6 & 3 \\
Angular motion of the TMs & 6 & 6 \\
\hline
Total & 20 & 14 \\
\hline
\hline
\end{tabular}
\end{table}

\ChangeFigFolder{3_dynamics}


\chapter{Controlled dynamics} \label{chap:dynamics}


As said in the Introduction, LPF is aimed at demonstrating the geodesic motion of TMs within a single SC reproducing a down-scaled version of the LISA arm. The previous chapter discussed the fundamental physics of the Doppler link and the way external forces can be measured by frequency shifts of photons exchanged between two TMs. The GW signal, non-gravitational disturbances and all noise sources can be effectively viewed as input equivalent accelerations to a differential time-delayed accelerometer. In a step by step discussion it was shown that many effects may corrupt the measurement and, among all, there is the fact that the link is actually implemented in a dynamical system of 3 extended bodies, whose relative motions are optically tracked with inevitable pointing inaccuracies and misalignments.

This chapter introduces a further concept: the control. In LISA the drag-free controller acts as a shield for the external disturbances. In the adopted scheme, the SCs are actuated to follow the free-falling TMs along the measurement axes, whereas the TMs are actuated along the degrees of freedom orthogonal to those axes. This concept is implemented and verified in LPF with a difference. In LISA each SC contains two TMs belonging to different measurement axes, the links to the faraway SCs. In LPF, as shown in \figref{fig:dynamics:lpf_scheme_subsystems}, there is only one measurement axis, therefore the SC can not follow both TMs independently. While the SC follows the reference TM, the other TM must be capacitively actuated to follow the reference TM. This is the target configuration named science mode.

\figuremacroW{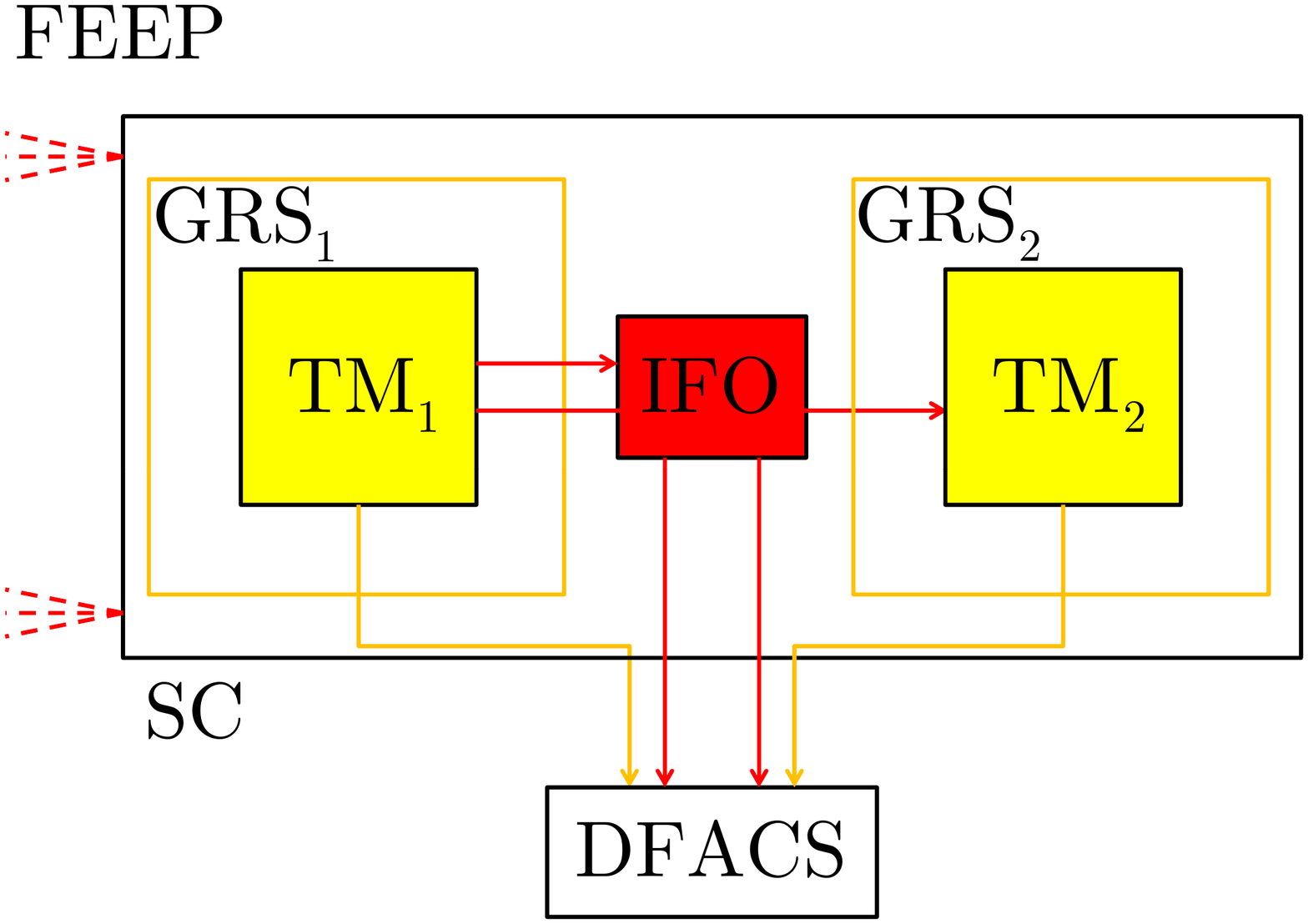}{Simplified scheme of \figref{fig:introduction:lpf_scheme_subsystems}. In spite of LISA, in LPF there is only one measurement axis. As the reference TM is in free fall and the SC is forced to follows it to compensate for external disturbances, the second TM is forced to follow the reference TM.}{fig:dynamics:lpf_scheme_subsystems}{0.8}

Scope of this chapter is to step into the details of the measurement scheme of LPF. A unified formalism is introduced to describe dynamics, sensing and control as a whole in view of defining a fundamental operator that:
\begin{enumerate}
  \item converts the sensed TM relative motion into total equivalent input relative acceleration;
  \item subtracts known force couplings, control forces and the cross-talk (sensing, dynamical and actuation);
  \item suppresses system transients.
\end{enumerate}
\sectref{sect:dynamics:matrix_formalism} introduces the formalism describing the closed-loop implementation of the LISA arm in LPF. \sectref{sect:dynamics:transients} discusses on the suppression of system transients in the total reconstructed equivalent acceleration noise as a natural consequence of the formalism. \sectref{sect:dynamics:model_along_x} discusses the first application: a dynamical model of LPF along the optical axis. \sectref{sect:dynamics:cross-talk_matrix} presents the mathematical description of the cross-talk from nominally orthogonal degrees of freedom to the optical axis. \sectref{sect:dynamics:model_xy_cross-talk} discusses the second application: an example of cross-talk.

\section{Closed-loop formalism} \label{sect:dynamics:matrix_formalism}

The formalism developed in this section is effective in mapping a complex dynamics into a simple equation, treating different aspects of the system at the same time as a whole, and allowing for the reconstruction of the total input differential acceleration from the interferometrically-sensed motion.

Like every physical dynamical system, LPF can be described by three main conceptual parts:
\begin{enumerate}
  \item free dynamics;
  \item sensing;
  \item control and actuation.
\end{enumerate}
The first one is the natural free evolution of the system. This gives the dynamical evolution of the TMs as they were left alone in their flight. However, small unwanted disturbances can take each TM away from the ideal geodesic, the \textit{reference} trajectory. On-ground measurements and models predict that to first order the TMs are electrostatically coupled with the SC through negative force gradients described by unstable oscillators. If the TMs were left to follow their free evolution, the system would exponentially destabilize in a very small timescale. Referring to \figref{fig:dynamics:lpf_scheme_subsystems}, in the main science mode the sensed motion between the TM and the interferometer and the sensed relative motion between the TMs is fed into the DFACS controller to command actuation on the SC and the second TM to both follow the reference TM. In this way, one would say that the controller utilizes the sensed relative motion to suppress the disturbances by ``pushing'' a body toward the reference trajectory, i.e., by actuating it along specifical degrees of freedom.

In turn, \sectref{sect:dynamics:coordinates} lists the relevant coordinates in LPF, the sensors, the control laws and the actuators for each degree of freedom; \sectref{sect:dynamics:controller} provides for a general description of the control philosophy; \sectref{sect:dynamics:eom} describes the generalized equation of motion for LPF.

\subsection{Coordinate definitions} \label{sect:dynamics:coordinates}

As pointed out at the end of the previous chapter, LPF is a 3-body dynamical system composed by a SC containing two TMs, whose relative motion is sensed by an interferometer and the capacitive sensors, as described in the Introduction. As LPF characterizes the relative motion between those bodies, the inertial acceleration of the SC is not sensed. Therefore, the degrees of freedom of the system are:
\begin{enumerate}
  \item the relative translations of the TM with respect to the SC, $3+3$;
  \item the relative attitudes of the TM with respect to the SC, $3+3$;
  \item additionally, the absolute (inertial) attitude of the SC with respect to the celestial frame, 3.
\end{enumerate}

The naming convention for the sensed coordinates in LPF in science mode can be found in \figref{fig:dynamics:coordinates}. There are 15 control laws implemented by the DFACS, 12 for the TM relative motions and 3 for the SC absolute attitude. A coordinate guiding the drag-free loop, i.e., a thruster actuation on the SC, is named \textit{drag-free coordinate}. Analogously, a coordinate guiding the electrostatic suspension loop, i.e., a capacitive actuation on the TMs, is named \textit{electrostatic suspension coordinate}. Finally, a coordinate guiding the attitude loop, i.e., a capacitive actuation on the TMs to maintain the inertial orientation, is named \textit{attitude coordinate}. The names of the control loops, the sensor readouts used as inputs to the control laws and the actuators are reported in \tabref{tab:dynamics:coordinates} for all controlled degrees of freedom in the main science mode.

\figuremacroW{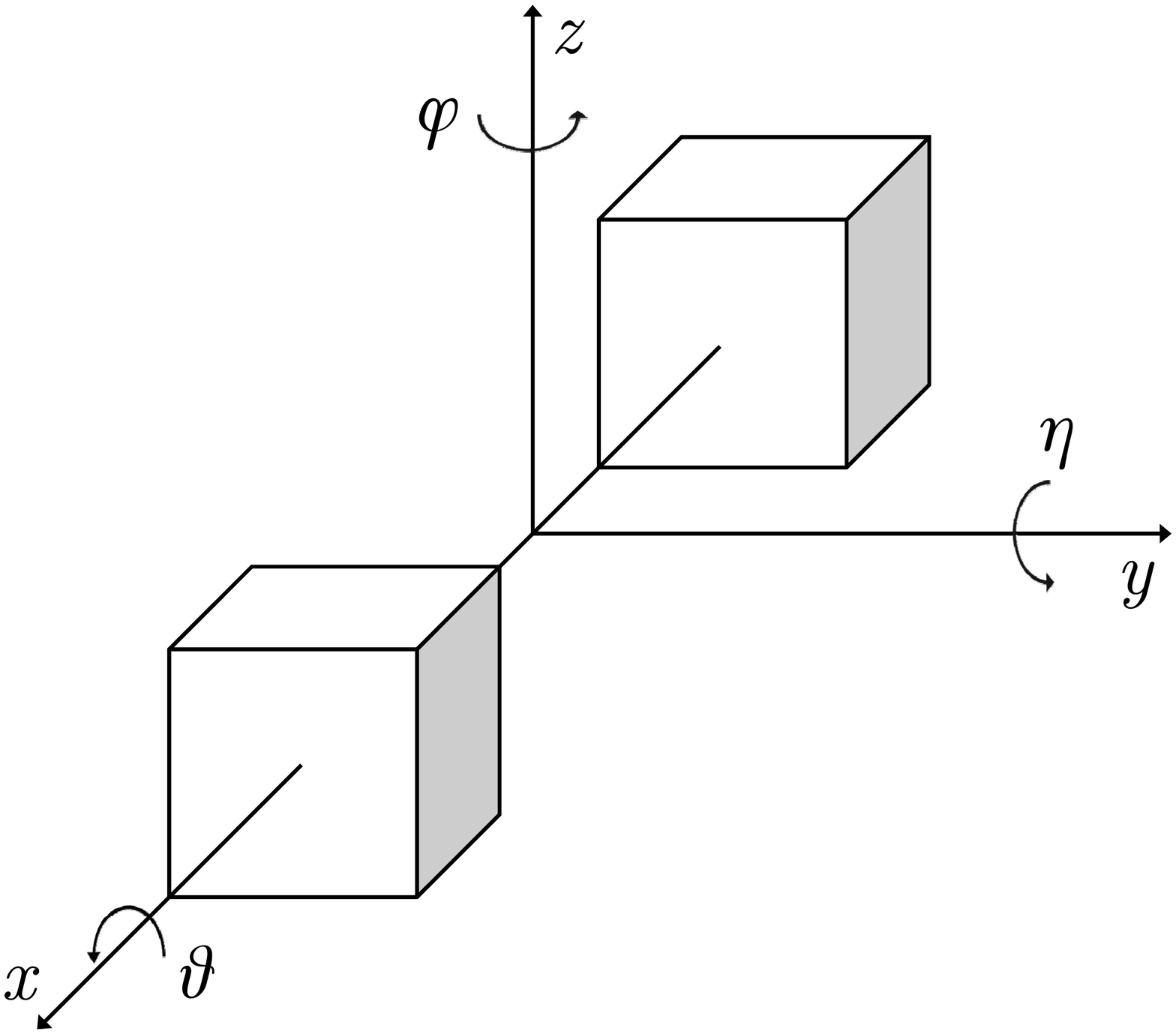}{Coordinate naming convention for the 3-body LPF system. The $x$-axis is the laser sensitive translational degree of freedom, as well as the $\eta$ and $\phi$ angles are optically detected. The $\theta$ angle is not interferometrically detectable. Other coordinates can be read out by capacitive sensors, especially along $y$ and $z$.}{fig:dynamics:coordinates}{0.6}

\begin{table}[!htbp]
\caption{\label{tab:dynamics:coordinates}\footnotesize{List of all controlled degrees of freedom for the LPF mission in the main science mode. The drag-free, electrostatic suspension and attitude control loops, together with the interferometer, capacitive and star-tracker sensors and the thruster and capacitive actuators are reported for each coordinate. Interferometric sensing is used in place of the capacitive whenever possible. Notice that the interferometer measures the relative linear and angular motion between the TMs, i.e., $x_{12}=x_2-x_1$, $\eta_{12}=\eta_2-\eta_1$ and $\phi_{12}=\phi_2-\phi_1$. The SC absolute position is not sensed.}}
\centering
\begin{tabular}{@{\hspace{10pt}}c@{\hspace{10pt}} @{\hspace{10pt}}c@{\hspace{10pt}} D{=}{\,=\,}{4.8} @{\hspace{10pt}}c@{\hspace{10pt}}}
\hline
\hline
Coordinate & Control & \multicolumn{1}{c}{Sensor} & Actuator \\
\hline
$x_1$ & Drag-free & o_1=\text{IFO}[x_1] & FEEP \\
$y_1$ & Drag-free & o_{y_1}=\text{GRS}[y_1] & FEEP \\
$z_1$ & Drag-free & o_{z_1}=\text{GRS}[z_1] & FEEP \\
$\theta_1$ & Drag-free & o_{\theta_1}=\text{GRS}[\theta_1] & FEEP \\
$\eta_1$ & Elect.\,suspension & o_{\eta_1}=\text{IFO}[\eta_1] & GRS \\
$\phi_1$ & Elect.\,suspension & o_{\phi_1}=\text{IFO}[\phi_1] & GRS \\
\hline
$x_2$ & Elect.\,suspension & o_{12}=\text{IFO}[x_{12}] & GRS \\
$y_2$ & Drag-free & o_{y_2}=\text{GRS}[y_2] & FEEP \\
$z_2$ & Drag-free & o_{z_2}=\text{GRS}[z_2] & FEEP \\
$\theta_2$ & Elect.\,suspension & o_{\theta_2}=\text{GRS}[\theta_2] & GRS \\
$\eta_2$ & Elect.\,suspension & o_{\eta_{12}}=\text{IFO}[\eta_{12}] & GRS \\
$\phi_2$ & Elect.\,suspension & o_{\phi_{12}}=\text{IFO}[\phi_{12}] & GRS \\
\hline
$\theta_\text{SC}$ & Attitude & o_{\theta_\text{SC}}=\text{ST}[\theta_\text{SC}] & GRS \\
$\eta_\text{SC}$ & Attitude & o_{\eta_\text{SC}}=\text{ST}[\eta_\text{SC}] & GRS \\
$\phi_\text{SC}$ & Attitude & o_{\phi_\text{SC}}=\text{ST}[\phi_\text{SC}] & GRS \\
\hline
\hline
\end{tabular}
\end{table}

Basically, in the main science mode all optical readings are used whenever possible and:
\begin{enumerate}
  \item along $x$: guided by the optical $x_1$, the SC is forced to follow the reference TM through thruster actuation; guided by the optical $x_{12}$ the second TM is forced to follow the reference TM through capacitive actuation;
  \item along orthogonal degrees of freedom: guided by the average linear motion of the TMs read out by the capacitive sensors, the SC is forced to follow both TMs through thruster actuation; guided by the star-tracker inertial attitude the TMs are oriented through capacitive actuation;
  \item along rotational degrees of freedom: guided by the differential linear motion of the TMs read out by the capacitive sensors, the SC is forced to follow both TMs through thruster actuation; guided by the optical TM attitudes both TMs are oriented through capacitive actuation.
\end{enumerate}


\subsection{Controller} \label{sect:dynamics:controller}

The controller is a dynamical system (see \figref{fig:dynamics:controller_diagram}), in general multidimensional, taking the difference between the measured and the reference trajectories as inputs and producing forces to be applied to the bodies as outputs. If $\vect{o}$ is the sensed motion, the \textit{error} signals for all controlled degrees of freedom are
\begin{equation}
\vect{e} = \vect{o} - \vect{o}_\text{i}~,
\end{equation}
where $\vect{o}_\text{i}$ are named \textit{reference set-point} signals or simply \textit{guidance} signals.



\figuremacroW{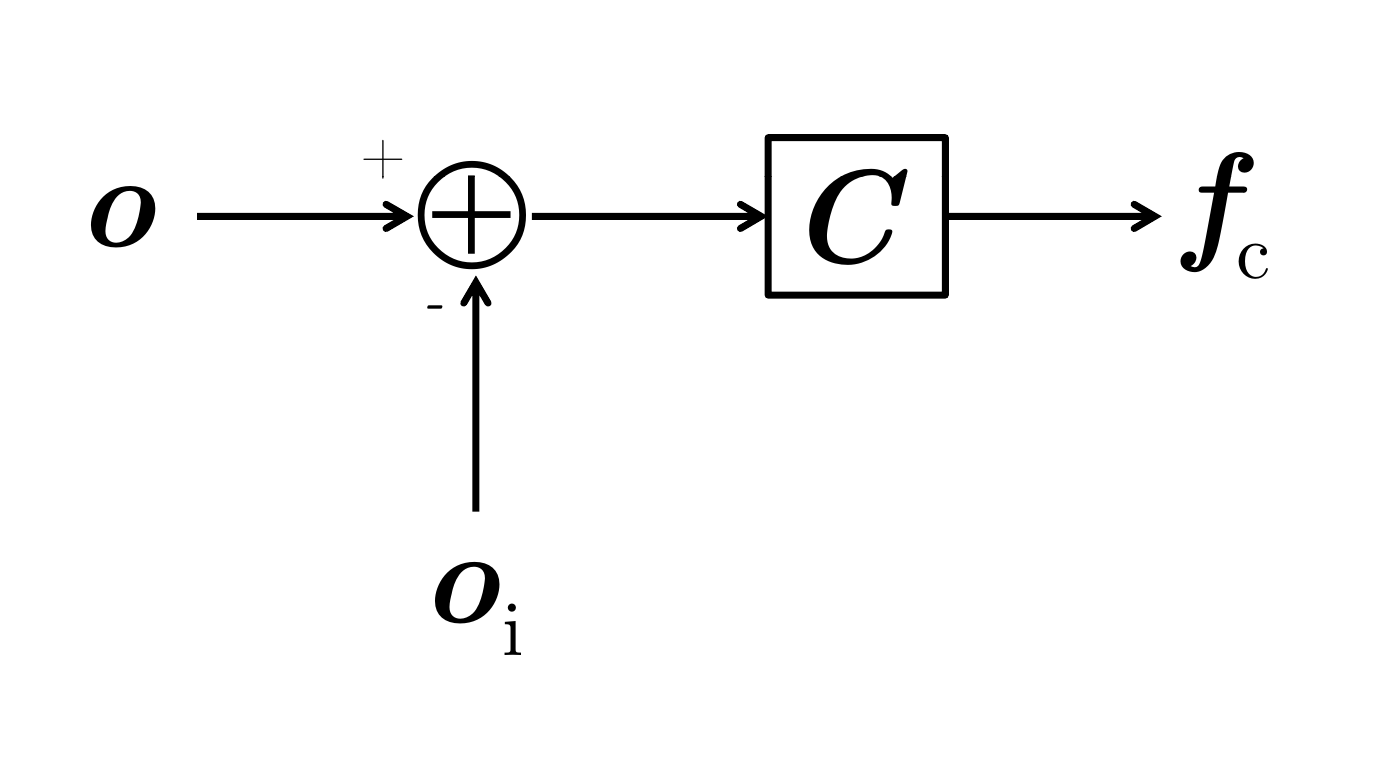}{Block diagram of the controller. It takes the differences between the measured coordinates $\vect{o}$ and the reference coordinates $\vect{o}_\text{i}$ and calculates control forces $\vect{f}_\text{c}$ to be applied to the SC and the TMs.}{fig:dynamics:controller_diagram}{0.45}


The DFACS is responsible of the minimization of the error signals. In this way, it compensates for negative force gradients and makes the system stable. It utilizes the sensed relative motion along different degrees of freedom, contained in the error signal, to compute actuation forces $\vect{f}_\text{c}$. The discrete implementation of the $n$-th value of the commanded force $f_{\text{c},n}$, for a generic control law in LPF \cite{S2-ASD-TN-2003} controlling a single degree of freedom, is a linear combination of the past values of the force $f_{\text{c},n-1},f_{\text{c},n-2},\ldots$ and the present and past values of the error signal (the innovations) $e_n,e_{n-1},\ldots$
\begin{equation}
f_{\text{c},n} = \sum_i q_i f_{\text{c},n-i} + \sum_j p_j e_{n-j}~,
\end{equation}
where $i=1,\ldots,N_q$ and $N_q$ is the order of the autoregressive filter; $j=0,\ldots,N_p$  and $N_p$ is the order of the moving average filter. The $z$-transform of the above gives the well-known autoregressive moving average model of the discrete control law
\begin{equation}
C(z) = \frac{\sum_j p_j z^{-j}}{1-\sum_i q_i z^{-i}}~.
\end{equation}

The control design assures: (i) the compensation of negative force gradients; (ii) the asymptotic stability; (iii) the mitigation of system resonances; (iii) the minimal-cost performance, i.e., the control computes the minimum actuation forces that allow the TMs to reach the reference signals to within the given accuracy of $\unit[5]{nm\,Hz^{-\nicefrac{1}{2}}}$ around $\unit[1]{mHz}$ (for the relative displacement control as reported in \tabref{tab:introduction:subsystems_requirements}), whose unsuppressed part contributes to the residual noise budget. ASTRIUM \cite{astrium} -- the main industry contractor of LPF -- has provided only the continuous representation of the controller as a rational function in the $s$-domain (of maximum order 6), used for system modeling, simulation and analysis shown in \chapref{chap:sys_identification}.

\subsection{Equation of motion} \label{sect:dynamics:eom}

This section describes the formalism on the basis of the modeling of the closed-loop LPF system. The most important assumption concerns on the linearity of the equations, i.e., that all physical quantities characterizing the motion enter linearly into the equations. Here is a list of the involved limitations:
\begin{enumerate}
  \item the force couplings between the TMs and the SC are mainly caused by electrostatics and SC self-gravity: those forces decay as the inverse of the distance at most; they are treated to first order as spring-like forces;
  \item the interferometric sensing involves reflections and transmissions through optical elements: even in geometric optics the equations must involve trigonometric expressions of the angles; it is assumed that trigonometric functions confuses with angles, whenever applicable;
  \item the angular motion of a rigid body is described by the Euler equations: they are non-linear with respect to the angular velocities; if the angular motion is small, non-linearities are second-order effect.
\end{enumerate}
Since the controller forces the motion around the reference trajectories, it also assures that the motion is small enough that all forces and non-linear terms can be expanded to first order with good approximation. In this way, the coupling forces are modeled as negative spring-like constants; the non-linearities due to optics and the angular motion can be effectively ignored. In general, the linearized equations of motion must contain terms to within the order of an imperfection multiplied by a noise contribution. In fact, other combinations like a noise contribution multiplied by another noise contribution are second-order effects and must be neglected. The accuracy to which linearity is achieved depends on: (i) the assumption that the controller does not itself introduce non-linearities in the system; (ii) the unsuppressed noisy motion in the error signals is to within the requirement figure of the controller.

With these assumptions, LPF is viewed as a closed-loop Multi-Input-Multi-Output (MIMO) linear time-invariant dynamical system described by vector equations with operators modeling dynamics, sensing and control.
The linearized equations
for LPF are \cite{S2-UTN-TN-3045}, \cite{antonucci2011b,armano2009} and more recently \cite{congedo2012}
\begin{subequations}\label{eq:dynamics:eom_matrix}
\begin{align}
\vect{D}\,\vect{q} & = \vect{g}~, \label{eq:dynamics:eom_matrix_a} \\
\vect{g} & = \vect{f}_\text{n}+\vect{A}\left[\vect{f}_\text{i}-\vect{C}\,(\vect{o}-\vect{o}_\text{i})\right]~,\label{eq:dynamics:eom_matrix_b} \\
\vect{o} & = \vect{S}\,\vect{q}+\vect{o}_\text{n}~. \label{eq:dynamics:eom_matrix_c}
\end{align}
\end{subequations}
The total forces (per unit mass) $\vect{g}$ produce the motion through the acting of the dynamics operator $\vect{D}$ onto the physical coordinates $\vect{q}$. The natural physical coordinates for LPF are given by the TM relative linear and angular motion. $\vect{D}$ is a differential operator containing time derivatives and the modeled coupling coefficients (the negative spring constants due to the linearization), and the dynamical cross-talk from other degrees of freedom to the sensitive axis as well. \sectref{sect:dynamics:cross-talk_matrix} generalizes this concept by decoupling the dynamics along the measurement axis (the nominal dynamics) from the dynamics along other degrees of freedom (the first-order perturbation). The external forces can be split into pure noise sources $\vect{f}_\text{n}$ -- mostly from the SC jitter and within the TM housings; applied biases $\vect{f}_\text{i}$ -- directly on each TM and the SC; applied biases through $\vect{o}_\text{i}$ -- the controller guidance signals already discussed in the preceding subsection. $\vect{C}$ is the operator containing the control laws. By changing the controller guidance signals, net forces on each body are commanded to the actuators
\begin{equation}\label{eq:dynamics:control_forces}
\vect{f}_\text{c} = -\vect{C}\,(\vect{o}-\vect{o}_\text{i})~,
\end{equation}
where $\vect{o}$ is the closed-loop measurement. Therefore, the application of biases in the controller guidance signals is equivalent to the application of explicit forces on the bodies. In this description, the application of the forces is modeled by an actuation operator $\vect{A}$. All force biases and control forces are fed into such an operator, responsible of the force dispatching on all bodies. In the main science mode, along the measurement axis, this implies a thruster actuation on the SC to follow the reference TM in free fall and a capacitive actuation on the second TM to follow the reference TM. Finally, the physical coordinates $\vect{q}$ are converted into the system readouts $\vect{o}$ (from interferometric and capacitive sensors) through the sensing operator $\vect{S}$, mostly diagonal, and corrupted by the readout noise $\vect{o}_\text{n}$. $\vect{S}$ is nominally an identity operator, but in reality there is a sensing cross-talk between different readout channels and miscalibrations as well. \figref{fig:dynamics:matrix_diagram} shows the block diagram of the closed-loop dynamics for LPF where all operators act on their own inputs and produce their outputs for the dynamical equations in \eqref{eq:dynamics:eom_matrix}; deterministic and stochastic inputs are also distinguished for clearness.


\figuremacroW{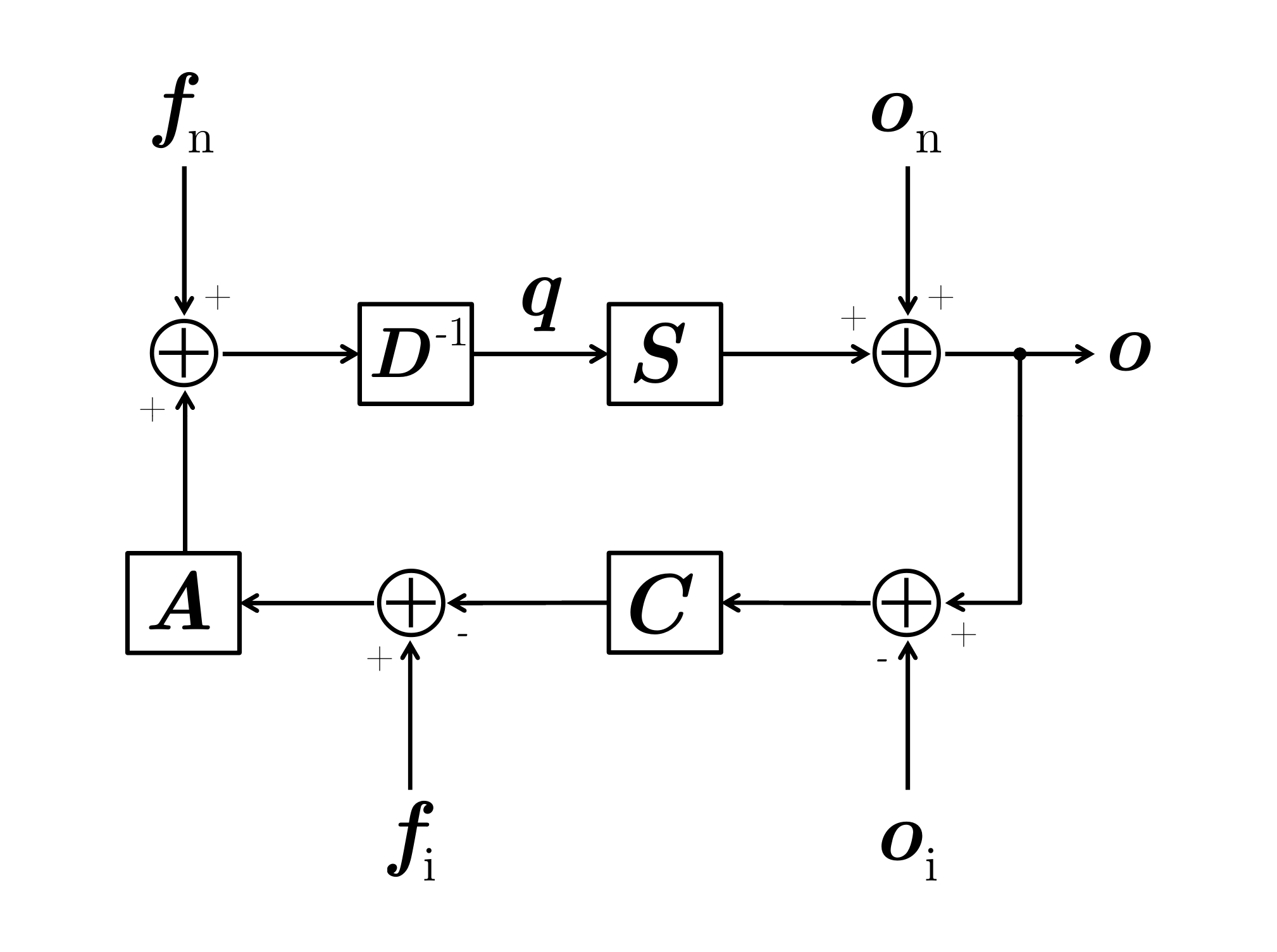}{Block diagram for the three main conceptual steps of LPF: dynamics, sensing and control. There are two different noise sources, $\vect{f}_\text{n}$ and $\vect{o}_\text{n}$, and biases to inject, $\vect{f}_\text{i}$ and $\vect{o}_\text{i}$. The open loops are defined by the transfers from forces to readouts. The forces produce the motion in the $\vect{q}$ coordinates through the inverse of $\vect{D}$. The coordinates are converted into sensed coordinates $\vect{o}$ through $\vect{S}$. The controller closes the loop in order to minimize the error signals, through $\vect{C}$ applied to the sensed coordinates. The calculated forces are then converted into actuation forces through $\vect{A}$.}{fig:dynamics:matrix_diagram}{0.7}

The full equation of motion in vector form and expressed in terms of the sensed relative coordinates, $\vect{o}$, can be obtained by manipulating the three equations in \eqref{eq:dynamics:eom_matrix}. The idea is to substitute \eqref{eq:dynamics:eom_matrix_c} and \eqref{eq:dynamics:eom_matrix_b} in \eqref{eq:dynamics:eom_matrix_a} and rearrange the equation so that the deterministic and stochastic inputs are on the right-hand side. The result is the \textit{equation of motion in the sensed coordinates}
\begin{equation}\label{eq:dynamics:eom_full}
\vect{\Delta}\,\vect{o} = \vect{f}_\text{n}+\vect{D}\,\vect{S}^{-1}\vect{o}_\text{n}+\vect{A}\left(\vect{f}_\text{i}+\vect{C}\,\vect{o}_\text{i}\right)~,
\end{equation}
where four terms are clearly recognized: force noise, readout noise, force bias and controller guidance bias, that all constitute the noise budget of LPF in terms of total equivalent acceleration. The second-order differential operator on the right-hand side is defined as
\begin{equation}
\vect{\Delta} = \vect{D}\,\vect{S}^{-1}+\vect{A}\,\vect{C}~. \label{eq:dynamics:diff_op}
\end{equation}
The deep meaning of the operator is that it allows for the reconstruction of the total equivalent input acceleration from the sensed relative motion and at the same time isolating and subtracting dynamics, sensing, control and actuation. Indeed, by looking at \figref{fig:dynamics:matrix_diagram}, $\vect{D}\,\vect{S}^{-1}$ is the \textit{open loop} from the sensed relative motion to input forces (inverting the direction of an arrow the corresponding operator must be inverted); whereas $\vect{A}\,\vect{C}$ is the \textit{control loop} consisting of all control laws commanding the force actuation.

In \eqref{eq:dynamics:eom_full} two transfer operators can be naturally identified
\begin{subequations}
\begin{align}
\vect{T}_{o\rightarrow f} & = \vect{\Delta}~, \label{eq:dynamics:ifo2acc} \\
\vect{T}_{o_\text{i}\rightarrow o} & = \vect{\Delta}^{-1}\vect{A}\,\vect{C}~. \label{eq:dynamics:ifo2ifo}
\end{align}
\end{subequations}
The second one solves the equation of motion for deterministic guidance signals and substituted into \eqref{eq:dynamics:control_forces} gives the following transfer operator
\begin{equation}\label{eq:dynamics:bias2control}
\vect{T}_{o_\text{i}\rightarrow f_\text{c}} = -\vect{C}\,(\vect{T}_{o_\text{i}\rightarrow o}-\vect{1})~,
\end{equation}
converting the bias injections $\vect{o}_\text{i}$ into the calculated control forces that the actuators must apply in order to stabilize the motion toward the reference signal.

The first transfer operator $\vect{T}_{o\rightarrow f}$ has fundamental relevance as it shows that the differential operator allows for the \textit{estimation of the total out-of-loop equivalent acceleration noise} \cite{monsky2009} on noisy interferometric data, i.e., when all explicit stimuli are set to zero, whose modeling in terms of force noise and readout noise is provided by the equation of motion \eqref{eq:dynamics:eom_full}. However, the evaluation requires the calibration of the dynamics $\vect{D}$, the sensing $\vect{S}$ and the actuation $\vect{A}$ operators overall depending on many system parameters. This critical procedure, named \textit{system identification}, which the performances of the LPF mission depend on, will be outlined in \chapref{chap:sys_identification}. It mainly consists on calibrating the second transfer operator $\vect{T}_{o_\text{i}\rightarrow o}$ and estimating all system parameters in dedicated experiments.

\section{Suppressing system transients} \label{sect:dynamics:transients}

As the main target of LPF is the estimation of the total equivalent input acceleration between the TMs, the transfer operator \eqref{eq:dynamics:ifo2acc}, once calibrated, allows for the compensation of the force gradients, but also for any system transients. Indeeed, the formalism of the previous section can be applied to understand in what sense system transients can be suppressed and the extent to which the suppression is effective.

In the approximation of small relative motion, the dynamical evolution of the TMs in LPF can be described by a linear differential equation with constant coefficients
\begin{equation}\label{eq:dynamics:operator_equation}
\vect{\Delta}\,\vect{o} = \vect{f}~,
\end{equation}
where the external forces $\vect{f}$ produce the motion in the sensed coordinates $\vect{o}$ through the acting of the second-order differential operator $\vect{\Delta}$. In LPF the operator also contains negative force gradients modeled as spring-like constants, the control laws, the actuation and the sensing conversion between physical coordinates and sensed coordinates. As it is shown in the next section this complex structure can be further described with the introduction of targeted parameters modeling the system. Those parameters may vary in time so that the equation has no longer constant coefficients: in principle, such behavior could be observable at very low frequency. Even in this case, there are theorems \cite{courant} ensuring the existence and uniqueness of solutions, at least locally, for reasonable conditions often met in practice.

The particular solution of \eqref{eq:dynamics:operator_equation} is provided to satisfy
\begin{equation}\label{eq:dynamics:operator_equation_part}
\vect{\Delta}\,\vect{o}_\text{s} = \vect{f}~,
\end{equation}
and gives the \textit{steady state} of the system where the evolution is completely driven by the external forces, noise in any form viewed as equivalent acceleration and any possible applied bias. The above equation is usually well-understood and easily solved, for example, in frequency domain.

On the contrary, the homogeneous solution of \eqref{eq:dynamics:operator_equation} is provided to satisfy
\begin{equation}\label{eq:dynamics:operator_equation_homog}
\vect{\Delta}\,\vect{o}_\text{t} = 0~,
\end{equation}
and gives the \textit{transient state} of the system. The transient state is not influenced by the steady state and vice-versa.


The operator \textit{kernel} is defined by the set of all solutions satisfying \eqref{eq:dynamics:operator_equation_homog}. It may be natural to question about the dimensionality of the kernel. There are two possible alternatives \cite{courant}:
\begin{enumerate}
  \item the kernel is trivial and the only possible homogeneous solution is the null solution;
  \item the kernel is non-trivial;
\end{enumerate}
Excluding the trivial case, whenever the operator is linear it is proved that the kernel itself is a vector space -- that in case can be provided with a norm or an inner dot -- where any combination of basis functions $\vect{\phi}_k$ in that space
\begin{equation}\label{eq:dynamics:transient_linear_decom}
\vect{o}_\text{t} = \sum_k c_k\,\vect{\phi}_k~,
\end{equation}
is still a solution of the homogeneous equation for $k$ running through the dimension of the space. $c_k$ are some constants depending on the initial (or boundary) conditions of the system.

So far, an extensive use of inversion operations -- in particular of the $\vect{\Delta}$ operator -- has been made in all calculations concerning dynamics. In general, this may not be completely allowed when the kernel is non-trivial. In this case the operator is singular by definition and can not be inverted. Equivalently, transients exist independently from the steady-state solution driven by the external forces.


The solution of the differential equation \eqref{eq:dynamics:operator_equation} exists and is unique for given input forces $\vect{f}$, initial conditions $c_k$ and suitable assumptions. The general solution is a sum of: (i) the steady state $\vect{o}_\text{s}$ proportional to the input forces; (ii) the transient state $\vect{o}_\text{t}$ set by the initial conditions. Hence, by applying the differential operator to both the steady state and the transient state, it holds
\begin{equation}
\begin{split}
\vect{\Delta}\,\vect{o} & = \vect{\Delta}\left(\vect{o}_\text{s} + \vect{o}_\text{t}\right) \\ 
& = \vect{\Delta}\,\vect{o}_\text{s} + \sum_k c_k\,\vect{\Delta}\,\vect{\phi}_k \\
& = \vect{f}~,
\end{split}
\end{equation}
where 
$k$ spans the kernel. Since $\vect{\Delta}\,\vect{\phi}_k=0$ for any $\vect{\phi}_k$ lying in the kernel, the operator automatically suppresses any system transients, if present.

For example, suppose that a system is modeled by a mono-dimensional harmonic oscillator, with $\Delta=\text{d}^2/\text{d}t^2-\omega_0^2$, where $\omega_0^2$ is the spring constant. An external force (per unit mass) $f(t)$ produces the sensed motion $o(t)$, from which the external force must be estimated. It is well-known that the transient solution is a combination of exponentials $o_\text{t}(t)=c_1\,\exp(-t/\tau)+c_2\,\exp(t/\tau)$, where $c_1$ and $c_2$ are constants depending on the initial conditions and $\tau=1/\omega_0^2$. By definition $\Delta o_\text{t}(t)=0$ and $\Delta o(t)$ provides an estimate of $f(t)$.

The accuracy to which the suppression of system transients is effective depends on the accuracy to which the system parameters have been previously calibrated. If $\vect{p}_\text{est}$ is the vector of parameter estimates approximating the ``true'' parameter values $\vect{p}_\text{true}$ modeling the system up to the inaccuracies $\delta\vect{p}$, i.e., $\vect{p}_\text{est}\simeq\vect{p}_\text{true}+\delta\vect{p}$, then to first order $\vect{\Delta}_\text{est}\simeq\vect{\Delta}_\text{true}+\delta\vect{\Delta}$, where $\vect{\Delta}_\text{true}=\vect{\Delta}(\vect{p}_\text{true})$ and $\vect{\Delta}_\text{est}=\vect{\Delta}(\vect{p}_\text{est})$. Therefore, it follows
\begin{equation}
\begin{split}
\vect{\Delta}_\text{est}\,\vect{o} & \simeq \left(\vect{\Delta}_\text{true}+\delta\vect{\Delta}\right)\left(\vect{o}_\text{s} + \vect{o}_\text{t}\right) \\
& = \vect{\Delta}_\text{true}\,\vect{o}_\text{s} + \vect{\Delta}_\text{true}\,\vect{o}_\text{t}
+ \delta\vect{\Delta}\left(\vect{o}_\text{s} + \vect{o}_\text{t}\right) \\
& = \vect{f}_\text{true} + \delta\vect{f}~,
\end{split}
\end{equation}
where the first term gives the true forces, the second is identically zero by definition and the last one gives the systematic errors in the reconstructed forces
\begin{equation}\label{eq:dynamics:transients_suppression}
\delta\vect{f} = \delta\vect{\Delta}\left(\vect{o}_\text{s} + \vect{o}_\text{t}\right)~.
\end{equation}
From the preceding equation, two cases can be distinguished:
\begin{enumerate}
  \item $\vect{o}_\text{t}\ll\vect{o}_\text{s}$, transients are negligible, but inaccuracies in the operator still produce systematic errors in the estimated total equivalent acceleration noise; a similar argument will be used in \sectref{sect:sys_identification:force_noise} to demonstrate that biases in the estimated parameters may produce biases in the estimated total equivalent acceleration noise;
  \item $\vect{o}_\text{t}\gg\vect{o}_\text{s}$, transients are important and inaccuracies in the operator makes impossible their complete suppression; much more important, biases in the estimated parameters may even amplify the transients.
\end{enumerate}

Once more, the suppression of transients is assured to the level by which the operator itself is calibrated. It is worth stressing the importance of the result. In LPF the main scientific target is the estimation of the total out-of-loop equivalent acceleration from the sensed relative motion.
The damping is fundamentally governed by the controller design, i.e., the efficiency to which the controller responds and stabilizes the system toward the zero-reference signal. As the controller is designed to mostly compensate the expected force gradients that are roughly $|\omega^2|\oforder\unit[1\e{-6}]{s^{-2}}$, this figure gives a typical timescale of $\tau\oforder\unit[6\e{3}]{s}$ (almost 2 hours) for the damping of initial transients. Since the mission operations are very time-constrained, it is not possible to wait for the steady state and the total equivalent acceleration noise must be estimated when the system is fully dominated by transients. The considerations enlightened in this section, together with the procedures of system identification described in \chapref{chap:sys_identification}, assure that in the estimation of the total out-of-loop equivalent acceleration noise, even though transients could almost certainly last for hours, they can be mitigated with good and reasonable confidence if the $\vect{\Delta}$ operator is calibrated on fiducial values of the system parameters.

\section{Dynamical model along $x$} \label{sect:dynamics:model_along_x}

In what follows a model for the LPF mission is elaborated in terms of the two main degrees of freedom along the optical axis: the relative motion of the reference TM to the optical bench and the differential motion between the TMs. In this formulation the relative motion is sensed with the interferometer -- the reference measurement for scientific operations -- while keeping in mind that the capacitive sensors could even be used in place of the interferometer as a backup option, even though such a measurement would be two orders of magnitude worse, especially at high frequency. However, along the other orthogonal axes the capacitive sensors are the only means by which the TM relative motion can be measured.

By tracing back the equations to \sectref{sect:dynamics:matrix_formalism}, the formalism developed so far allows for a straightforward description of LPF as a doubly closed-loop dynamical system in which the effect of the modeled couplings and control laws must be isolated and subtracted from the data when estimating the total equivalent acceleration noise. \figref{fig:dynamics:lpf_scheme_1d} shows a sketch of a LPF model, in the main science mode, along the optical axis that is discussed here in details.

\figuremacroW{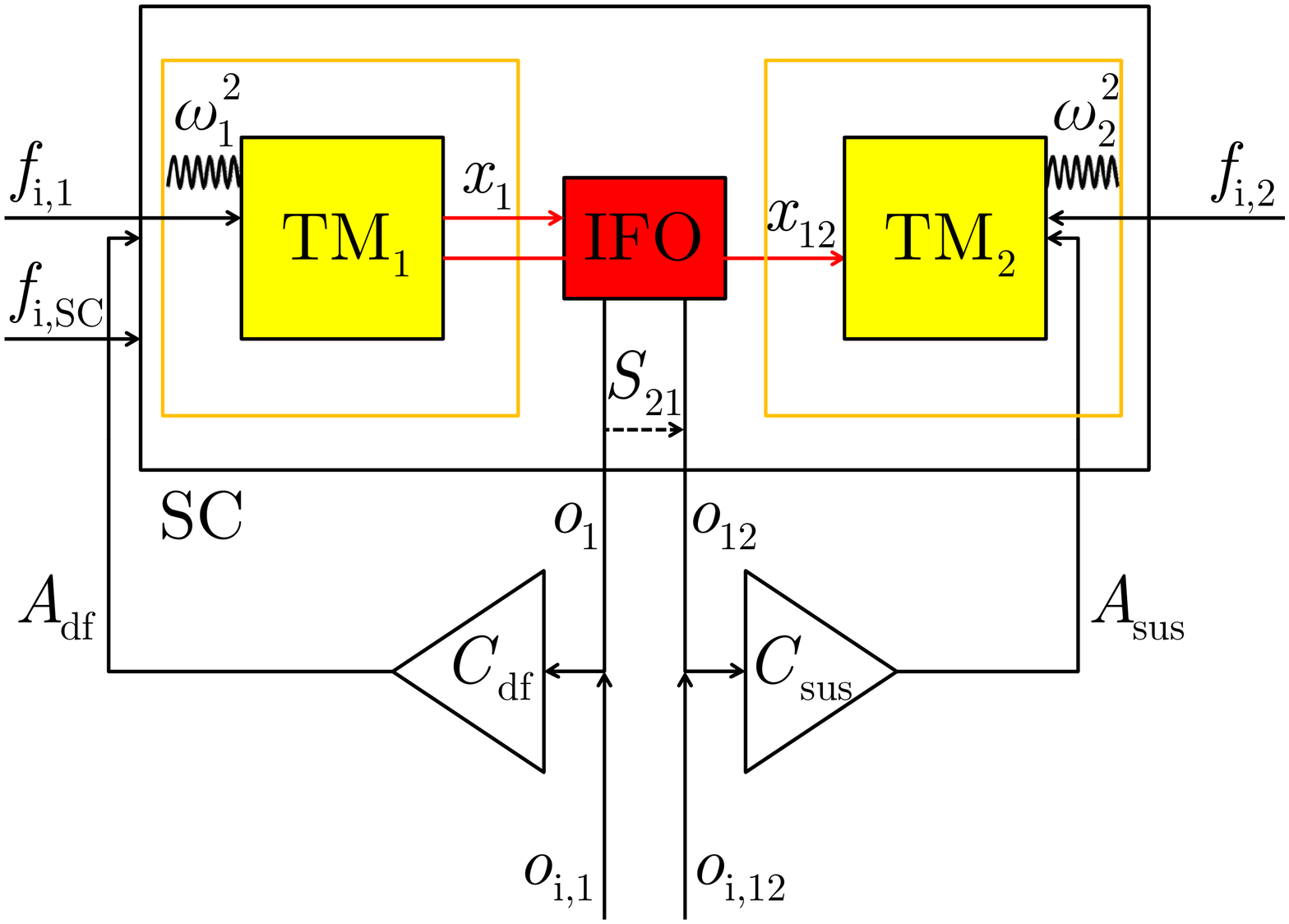}{Scheme of the LPF model along the optical axis in the main science mode. The first TM is in free fall along $x$ and its displacement to the optical bench ($o_1$) is sensed by the interferometer (IFO) and fed into the controller ($C_\text{df}$) to force the SC to follow the TM through thruster actuation (drag-free loop). Analogously, the sensed differential displacement between the two TMs ($o_{12}$) is fed into the controller ($C_\text{sus}$) to force the TM to follow the reference one through capacitive actuation (suspension loop). The critical system parameter are the TM spring-like couplings to the SC ($\omega_1^2$ and $\omega_2^2$), the sensing cross-talk ($S_{21}$) and the actuation gains ($A_\text{df}$ and $A_\text{sus}$). The system can be excited by injecting signals as direct forces on the masses ($f_{\text{i},1}$, $f_{\text{i},2}$ and $f_{\text{i},\text{SC}}$) or controller guidance signals ($o_{\text{i},1}$ and $o_{\text{i},12}$).}{fig:dynamics:lpf_scheme_1d}{0.8}

Referring to \figref{fig:dynamics:coordinates} and \figref{fig:dynamics:lpf_scheme_1d}, $x$ is the interferometric axis. $x_\text{SC}$ is the absolute SC position and $x_1$, $x_2$ are the relative TM positions with respect to the SC; $m_\text{SC}=\unit[422.7]{kg}$ and $m_1=m_2=\unit[1.96]{kg}$ are the respective masses; $\tilde{m}_1=\tilde{m}_2=5\e{-3}$ are the masses normalized to $m_\text{SC}$; $f_1$, $f_2$ and $f_\text{SC}$ are the total forces (per unit mass) containing noise in any form and applied biases.

In the linear approximation (small motion, small forces, as already discussed) the 3-body dynamics is described by a linear system of differential equations. In frequency domain and assuming null initial conditions the equations of motion are
\begin{subequations}
\begin{align}
s^2\,x_1+s^2\,x_\text{SC} + \omega_1^2\,x_1 + \Gamma_x\left(x_2-x_1\right) & = f_1~, \label{eq:dynamics:eom_1D_1} \\
s^2\,x_2+s^2\,x_\text{SC} + \omega_2^2\,x_2 - \Gamma_x\left(x_2-x_1\right) & = f_2 - C_\text{sus}(s)\,o_{12}~, \label{eq:dynamics:eom_1D_2} \\
\begin{split}\label{eq:dynamics:eom_1D_3}
s^2\,x_\text{SC} - \tilde{m}_1\,\omega_1^2\,x_1 - \tilde{m}_2\,\omega_2^2\,x_2 & = f_\text{SC} + C_\text{df}(s)\,o_1 \\
& \quad - \tilde{m}_1\,f_1 - \tilde{m}_2\,f_2 \\
& \quad + \tilde{m}_2\,C_\text{sus}(s)\,o_{12}~,
\end{split}
\end{align}
\end{subequations}
where $\omega_1^2\simeq\omega_2^2\oforder\unit[-1\e{-6}]{s^{-2}}$ are spring constants modeling oscillator-like force couplings between the TMs and the SC, named \textit{parasitic stiffness}. As the dominating part of such force gradients is due to electrostatics, the oscillators are unstable: that is the reason why a controller is employed. $\Gamma_x\oforder\unit[4\e{-9}]{s^{-2}}$ is the gravity gradient (per unit mass) between the TMs corresponding to a nominal separation of $\oforder\unit[38]{cm}$. All terms containing normalized masses are back-reactions that can be neglected to zeroth order.

In writing the dynamics the control in the science mode is implicitly assumed, where the SC is forced to follow a reference TM in free fall along the optical axis and the other TM is forced to follow the reference TM along the same axis. As declared by \tabref{tab:dynamics:coordinates}, the interferometric readout $o_1$ ($x_1$ coordinate) is a drag-free coordinate and is the input to the drag-free control law $C_\text{df}(s)$ assuring thruster actuation. The interferometric readout $o_{12}$ ($x_{12}=x_2-x_1$ coordinate) is an electrostatic suspension coordinate and is the input to the electrostatic suspension control law $C_\text{sus}(s)$ assuring capacitive actuation on the second TM.

The first step is to rearrange the equations so that to eliminate the unmeasurable absolute position $x_\text{SC}$ and rewrite them in terms of the two main degrees of freedom $x_1$ and $x_{12}$. In fact, by taking the difference between \eqref{eq:dynamics:eom_1D_2} and \eqref{eq:dynamics:eom_1D_1} the SC acceleration vanishes. Then, the SC acceleration in \eqref{eq:dynamics:eom_1D_3} is substituted in \eqref{eq:dynamics:eom_1D_1}. The structure of the equations suggests to define the \textit{differential forces} $f_{12} = f_2-f_1$ and the \textit{differential parasitic stiffness} $\omega_{12}^2 = \omega_2^2-\omega_1^2$. The equations can be finally condensed into the formalism of \eqref{eq:dynamics:eom_matrix_a}, where the dynamics operator has the following matrix representation
\begin{equation}\label{eq:dynamics:D_matrix_along_x}
\vect{D} =
\begin{pmatrix}
s^2+\left(1+\tilde{m}_1+\tilde{m}_2\right)\omega_1^2+\tilde{m}_2\,\omega_{12}^2 & \Gamma_x+\tilde{m}_2\left(\omega_1^2+\omega_{12}^2\right) \\
\omega_{12}^2 & s^2 + \omega_1^2 + \omega_{12}^2 - 2\,\Gamma_x
\end{pmatrix}~,
\end{equation}
that acting on the system coordinates
\begin{equation}
\vect{q} = \begin{pmatrix} x_1 \\ x_{12} \end{pmatrix}~,
\end{equation}
produces the external forces
\begin{equation}
\vect{g} =
\begin{pmatrix}
\left(1+\tilde{m}_1+\tilde{m}_2\right)f_1 + \tilde{m}_2\,f_{12} - f_\text{SC} - C_\text{df}(s)\,o_1  - \tilde{m}_2\,C_\text{sus}(s)\,o_{12} \\
f_{12} - C_\text{sus}(s)\,o_{12}
\end{pmatrix}~.
\end{equation}
The preceding contains force noise sources and injected biases. Neglecting all back-reactions it shows that the first degree of freedom $x_1$ is dominated by the thruster noise and the drag-free actuation; the second degree of freedom $x_{12}$ is dominated by the differential force noise and the capacitive actuation on the second TM. The identified control operator of \eqref{eq:dynamics:eom_matrix_b} is given by
\begin{equation}
\vect{C} =
\begin{pmatrix}
C_\text{df}(s) & \tilde{m}_2\,C_\text{sus}(s) \\
0 & C_\text{sus}(s)
\end{pmatrix}~,
\end{equation}
where the off-diagonal quantity is the back-reaction from the suspension to the drag-free loop.

The dynamical equations shown above assume a perfect actuation. This implies that $\vect{A}$ is an identity. Otherwise, actuation gains $A_\text{df}$ and $A_\text{sus}$ may be conveniently introduced to model the efficiency to which commanded forces are converted to actual applied forces by the corresponding loops.

The expression in \eqref{eq:dynamics:eom_matrix_c} gives the sensing conversion between the physical coordinates $\vect{q}$ and the interferometric readouts
\begin{equation}
\vect{o} = \begin{pmatrix} o_1 \\ o_{12} \end{pmatrix}~,
\end{equation}
being fed up into the controller. The perfect conversion is represented by an identity matrix. The imperfect conversion is due to both miscalibrations (the diagonal terms) or cross-talk contributions (the off-diagonal terms). The on-ground characterization and the theoretical modeling of the interferometer \cite{hechenblaikner2010,S2-ASD-TN-2024} suggests that the most relevant is the cross-talk from $o_1$ to $o_{12}$ that mixes the two nominally independent degrees of freedom in the following way
\begin{equation}
\vect{S} =
\begin{pmatrix}
1 & 0 \\
S_{21} & 1
\end{pmatrix}~.
\end{equation}
The cross-talk is explained by a tiny difference in the incidence angles with which light reflects on the TM surface for the two readings. The photon phase $\phi$ is built up by taking the difference between the averaged measurement of the 4 quadrants of each photodiode and the reference phase used for common-mode rejection of any residual optical path length variation. A generic displacement readout is proportional to the measured phase
\begin{equation}\label{eq:dynamics:displacement_phase}
o = \frac{\lambda}{4\pi\cos{\delta}}\phi~,
\end{equation}
where $\lambda=\unit[1.064]{\mu m}$ is the laser wavelength and $\delta\oforder\unit[4.5]{^\circ}$ is the nominal incidence angle. The sensing cross-talk due to a small mismatch in incidence angles is
\begin{equation}
S_{21} = \frac{\delta_2-\delta_1}{\delta_1}~.
\end{equation}
Therefore, a measured difference of $(\delta_2-\delta_1)\oforder\unit[5]{''}$ produces a sensing cross-talk as large as $S_{21}\oforder{3\e{-4}}$.

The model described in this section, with some slight improvements, has been extensively used for simulations and analysis -- and examples with references will be discussed in details in the next chapter -- to test the algorithms aimed at estimating the TM couplings, the sensing cross-talk and other relevant parameters needed for system calibration. Such calibration is also critical for the unbiased estimation of the total equivalent acceleration noise. The same model is planned to be employed during the identification experiments of the LPF mission.

\section{Cross-talk from degrees of freedom other than the optical axis} \label{sect:dynamics:cross-talk_matrix}

\sectref{sect:dynamics:matrix_formalism} has introduced a general formalism -- the three master equations of \eqref{eq:dynamics:eom_matrix} -- capable in describing the evolution of LPF as a physical system continuously subjected to a digital control. Subsequently, in \sectref{sect:dynamics:model_along_x} the formalism has been applied in the derivation of a model of LPF along the two main optical degrees of freedom. However, there are sources of non-idealities in this description. The knowledge of the operators $\vect{D}$, $\vect{A}$ and $\vect{S}$ might lack because of a poor calibration, misalignments, pointing inaccuracies as in \sectref{sect:metrology:fiducial_points}, force gradients and force noise along different axes. Such errors couple with the main interferometric axis and manifest themselves as a cross-talk from other degrees of freedom. Despite describing the system in its full complexity with all degrees of freedom at once, in the hypothesis of small motion, the absence of strong non-linearities assures that the contribution from other degrees of freedom is a small perturbation to the nominal dynamics along the optical axis.

For the sake of clarity, the approach used here in describing such effects is based on a first-order perturbation theory, where the zeroth-order dynamics, named the \textit{nominal} dynamics, is fully known. The readouts $\vect{o}$ can then be split into a nominal response of the system to the $x$-dynamics, say $\vect{o}_0$, and a small perturbation coming from other degrees of freedom ($y$, $z$ or some torsional angles), say $\delta\vect{o}$, therefore
\begin{equation}
\vect{o}\simeq\vect{o}_0+\delta\vect{o}~.
\end{equation}
The leading correcting terms can be embedded into the dynamics as imperfections to the operators introduced in the previous section. For example, if $\vect{D}_0$ is the nominal dynamics operator, $\delta\vect{D}$ is the relative imperfection.

To first order the dynamics is a generalization of \eqref{eq:dynamics:eom_matrix} \cite{S2-UTN-TN-3052}
\begin{subequations}\label{eq:dynamics:eom_matrix_crosstalk}
\begin{align}
(\vect{D}_0+\delta\vect{D})\,\vect{q} & = \vect{g}~, \\
\vect{g} & = \vect{f}_\text{n} + (\vect{A}_0+\delta\vect{A})\left[\vect{f}_\text{i}-\vect{C}_0\,(\vect{o}-\vect{o}_\text{i})\right]~, \\
\vect{o} & = (\vect{S}_0+\delta\vect{S})\,\vect{q}+\vect{o}_\text{n}~.
\end{align}
\end{subequations}
where the control laws in $\vect{C}_0$ are exactly known from the original design. \figref{fig:dynamics:matrix_diagram_crosstalk} shows a block diagram of the above set of equations, in where a dashed arrow stands for a cross-talk contribution. Every time a physical quantity must be passed throughout an operator, then the relative imperfection mixes it up among many degrees of freedom.


\figuremacroW{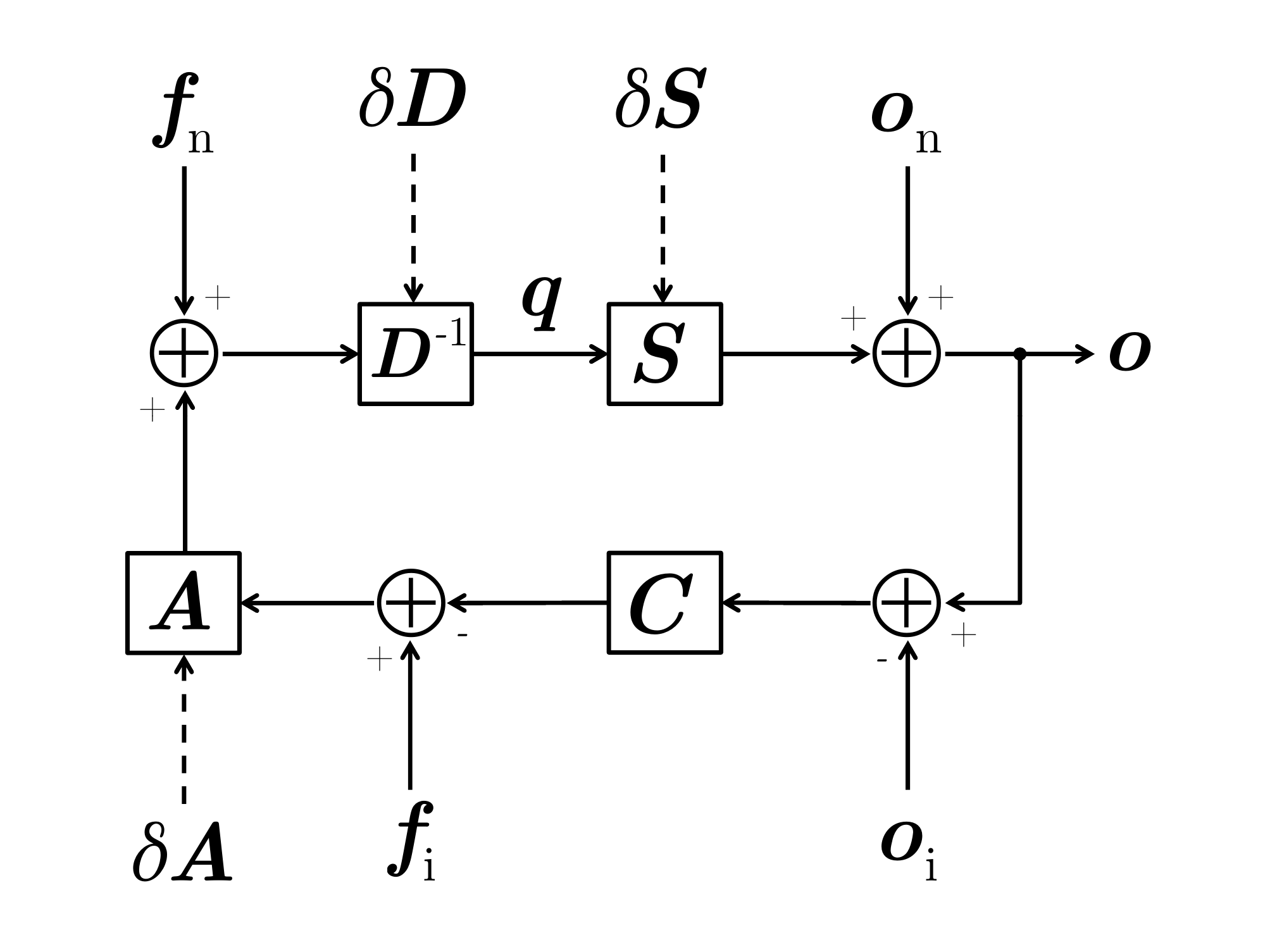}{An update of the block diagram shown in \figref{fig:dynamics:matrix_diagram} where the cross-talks (dynamics, sensing and actuation) are introduced as imperfections (dashed arrows) to the nominal operators. For example, $\vect{D}\simeq\vect{D}_0+\delta\vect{D}$.}{fig:dynamics:matrix_diagram_crosstalk}{0.7}

The nominal solution $\vect{o}_0$ is provided by \eqref{eq:dynamics:eom_full} in which all operators have the proper subscript indicating a zeroth-order term. The differential operator is obtained from the definition \eqref{eq:dynamics:diff_op} by means of simple algebra and using the inversion lemma \footnote{$\left(\vect{S}_0+\delta\vect{S}\right)^{-1} = \vect{S}_0^{-1}-\vect{S}_0^{-1}\,\delta\vect{S}\left(\vect{1}+\vect{S}_0^{-1}\,\delta\vect{S}\right)^{-1}\,\vect{S}_0^{-1} \simeq \vect{S}_0^{-1}-\vect{S}_0^{-1}\,\delta\vect{S}\,\vect{S}_0^{-1}$.}
\begin{equation}
\begin{split}
\vect{\Delta} & \simeq \left(\vect{D}_0+\delta\vect{D}\right)\left(\vect{S}_0+\delta\vect{S}\right)^{-1}+\left(\vect{A}_0+\delta\vect{A}\right)\vect{C}_0 \\
& \simeq \left(\vect{D}_0+\delta\vect{D}\right)\vect{S}_0^{-1}\left(\vect{1}-\delta\vect{S}\,\vect{S}_0^{-1}\right)+\left(\vect{A}_0+\delta\vect{A}\right)\vect{C}_0 \\
& \simeq \left(\vect{D}_0\,\vect{S}_0^{-1}+\vect{A}_0\,\vect{C}_0\right) + \left(\delta\vect{D}\,\vect{S}_0^{-1} - \vect{D}_0\,\vect{S}_0^{-1}\,\delta\vect{S}\,\vect{S}_0^{-1} + \delta\vect{A}\,\vect{C}_0\right)
\end{split}
\end{equation}
where the nominal operator and its imperfection are defined as
\begin{subequations}\label{eq:dynamics:diff_op_crosstalk}
\begin{align}
\vect{\Delta}_0 & = \vect{D}_0\,\vect{S}_0^{-1}+\vect{A}_0\,\vect{C}_0~, \label{eq:dynamics:diff_op_crosstalk_a} \\
\delta\vect{\Delta} & = \delta\vect{D}\,\vect{S}_0^{-1} - \vect{D}_0\,\vect{S}_0^{-1}\,\delta\vect{S}\,\vect{S}_0^{-1} + \delta\vect{A}\,\vect{C}_0~. \label{eq:dynamics:diff_op_crosstalk_b}
\end{align}
\end{subequations}

There are two ways to obtain the evolution of the first-order correction $\delta\vect{o}$. The first method involves a direct computation on \eqref{eq:dynamics:eom_matrix_crosstalk} similar to the reasoning of \sectref{sect:dynamics:matrix_formalism}. The idea is to combine the three equations, substitute back the zeroth-order and keep only first-order terms. The second method is more straightforward and is based on the elaboration of the equation of motion \eqref{eq:dynamics:eom_full}. Following this idea, the expansion of \eqref{eq:dynamics:eom_full} is
\begin{equation}\label{eq:dynamics:eom_full_crosstalk}
\begin{split}
\left(\vect{\Delta}_0+\delta\vect{\Delta}\right)\left(\vect{o}_0+\delta\vect{o}\right) & \simeq \vect{f}_\text{n} \\
& \quad +\left(\vect{D}_0+\delta\vect{D}\right)\left(\vect{S}_0+\delta\vect{S}\right)^{-1}\,\vect{o}_\text{n} \\
& \quad +\left(\vect{A}_0+\delta\vect{A}\right)\left(\vect{f}_\text{i}+\vect{C}_0\,\vect{o}_\text{i}\right)~,
\end{split}
\end{equation}
Now, to first order the equation becomes
\begin{equation}
\begin{split}
\vect{\Delta}_0\,\vect{o}_0+\vect{\Delta}_0\,\delta\vect{o}+\delta\vect{\Delta}\,\vect{o}_0 & \simeq \vect{f}_\text{n} \\
& \quad +\left(\vect{D}_0\,\vect{S}_0^{-1}+\delta\vect{D}_0\,\vect{S}_0^{-1}-\vect{D}_0\,\vect{S}_0^{-1}\,\delta\vect{S}\,\vect{S}_0^{-1}\right)\vect{o}_\text{n} \\
& \quad +\left(\vect{A}_0+\delta\vect{A}\right)\left(\vect{f}_\text{i}+\vect{C}_0\,\vect{o}_\text{i}\right)~,
\end{split}
\end{equation}
which is further simplified by assuming the knowledge of the nominal dynamics. Indeed, $\vect{\Delta}_0\,\vect{o}_0$ on the left-hand side is canceled out by $\vect{f}_\text{n}+\vect{D}_0\,\vect{S}_0^{-1}\,\vect{o}_\text{n}+\vect{A}_0\left(\vect{f}_\text{i}+\vect{C}_0\,\vect{o}_\text{i}\right)$ on the right-hand side. Therefore, the first-order dynamics is given by
\begin{equation}
\begin{split}
\vect{\Delta}_0\,\delta\vect{o} & \simeq -\delta\vect{\Delta}\,\vect{o}_0 + \left(\delta\vect{D}\,\vect{S}_0^{-1}-\vect{D}_0\,\vect{S}_0^{-1}\,\delta\vect{S}\,\vect{S}_0^{-1}\right)\vect{o}_\text{n} \\
& \quad +\delta\vect{A}\left(\vect{f}_\text{i}+\vect{C}_0\,\vect{o}_\text{i}\right)~,
\end{split}
\end{equation}
where the operator in front of $\vect{o}_\text{n}$ is exactly $\delta\vect{\Delta}-\delta\vect{A}\,\vect{C}_0$ (it can be checked by comparing to \eqref{eq:dynamics:diff_op_crosstalk_b}). The equation also shows that the first-order dynamics is given in terms of the zeroth-order $\vect{o}_0$. To write down an explicit form of the above formula, it is necessary to substitute the expression of $\vect{o}_0$ in \eqref{eq:dynamics:eom_full} and get
\begin{equation}
\begin{split}
\vect{\Delta}_0\,\delta\vect{o} & \simeq -\delta\vect{\Delta}\,\vect{\Delta}_0^{-1}\left[\vect{f}_\text{n}+\vect{D}_0\,\vect{S}_0^{-1}\,\vect{o}_\text{n}+\vect{A}_0\left(\vect{f}_\text{i}+\vect{C}_0\,\vect{o}_\text{i}\right)\right] \\
& \quad + \left(\delta\vect{\Delta}-\delta\vect{A}\,\vect{C}\right)\vect{o}_\text{n} + \delta\vect{A}\left(\vect{f}_\text{i}+\vect{C}_0\,\vect{o}_\text{i}\right) \\
& = -\delta\vect{\Delta}\,\vect{\Delta}_0^{-1}\,\vect{f}_\text{n} + \left(\delta\vect{\Delta}-\delta\vect{A}\,\vect{C}_0-\delta\vect{\Delta}\,\vect{\Delta}_0^{-1}\,\vect{D}_0\,\vect{S}_0^{-1}\right)\vect{o}_\text{n} \\
& \quad + \left(\delta\vect{A}-\delta\vect{\Delta}\,\vect{\Delta}_0^{-1}\,\vect{A}_0\right)\left(\vect{f}_\text{i} + \vect{C}_0\,\vect{o}_\text{i}\right)
~.
\end{split}
\end{equation}
The operator in front of $\vect{f}_\text{n}$ is the force noise cross-talk, whose transfer to total equivalent acceleration is denoted with
\begin{equation}
\delta\vect{T}_{f_\text{n}\rightarrow f} = -\delta\vect{\Delta}\,\vect{\Delta}_0^{-1}~;
\end{equation}
the operator in front of $\vect{o}_\text{n}$ is the readout noise cross-talk and can be rewritten as
\begin{equation}
\begin{split}
\delta\vect{T}_{o_\text{n}\rightarrow f} & = \delta\vect{\Delta}-\delta\vect{A}\,\vect{C}_0-\delta\vect{\Delta}\,\vect{\Delta}_0^{-1}\,\vect{D}_0\,\vect{S}_0^{-1} \\
& = \delta\vect{\Delta}-\delta\vect{A}\,\vect{C}_0-\delta\vect{\Delta}\,\vect{\Delta}_0^{-1}\left(\vect{\Delta}_0-\vect{A}_0\,\vect{C}_0\right) \\
& = \delta\vect{\Delta}\,\vect{\Delta}_0^{-1}\,\vect{A}_0\,\vect{C}_0-\delta\vect{A}\,\vect{C}_0 \\
& = -\left(\delta\vect{A}+\delta\vect{T}_{f_\text{n}\rightarrow f}\,\vect{A}_0\right)\vect{C}_0~;
\end{split}
\end{equation}
the operator in front of $\vect{f}_\text{i}$ is the force actuation cross-talk
\begin{equation}
\delta\vect{T}_{f_\text{i}\rightarrow f} = \delta\vect{A}+\delta\vect{T}_{f_\text{n}\rightarrow f}\,\vect{A}_0~;
\end{equation}
and the operator in front of $\vect{o}_\text{i}$ is the bias actuation cross-talk
\begin{equation}
\delta\vect{T}_{o_\text{i}\rightarrow f} = \left(\delta\vect{A}+\delta\vect{T}_{f_\text{n}\rightarrow f}\,\vect{A}_0\right)\vect{C}_0~,
\end{equation}
that is exactly the readout noise cross-talk modulo a sign. As reasonable to expect, the symmetry between the various cross-talk terms comes from the linearization of the problem.

The equation of motion for the first-order cross-talk is finally given by
\begin{equation}\label{eq:dynamics:eom_full_crosstalk_final}
\vect{\Delta}_0\,\delta\vect{o} \simeq \delta\vect{T}_{f_\text{n}\rightarrow f}\,\vect{f}_\text{n} +
\delta\vect{T}_{o_\text{n}\rightarrow f}\,\vect{o}_\text{n} +
\delta\vect{T}_{f_\text{i}\rightarrow f}\,\vect{f}_\text{i} +
\delta\vect{T}_{o_\text{i}\rightarrow f}\,\vect{o}_\text{i}~,
\end{equation}
that enlightens the various contributions to the overall cross-talk from other degrees of freedom. On one side, in a pure noise measurement during the LPF mission (no application of forces or biases) a noise cross-talk sums up to the nominal dynamics along the main degrees of freedom. On the other side, whatever a bias is applied to the system, a perturbation is produced along the sensitive axis. As it is clear, a non-negligible cross-talk at any level of the system (dynamics, sensing and actuation) actually breaks the nominal orthogonality between different degrees of freedom.

\section{Dynamical model for $xy$ cross-talk} \label{sect:dynamics:model_xy_cross-talk}

The application of the idea of the preceding section is a further development of the model along the optical axis described in \sectref{sect:dynamics:model_along_x}, i.e., the cross-talk from other nominally orthogonal degrees of freedom to the optical axis. The example discussed in this section is the cross-talk from the $y$ degree of freedom to $x$.

Referring to \figref{fig:dynamics:coordinates} for the usual coordinate naming convention, the $xy$ cross-talk can be viewed as a first-order perturbation of the dynamics in the $xy$ plane to the zeroth-order dynamics along $x$. Obviously, the dynamics in $xy$ contains the rotation $\phi$ about the $z$ axis.

The control design of the main science mode for this simplified model requires a minimal number of drag-free, electrostatic suspension and attitude coordinates as inputs to the same control loops. \tabref{tab:dynamics:coordinates_xy_cross-talk} presents a list of those coordinates relevant for the $xy$ cross-talk as taken from \tabref{tab:dynamics:coordinates}.
\begin{table}[htb]
\caption{\label{tab:dynamics:coordinates_xy_cross-talk}\footnotesize{
List of the 7 controlled degrees of freedom for the $xy$ cross-talk of the LPF mission in the main science mode. Refer to \tabref{tab:dynamics:coordinates} for a comprehensive description of all coordinates. Notice that $o_{\phi_2}=o_{\phi_1}+o_{\phi_{12}}$ is used in the equations for clearness and simplicity.}}
\centering
\begin{tabular}{@{\hspace{10pt}}c@{\hspace{10pt}} @{\hspace{10pt}}c@{\hspace{10pt}} D{=}{\,=\,}{4.8} @{\hspace{10pt}}c@{\hspace{10pt}}}
\hline
\hline
Coordinate & Control & \multicolumn{1}{c}{Sensor} & Actuator \\
\hline
$x_1$ & Drag-free & o_1=\text{IFO}[x_1] & FEEP \\
$y_1$ & Drag-free & o_{y_1}=\text{GRS}[y_1] & FEEP \\
$\phi_1$ & Elect.\,suspension & o_{\phi_1}=\text{IFO}[\phi_1] & GRS \\
\hline
$x_2$ & Elect.\,suspension & o_{12}=\text{IFO}[x_{12}] & GRS \\
$y_2$ & Drag-free & o_{y_2}=\text{GRS}[y_2] & FEEP \\
$\phi_2$ & Elect.\,suspension & o_{\phi_{12}}=\text{IFO}[\phi_{12}] & GRS \\
\hline
$\phi_\text{SC}$ & Attitude & o_{\phi_\text{SC}}=\text{ST}[\phi_\text{SC}] & GRS \\
\hline
\hline
\end{tabular}
\end{table}

The control is such that:
\begin{enumerate}
  \item along $x$: guided by the optical $x_1$, the SC is forced to follow the reference TM through thruster actuation; guided by the optical $x_{12}$, the second TM is forced to follow the reference TM through capacitive actuation;
  \item along $y$: guided by the capacitive $(y_1+y_2)/2$, the SC is forced to follow both TMs through thruster actuation; guided by the star-tracker $\phi_\text{SC}$, the TMs are oriented along $\phi$;
  \item along $\phi$: guided by the capacitive $(y_1-y_2)/2$, the SC is forced to follow both TMs through thruster actuation; guided by the optical $\phi_1$ and $\phi_2$, both TMs are oriented along $\phi$ through capacitive actuation.
\end{enumerate}

%

As already pointed out, the cross-talk can be described by a first-order perturbation of the nominal dynamics along $x$. Three different types of cross-talks can be identified:
\begin{enumerate}
  \item dynamical cross-talk;
  \item actuation cross-talk;
  \item sensing cross-talk.
\end{enumerate}
All equations must be written to within linear terms of an imperfection or noise contribution.

Concerning the dynamics along the $x$ axis, in the approximation of small motion, the stiffness constant has been introduced to model residual oscillator-like couplings between the TMs and the SC. The generalization in three dimensions is straightforward. Since there are electrodes all around the TMs and the most important coupling is indeed due to electrostatics, in place of a single oscillator, 6 coupled harmonic oscillators along the translational and rotational degrees of freedom must be considered. Therefore, the stiffness constant becomes a $6\times6$ quasi-diagonal matrix, the \textit{stiffness matrix}. For the $xy$ cross-talk, since it makes sense to inspect the cross-stiffness from $y$ to $x$ or $\phi$ to $x$, the structure of the matrix for the first TM is
\begin{equation}
\vect{\kappa}_1=
\begin{pmatrix}
m_1\,\omega_{1,x}^2 & \delta_{1,xy}\,m_1\,\omega_{1,y}^2 & \delta_{1,x\phi}\,I_{1,z}\,\omega_{1,\phi}^2 \\
0 & m_1\,\omega_{1,y}^2 & 0 \\
0 & 0 & I_{1,z}\,\omega_{1,\phi}^2
\end{pmatrix}~,
\end{equation}
where $m_1=\unit[1.96]{kg}$ is the TM mass, $I_{1,z}=\nicefrac{1}{6}\,m_1\,l^2\oforder\unit[7\e{-4}]{kg\,m^2}$ is the inertia matrix about $z$ and $l=\unit[46]{mm}$ is the TM size. A $\delta$-coefficient, with the number of the TM and the names of two coordinates as subscripts, denotes a dynamical cross-talk imperfection typically $\oforder1\e{-3}$, an assumption based on on-ground measurements and theoretical models of the GRS.

The second type is the actuation cross-talk due to misalignments in the setup of the thrusters and electrodes. The idea is that every time a force is actuated along a nominally orthogonal degree of freedom it couples with $x$. A $\delta$-coefficient, with the names of the actuation and the coordinate, denotes an actuation cross-talk imperfection.

Finally, the third type is the sensing cross-talk due to miscalibrations and misalignments in the sensors. For the capacitive sensing, misalignments in the electrodes produces a mixing in the sensed coordinates. For the optical sensing, misalignments in the optical elements produce an analogous result. As in \sectref{sect:dynamics:model_along_x}, the 7 physical coordinates
\begin{equation}
\vect{q} =
\scalebox{.8}{$
\begin{pmatrix}
x_1 \\ x_{12} \\ y_1 \\ y_2 \\ \phi_1 \\ \phi_2 \\ \phi_\text{SC}
\end{pmatrix}$}~,
\end{equation}
are converted into the sensed coordinates
\begin{equation}
\vect{o} =
\scalebox{.8}{$
\begin{pmatrix}
o_1 \\ o_{12} \\ o_{y_1} \\ o_{y_2} \\ o_{\phi_1} \\ o_{\phi_2} \\ o_{\phi_\text{SC}}
\end{pmatrix}$}~,
\end{equation}
with an inevitable mixing. Assuming a sensing operator which is nominally identity, the relative imperfection operator has the following matrix representation
\begin{equation}
\delta\vect{S} =
\left(\begin{smallmatrix}
0 & 0 & \delta_{S,1y_1} & \delta_{S,1y_2} & \frac{l}{2}\,\delta_{S,1\phi_1} & \frac{l}{2}\,\delta_{S,1\phi_2} & \frac{L}{2}\,\delta_{S,1\phi_\text{SC}} \\
S_{21} & 0 & \delta_{S,2y_1} & \delta_{S,2y_2} & \frac{l}{2}\,\delta_{S,2\phi_1} & \frac{l}{2}\,\delta_{S,2\phi_2} & \frac{L}{2}\,\delta_{S,2\phi_\text{SC}} \\
0 & 0 & 0 & 0 & 0 & 0 & 0 \\
0 & 0 & 0 & 0 & 0 & 0 & 0 \\
0 & 0 & 0 & 0 & 0 & 0 & 0 \\
0 & 0 & 0 & 0 & 0 & 0 & 0 \\
0 & 0 & 0 & 0 & 0 & 0 & 0
\end{smallmatrix}\right)~,
\end{equation}
where $l=\unit[46]{mm}$ is the TM size and $L=\unit[38]{cm}$ is the nominal separation between the TM centers of mass. $S_{21}$ is the sensing cross-talk between $o_1$ and $o_{12}$, already discussed, and a generic $\delta$-coefficient denotes a sensing cross-talk imperfection from an orthogonal degree of freedom to $x$. Since the target of the investigation is the cross-talk to the two optical degrees of freedom, no cross-talk to other coordinates is considered as this is second order.

In the derivation of the equations of motion the gravity/torque gradients between the TMs can be neglected without loss of generality, as well as the back-reactions, since they are all second order effects. The first set of equations is along the $x$ degree of freedom. With the same considerations of \sectref{sect:dynamics:model_along_x}, i.e., assuming small motion, small forces and null initial conditions, the linear equations of motion, per unit mass, in frequency domain are
\begin{subequations}\label{eq:dynamics:eom_XY_along x}
\begin{align}
\begin{split}
s^2\,x_1+s^2\,x_\text{SC} + \omega_{1,x}^2\,x_1 \quad & \\
\quad +\,\delta_{1,xy}\,\omega_{1,y}^2\,y_1 + \delta_{1,x\phi}\,\frac{l}{2}\,\omega_{1,\phi}^2\,\phi_1 & = f_{1,x} \\
& \quad + \delta_{\text{sus},\,y_1}\,\frac{L}{2}\,C_{\text{sus},\,\phi_\text{SC}}(s)\,o_{\phi_\text{SC}} \\
& \quad + \delta_{\text{sus},\,\phi_1}\,\frac{l}{2}\,C_{\text{sus},\,\phi_1}(s)\,o_{\phi_1}~,
\end{split} \\
\begin{split}
s^2\,x_2+s^2\,x_\text{SC} + \omega_{2,x}^2\,x_2 \quad & \\
\quad +\,\delta_{2,xy}\,\omega_{2,y}^2\,y_2 + \delta_{2,x\phi}\,\frac{l}{2}\,\omega_{2,\phi}^2\,\phi_2 & = f_{2,x} - C_{\text{sus},\,x}(s)\,o_{12} \\
& \quad - \delta_{\text{sus},\,y_2}\,\frac{L}{2}\,C_{\text{sus},\,\phi_\text{SC}}(s)\,o_{\phi_\text{SC}} \\
& \quad + \delta_{\text{sus},\,\phi_2}\,\frac{l}{2}\,C_{\text{sus},\,\phi_2}(s)\,o_{\phi_2}~,
\end{split} \\
\begin{split}
s^2\,x_\text{SC} & = f_{\text{SC},x} + C_{\text{df},\,x}(s)\,o_1 \\
& \quad + \delta_{\text{df},\,y_\text{SC}}\,\frac{1}{2}\left[C_{\text{df},\,y_1}(s)\,o_{y_1}+C_{\text{df},\,y_2}(s)\,o_{y_2}\right] \\
& \quad - \delta_{\text{df},\,\phi_{\text{SC}}}\,\frac{1}{2}\left[C_{\text{df},\,y_1}(s)\,o_{y_1}-C_{\text{df},\,y_2}(s)\,o_{y_2}\right]~,
\end{split}
\end{align}
\end{subequations}
The second row of the left-hand side of the equations for the TMs contain the dynamical cross-talk due to the stiffness matrix (imperfections $\delta_{1,xy}$, $\delta_{1,x\phi}$, $\delta_{2,xy}$ and $\delta_{2,x\phi}$). Instead, the last two rows of the right-hand side of the equations are actuation cross-talk. Notice that for the TMs there is a cross-talk from the SC inertial attitude control ($\delta_{\text{sus},\,y}$) and one from the TM attitude control ($\delta_{\text{sus},\,\phi}$). For the SC there is a cross-talk from the $y$ and $\phi$ drag-free actuation ($\delta_{\text{df},\,y_\text{SC}}$ and $\delta_{\text{df},\,\phi_{\text{SC}}}$). Other relevant quantities have been already defined in the previous section and are not further discussed here.

The second set of equations describes the dynamics along the nominally orthogonal $y$ axis
\begin{subequations}\label{eq:dynamics:eom_XY_along y}
\begin{align}
s^2\,y_1+s^2\,y_\text{SC} + \omega_{1,y}^2\,y_1 - \frac{L}{2}\,s^2\,\phi_\text{SC} & =
f_{1,y} + \frac{L}{2}\,C_{\text{sus},\,\phi_\text{SC}}(s)\,o_{\phi_\text{SC}}~, \\
s^2\,y_2+s^2\,y_\text{SC} + \omega_{2,y}^2\,y_2 + \frac{L}{2}\,s^2\,\phi_\text{SC} & =
f_{2,y} - \frac{L}{2}\,C_{\text{sus},\,\phi_\text{SC}}(s)\,o_{\phi_\text{SC}}~, \\
s^2\,y_\text{SC} & =
f_{\text{SC},y} + \frac{1}{2}\left[C_{\text{df},\,y_1}(s)\,o_{y_1} + C_{\text{df},\,y_2}(s)\,o_{y_2}\right]~.
\end{align}
\end{subequations}
The inertial attitude control is implemented as electrostatic suspension actuation on the TMs along $y$ (notice the opposite signs) through the $C_{\text{sus},\,\phi_\text{SC}}(s)$ control law. $C_{\text{df},\,y_1}(s)$ and $C_{\text{df},\,y_2}(s)$ are the two drag-free control laws along $y$. On the left-end side of the equations the SC absolute angular acceleration $s^2\,\phi_\text{SC}$ also appears as a strict consequence of the coupling between the $\phi$ motion with $y$.

The third set of equations describes the dynamics along the nominally orthogonal $\phi$ angle
\begin{subequations}\label{eq:dynamics:eom_XY_along phi}
\begin{align}
s^2\,\phi_1+s^2\,\phi_\text{SC} + \omega_{1,\phi}^2\,\phi_1 & =
\tau_{1,z} - C_{\text{sus},\,\phi_1}(s)\,o_{\phi_1}~, \\
s^2\,\phi_2+s^2\,\phi_\text{SC} + \omega_{2,\phi}^2\,\phi_2 & =
\tau_{2,z} - C_{\text{sus},\,\phi_2}(s)\,o_{\phi_2}~, \\
s^2\,\phi_\text{SC} & =
\tau_{\text{SC},z} + \frac{1}{L}\left[C_{\text{df},\,y_2}(s)\,o_{y_2} - C_{\text{df},\,y_1}(s)\,o_{y_1}\right]~,
\end{align}
\end{subequations}
where $\tau$ denotes a generic component of torque per unit of inertia $I_{\text{TM},z}\oforder\unit[7\e{-4}]{kg\,m^2}$ and $I_{\text{SC},z}\oforder\unit[1\e{3}]{kg\,m^2}$ \cite{S2-ASD-ICD-2011}. On the left-end side, the angular stiffness constants are evident. On the right-end side, there are the $C_{\text{sus},\,\phi}(s)$ electrostatic suspension control law and the $C_{\text{df},\,y}(s)$ drag-free control law. The SC attitude is controlled by actuating along the differential TM positions along $y$.

As said, along $y$ the SC follows the average $y$ motion of the TMs sensed with the capacitive $o_{y_1}$ and $o_{y_2}$; the TMs are oriented following the star-tracker $o_{\phi_\text{SC}}$. Analogously, along $\phi$ the SC is oriented following the differential $y$ motion of the TMs sensed with the capacitive $o_{y_1}$ and $o_{y_2}$; the TMs are oriented following the optical $o_{\phi_1}$ and $o_{\phi_2}$.

The SC absolute linear acceleration is not measurable, whereas the SC absolute attitude is measured by the ST with respect to the celestial inertial frame. Therefore, the system of 9 equations turns into 7 equations because the SC acceleration along $x$ and $y$ must be canceled out. By doing so in \eqref{eq:dynamics:eom_XY_along x} and defining the differential TM displacement, the dynamics along $x$ becomes
\begin{subequations}
\begin{align}
\begin{split}
s^2\,x_1 + \omega_{1,x}^2\,x_1 \quad & \\
\quad +\,\delta_{1,xy}\,\omega_{1,y}^2\,y_1 + \delta_{1,x\phi}\,\frac{l}{2}\,\omega_{1,\phi}^2\,\phi_1 & = f_{1,x} - f_{\text{SC},x} - C_{\text{df},\,x}(s)\,o_1 \\
& \quad + \delta_{\text{sus},\,y_1}\,\frac{L}{2}\,C_{\text{sus},\,\phi_\text{SC}}(s)\,o_{\phi_\text{SC}} \\
& \quad + \delta_{\text{sus},\,\phi_1}\,\frac{l}{2}\,C_{\text{sus},\,\phi_1}(s)\,o_{\phi_1} \\
& \quad - \delta_{\text{df},\,y_\text{SC}}\,\frac{1}{2}\left[C_{\text{df},\,y_1}(s)\,o_{y_1}+C_{\text{df},\,y_2}(s)\,o_{y_2}\right] \\
& \quad + \delta_{\text{df},\,\phi_{\text{SC}}}\,\frac{1}{2}\left[C_{\text{df},\,y_1}(s)\,o_{y_1}-C_{\text{df},\,y_2}(s)\,o_{y_2}\right]~,
\end{split} \\
\begin{split}
s^2\,x_{12} + \omega_{2,x}^2\,x_2 - \omega_{1,x}^2\,x_1 \quad & \\
\quad +\,\delta_{2,xy}\,\omega_{2,y}^2\,y_2 + \delta_{2,x\phi}\,\frac{l}{2}\,\omega_{2,\phi}^2\,\phi_2 \\
\quad -\,\delta_{1,xy}\,\omega_{1,y}^2\,y_1 - \delta_{1,x\phi}\,\frac{l}{2}\,\omega_{1,\phi}^2\,\phi_1 & = f_{2,x} - C_{\text{sus},\,x}(s)\,o_{12} \\
& \quad - \delta_{\text{sus},\,y_2}\,\frac{L}{2}\,C_{\text{sus},\,\phi_\text{SC}}(s)\,o_{\phi_\text{SC}} \\
& \quad + \delta_{\text{sus},\,\phi_2}\,\frac{l}{2}\,C_{\text{sus},\,\phi_2}(s)\,o_{\phi_2} \\
& \quad - \delta_{\text{sus},\,y_1}\,\frac{L}{2}\,C_{\text{sus},\,\phi_\text{SC}}(s)\,o_{\phi_\text{SC}} \\
& \quad - \delta_{\text{sus},\,\phi_1}\,\frac{l}{2}\,C_{\text{sus},\,\phi_1}(s)\,o_{\phi_1}~.
\end{split}
\end{align}
\end{subequations}
Analogously, substituting the SC acceleration along $y$ and $\phi$ into the dynamics along $y$
\begin{subequations}
\begin{align}
\begin{split}
s^2\,y_1 + \omega_{1,y}^2\,y_1 & =
f_{1,y} - f_{\text{SC},y} + \frac{L}{2}\tau_{\text{SC},z} + \frac{L}{2}\,C_{\text{sus},\,\phi_\text{SC}}(s)\,o_{\phi_\text{SC}} \\
& \quad - C_{\text{df},\,y_1}(s)\,o_{y_1}~,
\end{split} \\
\begin{split}
s^2\,y_2 + \omega_{2,y}^2\,y_2 & =
f_{2,y} - f_{\text{SC},y} - \frac{L}{2}\tau_{\text{SC},z} - \frac{L}{2}\,C_{\text{sus},\,\phi_\text{SC}}(s)\,o_{\phi_\text{SC}} \\
& \quad - C_{\text{df},\,y_2}(s)\,o_{y_2}~.
\end{split}
\end{align}
\end{subequations}
$y_1$ and $y_2$ are named drag-free coordinates as they guide a drag-free actuation; $\phi_\text{SC}$ is an electrostatic suspension coordinate as it guides a capacitive actuation. At the same time the dynamics along $\phi$ is
\begin{subequations}
\begin{align}
\begin{split}
s^2\,\phi_1 + \omega_{1,\phi}^2\,\phi_1 & =
\tau_{1,z} - \tau_{\text{SC},z} - C_{\text{sus},\,\phi_1}(s)\,o_{\phi_1} \\
& \quad - \frac{1}{L}\left[C_{\text{df},\,y_2}(s)\,o_{y_2} - C_{\text{df},\,y_1}(s)\,o_{y_1}\right]~,
\end{split} \\
\begin{split}
s^2\,\phi_2 + \omega_{2,\phi}^2\,\phi_2 & =
\tau_{2,z} - \tau_{\text{SC},z} - C_{\text{sus},\,\phi_2}(s)\,o_{\phi_2} \\
& \quad - \frac{1}{L}\left[C_{\text{df},\,y_2}(s)\,o_{y_2} - C_{\text{df},\,y_1}(s)\,o_{y_1}\right]~.
\end{split}
\end{align}
\end{subequations}
and the one for $\phi_\text{SC}$ which remains unchanged. $\phi_1$ and $\phi_2$ are electrostatic suspension coordinates as they guide a capacitive actuation.

The equations of motion presented above can now be mapped to the formalism of the previous section. Therefore, the $7\times7$ nominal dynamics operator has the following matrix representation
\begin{equation}
\vect{D}_0 =
\scalebox{.8}{$
\begin{pmatrix}
s^2+\omega_{1,x}^2 & 0 & 0 & 0 & 0 & 0 & 0 \\
\omega_{12,x}^2 & s^2 + \omega_{1,x}^2 + \omega_{12,x}^2 & 0 & 0 & 0 & 0 & 0 \\
0 & 0 & s^2 + \omega_{1,y}^2 & 0 & 0 & 0 & 0 \\
0 & 0 & 0 & s^2 + \omega_{2,y}^2 & 0 & 0 & 0 \\
0 & 0 & 0 & 0 & s^2 + \omega_{1,\phi}^2 & 0 & 0 \\
0 & 0 & 0 & 0 & 0 & s^2 + \omega_{2,\phi}^2 & 0 \\
0 & 0 & 0 & 0 & 0 & 0 & s^2
\end{pmatrix}$}~,
\end{equation}
which is a natural generalization of the same operator \eqref{eq:dynamics:D_matrix_along_x} written for the model along $x$. The first-order perturbation due to the dynamical cross-talk is given by
\begin{equation}
\delta\vect{D} =
\scalebox{.8}{$
\begin{pmatrix}
0 & 0 & \delta_{1,xy}\,\omega_{1,y}^2 & 0 & \delta_{1,x\phi}\,\frac{l}{2}\,\omega_{1,\phi}^2 & 0 & 0 \\
0 & 0 & -\delta_{1,xy}\,\omega_{1,y}^2 & \delta_{2,xy}\,\omega_{2,y}^2 & -\delta_{1,x\phi}\,\frac{l}{2}\,\omega_{1,\phi}^2 & \delta_{2,x\phi}\,\frac{l}{2}\,\omega_{2,\phi}^2 & 0 \\
0 & 0 & 0 & 0 & 0 & 0 & 0 \\
0 & 0 & 0 & 0 & 0 & 0 & 0 \\
0 & 0 & 0 & 0 & 0 & 0 & 0 \\
0 & 0 & 0 & 0 & 0 & 0 & 0 \\
0 & 0 & 0 & 0 & 0 & 0 & 0
\end{pmatrix}$}~.
\end{equation}
Analogously, the control operator can be identified as
\begin{equation}
\vect{C}_0 =
\scalebox{.8}{$
\begin{pmatrix}
C_{\text{df},\,x}(s) & 0 & 0 & 0 & 0 & 0 & 0 \\
0 & C_{\text{sus},\,x}(s) & 0 & 0 & 0 & 0 & 0 \\
0 & 0 & C_{\text{df},\,y_1}(s) & 0 & 0 & 0 & -\frac{L}{2}\,C_{\text{sus},\,\phi_\text{SC}}(s) \\
0 & 0 & 0 & C_{\text{df},\,y_2}(s) & 0 & 0 & \frac{L}{2}\,C_{\text{sus},\,\phi_\text{SC}}(s) \\
0 & 0 & -\frac{1}{L}\,C_{\text{df},\,y_1}(s) & \frac{1}{L}\,C_{\text{df},\,y_2}(s) & C_{\text{sus},\,\phi_1}(s) & 0 & 0 \\
0 & 0 & -\frac{1}{L}\,C_{\text{df},\,y_1}(s) & \frac{1}{L}\,C_{\text{df},\,y_2}(s) & 0 & C_{\text{sus},\,\phi_2}(s) & 0 \\
0 & 0 & -\frac{1}{L}\,C_{\text{df},\,y_1}(s) & \frac{1}{L}\,C_{\text{df},\,y_2}(s) & 0 & 0 & 0
\end{pmatrix}$}~,
\end{equation}
where possible actuation gains can be intended as multiplicative factor of each single control law. The first-order perturbation due to the control actuation cross-talk is given by
\begin{equation}
\delta\vect{A}\,\vect{C}_0 = \frac{1}{2}
\scalebox{.7}{$
\begin{pmatrix}
0 & 0 & \delta_\text{df}^-\,C_{\text{df},\,y_1}(s) & \delta_\text{df}^+\,C_{\text{df},\,y_2}(s) & -\delta_{\text{sus},\,\phi_1}\,l\,C_{\text{sus},\,\phi_1}(s) & 0 & -\delta_{\text{sus},\,y_1}\,L\,C_{\text{sus},\,\phi_\text{SC}}(s) \\
0 & 0 & 0 & 0 & \delta_{\text{sus},\,\phi_1}\,l\,C_{\text{sus},\,\phi_1}(s) & -\delta_{\text{sus},\,\phi_2}\,l\,C_{\text{sus},\,\phi_2}(s) & \delta_\text{sus}^+\,L\,C_{\text{sus},\,\phi_\text{SC}}(s) \\
0 & 0 & 0 & 0 & 0 & 0 & 0 \\
0 & 0 & 0 & 0 & 0 & 0 & 0 \\
0 & 0 & 0 & 0 & 0 & 0 & 0 \\
0 & 0 & 0 & 0 & 0 & 0 & 0 \\
0 & 0 & 0 & 0 & 0 & 0 & 0
\end{pmatrix}$}~,
\end{equation}
where $\delta_\text{df}^- = \left(\delta_{\text{df},\,y_\text{SC}}-\delta_{\text{df},\,\phi_{\text{SC}}}\right)$, $\delta_\text{df}^+ = \left(\delta_{\text{df},\,y_\text{SC}} + \delta_{\text{df},\,\phi_{\text{SC}}}\right)$ and $\delta_\text{sus}^+ = \delta_{\text{sus},\,y_1}+\delta_{\text{sus},\,y_2}$ are three new definitions of effective cross-talk coefficients.

The imperfection matrices $\delta\vect{D}$, $\delta\vect{A}\,\vect{C}_0$ and $\delta\vect{S}$, together with the nominal matrices $\vect{D}_0$, $\vect{C}_0$ and $\vect{S}_0=\vect{1}$, allows for a simplification of the nominal differential operator and its imperfection in \eqref{eq:dynamics:diff_op_crosstalk}
\begin{subequations}
\begin{align}
\vect{\Delta}_0 & = \vect{D}_0+\vect{C}_0~, \\
\delta\vect{\Delta} & = \delta\vect{D} - \vect{D}_0\,\delta\vect{S} + \delta\vect{A}\,\vect{C}_0~.
\end{align}
\end{subequations}
The second equation explains the fact that the imperfection to the differential operator converting sensed coordinates into total equivalent acceleration is given by three terms: the dynamical cross-talk, the sensing cross-talk and the control actuation cross-talk. The application of the first matrix to the first-order correction to the sensed coordinates finally gives the various cross-talk contributions in the equation of motion \eqref{eq:dynamics:eom_full_crosstalk_final}. Inverting $\vect{\Delta}_0$, and applying it to the various cross-talk contributions of \eqref{eq:dynamics:eom_full_crosstalk_final}, it allows for a modeling of the response of the system along the optical axis to noise sources affecting the nominally orthogonal degrees of freedom.





\ChangeFigFolder{4_sys_identification}


\chapter{System identification} \label{chap:sys_identification}


This chapter focuses on a topic that can be considered the core of the whole LPF mission in view of characterizing the total equivalent acceleration noise affecting each single LISA arm. In system identification LPF is modeled as a matrix of parametric transfer functions. Targeted experiments where the system is stimulated on each degree of freedom can be used to infer the values of the critical parameters contained in those functions.

The preceding chapter described the closed-loop dynamics underlying LPF, the methods to handle and subtract the applied control forces, the sensing and the dynamical couplings between the TMs and the SC, the extent to which system transients can be suppressed and the estimation of the equivalent out-of-loop acceleration noise can be made possible.

This chapter shows an application of the ideas in a mission-like fashion with numerical applications. It assumes a model for LPF along the optical axis, which gives the description of the dynamics to first approximation. The aim is to simulate and analyze the data as they will be released during the mission. To simplify the discussion, only two experiments are considered, allowing for a complete identification of the system along the optical axis. As the methods developed in this chapter are general, they can also be applied to the study of more sophisticated experiments. Examples are the cross-talk experiments from orthogonal degrees of freedom to the optical axis: the modeled transfer functions are different, the dimensionality of the system is different, but the approach is the same. In the end, all experiments analyzed with the methods described in this chapter will hopefully provide a coherent understanding of the system, contributing to the final success of the LPF mission.

In turn, this chapter discusses: the dynamical model assumed for simulations and analysis; the noise characterization of the system; the simulated identification experiments; the parameter estimation method, the validation and the robustness to non-standard scenarios. Finally, it demonstrates the impact of system identification on the estimation of the residual equivalent acceleration noise and the suppression of transients in data produced by a simulator provided by ESA.

\section{Dynamical model} \label{sect:sys_identification:dynamical_model}

\sectref{sect:dynamics:model_along_x} provided a model of LPF along $x$, the optical axis. In the main science mode, the reference TM is in free fall along $x$. The other TM and the SC are, respectively, forced by capacitive and thruster actuation to follow the reference TM along the same axis. The interferometer keeps track of the relative motion between the reference TM and the SC ($o_1$) and between the two TMs ($o_{12}$). The two readouts expressed in displacement are fed into the DFACS to command force actuation, hence minimizing the relative motion.

\figref{fig:sys_identification:Cdf} and \figref{fig:sys_identification:Csus} show the frequency dependence of the two control laws converting the sensed displacements to commanded forces to the thrusters (drag-free loop) and the electrostatic suspensions (electrostatic suspension loop), respectively. At low frequency, the drag-free gain is very high due to the need for suppressing the SC jitter that dominates $o_1$. Instead, the electrostatic suspension gain is designed to suppress the force couplings between the TMs and the SC that dominate $o_{12}$. The control laws used in this thesis are provided by ASTRIUM \cite{S2-ASD-TN-2003} -- the main industry contractor of LPF.

\figuremacroW{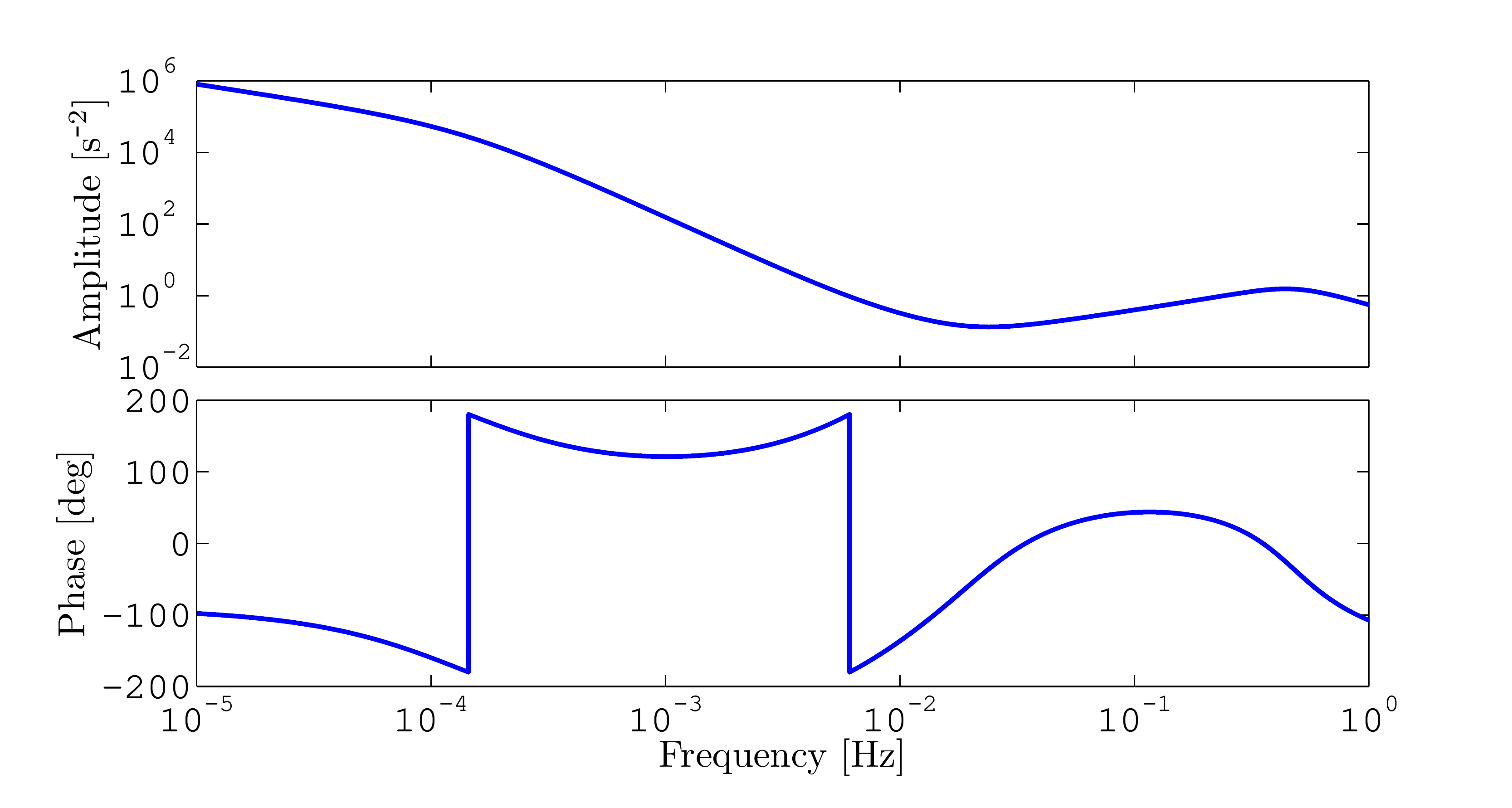}{Frequency dependence of the drag-free loop controller per unit SC mass. The very high gain at low frequency is explained by the need for removing the thruster noise. Following \eqref{eq:dynamics:bias2control}, $\unit[1]{\mu m}$ sensed displacement of the first TM relative to the SC produces a thruster actuation of $\oforder\unit[0.02]{\mu N}$ at $\unit[1]{mHz}$.}{fig:sys_identification:Cdf}{0.8}

\figuremacroW{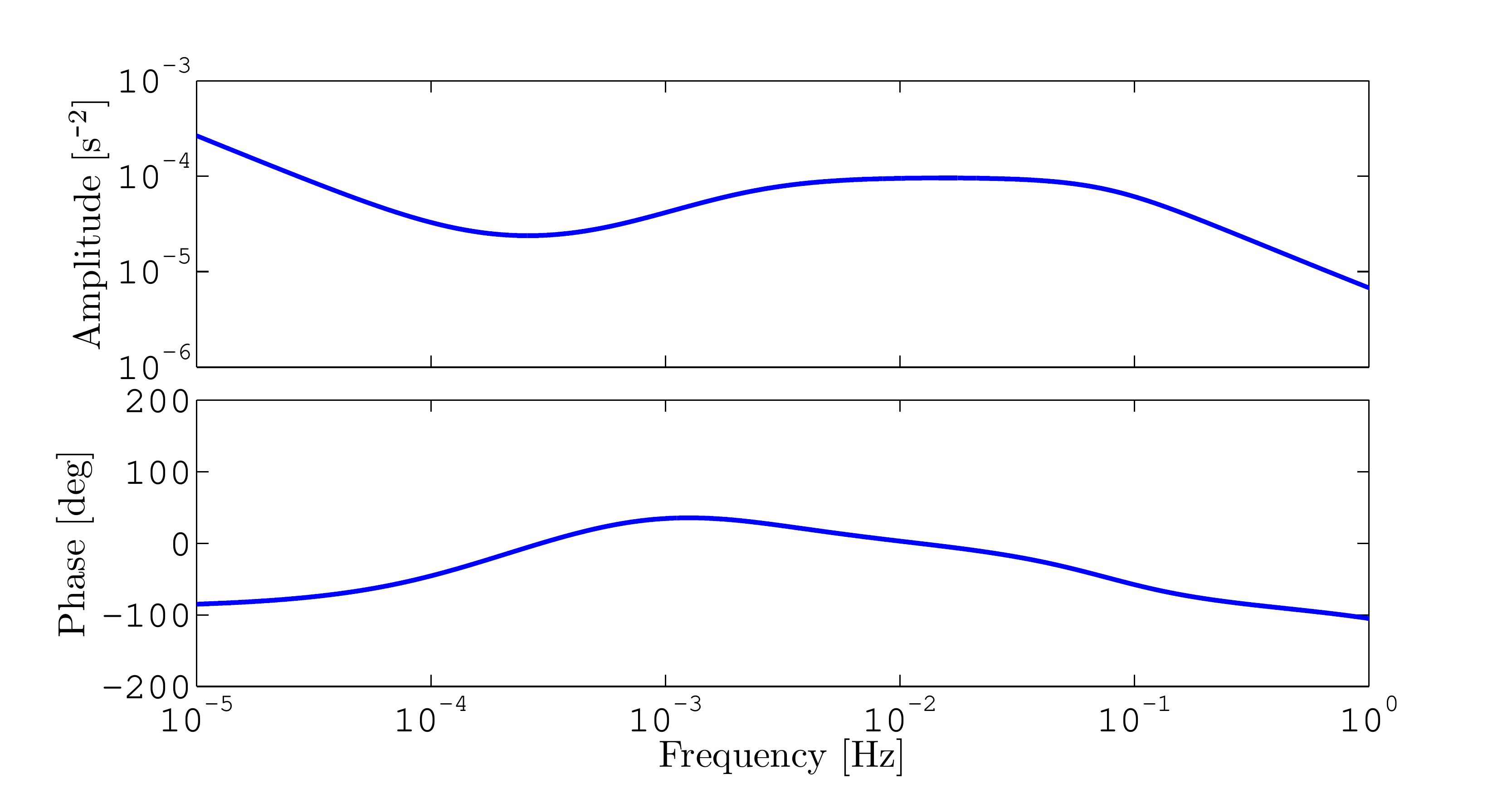}{Frequency dependence of the electrostatic suspension loop controller per unit TM mass. Here there is no such a huge variation in the order of magnitude as for the drag-free controller in \figref{fig:sys_identification:Cdf}. The control law is designed in particular to suppress the force couplings between the TMs and the SC at low frequency. Following \eqref{eq:dynamics:bias2control}, $\unit[1]{\mu m}$ sensed displacement of the second TM relative to the first one produces a capacitive actuation of $\oforder\unit[0.1]{nN}$ at $\unit[1]{mHz}$.}{fig:sys_identification:Csus}{0.8}

As described in \sectref{sect:dynamics:model_along_x}, the system can be modeled by the operators $\vect{D}$ (dynamics), $\vect{S}$ (sensing) and $\vect{A}$ (actuation) representing different non-idealities in the practical implementation of the closed-loop LISA arm. The operators contain all system parameters describing the dynamics along the optical axis. One last source of indetermination introduced here is a delay in the application of the guidance signals
\begin{equation}
\vect{T} =
\begin{pmatrix}
e^{- s\,\Delta t_1} & 0 \\
0 & e^{- s\,\Delta t_2}
\end{pmatrix}~,
\end{equation}
whose possible causes may be either due to the digitalization of the continuous control laws or to bus delays, a possibility not considered in a previous model \cite{nofrarias2010}. With the introduction of the delays, the model \eqref{eq:dynamics:ifo2ifo} becomes now
\begin{equation}\label{eq:sys_identification:ifo2ifo}
\vect{T}_{o_\text{i}\rightarrow o} = \vect{\Delta}^{-1}\vect{A}\,\vect{C}\,\vect{T}~,
\end{equation}
where the differential operator $\vect{\Delta}$, defined in \eqref{eq:dynamics:ifo2acc}, converts the sensed motion into total equivalent acceleration.

\figref{fig:sys_identification:ifo2ifo} shows the transfer gains of the model $\vect{T}_{o_\text{i}\rightarrow o}$, whereas the dynamical cross-talk from the differential channel to the first one is definitely negligible with peak gain of about $4\e{-6}$ at $\unit[30]{mHz}$. The diagonal elements have respectively peak gains of almost 3 at $\unit[0.1]{Hz}$ and about 2 at $\unit[0.8]{mHz}$. The dynamical cross-talk from the first channel to the differential one has peak gain of about $5\e{-2}$ at $\unit[0.5]{mHz}$. The above transfer matrix is used to both model the outputs of the system subjected to bias injections and perform system identification.

\figuremacroW{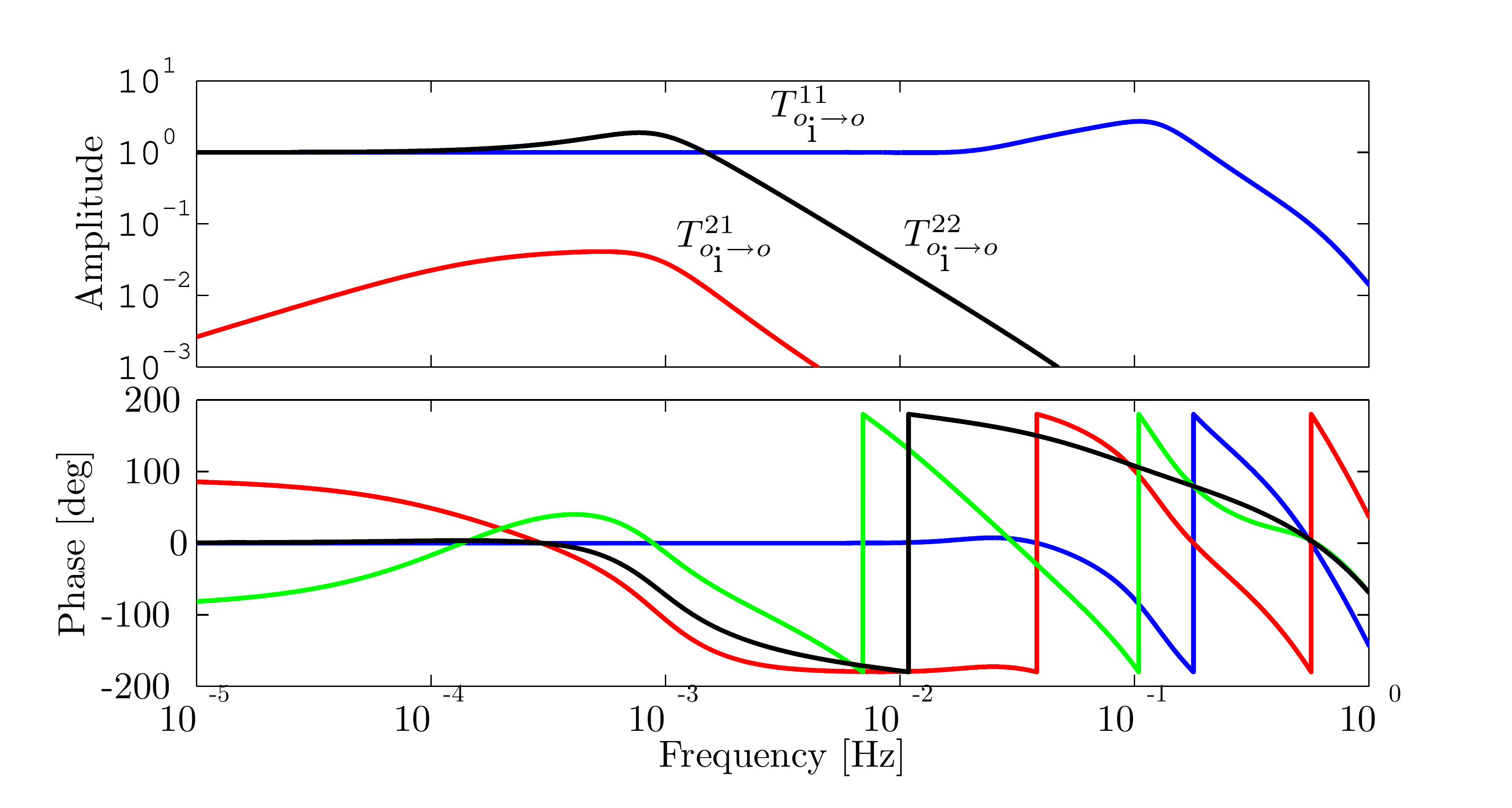}{Frequency dependence of the transfer matrix $\vect{T}_{o_\text{i}\rightarrow o}$ used for system identification. The transfer function $T^{11}_{o_\text{i}\rightarrow o}=T_{o_\text{i,1}\rightarrow o_1}$ has peak gain of almost 3 at $\unit[0.1]{Hz}$. The transfer function $T^{22}_{o_\text{i}\rightarrow o}=T_{o_\text{i,12}\rightarrow o_{12}}$ has peak gain of about 2 at $\unit[0.8]{mHz}$, then it quickly decays. The dynamical cross-talk $T^{12}_{o_\text{i}\rightarrow o}=T_{o_\text{i,1}\rightarrow o_{12}}$ has peak gain of about $5\e{-2}$ at $\unit[0.5]{mHz}$. The other dynamical cross-talk is negligible since has peak gain of about $4\e{-6}$ at $\unit[30]{mHz}$.}{fig:sys_identification:ifo2ifo}{0.8}

Throughout this chapter bias injections at the level of controller guidance signals $\vect{o}_\text{i}$ \footnote{Following \eqref{eq:dynamics:bias2control}, a bias in the guidance signals is equivalent to a commanded force bias directly applied onto the SC through thruster actuation and the second TM through capacitive actuation.} are considered and the transfer matrix in \eqref{eq:sys_identification:ifo2ifo} models the response of the system to those signals. As the modeled system parameters appear in the operators, $\vect{T}_{o_\text{i}\rightarrow o}$ is parameter-dependent. The modeled system response is then parameter-dependent. The parameters can be arranged in a vector that will be abstractly referred to $\vect{p}$
\begin{equation}
\vect{p} =
\scalebox{.8}{$
\begin{pmatrix}
\omega_1^2 \\ \omega_{12}^2 \\ S_{21} \\ A_\text{df} \\ A_\text{sus} \\ \Delta t_1 \\ \Delta t_2 \\
\end{pmatrix}$}~,
\end{equation}
where \tabref{tab:sys-identification:parameters} provides a description of the above system parameter with initial plausible estimates coming from on-ground measurements and theoretical modeling.
\begin{table}[!htbp]
\caption{\label{tab:sys-identification:parameters}\footnotesize{List of the modeled system parameters, introduced in \sectref{sect:dynamics:model_along_x}, except for $\Delta t_1$ and $\Delta t_2$, with descriptions and initial estimates. The parameters that are fitted to data are $\omega_1^2$, $\omega_{12}^2$, $S_{21}$, $A_\text{df}$, $A_\text{sus}$, $\Delta t_1$, $\Delta t_2$.}}
\centering
\scalebox{.9}{
\begin{tabular}{p{2.5cm}p{5cm}p{3.5cm}p{3cm}}
\hline
\hline
Parameter & Description & Note & Estimate \\
\hline
$\omega_1^2$, $\omega_{12}^2$ & parasitic stiffness constants modeling residual oscillator-like couplings between the SC and the reference TM and between the two TMs & must be estimated from experiments & $\oforder-\unit[1\e{-6}]{s^{-2}}$ \\
$S_{21}$ & sensing cross-talk between $o_1$ and $o_{12}$ interferometric readouts & must be estimated from experiments & $\oforder 1\e{-4}$ \\
$A_\text{df}$, $A_\text{sus}$ & actuation gains for the application of forces by the thrusters and the electrostatic suspensions & must be estimated from experiments & $\oforder 1$ \\
$\Delta t_1$, $\Delta t_2$ & delays in the application of biases to the controller computing the actuation & must be estimated from experiments & $\lesssim\unit[1]{s}$ \\
$\Gamma_x$ & gravity gradient between the two TMs & could be estimated from experiments with different actuation stiffness but difficult, considered fixed & $\oforder\unit[4\e{-9}]{s^{-2}}$ \\
$m_1$, $m_2$, $m_\text{SC}$ & masses of TMs and SC & considered fixed & $\unit[1.96]{kg}$, $\unit[422.7]{kg}$ \\
\hline
\hline
\end{tabular}
}
\end{table}

The aim of system identification, as thoroughly described in this chapter, is the estimation of these system parameters with targeted experiments.

\subsection{Anelasticity and damping}

The parameters defined above are implicitly assumed to be independent from frequency.
For example, the parasitic stiffness constant may show a frequency dependence due to anelasticity (an ``internal'' dissipation of the string constant) or a damping effect.

For the sake of clarity, $\omega_0^2$ is the (negative) stiffness constant not to be confused with the Fourier angular frequency $\omega$. An anelasticity can be modeled as a frequency dependence in the imaginary part \cite{saulson1990} of a complex stiffness constant
\begin{equation}
\tilde{\omega}^2(\omega) = \omega_0^2\left[1+i\phi(\omega)\right]~,
\end{equation}
where $\phi$ is named the \textit{loss angle} modeling the dissipation. Sources of dissipation are the dielectric losses in the surface of the electrodes facing the TM that can be modeled by a constant
\begin{equation}
\phi_\epsilon(\omega) = -\delta_\epsilon~,
\end{equation}
such that it produces a force proportional to displacement and in phase with velocity \footnote{Thanks to the imaginary unit. The minus sign is due to the fact that the stiffness constant is usually negative.}. The other source of dissipation is the residual gas damping that can be modeled by a function proportional to frequency
\begin{equation}
\phi_\text{g}(\omega) = \frac{\omega}{\omega_0^2\tau}~,
\end{equation}
where $\tau$ is the damping characteristic time. The above produces a force proportional to velocity \footnote{In fact, the damped harmonic oscillator in frequency domain is $(-\omega^2+i\gamma\omega+\omega_0^2)x=f$, where $\gamma=1/\tau$ is the damping coefficient. Then, the complex stiffness constant is given by $\tilde{\omega}^2 = \omega_0^2+i\gamma\omega = \omega_0^2+i\omega/\tau$.}.

The loss angle function may show other interesting features beyond the ones reported here. To first approximation, the following analysis assumes that all parameters are independent from frequency, at least within the frequency band of interest.

\section{Noise characterization} \label{sect:sys_identification:ifo_noise}

One of the objectives of the LPF mission is to provide a full noise projection of the total equivalent differential acceleration noise between the TMs. As this is well beyond the scope of this thesis, the following presents a hint of the problem. Moreover, a theoretical projection of the observed displacement noise is needed in advance in order to identify the dominant effects in the noise and produce the generating filters used for all simulations. The noise projections shown in this section are given by plausible noise shapes implemented in the simulator provided by ESA (that will be specifically introduced in the first paragraph of \sectref{sect:sys_identification:simulator}).

\figref{fig:sys_identification:a1_noise_proj} shows a theoretical noise projection of the equivalent acceleration noise affecting the $x_1$ degree of freedom. Evidently, the thruster actuation noise dominates the total noise budget in the frequency band of interest. Other important noise sources are the infrared thermal emission of the SC external surface and the $o_1$ sensing noise.

\figuremacroW{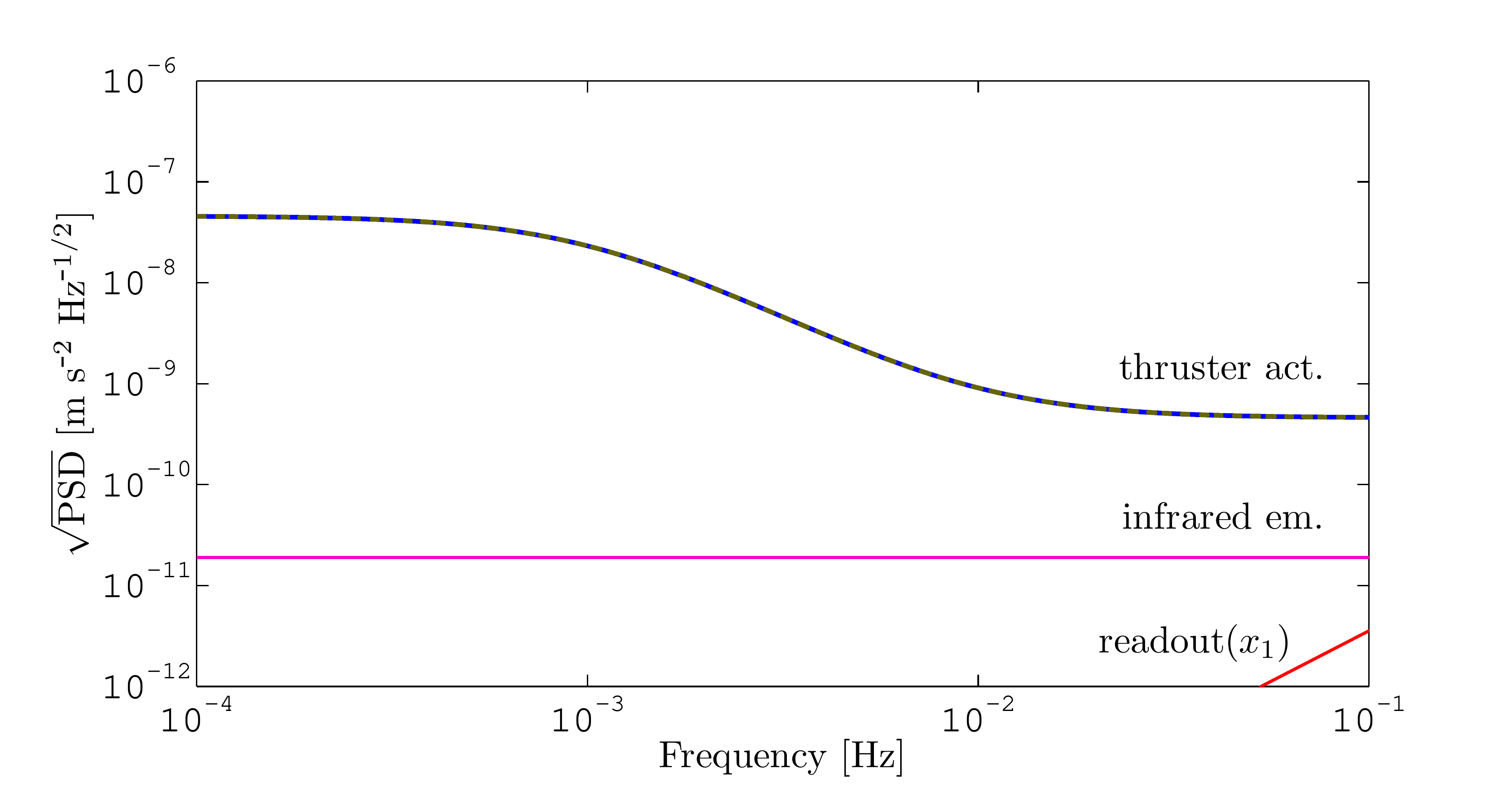}{Theoretical noise projection of the residual equivalent acceleration noise of the relative motion between the SC and the first TM for the nominal dynamics along $x$. The thruster actuation noise dominates the total noise budget (dashed line) in the frequency band of interest. Other important noise sources are the infrared thermal emission of the SC external surface and the $o_1$ sensing noise.}{fig:sys_identification:a1_noise_proj}{0.8}

\figref{fig:sys_identification:a12_noise_proj} shows the second and most important projection of the equivalent differential acceleration noise affecting the $x_{12}$ degree of freedom. A turning point around $\unit[6]{mHz}$ between two regimes is clearly evident. At high frequency, the $o_1$ sensing noise dominates the total noise budget. At low frequency, $\nicefrac{2}{3}$ of the total noise budget (in units of $\sqrt{\text{PSD}}$) is due to force couplings between the SC and the TMs. Other important noise sources, intervening at low frequency, are the capacitive actuation noise on the second TM, forces on the TMs coming from outside the SC and the $o_{12}$ and $o_1$ sensing noises.

\figuremacroW{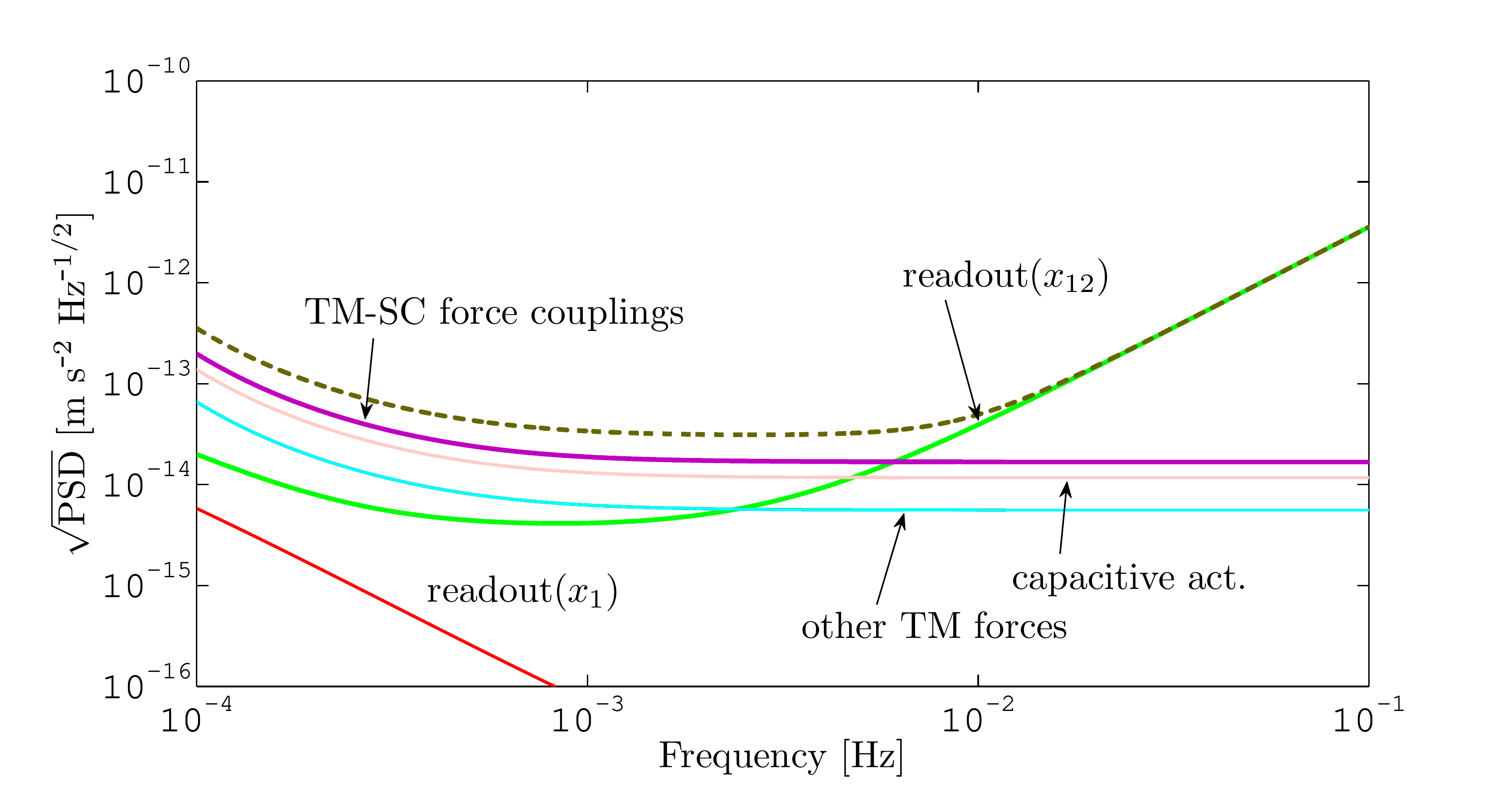}{Theoretical noise projection of the residual equivalent acceleration noise of the relative motion between the TMs for the nominal dynamics along $x$. At high frequency, the $o_{12}$ sensing noise dominates the total noise budget (dashed line). At low frequency, $\nicefrac{2}{3}$ of the total noise budget is due to force couplings between the SC and the TMs. Other important noise sources are the capacitive actuation noise on the second TM, forces on the TMs coming from outside the SC and the $o_{12}$ and $o_1$ sensing noises.}{fig:sys_identification:a12_noise_proj}{0.8}


The above acceleration noise projections are the equivalent acceleration inputs to LPF coming from reasonable noise shapes, producing a characteristic output in the interferometric readouts. \figref{fig:sys_identification:o1_noise_proj} and \figref{fig:sys_identification:o12_noise_proj} contain the relative projections for the two interferometric readouts, $o_1$ and $o_{12}$ along $x$, produced with a plausible transfer model. Analogously to the equivalent acceleration noise, for $o_1$ the thruster actuation noise dominates the total noise budget in the frequency band of interest. As previously pointed out, for $o_{12}$ there is a turning point around $\unit[6]{mHz}$. At high frequency, the $o_1$ sensing noise dominates the total noise budget. At low frequency, $\nicefrac{2}{3}$ of the total noise budget is due to force couplings between the SC and the TMs. Secondary sources, intervening at low frequency, are the capacitive actuation noise on the second TM, forces on the TMs coming from outside the SC, the $o_{12}$ and $o_1$ sensing noises and the thruster actuation noise. 

\figuremacroW{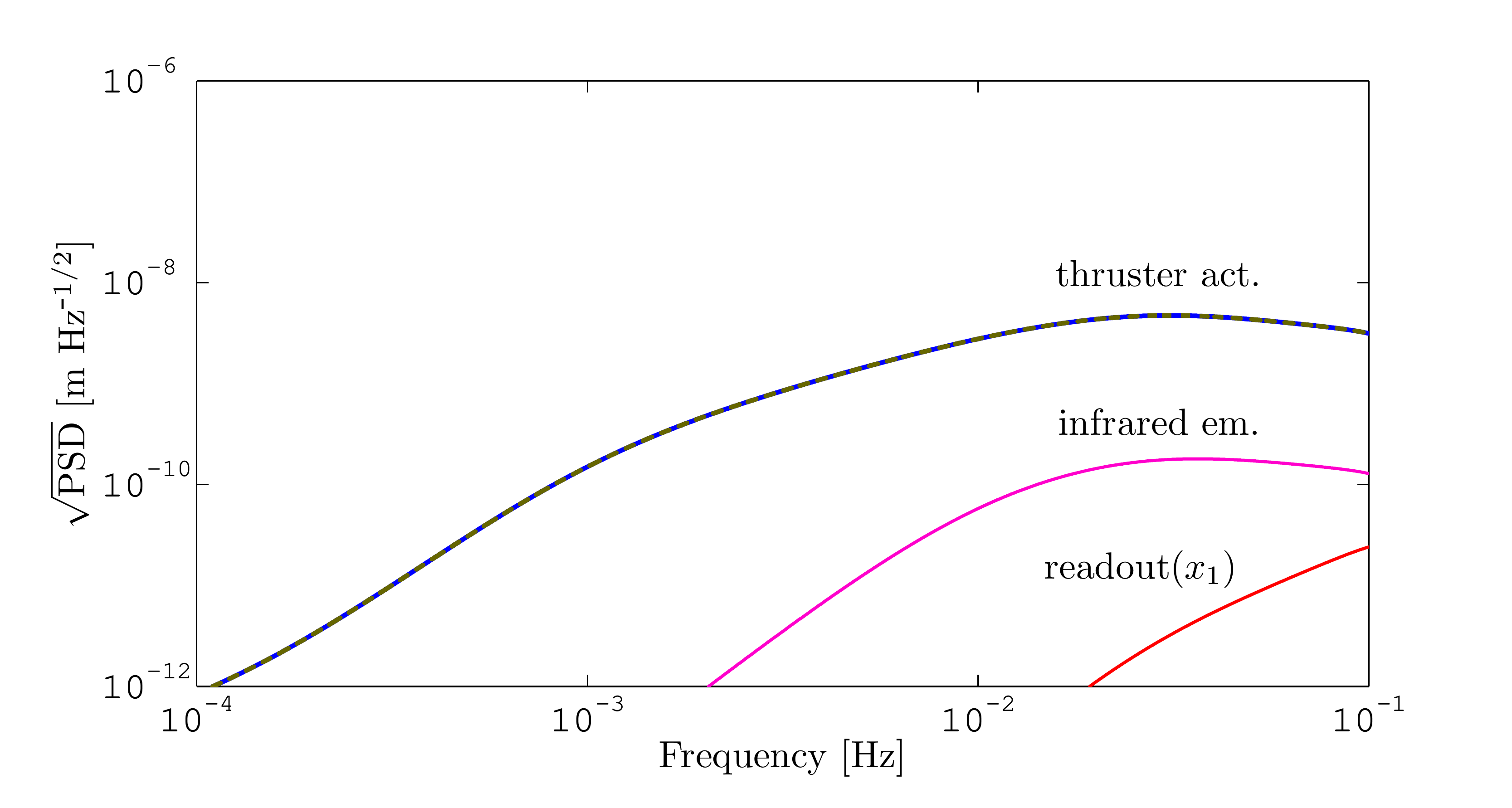}{Theoretical noise projection of the $o_1$ data channel sensing the relative motion between the SC and the first TM for the nominal dynamics along $x$. The thruster actuation noise dominates the total noise budget (dashed line) in the frequency band of interest. Secondary sources are the infrared thermal emission of the SC external surface and the $o_1$ sensing noise.}{fig:sys_identification:o1_noise_proj}{0.8}

\figuremacroW{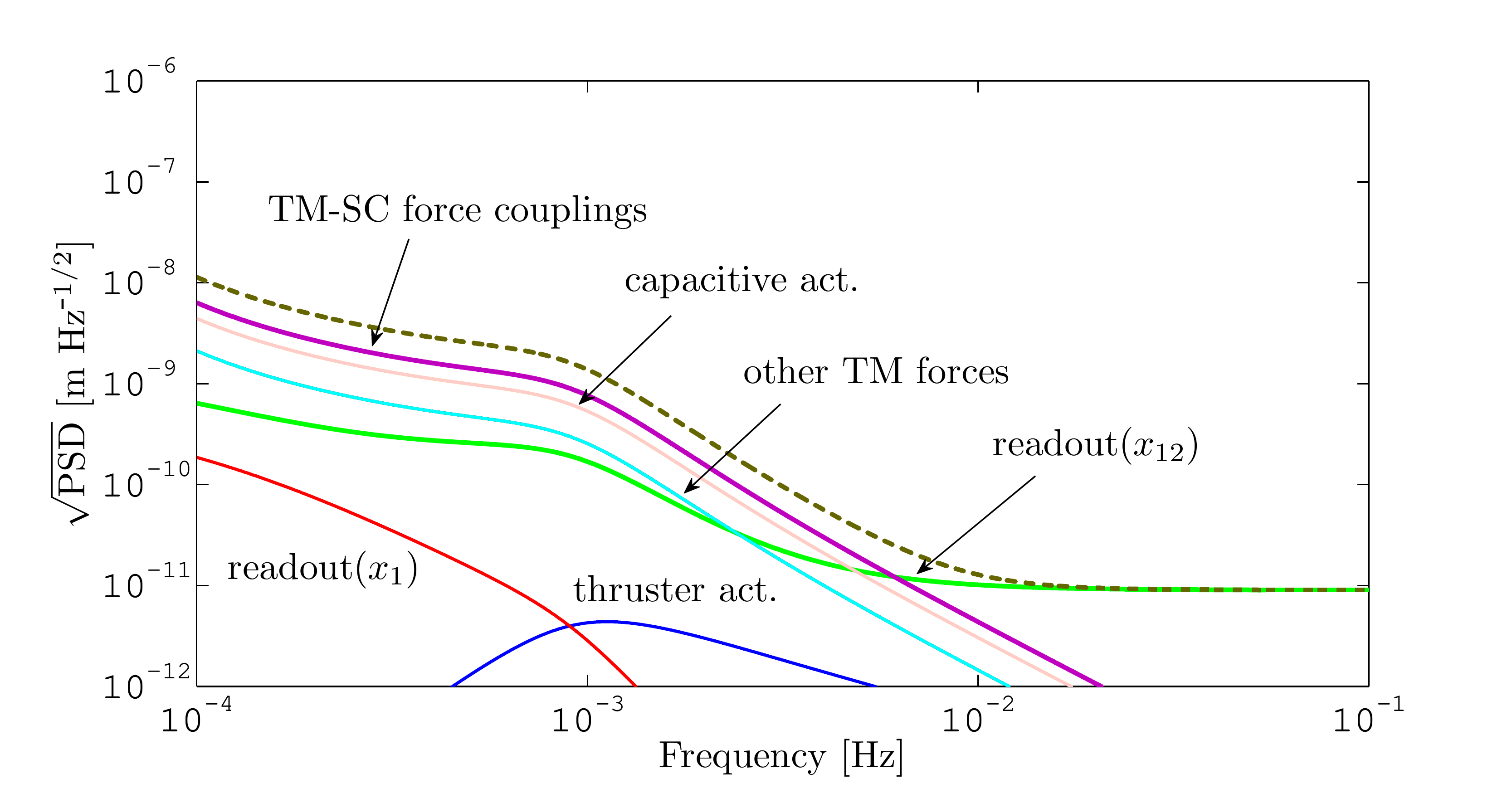}{Theoretical noise projection of the $o_{12}$ data channel sensing the relative motion between the TMs for the nominal dynamics along $x$. At high frequency, the $o_{12}$ sensing noise dominates the total noise budget (dashed line). At low frequency, $\nicefrac{2}{3}$ of the total noise budget is due to force couplings between the SC and the TMs. Secondary sources are the capacitive actuation noise on the second TM, forces on the TMs coming from outside the SC, the $o_{12}$, $o_1$ sensing noises and the thruster actuation noise.}{fig:sys_identification:o12_noise_proj}{0.8}

The noise shapes of the interferometric readouts (with their cross-correlation) are also used for simulation purposes. From those models, noise shaping filters are derived and integrated into a multi-channel cross-correlated noise generator \cite{ferraioli2010}. \figref{fig:sys_identification:noiserun} reports an example of a noise run lasting 12 hours and obtained by coloring an input zero-mean $\delta$-correlated (white) Gaussian noise with those filters. $o_{12}$ shows a huge red component caused by the increase of the PSD at low frequency, due to forces on the TMs, as predicted by \figref{fig:sys_identification:o12_noise_proj}. While $o_1$ is dominated by the thruster jitter, $o_{12}$ becomes much less noisy at high frequency, being dominated by readout noise only. The red noise shape of $o_{12}$ is an expected feature during the experiments of the LPF mission.

\figuremacroW{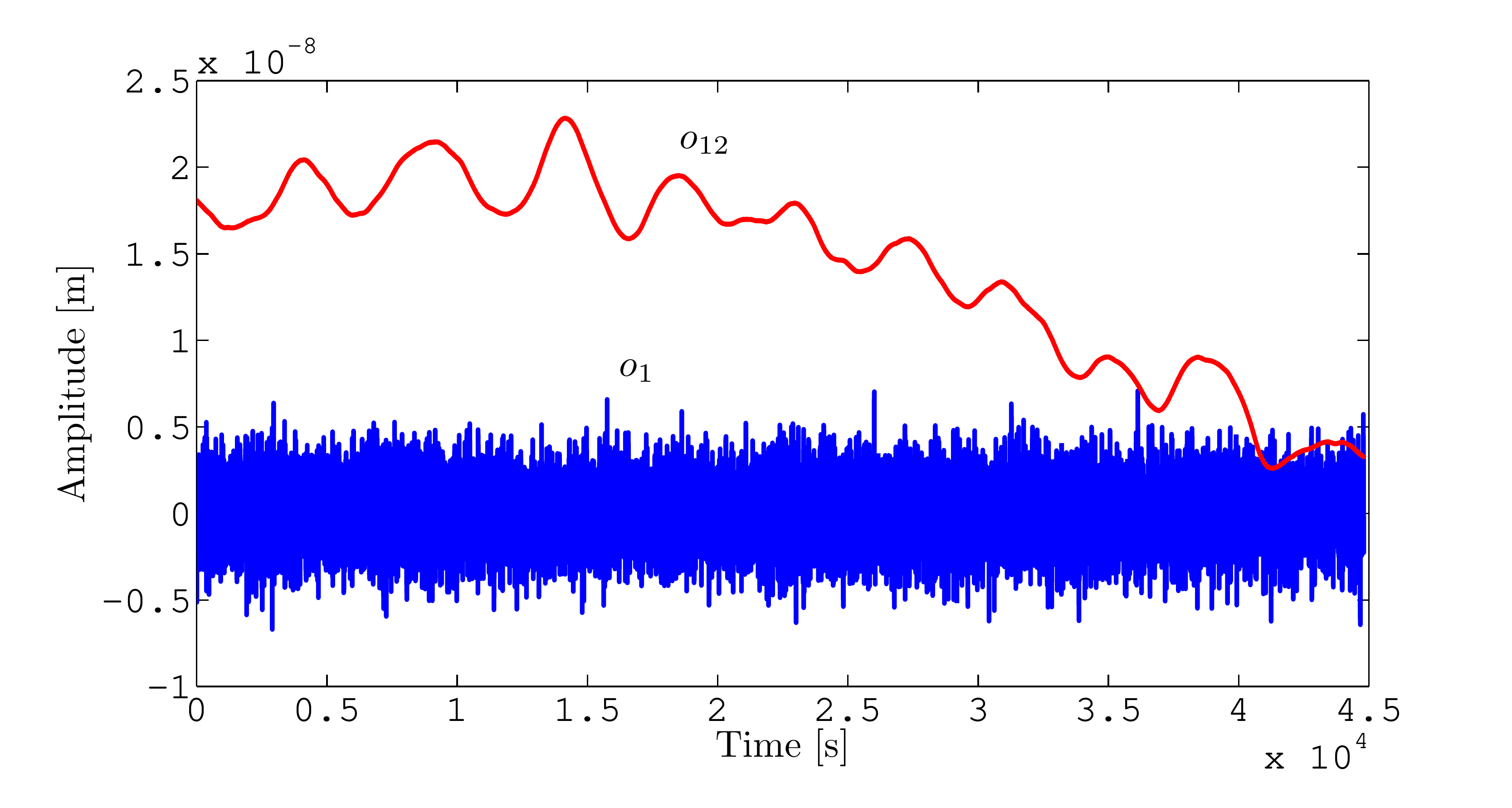}{A simulated noise run of about 12 hours. $o_1$ and $o_{12}$ are the two interferometer readings. Notice the behavior of $o_{12}$ at low frequency -- an expected feature during the LPF mission -- showing a huge red component caused by force couplings between the TMs and the SC. At high frequency, $o_{12}$ becomes much less noisy than $o_1$, the former being dominated by only interferometer readout noise and the latter by thruster noise.}{fig:sys_identification:noiserun}{0.8}

\section{Identification experiments} \label{sect:sys_identification:experiments}

Among the series of experiments characterizing the LPF mission, a few of capital importance will tackle system identification. This thesis considers two identification experiments allowing for a complete identification of the 7 most important system parameters introduced in \sectref{sect:sys_identification:dynamical_model}. As said, considering bias injections at the level of controller guidance signals is completely equivalent to applying direct force stimuli through the equivalence given by \eqref{eq:dynamics:control_forces}. In the nominal $x$-dynamics two experiments are defined:
\begin{enumerate}
\item an injection into the controller guidance of the $o_1$ channel, namely $o_{\text{i},1}$, producing forces on the SC through thruster actuation;
\item an injection into the controller guidance of the $o_{12}$ channel, namely $o_{\text{i},12}$, producing forces on the second TM through capacitive actuation.
\end{enumerate}

To naively understand how the parameters can be determined from the above experiments and the model described in \sectref{sect:sys_identification:dynamical_model}, it is useful to make a projection of the differential operator, whose inverse enters into the transfer matrix through \eqref{eq:sys_identification:ifo2ifo}.

\figref{fig:sys_identification:delta11_proj} contains the projection of the differential operator from the first channel to equivalent acceleration in terms of: (i) dynamics and sensing; (ii) control. Clearly, the control dominates the transfer for almost the entire frequency band, in order to attenuate the SC jitter. For this reason, injecting a signal into the first controller guidance (i.e., applying a thruster actuation on the SC) allows for the identification, in turn, of: the actuation gain, $A_\text{df}$, the first TM coupling to the SC, $\omega_1^2$, as well as a possible delay in the application of the same bias, $\Delta t_1$.

\figuremacroW{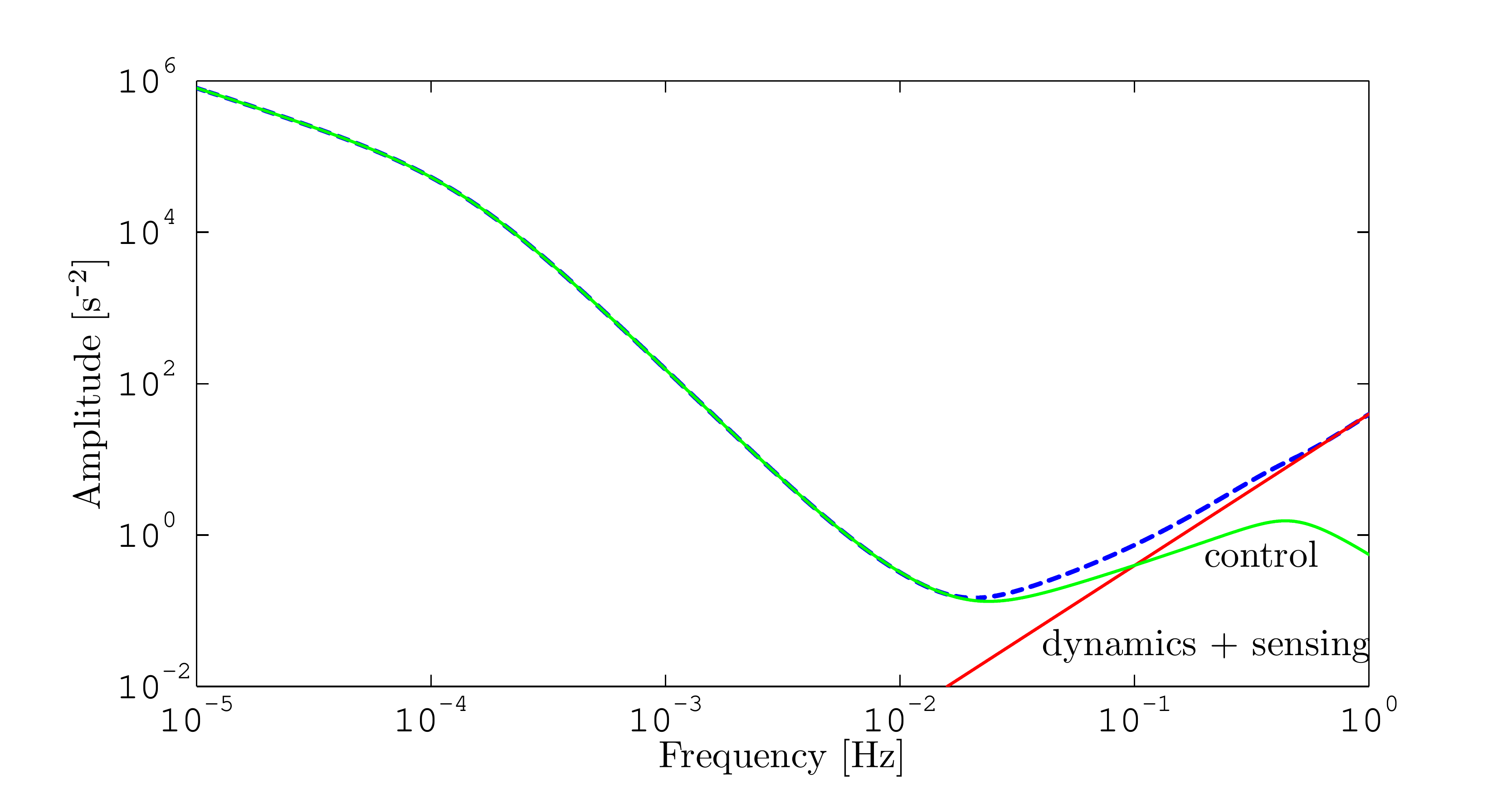}{Frequency dependence of the differential operator for the transfer from $o_1$ to equivalent acceleration. The control dominates the transfer for almost the entire frequency band, in order to attenuate the SC jitter.}{fig:sys_identification:delta11_proj}{0.8}

Analogously, \figref{fig:sys_identification:delta22_proj} contains the projection of the differential operator from the differential channel to equivalent acceleration. Below $\unit[1]{mHz}$, the control dominates the transfer in order to compensate the differential force disturbances. Above $\unit[1]{mHz}$, dynamics and sensing dominate the transfer. For this reason, injecting a signal into the second controller guidance (i.e., applying a capacitive actuation on the second TM) allows for the identification, in turn, of: the actuation gain, $A_\text{sus}$, the differential coupling between the TMs, $\omega_{12}^2$, as well as a possible delay in the application of the same bias, $\Delta t_2$. Given the cross-talk elucidated in \figref{fig:sys_identification:ifo2ifo} at low frequency, the sensing cross-talk, $S_{21}$, can also be determined.

\figuremacroW{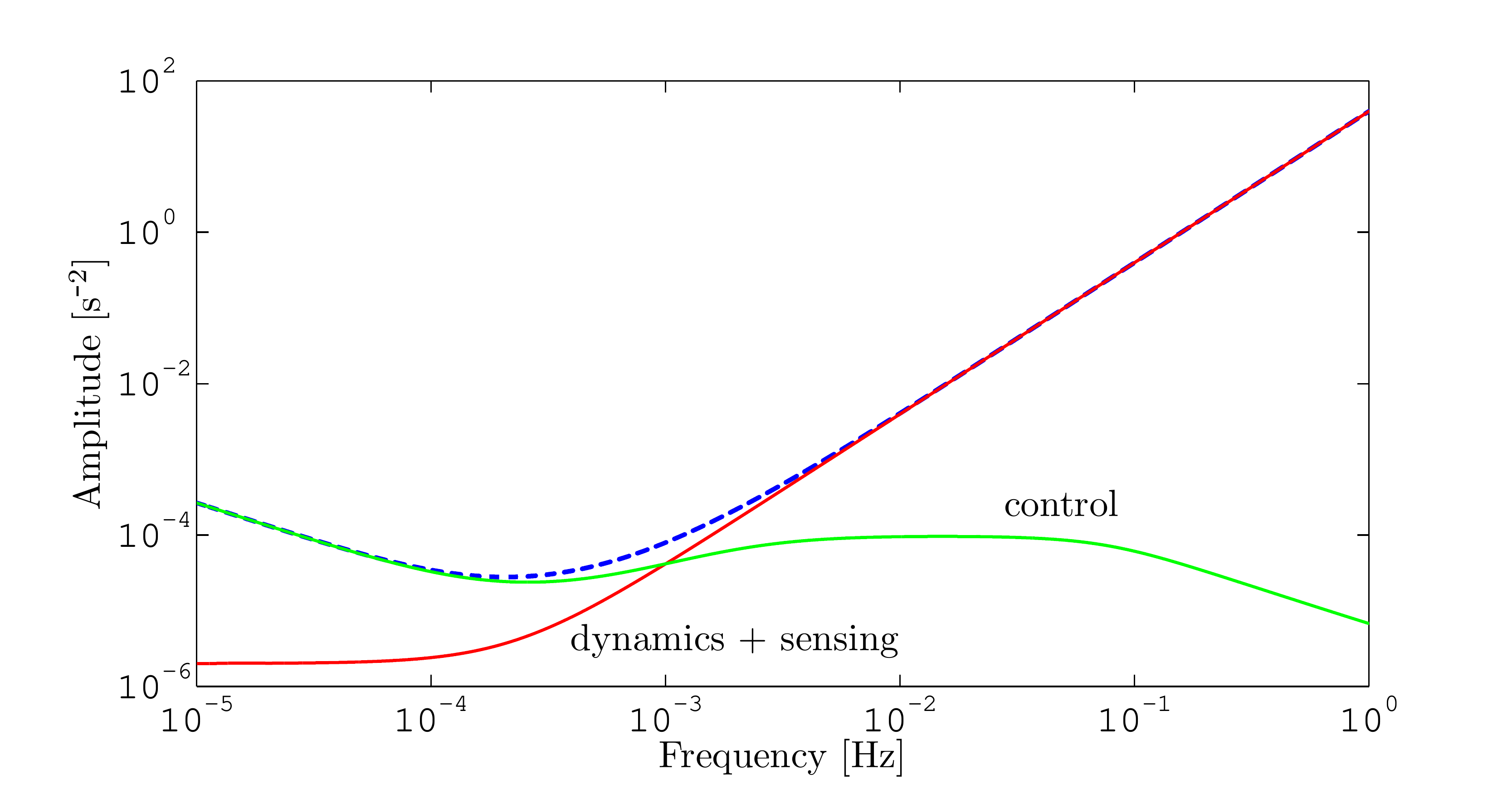}{Frequency dependence of the differential operator for the transfer from $o_{12}$ to equivalent acceleration. The control dominates the transfer at low frequency, in order to compensate the differential force disturbances.}{fig:sys_identification:delta22_proj}{0.8}

To conclude the discussion on the projection of the differential operator, it is worth noting that the off-diagonal terms contribute with a figure of $\unit[1\e{-7}]{s^{-2}}$. In particular, even if not shown in any figures, the transfer from the first channel to the equivalent differential acceleration is dominated by dynamics and sensing; the other by control at low frequency and dynamics and sensing at high frequency. As the SC motion is common-mode and the first and differential channel are correlated, the estimation of the differential acceleration noise can not be performed independently of the first channel, which is the only means by which the SC jitter can be measured and subtracted. The details of such an estimation will be given in \sectref{sect:sys_identification:force_noise}.

The next section is devoted to the estimation of the 7 system parameters by means of a MIMO approach that maximizes the overall information. The identification experiments defined at the beginning of this section are simulated for a total duration of almost 3 hours each -- a suitable timescale for the mission -- by injecting stimulating biases. The following facts are assumed:
\begin{enumerate}
\item the noise $\vect{o}_\text{n}$ is generated as in \sectref{sect:sys_identification:ifo_noise}, independently from the noise-only run which is used for noise characterization, and is Gaussian and stationary;
\item the signals $\vect{o}_\text{s}$ are simulated in time domain with a MIMO approach by means of \eqref{eq:dynamics:ifo2ifo}, i.e., by anti-Fourier transforming \footnote{The numerical implementation of the direct and inverse Fourier transform are the Fast Fourier Transform (FFT) and Inverse FFT (IFFT). Being circular operations, the input time-series needs to be zero-padded to avoid systematic errors caused by wrapped-around data \cite{press}. A conservative default value of one data length is assumed.} with $\mathcal{F}^{-1}$ the deterministic input signals
    \begin{equation}\label{fig:sys_identification:signals}
    \vect{o}_\text{s}(t,\vect{p}_\text{true})=\mathcal{F}^{-1}\left[\vect{T}_{o_\text{i}\rightarrow o}(\omega,\vect{p}_\text{true})\,\vect{o}_\text{i}(\omega)\right](t)~,
    \end{equation}
    where $\vect{p}_\text{true}$ is the set of assumed \textit{true} system parameter values to be estimated from the analysis and which the estimation of residual equivalent acceleration noise depends on;
\item the superposition principle of signals and noise holds true in the hypothesis of small motion and in absence of non-linearities in the system, so that the ``experimental'' data are simulated by
    \begin{equation}
    \vect{o}_\text{exp}=\vect{o}_\text{s}+\vect{o}_\text{n}~.
    \end{equation}
\end{enumerate}

The underlying idea in parameter estimation is to excite the system with proper high SNR signals so that the modeled parameters can be measured. A typical injected bias is a series of sine waves of logarithmically increasing frequency, with integer number of cycles, divided by gaps of $\unit[150]{s}$ to allow for system relaxation. The sine stretches last $1200\,\unit{s}$ each. The amplitudes are conservatively selected not to exceed $1\%$ of the operating range of the interferometer, corresponding to a maximum sensed displacement of $\unit[1]{\mu m}$, and $10\%$ of the maximum allowed force authority, corresponding to $\unit[10]{\mu N}$ of thruster actuation and $\unit[0.25]{nN}$ of capacitive actuation. The biases are parameterized in \tabref{tab:sys_identification:biases} and referred to the \textit{standard input signals} used for the rest of the analysis. Instead, \chapref{chap:optimal_design} will focus on the optimization of the same input signals.
\begin{table}[!htbp]
\caption{\label{tab:sys_identification:biases}\footnotesize{Controller guidance signals injected as biases for system identification. The sine stretches last $1200\,\unit{s}$ each and are separated by gaps of $\unit[150]{s}$. The sine waves perform an integer number of cycles, from 1 to 64. The amplitudes are selected to not exceed $1\%$ of the operating range of the interferometer and $10\%$ of the maximum force authority.}}
\centering
\begin{tabular}{D{.}{.}{5.5} D{.}{.}{5.5} D{.}{.}{5.5} D{.}{.}{5.5}}
\hline
\hline
\multicolumn{2}{c}{$o_{\text{i},1}$ for Exp.\,1} & \multicolumn{2}{c}{$o_{\text{i},12}$ for Exp.\,2} \\
\multicolumn{1}{c}{$\unit[f]{[mHz]}$} & \multicolumn{1}{c}{$\unit[a]{[\mu m]}$} & \multicolumn{1}{c}{$\unit[f]{[mHz]}$} & \multicolumn{1}{c}{$\unit[a]{[\mu m]}$} \\
\hline
0.83 & 1.0 & 0.83 & 0.80 \\
1.7 & 1.0 & 1.7 & 0.48 \\
3.3 & 1.0 & 3.3 & 0.19 \\
6.6 & 1.0 & 6.6 & 0.088 \\
13 & 0.59 & 13 & 0.096 \\
27 & 0.28 & 27 & 0.18 \\
53 & 0.14 & 53 & 0.46 \\
\hline
\hline
\end{tabular}
\end{table}

Data are simulated at $\unit[10]{Hz}$ and decimated to $\unit[1]{Hz}$ to ease data processing. During the mission, data will be collected at a sample rate between $1$ and $\unit[10]{Hz}$, depending on the experiment and available down-link bandwidth. The simulation of the first experiment, with injection of the $o_{i,1}$ signal of \tabref{tab:sys_identification:biases}, is shown in \figref{fig:sys_identification:data_exp1}. The response of the system in $o_1$ is approximately equal to $o_{i,1}$, except at high frequency where there is a modest gain due to the particular shape of the first diagonal element of the transfer function at that frequency. A residual signal in $o_{12}$ of absolute peak $\oforder\unit[4\e{-8}]{m}$ is also visible and due to dynamical cross-talk. As said before, the gaps allow for system relaxation, particularly at high frequency.

\figuremacroW{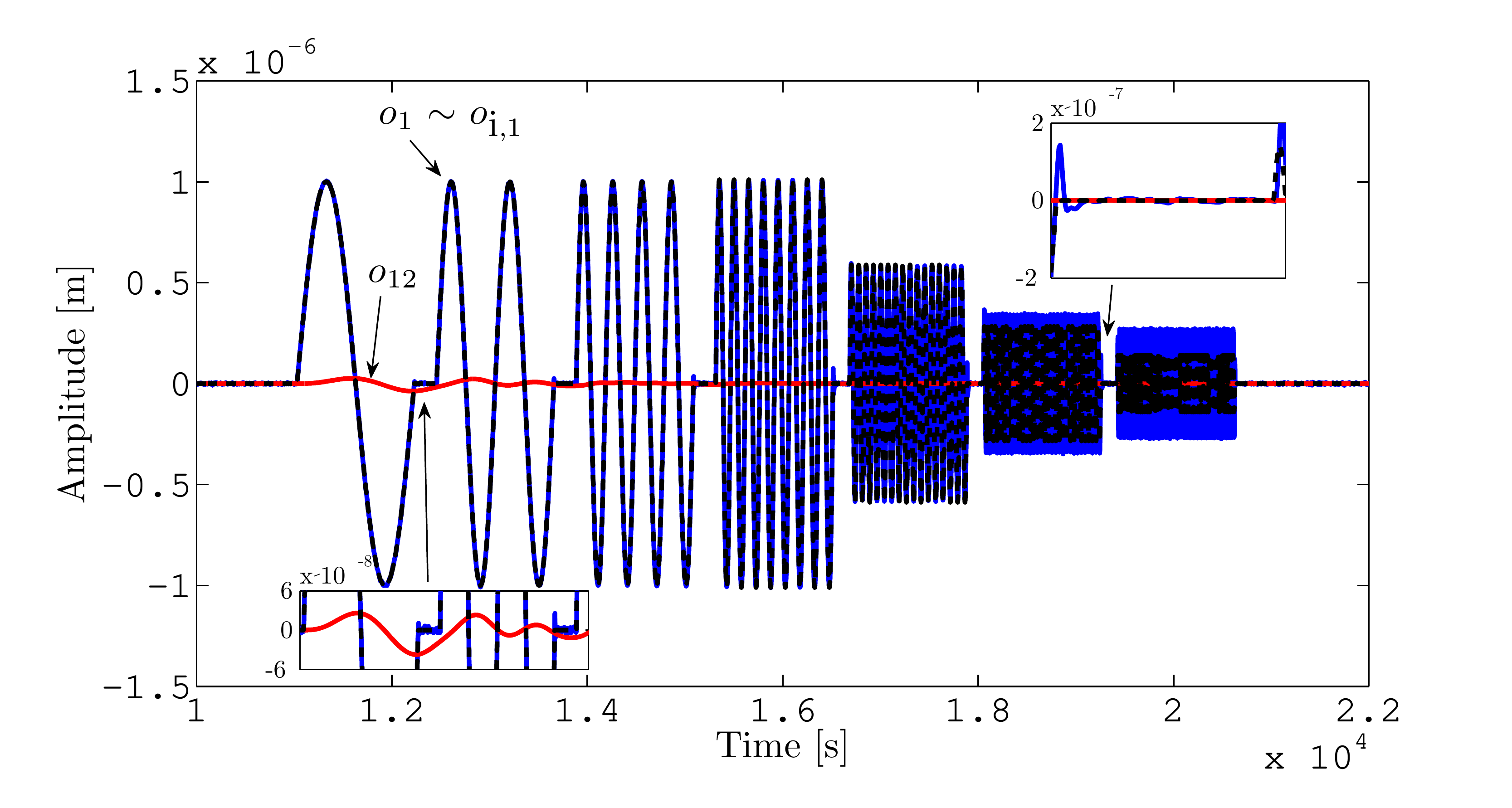}{Exp.\,1 synthetic data. An injection of sine-wave signals lasting for almost 3 hours into the first controller guidance $o_{\text{i},1}$ produces a different response in the two interferometer readings. The response in $o_1$ is approximately equal to $o_{\text{i},1}$ (dashed line), except at high frequency where there is a modest gain. A residual signal in $o_{12}$ of absolute peak $\oforder\unit[4\e{-8}]{m}$ is due to dynamical cross-talk (see inset at the left bottom side). Gaps between two cycles of injection allow for system relaxation (see inset at the right top side).}{fig:sys_identification:data_exp1}{0.8}

The simulation of the second experiment, with injection of the $o_{i,12}$ signal of \tabref{tab:sys_identification:biases}, is shown in \figref{fig:sys_identification:data_exp2}. The response of the system in $o_{12}$ is evidently phase delayed to $o_{\text{i},12}$. At high frequency, the very low gain of the transfer function almost suppresses the signal. Since the transfer from $o_{i,12}$ to $o_1$ is negligible, in this experiment $o_1$ has signal contribution completely hidden by noise. For this reason, during the mission the $o_1$ readout will serve as a useful sanity check for a first understanding of the model.

\figuremacroW{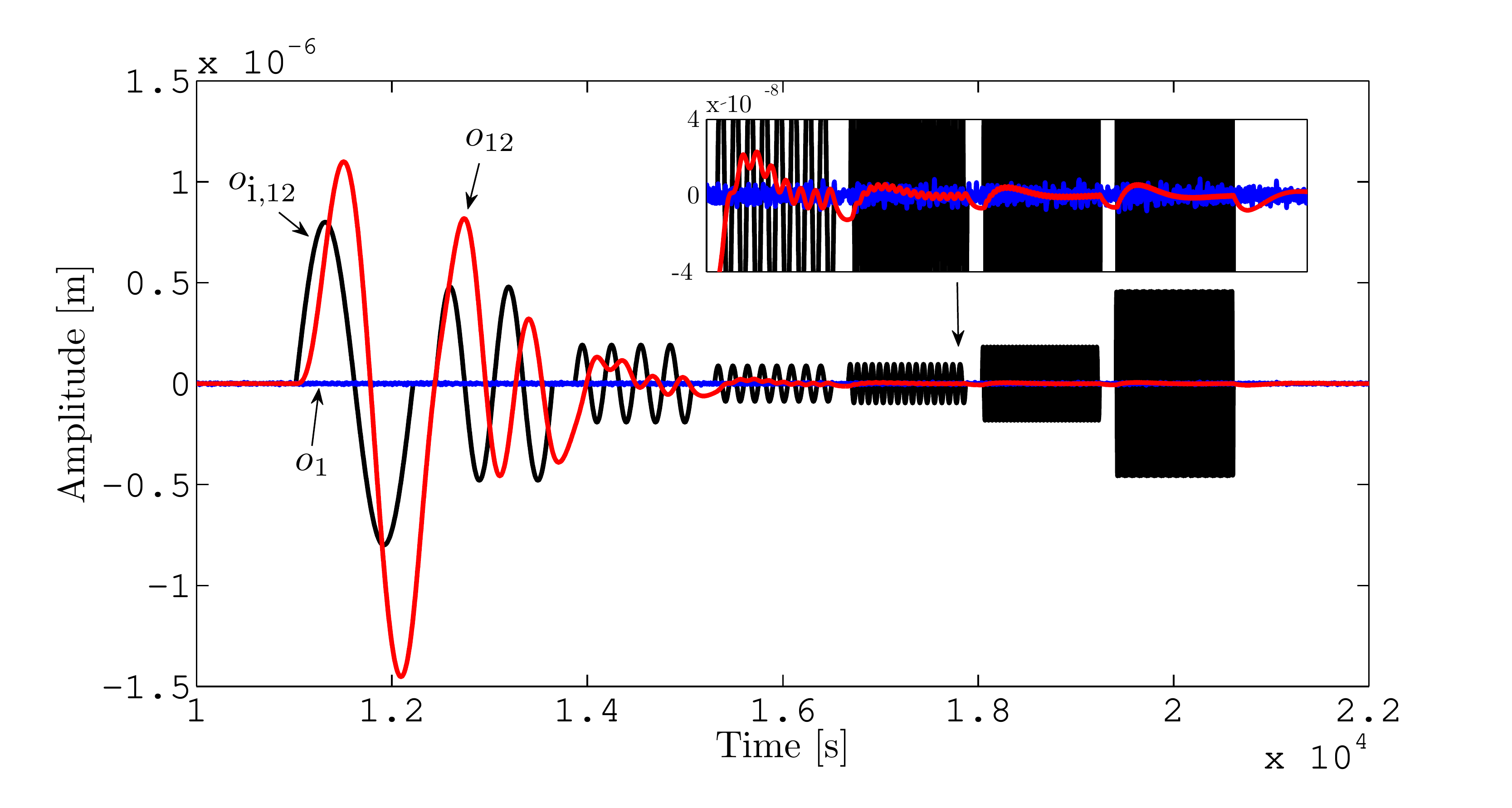}{Exp.\,2 synthetic data. An injection of sine-wave signals lasting for almost 3 hours into the second controller guidance $o_{\text{i},12}$ produces a different response in the two interferometer readings. The response in $o_{12}$ is evidently phase delayed to $o_{\text{i},12}$. At high frequency, the very low gain of the transfer function almost suppresses the signal (see inset). The $o_1$ data channel has negligible contribution hidden by the noise.}{fig:sys_identification:data_exp2}{0.8}

\section{Parameter estimation}

During the mission, noise runs will be used to characterize the noise itself and estimate the total equivalent input acceleration. The estimation of the total equivalent acceleration is possible if LPF is properly modeled. For this in the various experiments, signals will be injected along different degrees of freedom to study the response of the system. Along $x$, LPF will be characterized giving, as a first approximation, the nominal dynamics. Instead, along others degrees of freedom, LPF will be characterized in terms of the many cross-talk contributions arising from the dynamical couplings, the imperfections in the sensing conversion and the imperfections in the actuation.

This section handles the general problem of estimating the LPF parameters modeled as a MIMO dynamical system, where different inputs enters into the system and produce a response in different outputs. For the sake of simplicity, for the rest only the two experiments introduced above -- the characterization of the nominal dynamics along the optical axis -- are considered, bearing in mind that the method is general enough to handle more complex experiments. An example would be the identification of the $xy$ cross-talk, in where guidance or force bias signals are injected, in turn, along $y_1$, $y_2$, $\phi_1$, $\phi_2$ and $\phi_\text{SC}$ to study the response along the optical axis.

Finally, this section develops and validates the estimation procedures on the two most important experiments described in the previous section. It also shows the application to a couple of non-standard scenarios that may happen during the real LPF mission.

\subsection{Review of the problem}

The experimental data (either simulated or from the mission) can be modeled superimposing deterministic signals with noise
\begin{equation}
\vect{o}_\text{exp}=\vect{o}_\text{s}+\vect{o}_\text{n}~,
\end{equation}
where $\vect{o}_\text{n}$ is the output noise with cross PSD matrix $\vect{S}_\text{n}$ and
\begin{equation}\label{eq:sys_identification:templates}
\vect{o}_\text{s}(t,\vect{p})=\mathcal{F}^{-1}\left[\vect{T}_{o_\text{i}\rightarrow o}(\omega,\vect{p})\,\vect{o}_\text{i}(\omega)\right](t)~,
\end{equation}
are the so-called \textit{template} signals obtained by injecting bias guidance signals $\vect{o}_\text{i}$ into the system modeled by the transfer matrix $\vect{T}_{o_\text{i}\rightarrow o}$.

It is useful to think that the experimental data depends on the true parameter values
\begin{equation}
\vect{o}_\text{exp}=\vect{o}_\text{exp}(t,\vect{p}_\text{true})~,
\end{equation}
that need to be estimated from fitting procedures. In the case of simulated experiments, the true values are exactly those used in data generation. In the case of real mission experiments, the true values are actually those giving the best possible description of the data, the one that perfectly subtracts the deterministic signals, hence recovering the instrumental noise shapes.

In the same way, the observed noise (either simulated or from the mission) depends on the parameter values
\begin{equation}
\vect{o}_\text{n}=\vect{o}_\text{n}(t,\vect{p}_\text{true})~;
\end{equation}
but can be considered constant with respect to the parameter values for the timescale of an identification experiment where only high SNR signals will be injected.

The scope of parameter estimation is to recover the best possible description of the experimental data. If the \textit{residuals} between the experimental data and the modeled template signals are defined by
\begin{equation}
\vect{o}_\text{r}=\vect{o}_\text{exp} - \vect{o}_\text{s}~,
\end{equation}
the best possible description of the experimental data is given by
\begin{equation}
\vect{o}_\text{r}(t,\vect{p}_\text{est})\simeq\vect{o}_\text{n}(t,\vect{p}_\text{true})~,
\end{equation}
implying that the residuals evaluated at the estimated parameter values $\vect{p}_\text{est}$ recover the true instrumental noise.

\subsection{Estimation method}

LPF is a MIMO dynamical system for which each experiment has a unique set of meaningful parameters. Hence, for two generic experiments two sets of parameters can be independently determined. Sometimes a subset may be shared between the two; sometimes there could be parameters that can be estimated by only a particular experiment. Moreover, each experiment has multiple readouts sensitive to different parameters. \sectref{sect:sys_identification:experiments} has given an intuitive hint of such an idea.

The first approach is to build an information-weighted average \cite{congedo2012,nofrarias2010} of different parameter estimates coming from all readouts and experiments. If $\vect{p}_{ij}$ are the parameter estimates of the $i$-th experiment and $j$-th readout, the corresponding Fisher information matrix \cite{finn1992}
\begin{equation}
\vect{\mathcal{I}}_{ij}=
\int
\ctranspose{\nabla_{\vect{p}} \vect{o}_\text{r}^{(ij)}(\omega,\vect{p}_\text{est})} \,
\vect{S}_\text{n}^{(ij)}(\omega)^{-1} \,
\nabla_{\vect{p}} \vect{o}_\text{r}^{(ij)}(\omega,\vect{p}_\text{est}) \,\text{d}\omega~,
\end{equation}
where $\vect{S}_\text{n}^{(ij)}$ is the noise PSD of $i$-th experiment and $j$-th readout, $\vect{o}_\text{r}^{(ij)}$ is the corresponding vector of residuals, $\nabla_{\vect{p}}$ is the gradient with respect to the parameters and $\ctranspose{}$ is the conjugate transpose. The final combined parameter estimates are given by
\begin{equation}\label{eq:sys_identification:comb_estimates}
\vect{p}=\vect{\mathcal{I}}^{-1}\sum_{i=1}^{N_\text{exp}}\sum_{j=1}^{N_o}\vect{\mathcal{I}}_{ij}\,\vect{p}_{ij}~,
\end{equation}
where $N_\text{exp}$ is the number of experiments and $N_o$ the number of readouts per experiment assumed the same across the experiments. The combined Fisher information matrix is
\begin{equation}\label{eq:sys_identification:comb_info_matrix}
\vect{\mathcal{I}}=\sum_{i=1}^{N_\text{exp}}\sum_{j=1}^{N_o}\vect{\mathcal{I}}_{ij}~.
\end{equation}
Notice that the estimates $\vect{p}_{ij}$ may have different dimension depending of the $i$-th experiment and $j$-th interferometric readout; the same happens for the corresponding information matrices. The issue can be easily solved by inserting zeros where there is no information.

An example can readily show that the definition of \eqref{eq:sys_identification:comb_estimates} is not robust. In fact, suppose that the estimation of the system parameters is performed independently on each readout and one of those parameters has a biased value for an inaccuracy of the transfer matrix model. Therefore, the information matrix for that estimate is biased and the combined one in \eqref{eq:sys_identification:comb_info_matrix} as well. The numerical inversion in \eqref{eq:sys_identification:comb_estimates} inexorably amplifies that bias to the combined parameter estimates. To overcome the problem, one could try removing the failing estimates (which is possible only if one has good indication of what the real values are, for example, from ground measurements or previous independent experiments), but in doing so information and precision would definitely be lost.

The only solution is to attack the problem by a complete MIMO approach where the poor information coming from the biased model of a readout is continuously compensated by the others as the optimization goes on. One other advantage is that a joint information can likely remove or, at least, reduce the effect of parameter degeneracies.

The MIMO-Multi-Experiment joint log-likelihood of the system is a generalization of the standard definition \cite{finn1992} and is given by
\begin{equation}\label{eq:sys_identification:log-likelihood}
\chi^2(\vect{p})=
\int
\ctranspose{\vect{o}_\text{r}(\omega,\vect{p})} \,
\vect{S}_\text{n}(\omega)^{-1} \,
\vect{o}_\text{r}(\omega,\vect{p})\,\text{d}\omega~,
\end{equation}
where
\begin{equation}
\vect{o}_\text{r}(\omega,\vect{p})=\vect{o}_\text{exp}(\omega) - \vect{T}_{o_\text{i}\rightarrow o}(\omega,\vect{p}) \, \vect{o}_\text{i}(\omega)~,
\end{equation}
are the residuals between the experimental data $\vect{o}_\text{exp}$ and the modeled system response. $\vect{o}_\text{i}$ are the controller biases, $\vect{T}_{o_\text{i}\rightarrow o}$ the transfer matrix depending on all system parameters $\vect{p}$ (stiffness constants, sensing cross-talk, etc.), $\vect{S}_\text{n}$ the cross output noise PSD matrix assumed constant to the system parameters. For two experiments and two interferometric readouts each, $\vect{o}_\text{i}$ is a 4-vector, null in the second and third element, since the injection is in $o_{\text{i},1}$ (first experiment) and $o_{\text{i},12}$ (second experiment); $\vect{T}_{o_\text{i}\rightarrow o}$ is a block diagonal $4\times4$-matrix replicating the same $2\times2$-matrix; $\vect{S}_\text{n}$ is a $4\times4$-matrix of cross PSDs between different readouts and experiments; $\vect{o}_\text{exp}$ is a 4-vector of all experimental readouts.

Assuming that all readouts are sampled at the same rate and last for the same duration, the overall number $\nu$ of degrees of freedom for the problem is defined as
\begin{equation}
\nu=N_\text{exp} \times N_o \times N_\text{data} - N_p~,
\end{equation}
where $N_\text{exp}$ is the number of experiments; $N_o$ is the number of readouts per experiment (assumed the same across the experiments); $N_\text{data}$ is the number of data points per readout; $N_p$ is the dimension of the parameter space. For example, $\nu\oforder4\e{4}$ for two experiments, two readouts each, lasting for about 3 hours and sampled at $\unit[1]{Hz}$. For the rest, if not otherwise stated, the reduced log-likelihood $\chi^2/\nu$ will be used in place of the standard definition, as its expectation value is 1.

Notice that system identification may be also implemented in the domain of equivalent acceleration. If the $\vect{\Delta}$ operator is invertible, the two approaches -- identification in acceleration and displacement -- are completely equivalent. In fact,
\begin{equation}
\begin{split}
\chi^2 & = \int\ctranspose{\left(\vect{f}_\text{exp}-\vect{f}_\text{mdl}\right)}\,\vect{S}_{\text{n},f}^{-1}\left(\vect{f}_\text{exp}-\vect{f}_\text{mdl}\right)\,\text{d}\omega \\
& = \int\ctranspose{\left(\vect{o}_\text{exp}-\vect{o}_\text{mdl}\right)}\,\ctranspose{\vect{\Delta}}
\left(\left.\ctranspose{\vect{\Delta}}\right.^{-1}\vect{S}_{\text{n},o}^{-1}\,\vect{\Delta}^{-1}\right)
\vect{\Delta}\left(\vect{o}_\text{exp}-\vect{o}_\text{mdl}\right)\,\text{d}\omega \\
& = \int\ctranspose{\left(\vect{o}_\text{exp}-\vect{o}_\text{mdl}\right)}\,\vect{S}_{\text{n},o}^{-1}\,\left(\vect{o}_\text{exp}-\vect{o}_\text{mdl}\right)\;\text{d}\omega ~.
\end{split}
\end{equation}
where $\vect{f}_\text{mdl}$ and $\vect{f}_\text{exp}$ are the modeled and experimental equivalent accelerations; $\vect{o}_\text{mdl}$ and $\vect{o}_\text{exp}$ are the modeled and experimental displacement readouts. In the preceding equation, $\vect{\Delta}$ and $\vect{\Delta}^{-1}$ are used to transform the sensed relative motion into equivalent acceleration (and vice-versa) and contain the dependence on the modeled parameters. The main benefit of working with accelerations is the automatic subtraction of system transients as described in \sectref{sect:dynamics:transients} and that is numerically demonstrated at the end of this chapter. Despite the identification in displacement where the parameters are explicit in the modeled template signal, in the identification in acceleration the parameters are implicit in the estimated acceleration. Even though there is no real experimental acceleration because this must be estimated from the displacement readouts, system identification in acceleration domain can be still carried out numerically with a non-standard approach based upon a closed-loop optimization over the estimated acceleration data, whereas the modeled forces are the injected bias signals. For the rest, the following discussion employs the estimation in the domain of displacement readouts, as the other approach is currently under investigation.


The MIMO-Multi-Experiment Fisher information matrix for the parameter estimates $\vect{p}_\text{est}$ is the local curvature of the log-likelihood surface around the minimum and is given by
\begin{equation}\label{eq:sys_identification:info_matrix}
\vect{\mathcal{I}}=
\int
\ctranspose{\vect{o}_\text{i}(\omega)} \, \ctranspose{\nabla_{\vect{p}} \vect{T}_{o_\text{i}\rightarrow o}(\omega,\vect{p}_\text{est})} \,
\vect{S}_\text{n}(\omega)^{-1} \,
\nabla_{\vect{p}} \vect{T}_{o_\text{i}\rightarrow o}(\omega,\vect{p}_\text{est}) \, \vect{o}_\text{i}(\omega)\,\text{d}\omega~,
\end{equation}
where $\nabla_{\vect{p}}$ is the gradient with respect to all 7 system parameters. As above, if $\vect{T}_{o_\text{i}\rightarrow o}$ is a $4\times4$-matrix, then $\nabla_{\vect{p}}\vect{T}_{o_\text{i}\rightarrow o}$ is a $7\times4\times4$-tensor and the information is a $7\times7$-matrix as required. The very high SNR regime of the signals in \figref{fig:sys_identification:data_exp1} and \figref{fig:sys_identification:data_exp2} assures that the linear approximation of \eqref{eq:sys_identification:info_matrix} holds true and no corrective terms arise as pointed out by \cite{vallisneri2008} and more recently by \cite{vallisneri2011}. As the inverse of the information matrix provides the estimated covariance matrix, the validity of the linear approximation is checked a posteriori in \sectref{sect:sys_identification:montecarlo} by inspecting the statistics of a Monte Carlo simulation.

\subsection{Whitening} \label{sect:sys_identification:whitening}

The colored noise behavior of a typical LPF run makes mandatory to decorrelate the data used for system identification in order for a generic statistical estimator be unbiased. Consider for example a stationary noisy time-series $o(t)$ with noise PSD $S_{\text{n}}(\omega)$. The SNR of the signal \cite{finn1992} can be recast as
\begin{equation}
\begin{split}
\rho^2 & = \int\frac{\ctranspose{o}(\omega)\,o(\omega)}{S_{\text{n}}(\omega)}\,\text{d}\omega \\
& = \int o^*_\text{w}(\omega)\,o_\text{w}(\omega)\,\text{d}\omega~,
\end{split}
\end{equation}
which can be viewed as the acting of the \textit{whitening filter} $W(\omega)=1/\sqrt{S_{\text{n}}(\omega)}$ on $o(\omega)$ to produce the whitened series
\begin{equation}
o_\text{w}(\omega)=W(\omega)\,o(\omega)~.
\end{equation}
Here ``whitened'' is equivalent to saying that the noise PSD of the filtered series is approximately frequency-independent. The discrete time-domain version of the preceding involves the noise covariance matrix $\vect{C}_\text{n}$
\begin{equation}\label{eq:sys_identification:snr_time_domain}
\begin{split}
\rho^2 & = \transpose{\vect{o}}\,\vect{C}_\text{n}^{-1}\vect{o} \\
& = \transpose{\vect{o}}_\text{w}\,\vect{\Lambda}_\text{n}^{-1}\vect{o}_\text{w}~,
\end{split}
\end{equation}
which again can be viewed as the acting of the whitening filter $\vect{W}$, an orthogonal matrix satisfying $\vect{C}_\text{n}^{-1}=\transpose{\vect{W}}\vect{\Lambda}_\text{n}^{-1}\vect{W}$ \footnote{In fact, if $\vect{C}_\text{n}=\transpose{\vect{U}}\vect{\Lambda}_\text{n}\vect{U}$ where $\vect{U}$ is an orthogonal matrix and $\vect{\Lambda}_\text{n}$ is the eigen-decomposition of $\vect{C}_\text{n}$, then it turns out that $\vect{U}^{-1}=\transpose{\vect{W}}$.}, on $\vect{o}$ to produce the whitened unit-variance series
\begin{equation}
\vect{o}_\text{w}=\vect{W}\vect{o}~.
\end{equation}
As above, ``whitened'' means that the process diagonalizes the covariance matrix, so that $\vect{\Lambda}_\text{n}$ effectively becomes an identity matrix.

For simulation and analysis purposes, whitening a time-series is formally the inverse process of noise generation. Whitening filters are obtained by performing a fit in the $z$-domain to the inverse of the estimated PSD \footnote{Throughout this thesis, if not otherwise stated, it is assumed that a PSD is estimated by means of the Welch (modified periodogram) method \cite{welch1967} using a 4-sample 92-dB Blackman-Harris window \cite{harris1978}, 16-segments averaged, $66\%$ overlap and mean detrended.\label{foot:sys_identification:psd}}. \figref{fig:sys_identification:whitening} reports an example of whitening \footnote{Data filtering can produce fake transients at the beginning of the filtered time-series. To avoid this possibility, an initial segment of data is usually cut away.} a typical 28-hour run of interferometric noise. The effect of the whitening filters, as required, is to flatten the noise shapes, i.e., to decorrelate the time-series.

\figuremacroW{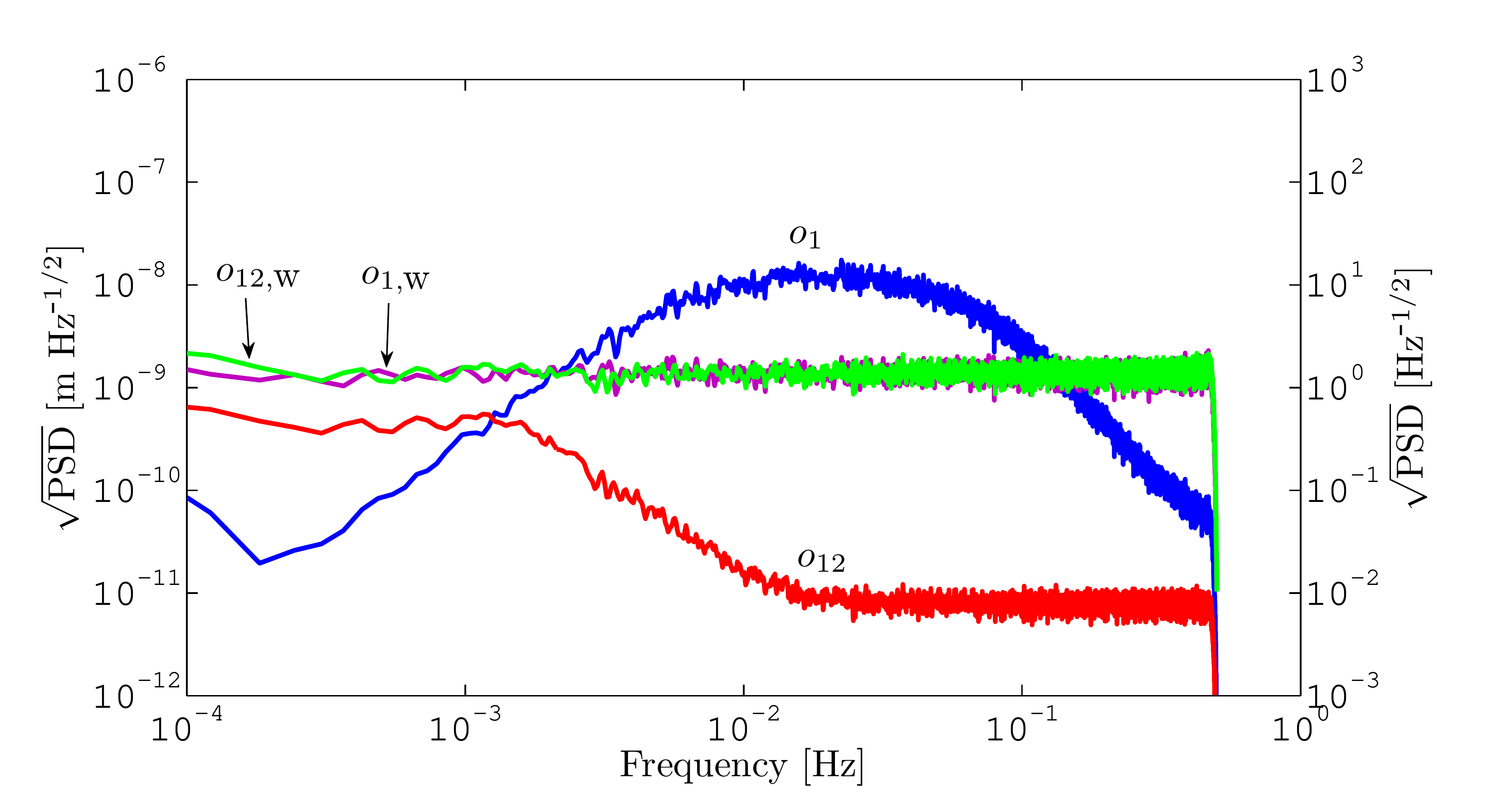}{Whitening of a simulated noise run. $o_1$ and $o_{12}$ are the two interferometer readings with PSD reported on the basis of the scale on the left end side. $o_{1,\text{w}}$ and $o_{12,\text{w}}$ are the whitened counterparts with PSD reported on the basis of the scale at the right end side. They show how the whitening filters can flatten the noise shapes. The convolution with a low-pass filter of the data resampling from 10 to $\unit[1]{Hz}$ is the cause of the drop around $\unit[0.5]{Hz}$.}{fig:sys_identification:whitening}{0.8}

Despite the PSD shapes which seem reasonably good at first sight, a residual red component still persists. \tabref{tab:sys_identification:whitening} reports two higher-order moments (skewness and excess kurtosis) of the empirical distribution together with their uncertainties \cite{press}. By inspecting the values, it turns out that the sample mean of the differential channel $o_{12}$ is not compatible with zero, as one would expect. Usually, a first or second order polynomial fit is necessary to subtract that residual component. The result is not surprising: the intrinsical difficulty is that the whitening process is performed on a restricted frequency band (the one of the estimated PSD) and low-frequency components may survive after the filtering.
\begin{table}[!htbp]
\caption{\label{tab:sys_identification:whitening}\footnotesize{Sample mean $\mu$, standard deviation $\sigma$ and higher moments, the sample skewness $\gamma_1$ and the excess kurtosis $\gamma_2$, for the whitened data channels $o_1$ and $o_{12}$. Assuming Gaussian-distributed data, the approximate standard deviations are $\sigma_\mu\simeq\sigma/\sqrt{N}$, $\sigma_\sigma\simeq\sigma/\sqrt{2N}$, $\sigma_{\gamma_1}\simeq\sqrt{\nicefrac{6}{N}}$, $\sigma_{\gamma_2}\simeq\sqrt{\nicefrac{24}{N}}$, with $N$ the number of data samples.}}
\centering
\begin{tabular}{ l D{!}{\,\pm\,}{6.5} D{!}{\,\pm\,}{5.5} D{!}{\,\pm\,}{3.7} D{!}{\,\pm\,}{2.7}}
\hline
\hline
\multicolumn{1}{c}{Data} & \multicolumn{1}{c}{$\mu$} & \multicolumn{1}{c}{$\sigma$}  & \multicolumn{1}{c}{$\gamma_1$} & \multicolumn{1}{c}{$\gamma_2$} \\
\hline
$o_{1,\text{w}}$               & 0.008 ! 0.003         & 0.970 ! 0.002     & (-5 ! 8)$\e{-3}$    & (0 ! 2)$\e{-2}$ \\
$o_{12,\text{w}}$            & -0.254 ! 0.003   & 1.002 ! 0.002     & (0 ! 8)$\e{-3}$          & (3 ! 2)$\e{-2}$ \\
\hline
\hline
\end{tabular}
\end{table}

The extent to which the idea of this section holds true depends on the assumption of stationarity and Gaussianity. Even though for LPF the interferometric noise is not explicitly dependent on the system parameters, it may depend implicitly through the coupling between the external force noise and the system response. Yet, as it will be discussed later in this chapter, the estimated equivalent acceleration noise depends explicitly on the system parameters through the transfer matrix given by the differential operator.

As a matter of fact, a non-stationarity in any of the system parameters implies a non-stationarity in the noise. In fact, if $o=o\left(t,p(t)\right)$ is a generic interferometer readout depending, for simplicity, on just one parameter fluctuating of $\delta p$ around the nominal value $p_0$, then to first order $o \simeq o_0+o'\,\delta p$, where $o_0=o(t,p_0)$
and $o'=\left.\partial o(t,p)/\partial o\right|_{p_0}$. For a zero-mean process the total variance is
\begin{equation}\label{eq:sys_identification_noise_variance}
\var[o] \simeq \var[o_0]+\var'[o_0]\,\delta p+\var[o']\,\delta p^2~,
\end{equation}
where the linear and quadratic terms come from the covariance between $o_0$ and $o'$
and the variance of $o'$ itself (see Appendix \ref{sect:appendix:non_stationary_noise} for details). Therefore, if any of the system parameters fluctuates, noise is likely to become non-stationary. In LPF all PSDs must be estimated piecewise along data segments approximately stationary on a timescale given by the one of the fluctuating parameter. The converse, i.e., a non-stationarity in the noise implies a non-stationarity in any of the parameters is not assured, since other effects, independent from those parameters, may still be relevant. For example, \sectref{sect:sys_identification:non_gaussianities} describes the possible existence of glitches, a non-stationary behavior in the noise, and its impact to system identification. Instead, Appendix \ref{sect:sys_identification:non-stationary_noise_time_freq} introduces the time-frequency approach in the study of non-stationarity noise.

\subsection{Search algorithm}

The joint log-likelihood \eqref{eq:sys_identification:log-likelihood} for two experiments, two readouts each, is implemented in time domain by means of FFT/IFFT the whitened time-series. The relevant iteration steps of the process taking to the final estimates of the system parameters, in loop of increasing accuracy, are:
\begin{enumerate}
  \item the whitening filters are estimated on a long noise run, as in \sectref{sect:sys_identification:whitening};
  \item the interferometric readouts of each experiments are whitened;
  \item the templates are generated according to \eqref{eq:sys_identification:templates} for the current parameter values;
  \item the templates are whitened;
  \item the log-likelihood is evaluated, i.e., ``models fit the data'', for the current parameter values;
  \item the parameter values are updated according to the adopted optimization scheme.
\end{enumerate}
From the optimization viewpoint, the log-likelihood is named the \textit{merit function}, i.e., the one being minimized as the parameter values are updated. \figref{fig:sys_identification:sys_identification_scheme} shows a sketch of the whole process of system identification. The data production provides for the noise run and the experiments, with both interferometric readouts and injected biases. Instead, the modeling provides for the proper transfer matrix being used for simulating the template signals. Finally, the data analysis concerns the estimation of the whitening filters and the algorithm for the log-likelihood minimization.

\figuremacroW{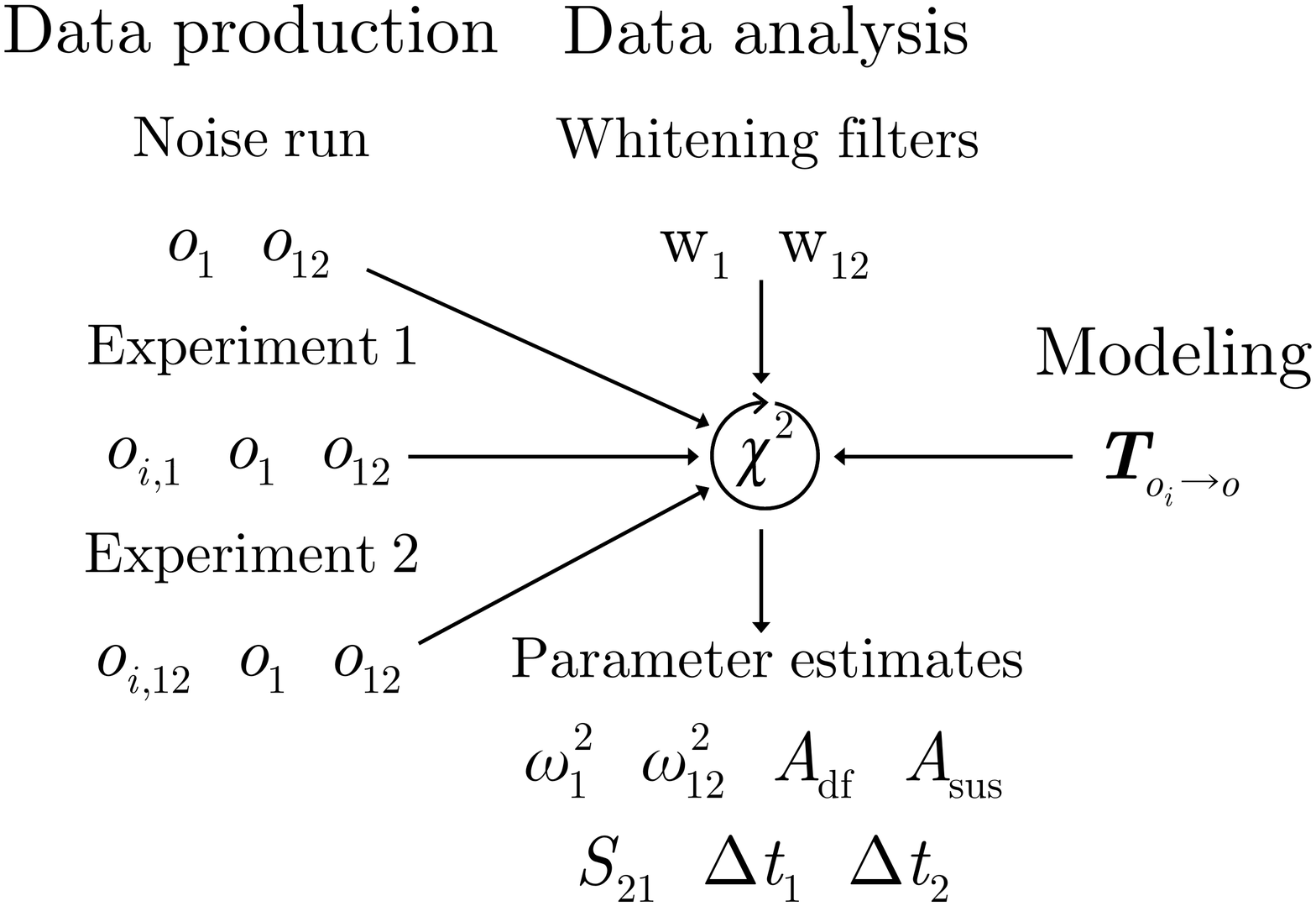}{Sketch of the system identification process for the two simulated experiments along the optical axis. Noise run and experiments pertain to data production. The modeling provides for the transfer matrix being used for simulation and analysis. For system identification, data analysis comprises the estimation of whitening filters and the log-likelihood optimization. The estimated parameters are output together with their covariance matrix.}{fig:sys_identification:sys_identification_scheme}{0.7}

The algorithm performs a log-likelihood minimization by taking advantage of the most recent developments in numerical non-linear optimization \cite{press}. During the work for this thesis, an investigation of different optimization algorithms was carried out. Non-standard schemes like the \textit{simulated annealing}, \textit{genetic algorithms} and the \textit{pattern search}, with or without a multi start (an initial Monte-Carlo-like exploration of the parameter space in which the initial most likely points are taken into account for further processing), were considered for the purposes of system identification. They were also compared to a mixed strategy employing more standard and widely-used optimization algorithms applied in sequence:
\begin{enumerate}
  \item the \textit{preconditioned conjugate gradient search} (alternatively, the quasi-Newton method) explores the parameter space to large scales;
  \item the \textit{derivative-free simplex} allows to reach the required numerical accuracy.
\end{enumerate}
The key advantage of mixing different approaches is that the global structure of the parameter space can be explored while keeping the numerical accuracy. Such an investigation proved that for the LPF system identification non-standard schemes have comparable performances with respect to the one proposed above which is assumed for the rest. The optimization is numerically controlled and stopped until either the function tolerance or the average parameter tolerance meets the requirement of $1\e{-4}$. The final parameter estimates are output from the fitting tool, together with the estimated covariance matrix, obtained by inverting the Fisher information matrix \eqref{eq:sys_identification:info_matrix} around the minimum.


Before showing the application to data simulated specifically for LPF, the tool was checked against more simpler cases like the linear fit (which is analytically solvable to any order), the chirped sine and the harmonic oscillator \cite{S2-UTN-TN-3071,S2-UTN-TN-3072}. The result is that there are no systematic errors and parameter uncertainties are in accordance to a Monte Carlo simulation. A similar check for LPF is discussed in \sectref{sect:sys_identification:montecarlo}.

\subsection{ESA simulator} \label{sect:sys_identification:simulator}

A very important test-bench on both system modeling and validation of the estimation techniques is the analysis of realistic data, closer to the actual LPF mission than the ones simulated and shown in this work. A real LPF simulator, named Off-line Simulation Environment (OSE), provided by ESA and written by ASTRIUM has given the chance to promptly analyze the data as they were realistically produced during the mission. The OSE is a state-space representation of a 3-dimensional LPF model written under the MATLAB$^\circledR$ and Simulink$^\circledR$ \cite{simulink} environments. It contains the most relevant disturbances and noise sources, the same actuation algorithms for drag-free, electrostatic suspension and attitude controls (DFACS) embedded in LPF, all couplings within the dynamics along the optical axis and between different degrees of freedom. The OSE was written to mainly check all procedures, the mission timeline, the experiments and validate the noise budget.

Several extended data analysis operational exercises were called in the past 2 years, where parameter estimation had a pivotal role, and therefore very similar to a mock data challenge, where data production is strictly separated from data analysis. The operational exercises culminated with the sixth one targeted to parameter estimation using a linear fit with singular value decomposition, a Markov-chain Monte Carlo method and the one described in this thesis. The first application of system identification on that operational exercise is contained in \cite{congedo2011}. The final conclusion of the activity on the same exercise is recently described in \cite{nofrarias2011}. The three methods are apparently in good agreement to each other, particularly the first and third approaches, but an investigation of the fit residuals, like the one in \figref{sect:sys_identification:initial_guess}, shows a mismatch in the first experiment between data and model at high frequency. The fact is confirmed by a statistical comparison between the residual PSDs to a noise-only measurement with a very general and model-independent method based on the Kolmogorov-Smirnov test \cite{ferraioli2012}. The explanation of such a mismatch will be given in the near future with further operational exercises and much more detailed knowledge of the simulator.

\subsection{Monte Carlo validation} \label{sect:sys_identification:montecarlo}

The aim of this section is to statistically validate the estimation method presented so far. A Monte Carlo simulation of $1000$ different noise realizations is used to check for consistency of the method. The estimation is identically repeated at each step, enabling fine tuning and the study of the statistics for every system parameter.

\tabref{tab:sys_identification:montecarlo} reports on the comparison between the mean best-fit values and the true values: the accordance is at the level of at most 2 standard deviations and demonstrates that the estimation method is statistically unbiased. Secondarily, it shows the best-fit standard deviations, i.e., the parameter fluctuations due to noise, compared to the mean expected standard deviations (the mean fit errors).
\begin{table}[!htbp]
\caption{\label{tab:sys_identification:montecarlo}\footnotesize{Monte Carlo validation of 1000 independent noise realizations on which parameter estimation is repeated identically at each step. The mean best-fit values are compatible with the true values within 2 standard deviations. The terms in brackets are the error relative to the rightmost digit. The mean expected standard deviations (estimated from the fit) and the best-fit standard deviations are approximately the same order of magnitude. The mean log-likelihood is $\chi^2=0.96$ with $\nu=79993$.}}
\centering
\begin{tabular}{l D{.}{.}{3.5} D{.}{.}{3.9} D{!}{\times}{4.5} D{!}{\times}{3.5}}
\hline
\hline
\multicolumn{1}{l}{\multirow{2}{*}{Parameter}} & \multicolumn{1}{c}{\multirow{2}{*}{True}} & \multicolumn{1}{c}{Mean} & \multicolumn{1}{c}{Best-fit} & \multicolumn{1}{c}{Mean} \\
& & \multicolumn{1}{c}{best-fit} & \multicolumn{1}{c}{st.\,dev.} &  \multicolumn{1}{c}{exp.\,st.\,dev.} \\
\hline
$\omega_1^2\,[10^{-6}\,\text{s}^{-2}]$              & -1.303         & -1.303006(7)         & 2!10^{-4}  & 1!10^{-3} \\
$\omega_{12}^2\,[10^{-6}\,\text{s}^{-2}]$           & -0.698         & -0.697998(6)         & 2!10^{-4}  & 5!10^{-4} \\
$S_{21}\,[10^{-4}]$                                 & 0.9            & 0.90004(9)           & 3!10^{-3}  & 4!10^{-3} \\
$A_\text{df}$                                     & 1.003          & 1.00297(1)           & 4!10^{-4}  & 4!10^{-4} \\
$A_\text{sus}$                                    & 0.9999         & 0.9999001(1)         & 4!10^{-6}  & 2!10^{-5} \\
$\Delta t_1\,[\text{s}]$                            & 0.06           & 0.059995(3)          & 9!10^{-5}  & 3!10^{-4} \\
$\Delta t_{12}\,[\text{s}]$                         & 0.05           & 0.05000(3)           & 8!10^{-4}  & 1!10^{-3} \\
\hline
\hline
\end{tabular}
\end{table}

\begin{figure}[!htbp]
\centering
\begin{tabular}{*{3}{@{\hspace{-6pt}}c@{\hspace{-6pt}}}}
\includegraphics[width=0.27\textwidth]{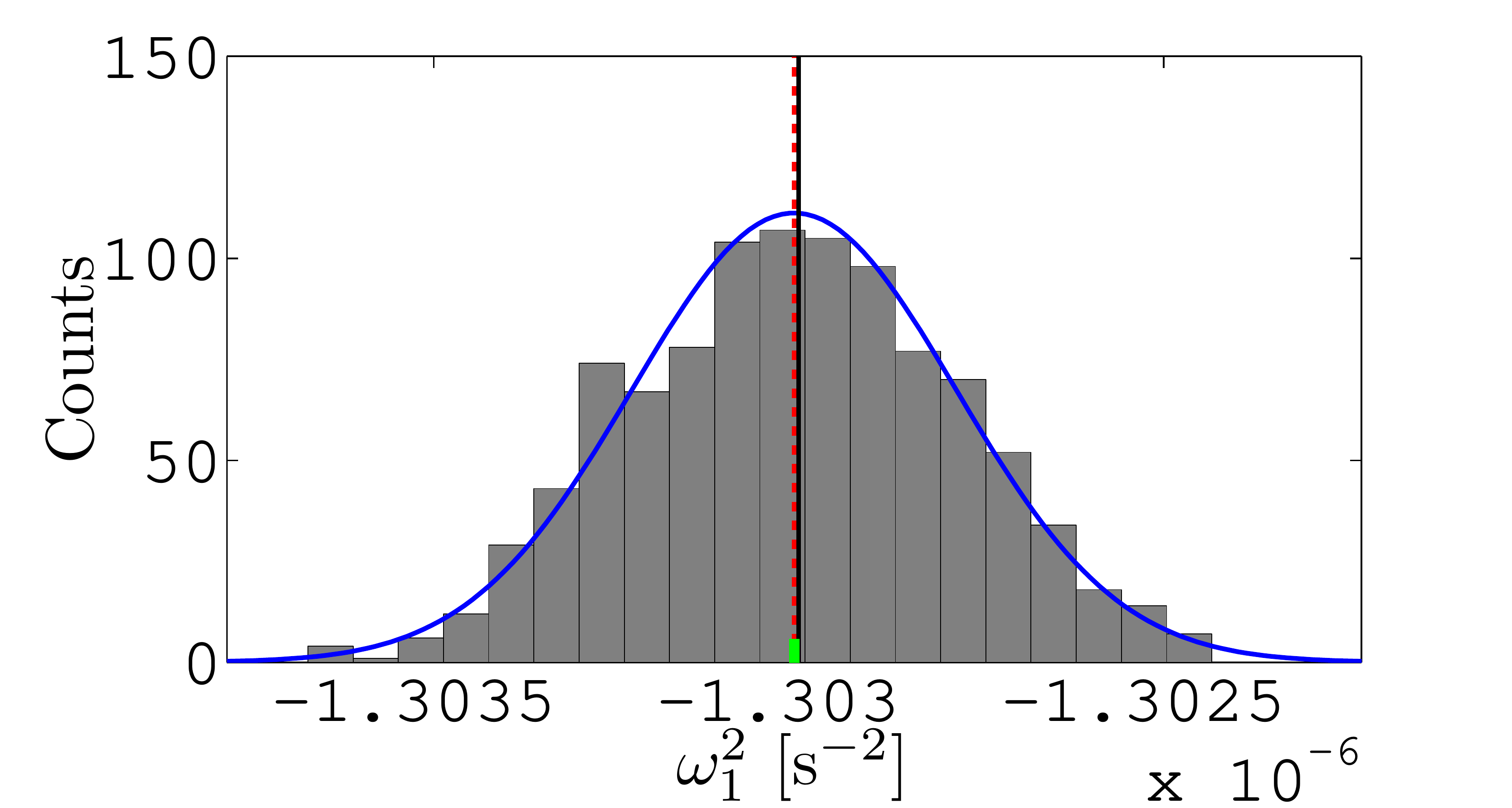} &
\includegraphics[width=0.27\textwidth]{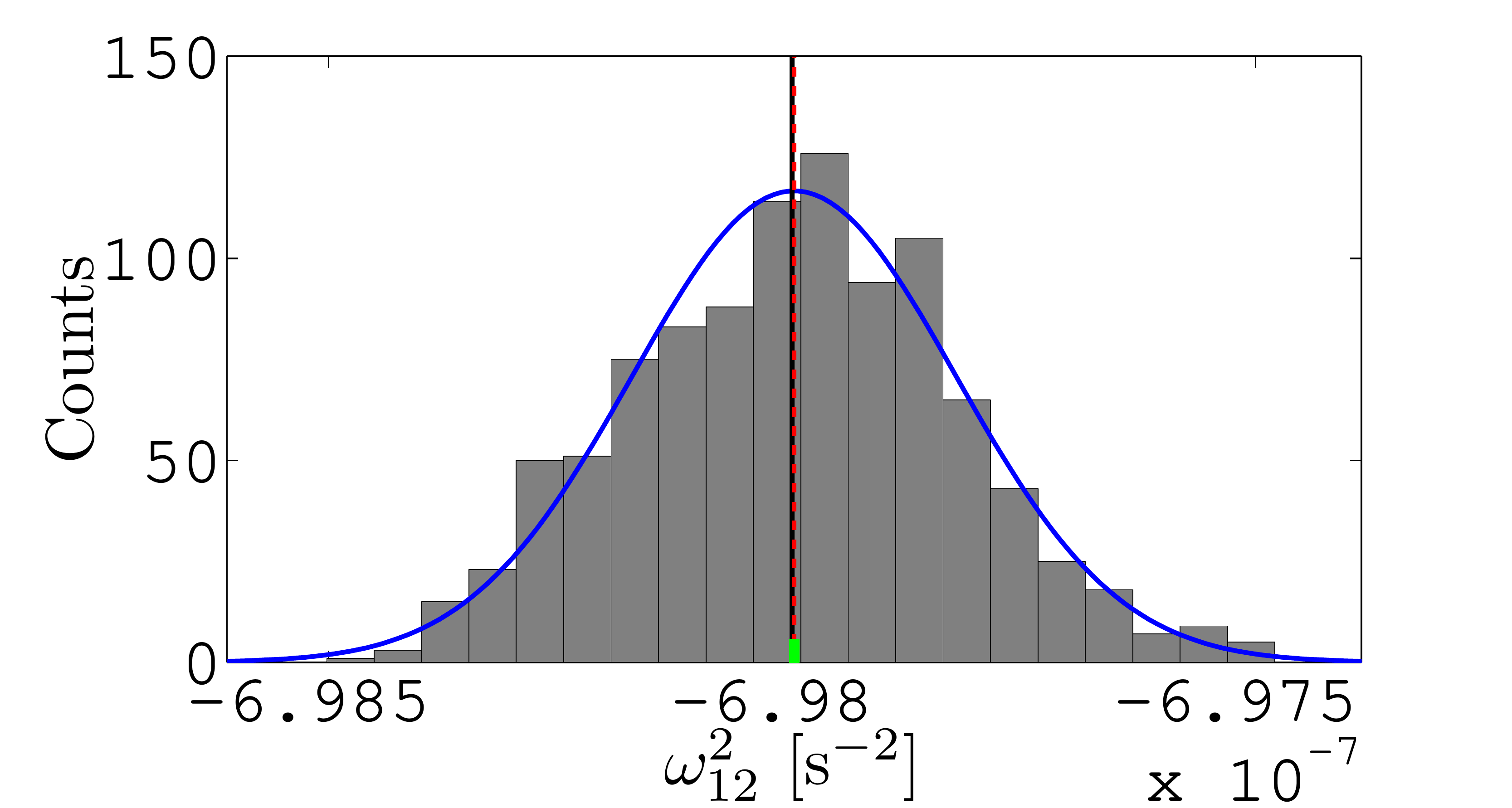} &
\includegraphics[width=0.27\textwidth]{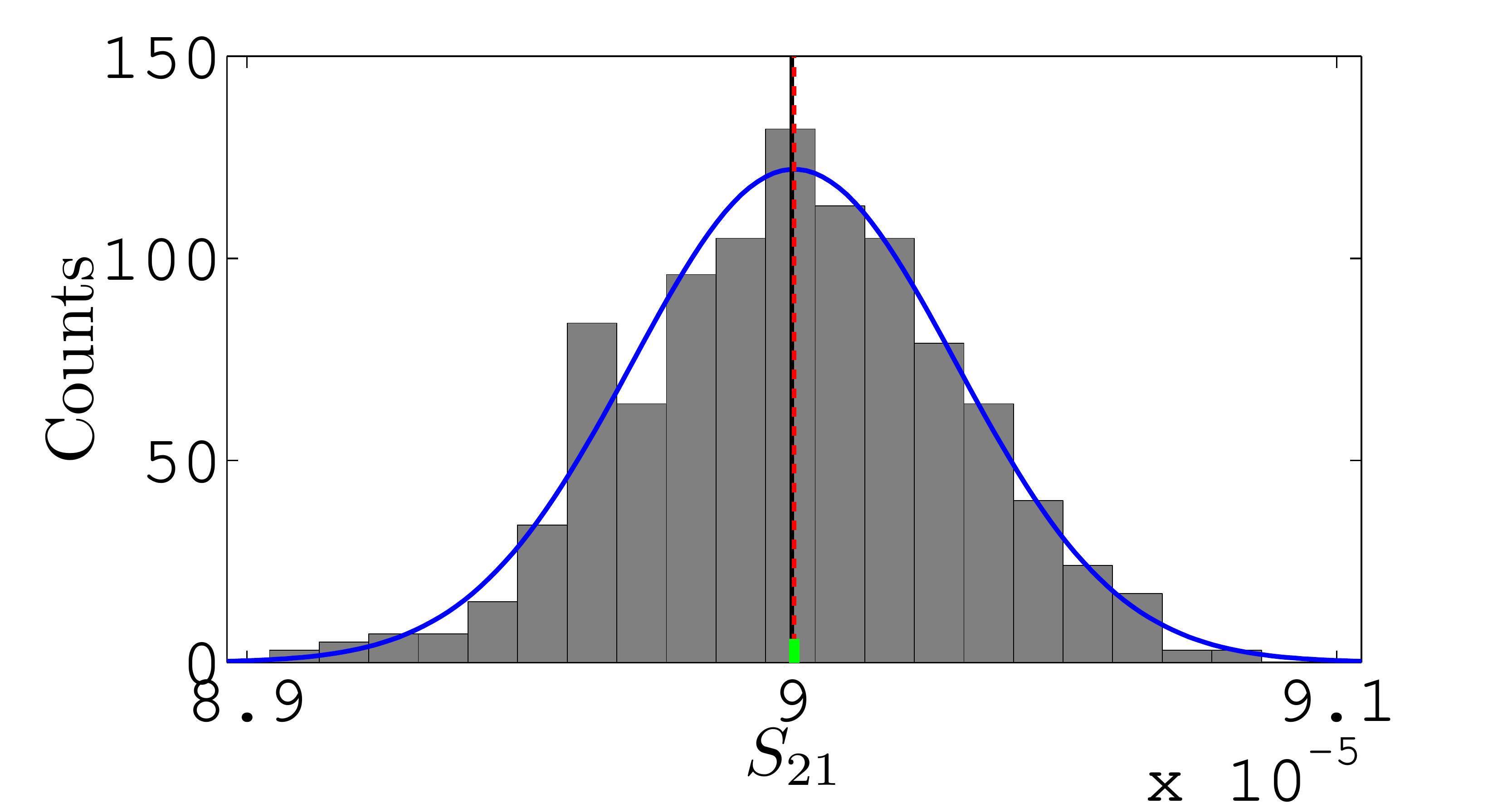} \\
\footnotesize{(a)} &
\footnotesize{(b)} &
\footnotesize{(c)} \\
\end{tabular}
\begin{tabular}{*{4}{@{\hspace{-6pt}}c@{\hspace{-6pt}}}}
\includegraphics[width=0.27\textwidth]{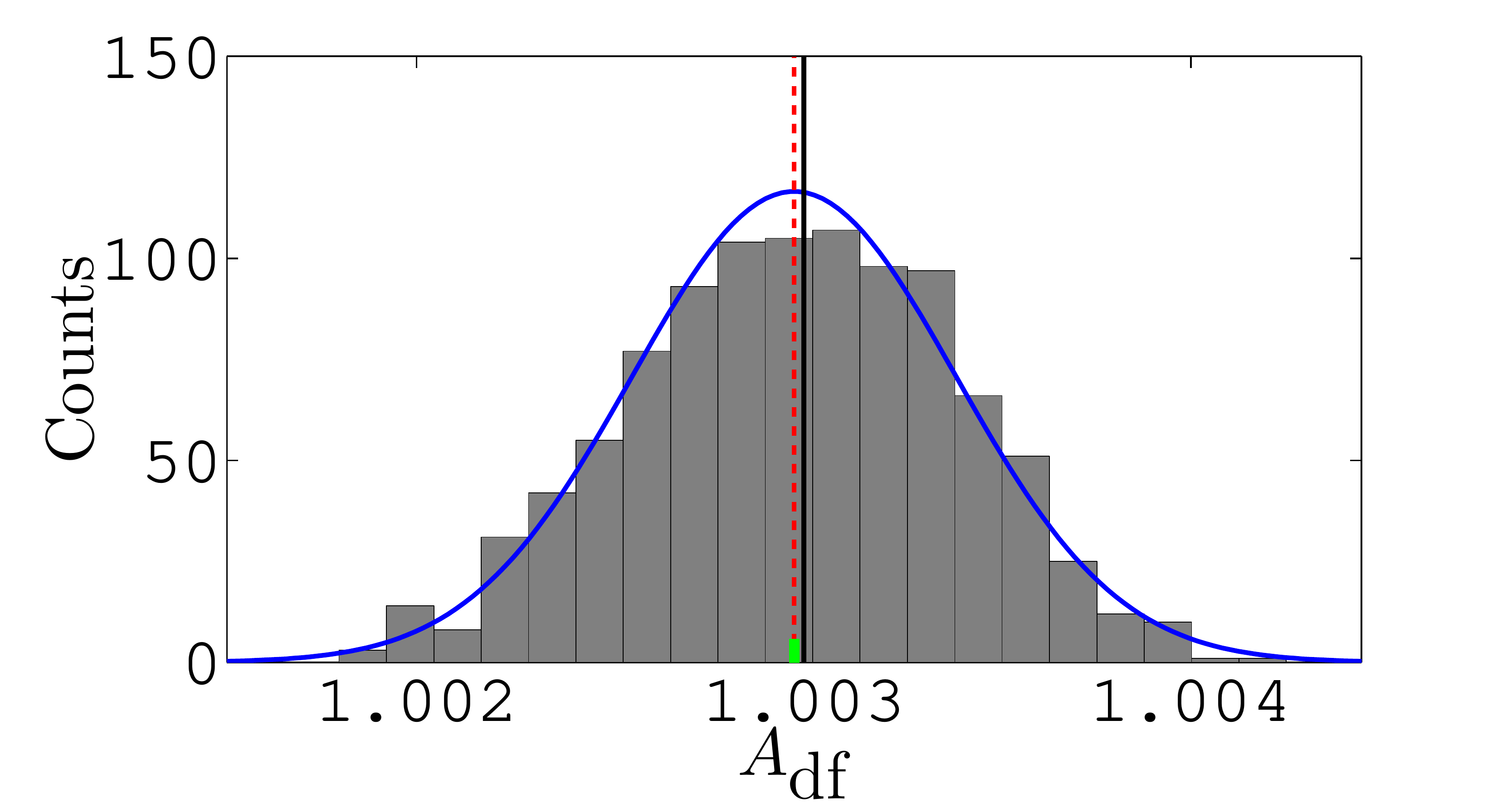} &
\includegraphics[width=0.27\textwidth]{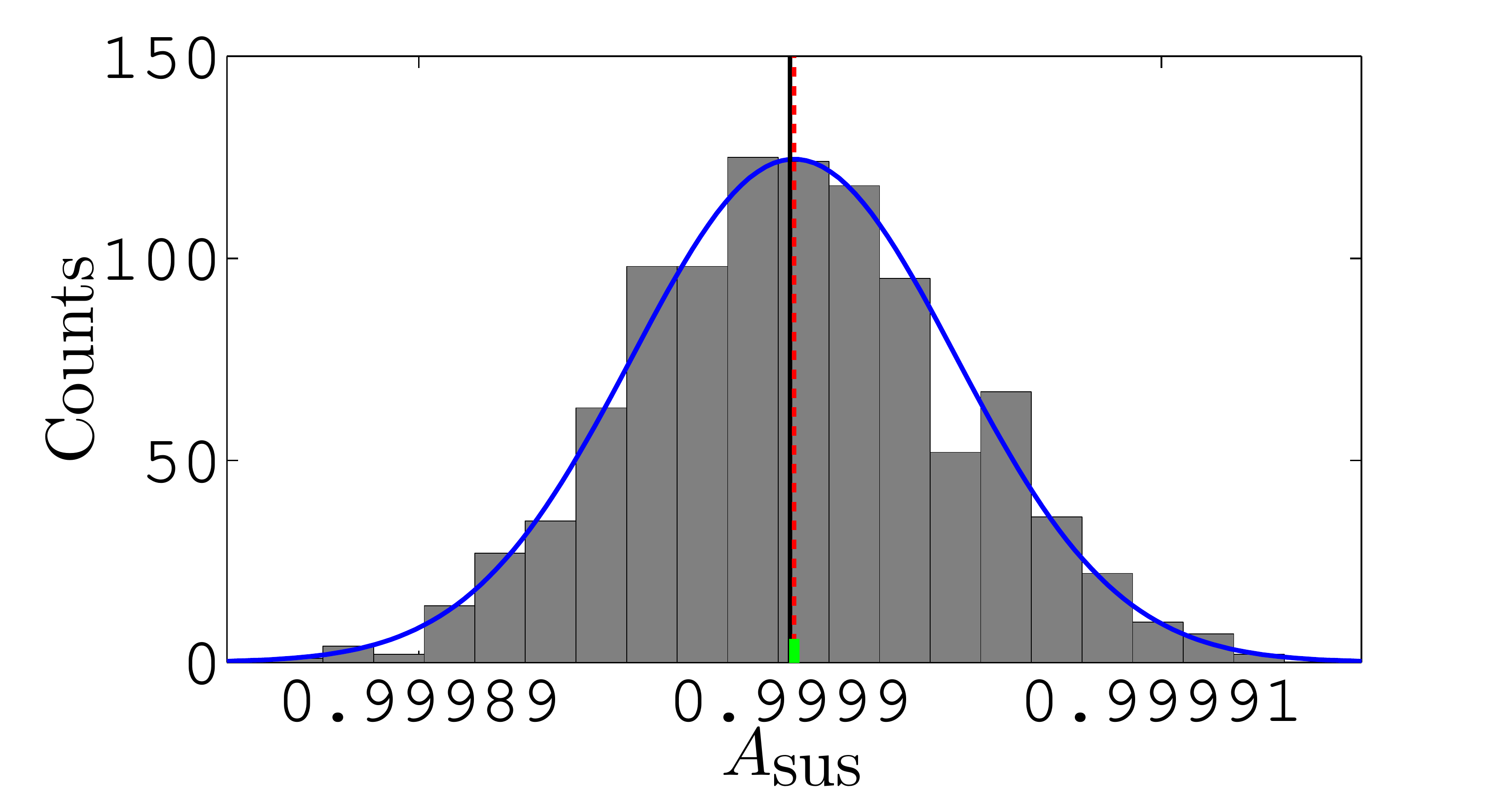} &
\includegraphics[width=0.27\textwidth]{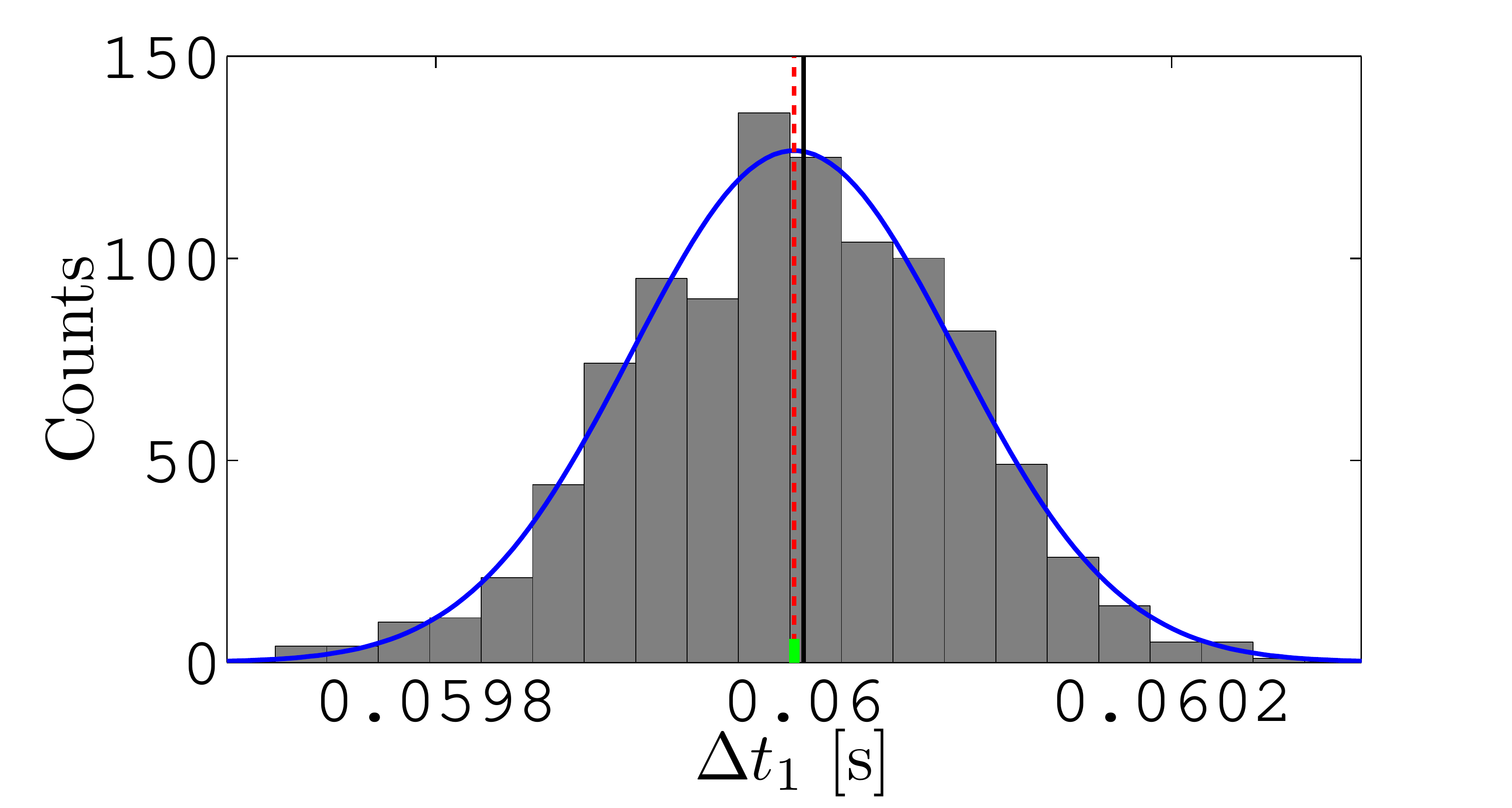} &
\includegraphics[width=0.27\textwidth]{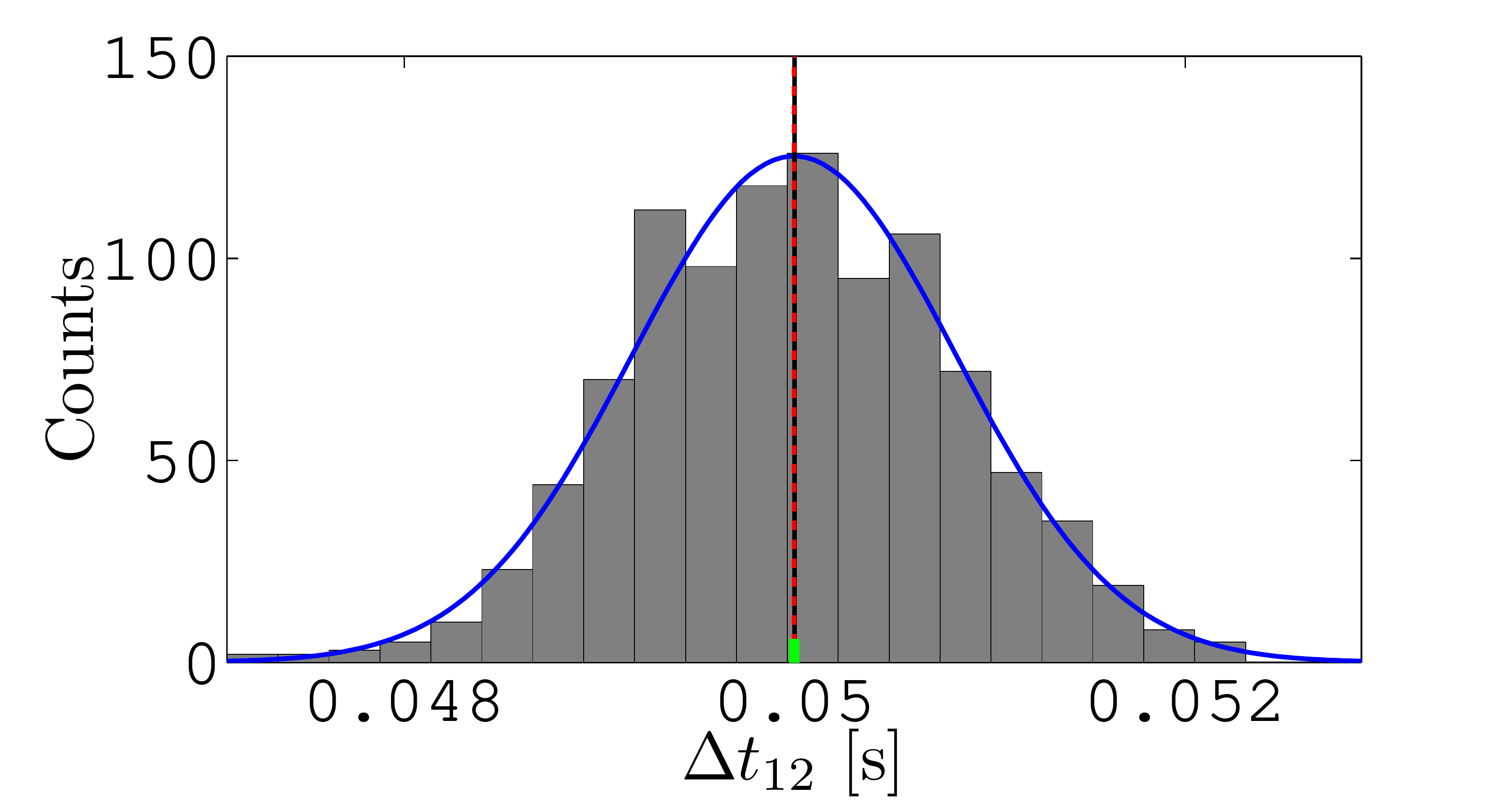} \\
\footnotesize{(d)} &
\footnotesize{(e)} &
\footnotesize{(f)} &
\footnotesize{(g)} \\
\end{tabular}
\caption{\footnotesize{Monte Carlo validation of 1000 independent noise realizations on which parameter estimation is repeated identically at each step. The plots show the statistics for all parameter estimates (a)-(g). The scaled Gaussian PDF is evaluated at the sample mean (dashed vertical lines) and sample standard deviation (half horizontal bars), which are compared to the true values (solid vertical lines).}}
\label{fig:sys_identification:montecarlo_parameters}
\end{figure}

\figref{fig:sys_identification:montecarlo_parameters} shows a more in-depth analysis of all parameter statistics. The accordance between the sample statistics of the Monte Carlo simulation and the scaled theoretical Gaussian Probability Density Function (PDF) (evaluated at the sample mean and standard deviation) is self-evident and demonstrates that: (i) the estimation is statistically unbiased; (ii) the parameters are Gaussian distributed.

\begin{figure}[!htbp]
\centering
\begin{tabular}{*{3}{@{\hspace{-6pt}}c@{\hspace{-6pt}}}}
\includegraphics[width=0.27\textwidth]{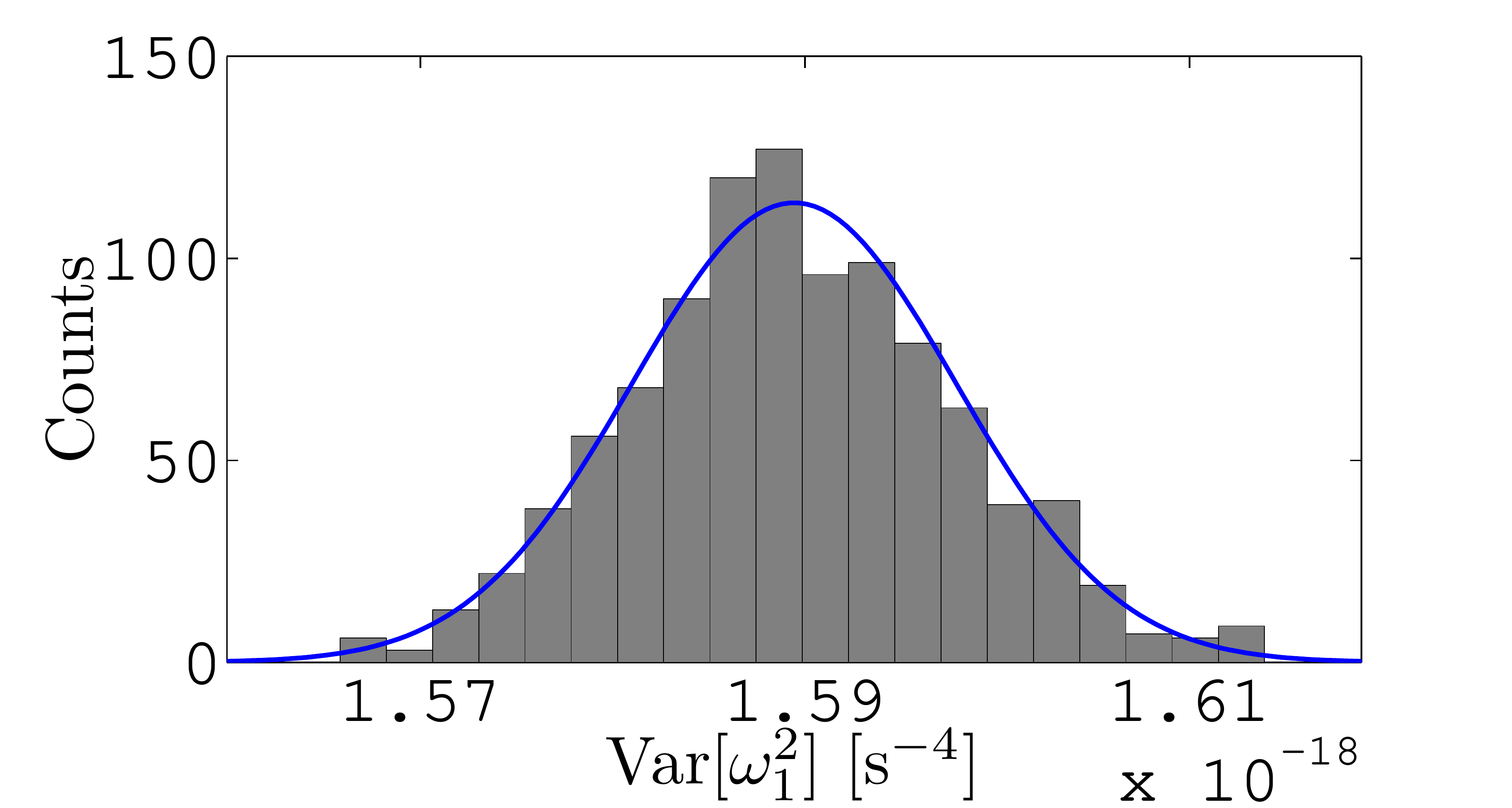} &
\includegraphics[width=0.27\textwidth]{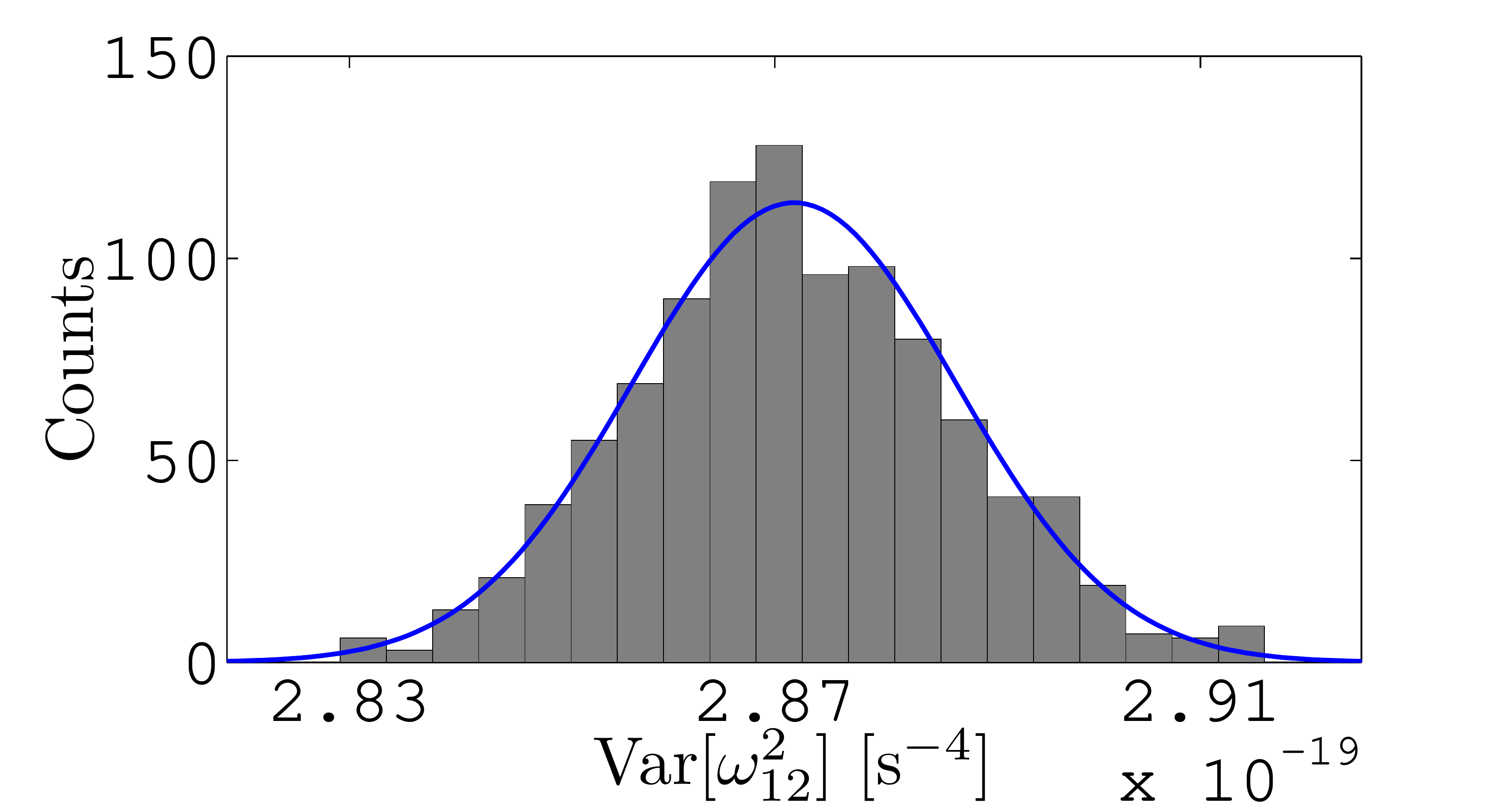} &
\includegraphics[width=0.27\textwidth]{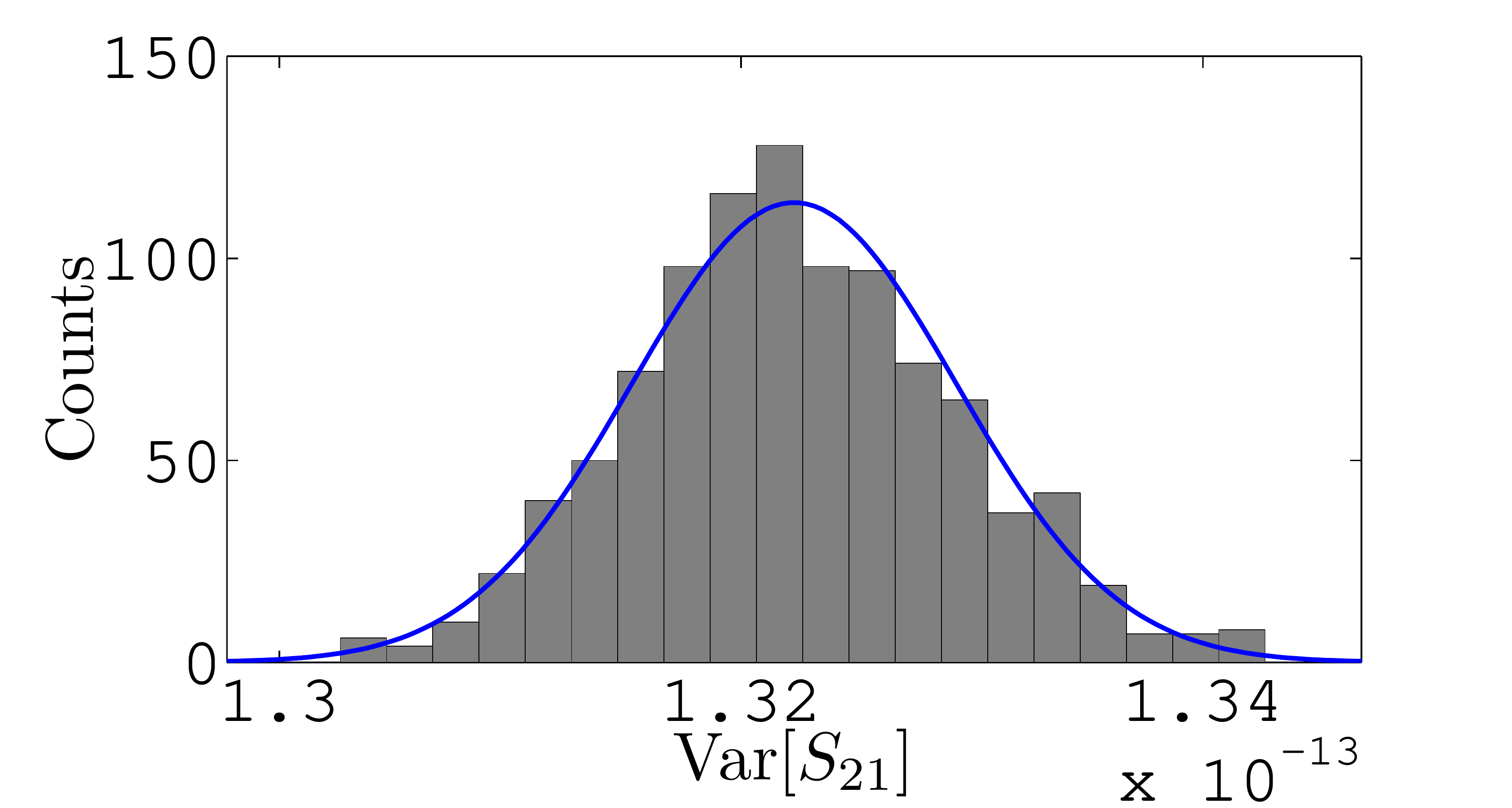} \\
\footnotesize{(a)} &
\footnotesize{(b)} &
\footnotesize{(c)} \\
\end{tabular}
\begin{tabular}{*{4}{@{\hspace{-6pt}}c@{\hspace{-6pt}}}}
\includegraphics[width=0.27\textwidth]{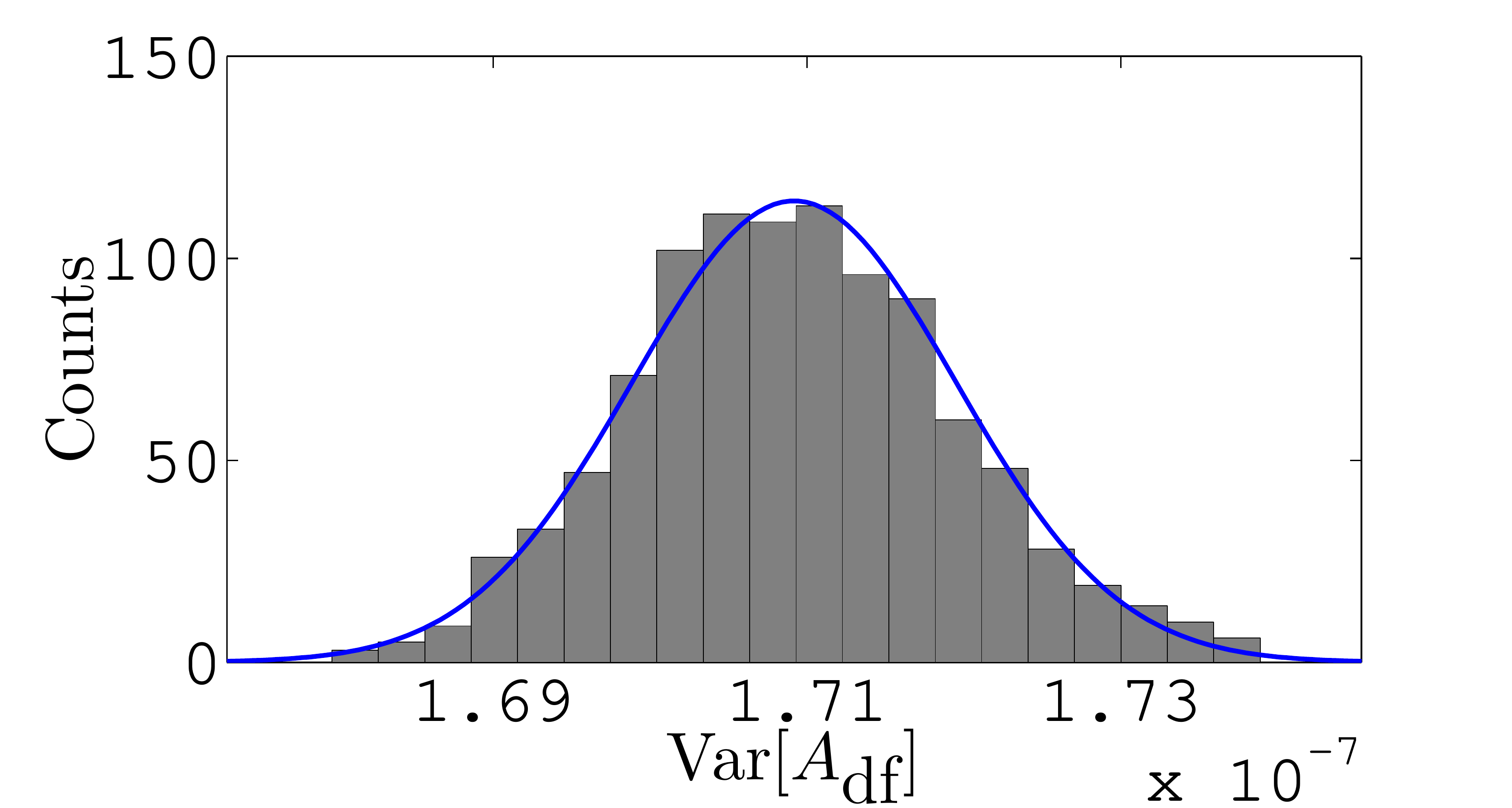} &
\includegraphics[width=0.27\textwidth]{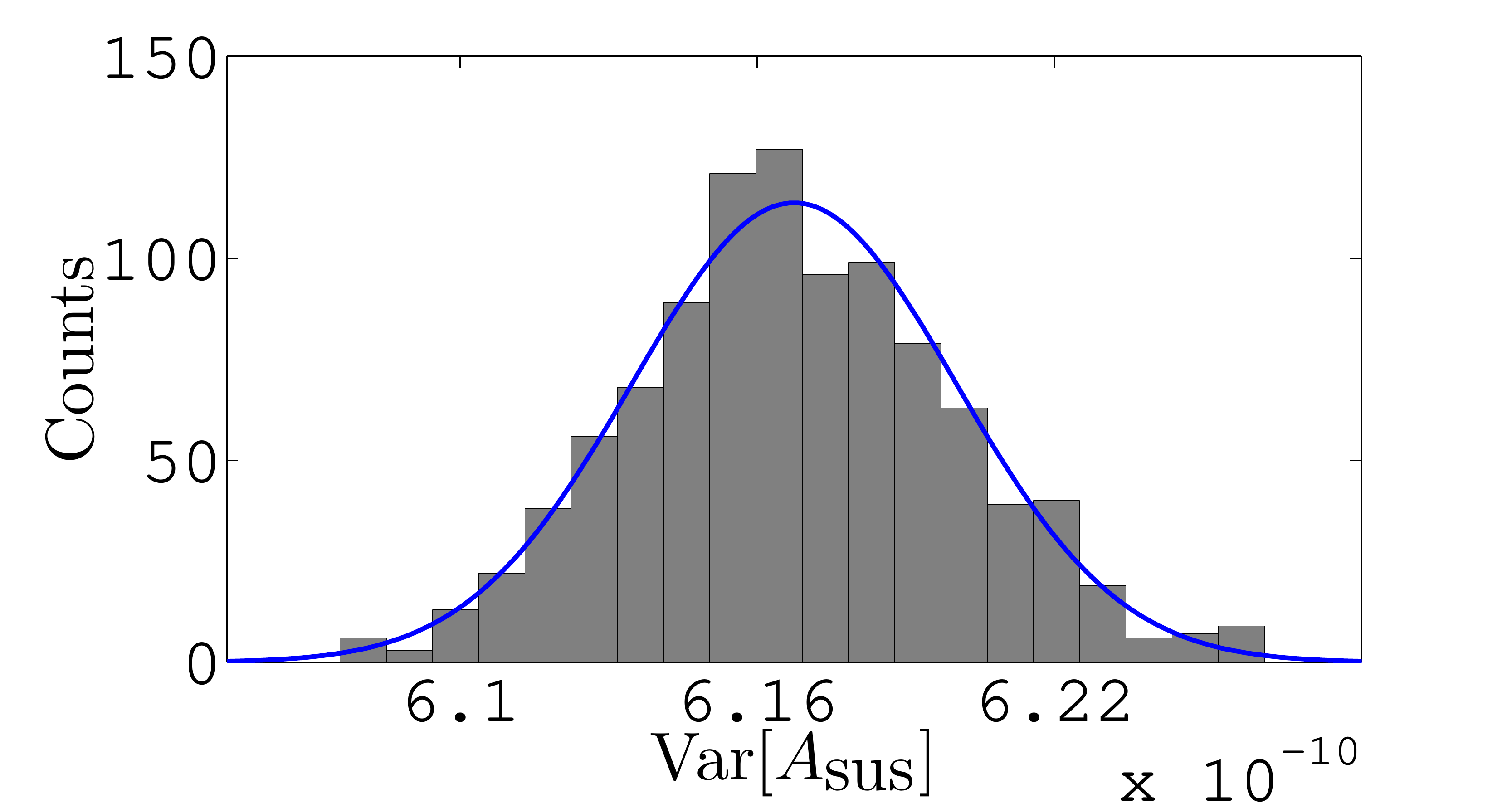} &
\includegraphics[width=0.27\textwidth]{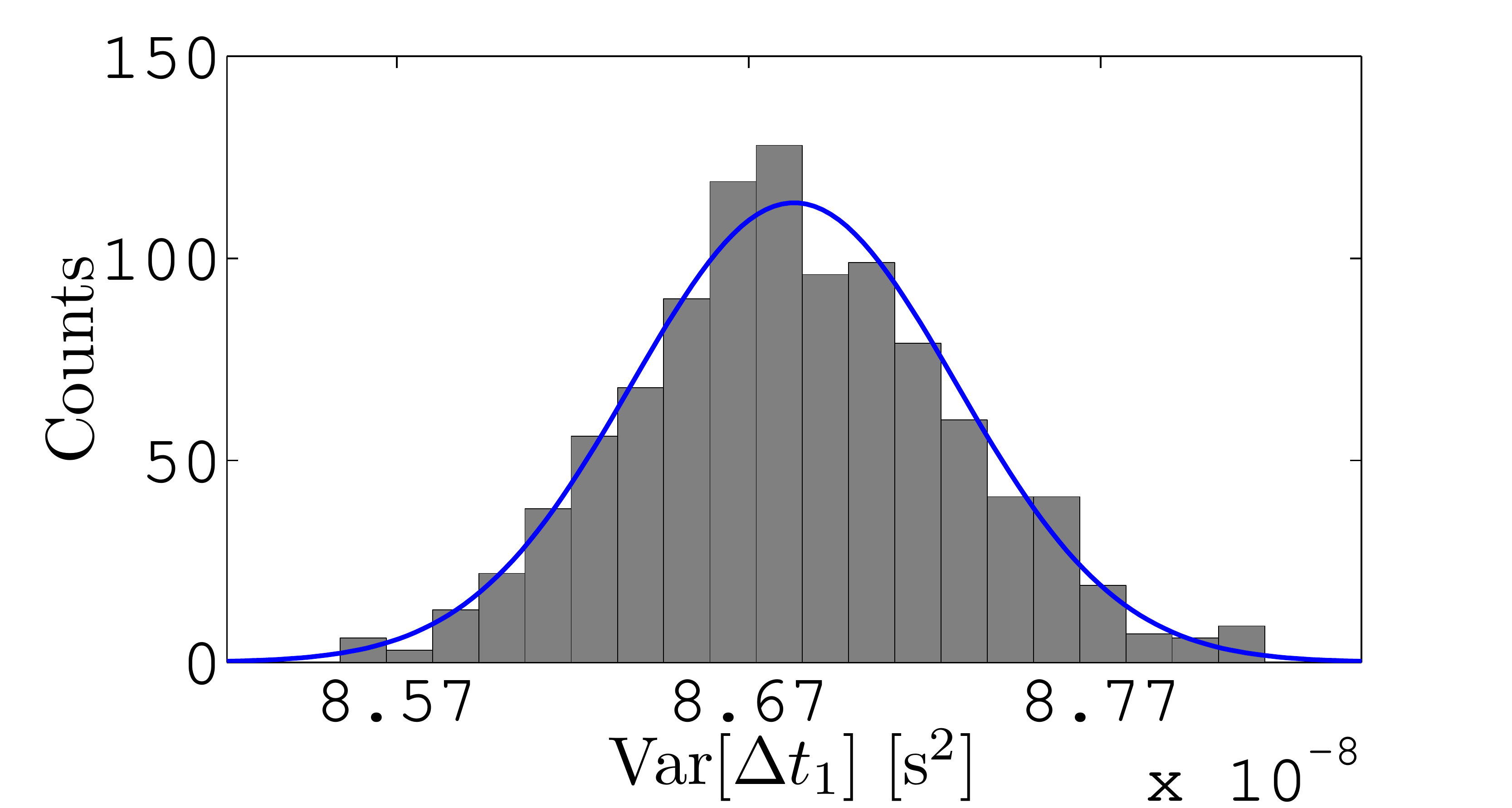} &
\includegraphics[width=0.27\textwidth]{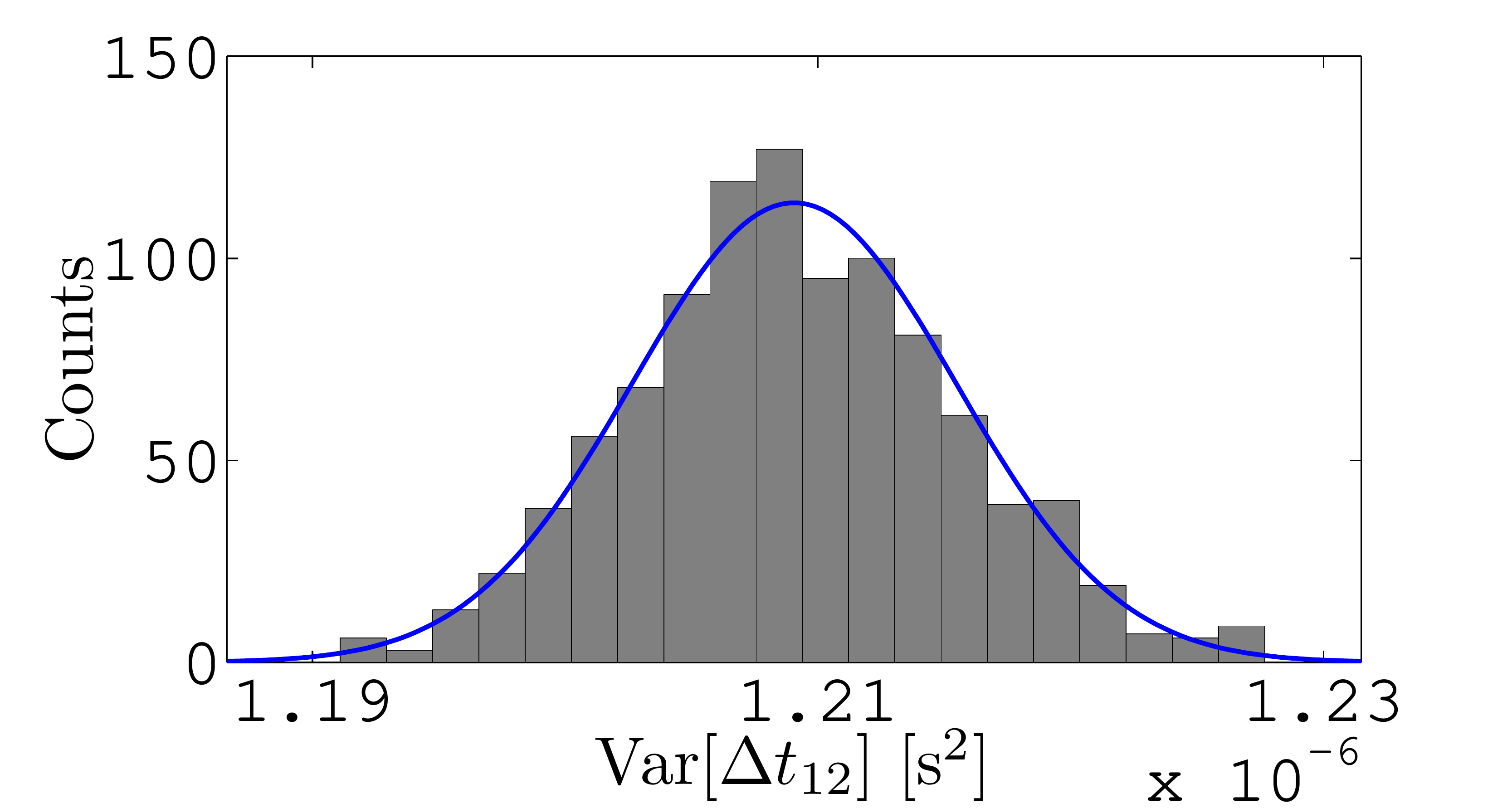} \\
\footnotesize{(d)} &
\footnotesize{(e)} &
\footnotesize{(f)} &
\footnotesize{(g)} \\
\end{tabular}
\caption{\footnotesize{Monte Carlo validation of 1000 independent noise realizations on which parameter estimation is repeated identically at each step. The plots show the statistics for all parameter variances (a)-(g). The scaled Gaussian PDF is evaluated at the sample mean and standard deviation.}}
\label{fig:sys_identification:montecarlo_uncertainties}
\end{figure}

Analogously, \figref{fig:sys_identification:montecarlo_uncertainties} shows the statistics for the estimated variances. Theory prescribes that the variance must be $\chi^2$ distributed, but for $\nu=79993$ the $\chi^2$ distribution tends to a Gaussian distribution with very good approximation, as is clear from the plots.

Appendix \ref{sect:appendix:montecarlo} also discusses some other interesting features of the Monte Carlo statistics, like the parameter correlation, related to the rotation of the log-likelihood paraboloid principal axis around the minimum, and the scatter of the estimation chains due to the noise fluctuation.

\begin{figure}[!htbp]
\centering
\begin{tabular}{*{2}{@{\hspace{-10pt}}c@{}}}
\includegraphics[width=0.5\columnwidth]{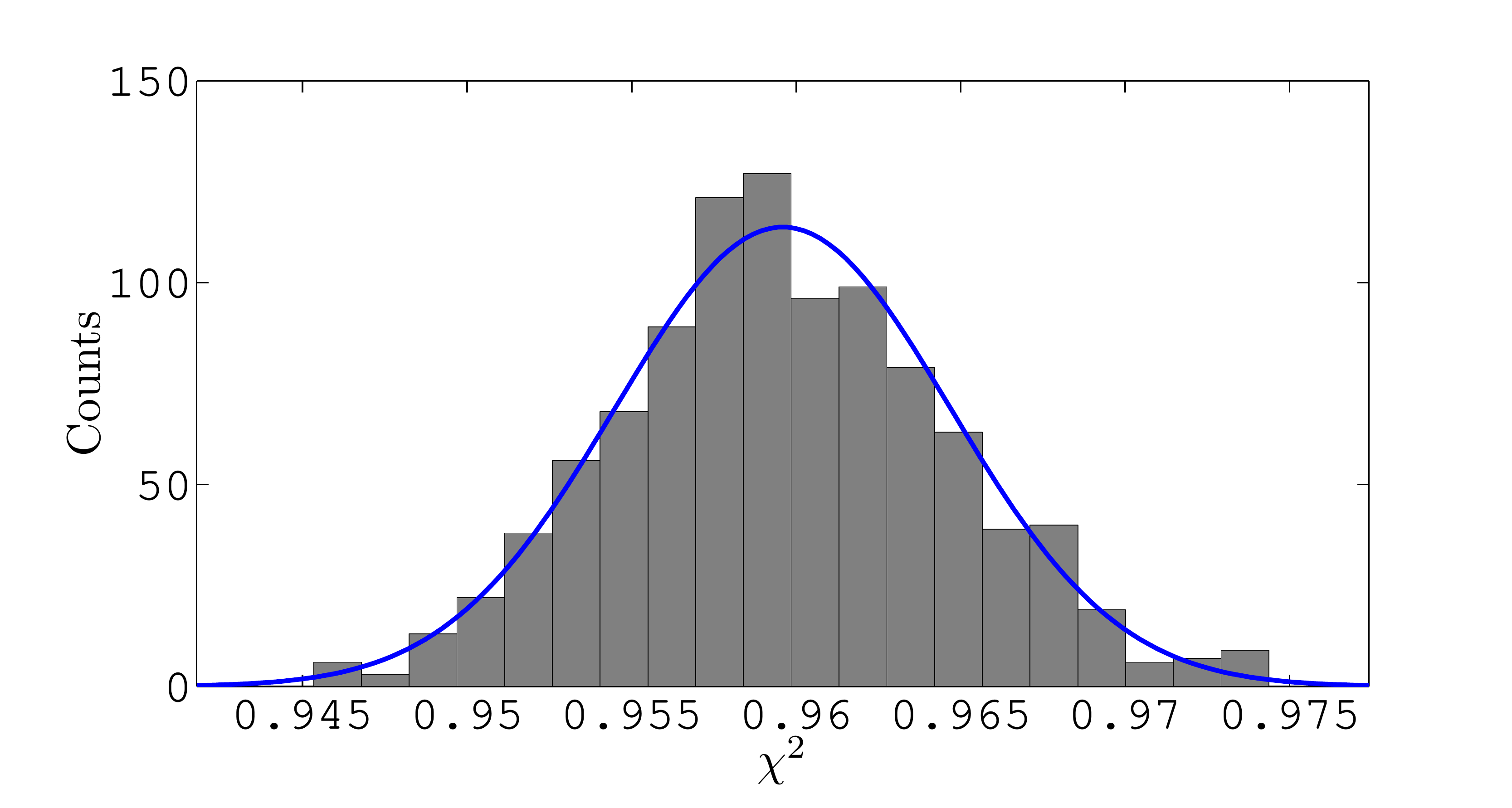} &
\includegraphics[width=0.5\columnwidth]{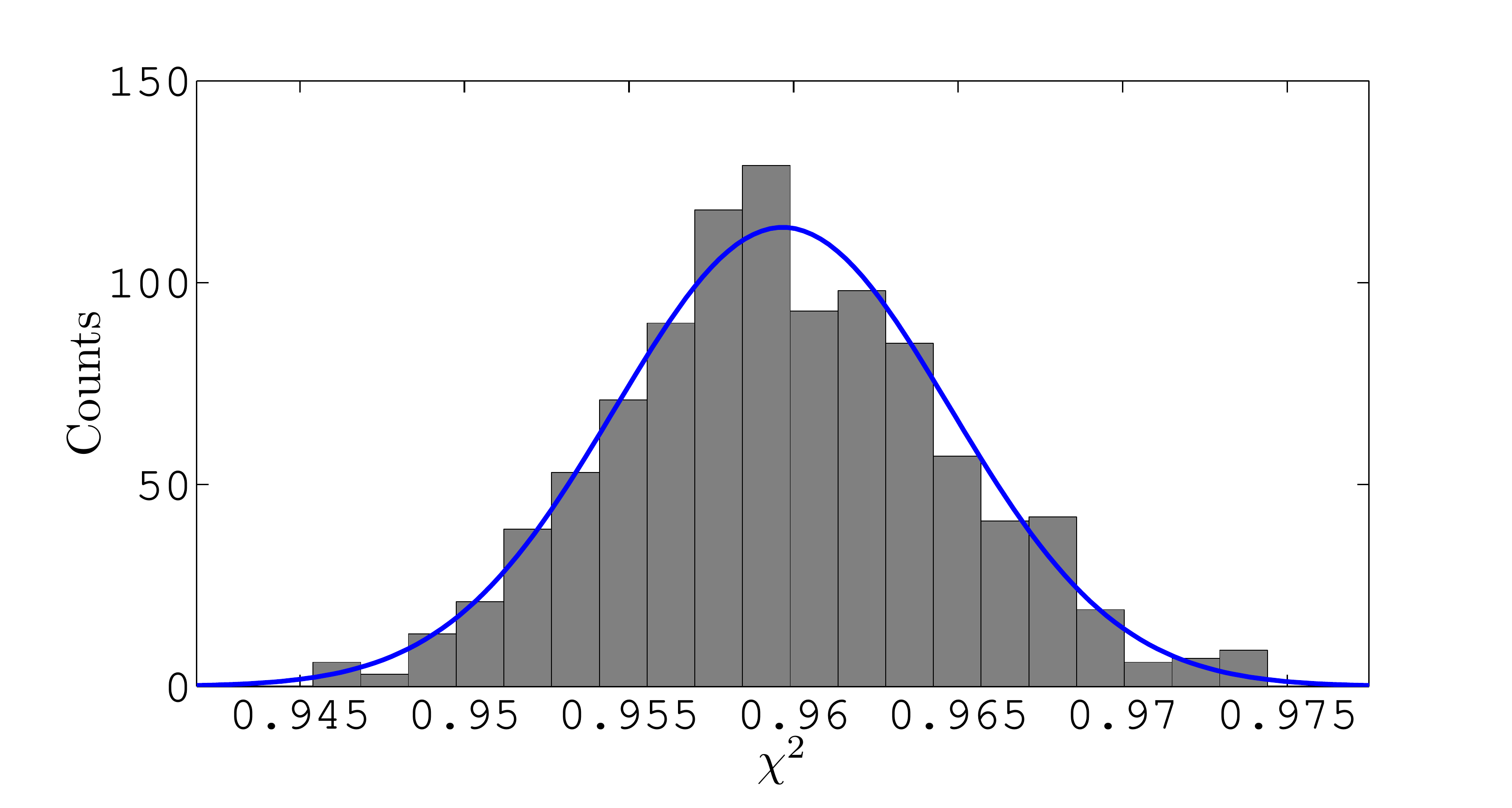} \\
\hspace{10pt}\footnotesize{(a)} &
\hspace{10pt}\footnotesize{(b)}
\end{tabular}
\caption{\footnotesize{Monte Carlo validation of 1000 independent noise realizations on which parameter estimation is repeated identically at each step. The plots show the statistic for (a) the fit $\chi^2$ log-likelihood and (b) the noise $\chi^2$ log-likelihood. The agreement between the two demonstrates that the deterministic signals are statistically suppressed out of the data.}}
\label{fig:sys_identification:montecarlo_chi2}
\end{figure}

The final, and most remarkable check, is the comparison between the fit $\chi^2$ log-likelihood and the one calculated on pure noise data contained in \figref{fig:sys_identification:montecarlo_chi2}. It is worth stressing that both the fit and the noise $\chi^2$ showed agreement between each other, but they were both positively skewed in a preliminary Monte Carlo simulation. The following facts explain why. \sectref{sect:sys_identification:whitening} has discussed the practical method to implement the diagonalization of the noise covariance matrix with its main limitation. This consists in the impossibility of filtering out the lowest frequencies, due to the finiteness of the data stretches from which whitening filters are derived and which causes the skewness. Transparently, the application of a high-pass filter to the data has solved the issue. The comparison in the plot provides for an important twofold test: on one side, the parameter variances are statistically distributed as the fit $\chi^2$ log-likelihood, as required; on the other, the fit $\chi^2$ log-likelihood is in agreement with the noise $\chi^2$ log-likelihood, showing that the estimation method has statistically suppressed the deterministic signals and recovered the noise statistic with no extra bias.

\subsection{Non-standard scenario: under-performing actuators and under-estimated couplings} \label{sect:sys_identification:initial_guess}

System identification has a key role in compensating the SC jitter and the TM couplings. Even in the unlikely (but possible) situation of under-performing actuators or under-estimated force couplings, it is still possible to retrieve the actual parameter values and allow for a precise estimation of the total equivalent acceleration noise without loosing sensibility and getting into systematic errors. The impact on the estimation of the total equivalent acceleration noise will be illustrated in \sectref{sect:sys_identification:force_noise}.

To introduce the problem, suppose that the predicted TM couplings are $\omega_1^2=\unit[-1.3\e{-6}]{s^{-2}}$ and $\omega_{12}^2=\unit[-0.7\e{-6}]{s^{-2}}$ and during the LPF mission:
\begin{enumerate}
  \item the actual TM couplings are about two times the predicted ones, due to unexpected/unmodeled stronger forces, like $\omega_1^2=\unit[-3\e{-6}]{s^{-2}}$ and $\omega_{12}^2=\unit[-2\e{-6}]{s^{-2}}$;
  \item the thruster and capacitive actuators unfortunately misfunction, due to both a breakdown of one or more thruster clusters and a loss of efficiency in the capacitive actuators on the second TM; this situation can be described by gains sensitively lower than one, like $A_\text{df}=0.62$ and $A_\text{sus}=0.6$;
  \item the interferometer introduces an extra cross-talk, $S_{21}=1.5\e{-3}$, ten times the expected one $S_{21}\oforder1\e{-4}$.
\end{enumerate}
In this very unfortunate situation, system identification, see \tabref{tab:sys_identification:robust2guess}, allows for the estimation of the true values within 1 standard deviation from the true values, so maintaining precision, even though the optimizations starts from initial guesses which are typically $\oforder10^3$ standard deviations away, so guaranteeing accuracy too.
\begin{table}[!htbp]
\caption{\label{tab:sys_identification:robust2guess}\footnotesize{Robustness to a non-standard scenario: under-performing actuators / under-estimated couplings. Initial estimates (guess) at $\chi^2=1.3\e5$, $\nu=79193$; best-fit values at $\chi^2=0.99$. The term in brackets is the error relative to the rightmost digit. In curly brackets the bias (absolute deviation from the real value in units of standard deviation) for each estimate.}}
\centering
\begin{tabular}{l D{.}{.}{5.5} D{.}{.}{3.7} D{.}{.}{3.3} D{.}{.}{4.1} D{!}{\times}{4.4}}
\hline
\hline
\multicolumn{1}{l}{Parameter} & \multicolumn{1}{c}{True} & \multicolumn{2}{c}{Best-fit} & \multicolumn{2}{c}{Guess} \\
\hline
$\omega_1^2\,[10^{-6}\,\text{s}^{-2}]$      & -3        & -2.9998(2)        & \{1.1\}     & -1.3 & \{7.8!10^3\}    \\
$\omega_{12}^2\,[10^{-6}\,\text{s}^{-2}]$   & -2        & -2.0000(1)        & \{0.32\}    & -0.7 & \{1.0!10^{4}\}    \\
$S_{21}\,[10^{-3}]$                         & -1.5      & -1.4998(1)        & \{0.55\}    & 0    & \{4.7!10^{3}\}    \\
$A_\text{df}$                               & 0.62      & 0.61994(8)        & \{0.77\}    & 1    & \{4.9!10^{3}\}    \\
$A_\text{sus}$                              & 0.6       & 0.599990(8)       & \{1.3\}     & 1    & \{5.1!10^{4}\}    \\
$\Delta t_1\,[\text{s}]$                    & 0.6       & 0.6013(7)         & \{1.8\}     & 0    & \{8.4!10^{2}\}    \\
$\Delta t_{12}\,[\text{s}]$                 & 0.4       & 0.398(2)          & \{0.95\}    & 0    & \{2.3!10^{2}\}    \\
\hline
\hline
\end{tabular}
\end{table}

\begin{figure}[!htbp]
\centering
\begin{tabular}{*{2}{@{\hspace{-10pt}}c@{}}}
\multicolumn{2}{c}{\includegraphics[width=0.5\columnwidth]{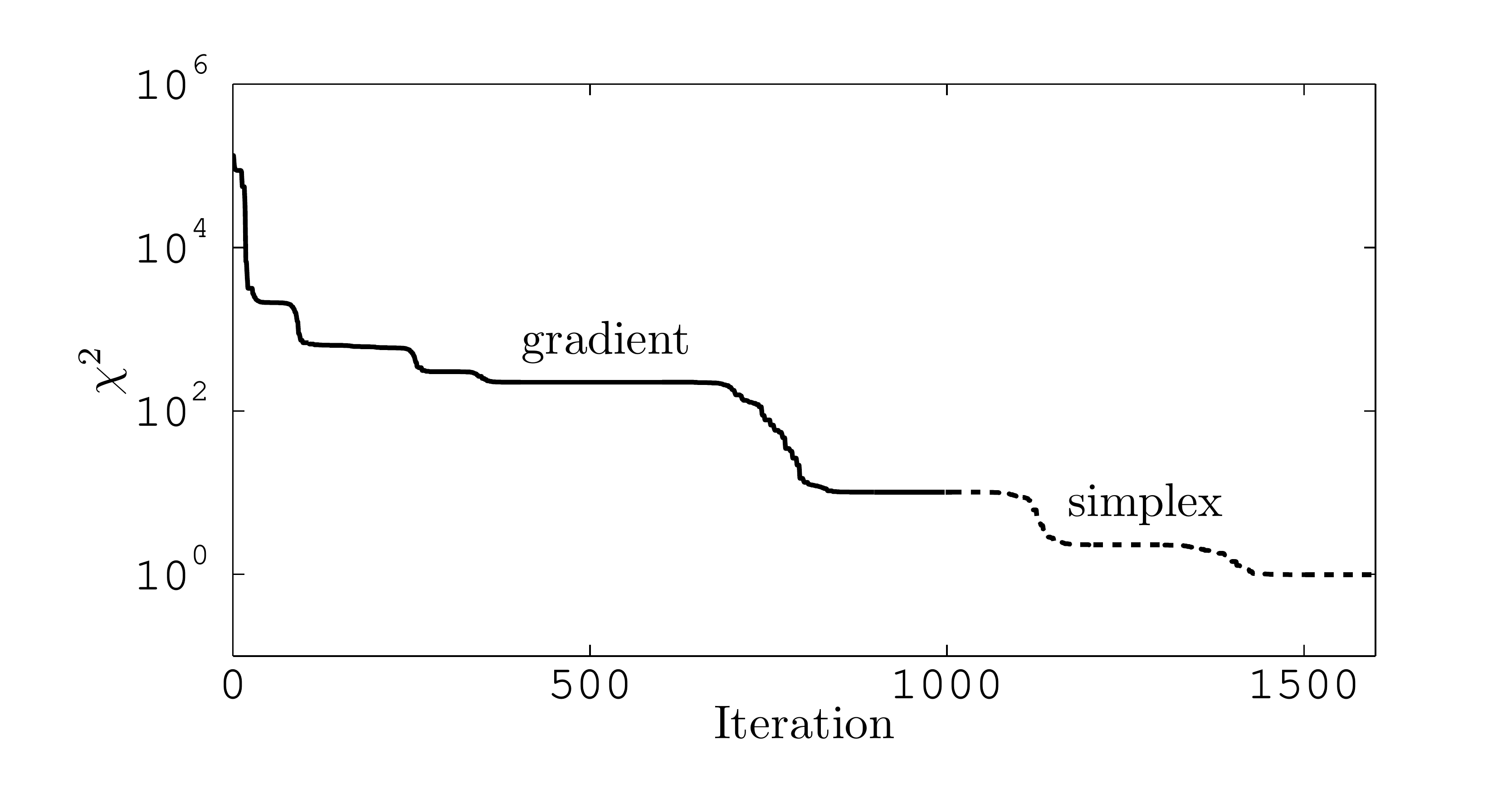}} \\
\multicolumn{2}{c}{\hspace{10pt}\footnotesize{(a)}} \\
\includegraphics[width=0.5\columnwidth]{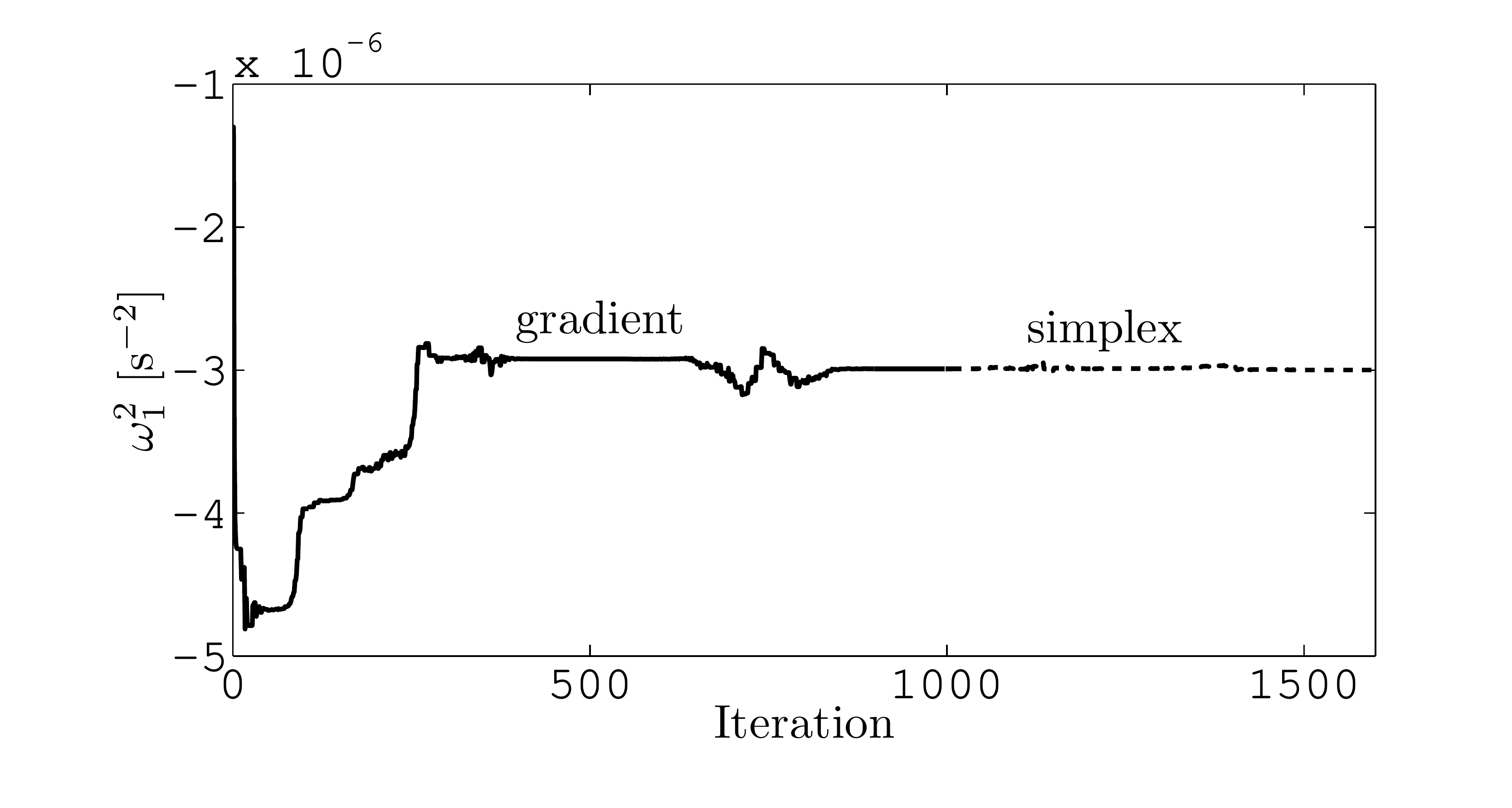} &
\includegraphics[width=0.5\columnwidth]{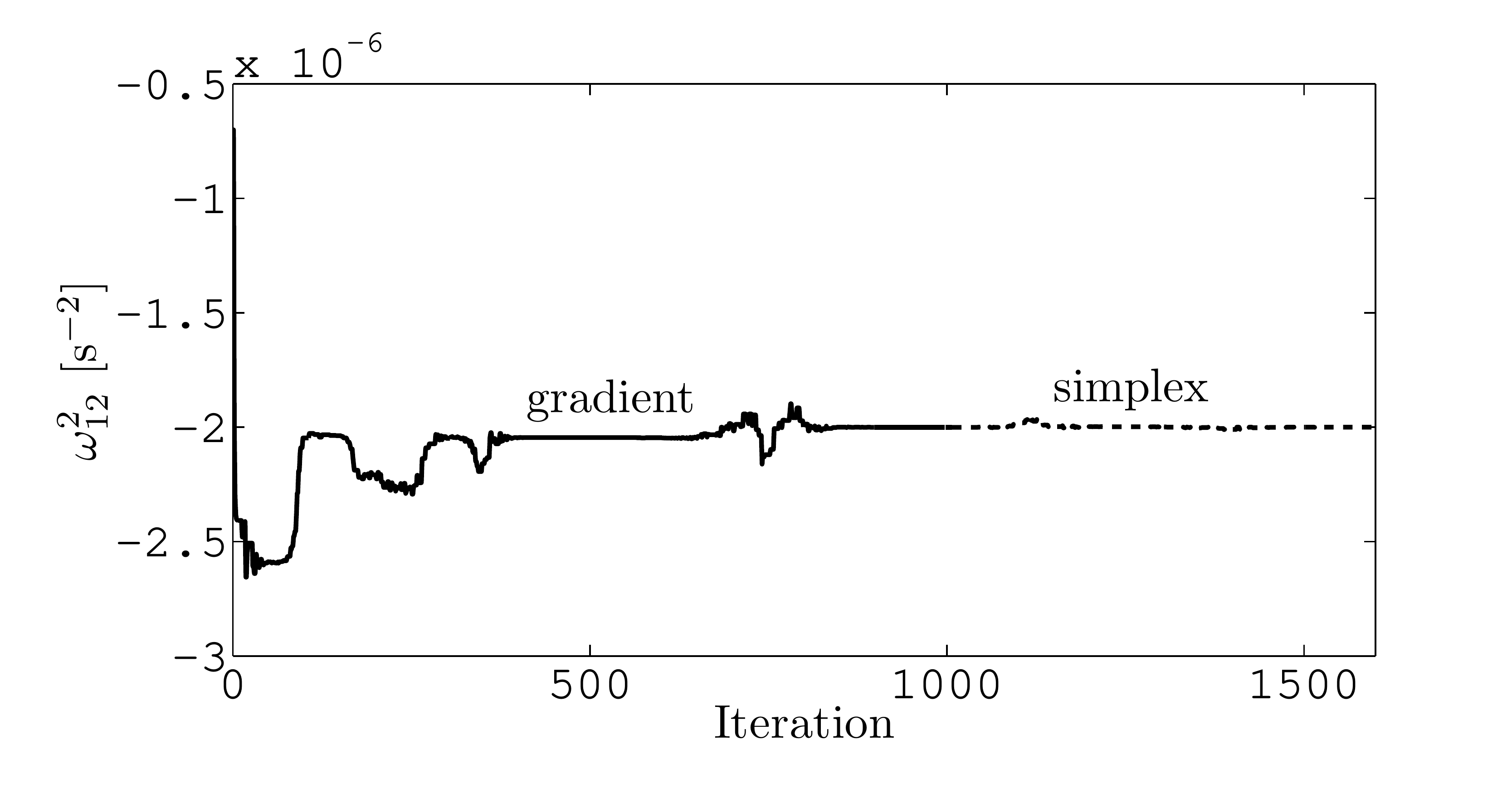} \\
\footnotesize{\hspace{10pt}(b)} &
\footnotesize{\hspace{10pt}(c)}
\end{tabular}
\caption{\footnotesize{Robustness to a non-standard scenario: under-performing actuators / under-estimated couplings. The estimation performances relative to the log-likelihood minimization (a) from $\oforder1\e{5}$ to the optimum $\oforder1$ and two examples of estimation chains for (b) $\omega_1^2$ and (c) $\omega_{12}^2$ showing the correlation with the big jumps in the $\chi^2$ chain. A preliminary global gradient search is followed by a local simplex. The process lasts for 1636 iterations and stops when the required tolerance is met.}}
\label{fig:sys_identification:robust2guess_performances}
\end{figure}

\figref{fig:sys_identification:robust2guess_performances} elucidates much more the results, showing the overall performances of the estimation. The $\chi^2$ is reduced from $1\e{5}$ to $\oforder1$ -- the required optimum -- within the given tolerances (set to $1\e{-4}$ in both log-likelihood and parameter values), while keeping both accuracy and precision. The figure reports two examples of estimation chains (for $\omega_1^2$ and $\omega_{12}^2$), showing the correlation with the big jumps in the $\chi^2$ chain and how the parameters saturate to the optimum values. The estimation, as already said, is divided into two phases: a gradient-based search, spanning the global structure of the parameter space, and a simplex search, improving the final accuracy.

\begin{figure}[!htbp]
\centering
\begin{tabular}{*{2}{@{\hspace{-10pt}}c@{}}}
\includegraphics[width=0.5\columnwidth]{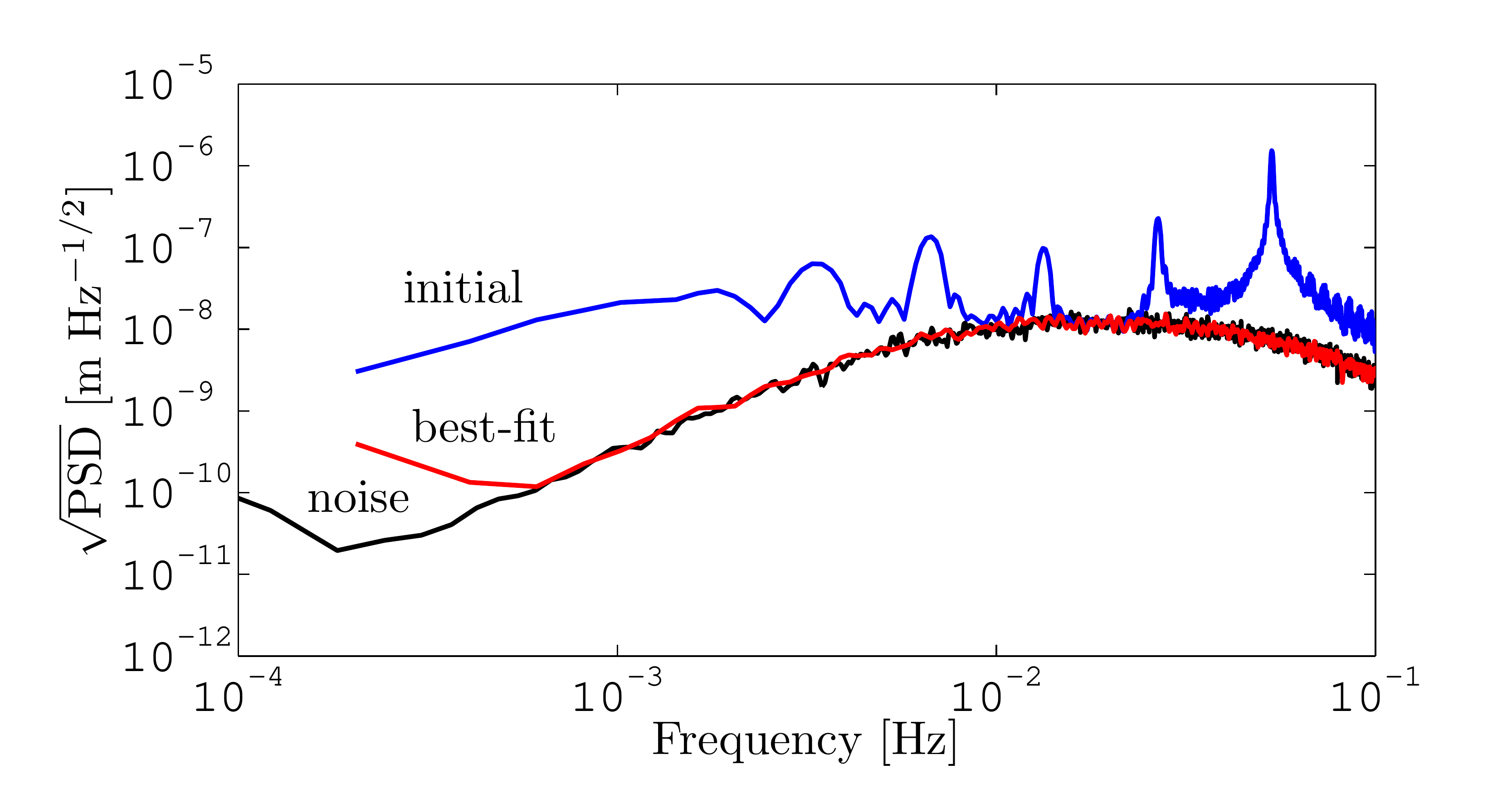} &
\includegraphics[width=0.5\columnwidth]{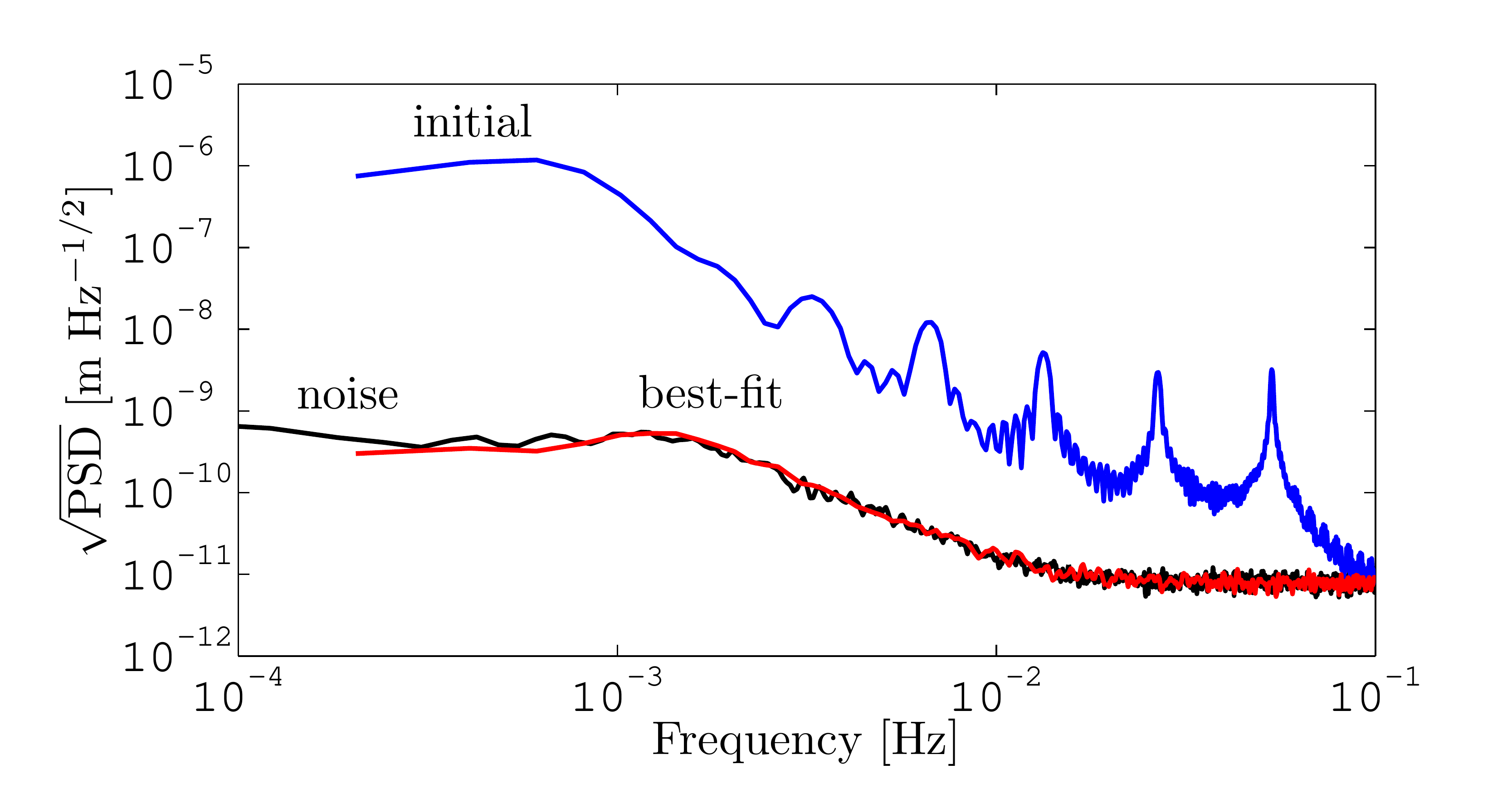} \\
\hspace{10pt}\footnotesize{(a)} &
\hspace{10pt}\footnotesize{(b)} \\
\includegraphics[width=0.5\columnwidth]{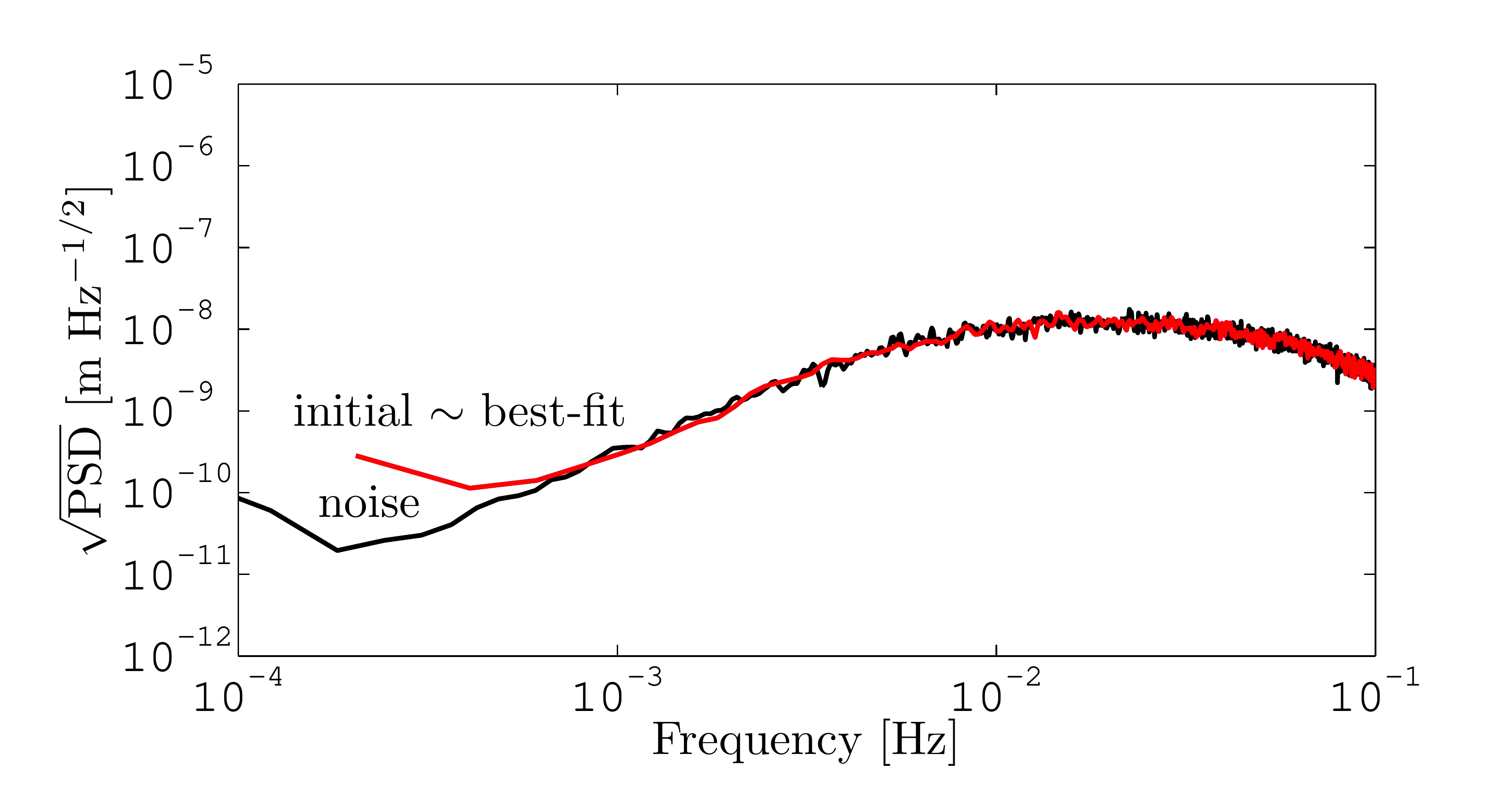} &
\includegraphics[width=0.5\columnwidth]{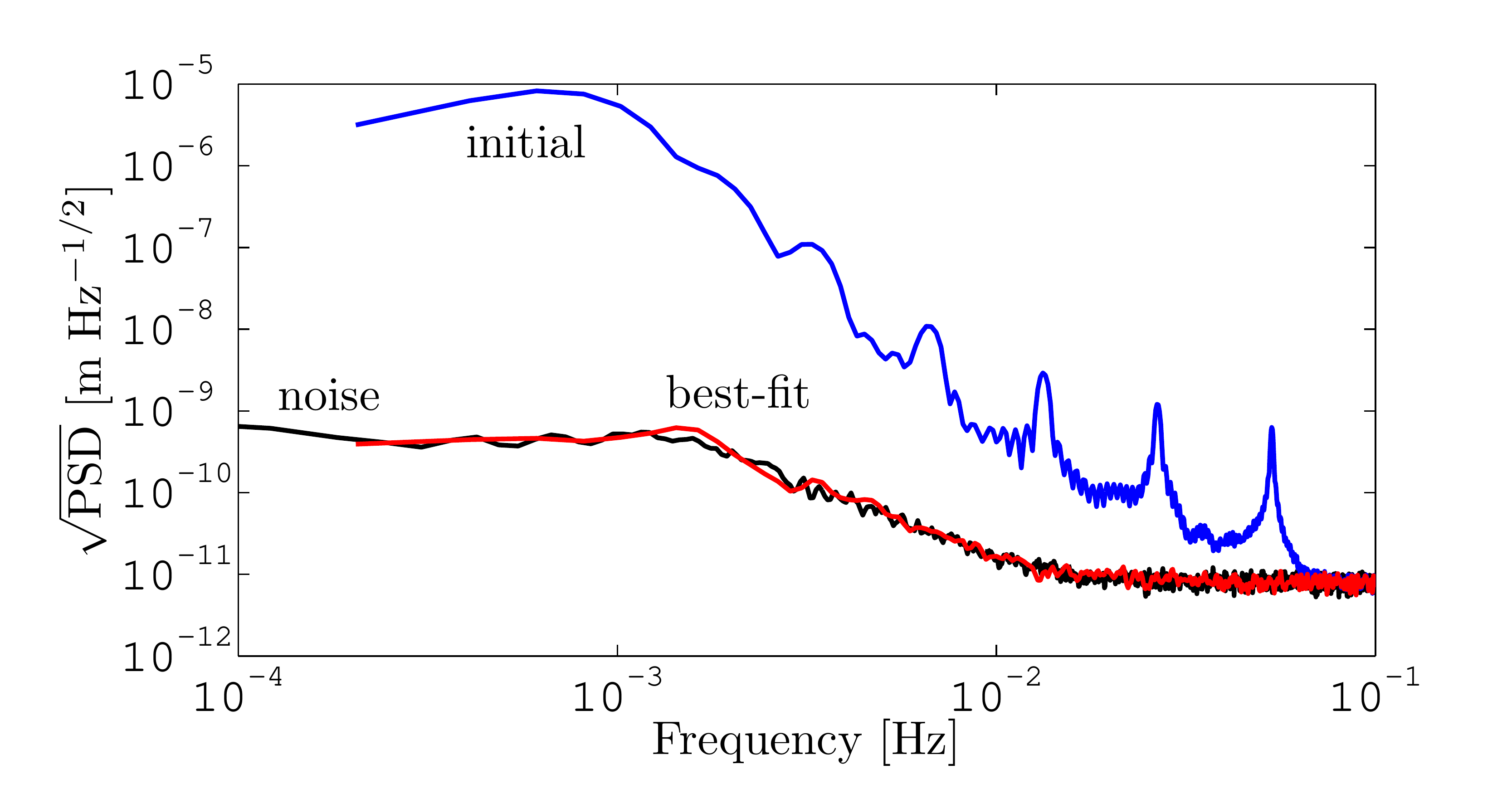} \\
\footnotesize{\hspace{10pt}(c)} &
\footnotesize{\hspace{10pt}(d)}
\end{tabular}
\caption{\label{fig:sys_identification:robust2guess_residuals}\footnotesize{Robustness to a non-standard scenario: under-performing actuators / under-estimated couplings. Analysis of residuals for all simulated identification experiments and interferometric readouts. Initial and best-fit residuals are compared to the expected noise shapes estimated from an independent run. For $o_{12}$ the improvement in both experiments (b) and (d) is of $\oforder 3$ orders of magnitude around 1 mHz; (a) for $o_1$ in the first experiment is $\oforder 2$ orders of magnitude; (c) contains no signal.}}
\end{figure}

The final and most important discussion is the analysis of residuals summarized in \figref{fig:sys_identification:robust2guess_residuals} for both identification experiments and interferometric readouts. The estimated PSDs of both initial and best-fit residuals are compared to the PSDs of an independent noise run. It is clear that the deterministic signals are completely subtracted from the data, hence recovering the noise shapes for all experiments and readouts. The improvement is mostly evident at low frequency: for $o_{12}$ the residuals are suppressed by $\oforder 3$ orders of magnitude around $\unit[1]{mHz}$. The same happens for $o_1$ in the first experiment where the improvement is $\oforder 2$ orders of magnitude. Only $o_1$ in the second experiment shows no improvement for the reason already discussed in \sectref{sect:sys_identification:experiments} (the signal is negligible).

\subsection{Non-standard scenario: non-Gaussianities} \label{sect:sys_identification:non_gaussianities}

This section is devoted to showing the impact of non-Gaussianities in the noise to parameter estimation. The main realistic behavior of experimental noise is the possible presence of outliers: consequently, the sampling distribution of the data may show some prominent tails. An example of such outliers is the manifestation of glitches, very short noise transients due to anomalous response in the readout/circuitry.

Given the non-Gaussian components in the noise, the log-likelihood defined so far is no longer well-behaved. Because of the intrinsical assumption of Gaussianity, it usually overweighs the outliers, and a systematic error may arise. A standard approach, named \textit{local L-estimate} \cite{press} \footnote{``L'' stands for ``likelihood''.}, requires the generalization of the definition of log-likelihood. The idea is to properly take care of the outliers by regularizing the usual square of whitened residuals with other similar definitions by means of a weighting function $\rho$
\begin{equation}
\chi^2 = \sum_i \rho(r_{\text{w},i})~,
\end{equation}
where, as an example, three possible choices, the squared, absolute and logarithmic deviations, are considered
\begin{equation}\label{eq:sys_identification:L_estimates}
\rho(r_{\text{w},i}) =
\begin{cases}
r_{\text{w},i}^2            & \text{mean squared dev.} \\
\left|r_{\text{w},i}\right| & \text{mean absolute dev.} \\
\log(1+r_{\text{w},i}^2)    & \text{mean logarithmic dev.}
\end{cases}~,
\end{equation}
corresponding to the cases of data distributed according to Gaussian, log-normal and Lorentzian distribution, respectively. The subscript $i$ is a generalized index counting the data available from all experiments and interferometric readouts and $r_{\text{w},i}$ is the whitened time-series of residuals. \figref{fig:sys_identification:robust2glitches_weighting} compares the three weighing functions for residuals out to 5 standard deviations. As is clear, the squared deviation overweighs the outliers. The absolute deviation gives a slightly better weight at high deviations, but performs poorly at low deviations. The logarithmic deviation has much more flexibility as it behaves like the squared deviation at low deviations and performs better than the absolute deviation.

\figuremacroW{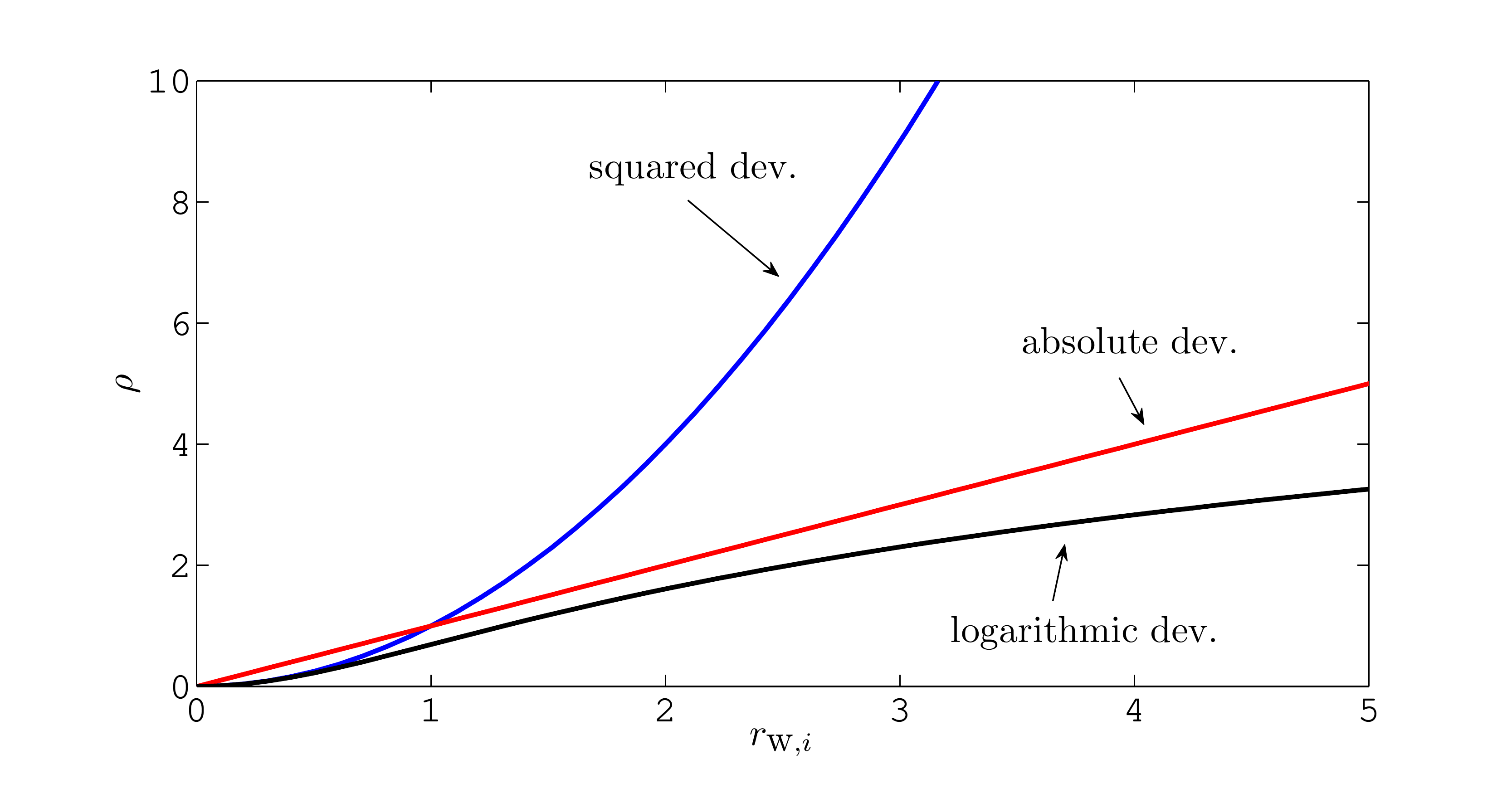}{Comparison of the three weighing functions of \eqref{eq:sys_identification:L_estimates} for the the proper weighing of outliers in the data. The logarithmic deviation is the most accurate as it behaves like the squared deviation at low deviations and performs better than the absolute deviation.}{fig:sys_identification:robust2glitches_weighting}{0.8}

The method can be successfully applied to data with glitches. Noise glitches are unpredictable high-frequency noise transients mostly
due to failures in the circuitry. Such outliers usually fall well beyond 3 standard deviations and produce an excess at the tails of the statistic. Since the output of the interferometer might be subject to similar phenomena, this section presents the results of the investigation of a realistic experiment containing glitches. Such transients are modeled as sine-Gaussian functions
\begin{equation}
o_\text{gl}(t) = a\sin\left[2\pi f_0(t-t_0)\right]\exp\left[-\frac{(t-t_0)^2}{\tau^2}\right]~,
\end{equation}
where the glitch parameters span a wide (uniformly distributed) range of values. In particular, the glitch frequency, $f_0$, covers the whole bandwidth $\unit[(10^{-4}\text{--}0.45)]{Hz}$; the injection time, $t_0$, is distributed all along the time-series; the characteristic time, $\tau$, giving the typical duration of the pulse is $\unit[(1\text{--}2)]{s}$; the amplitude, $a$, falls outside the Gaussian statistic by $(3\text{--}20)$ noise standard deviations. Moreover, the number of glitch injections is fixed as a fractional part of the whole data series, conventionally choosing $f_\text{gl}=N_\text{gl}/N_\text{data}=1\%$, since higher values are very unlikely. Notice that this value represents only the number of injections: the actual fraction of corrupted data is the order of $3\,\text{E}[\tau]\,f_\text{gl}\simeq5\%$.

\begin{figure}[!htbp]
\centering
\begin{tabular}{*{2}{@{\hspace{-10pt}}c@{}}}
\includegraphics[width=0.5\columnwidth]{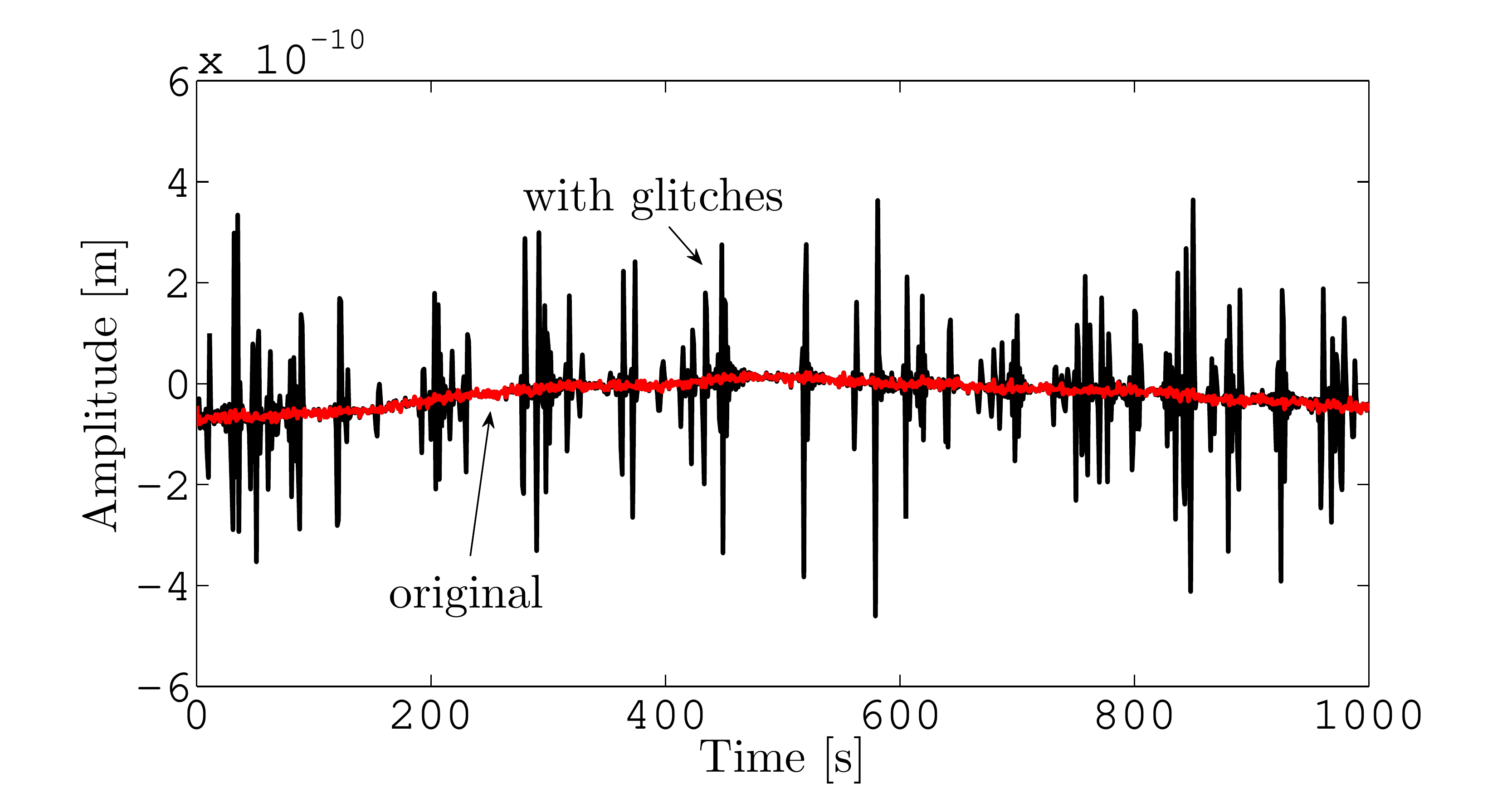} &
\includegraphics[width=0.5\columnwidth]{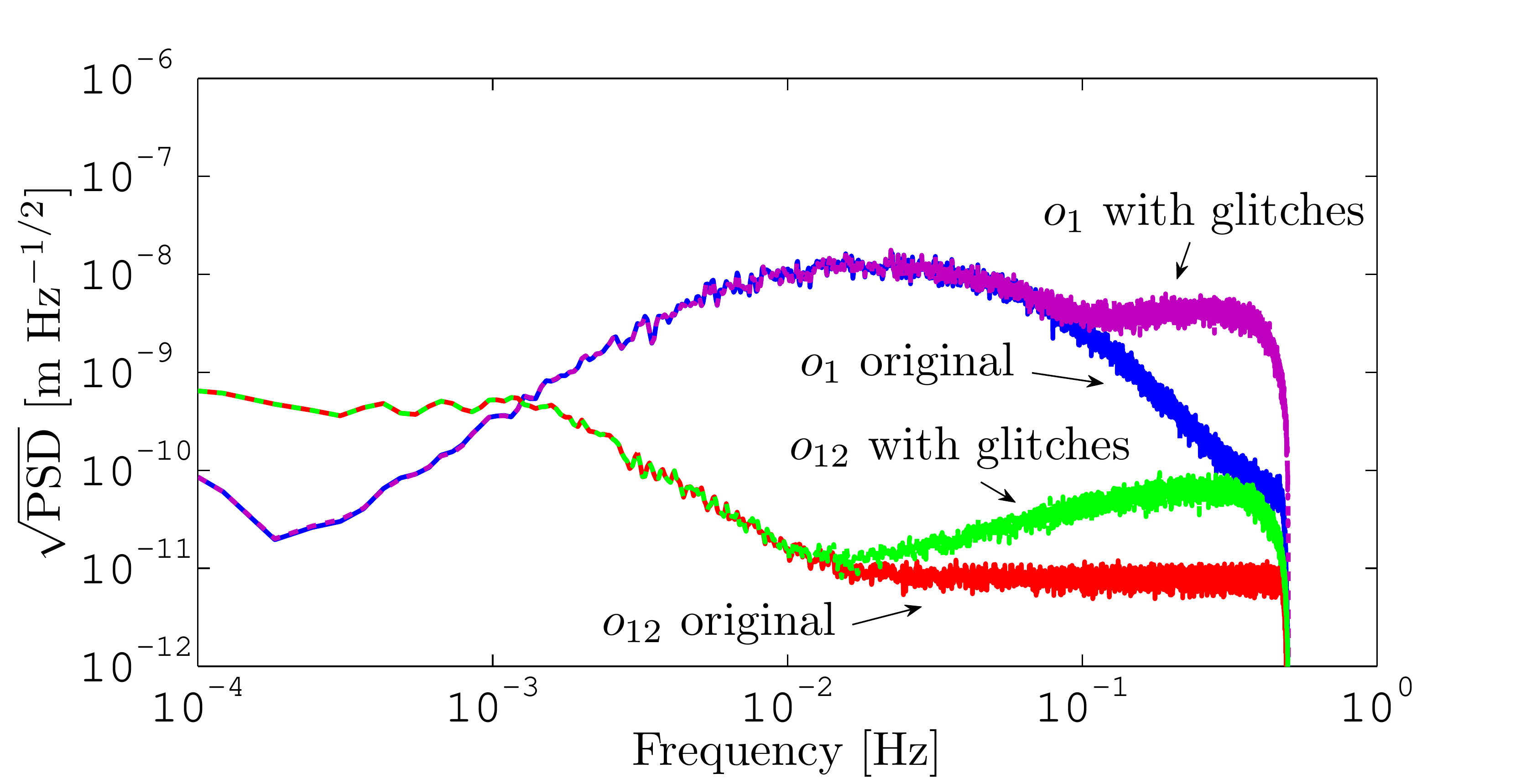} \\
\hspace{10pt}\footnotesize{(a)} &
\hspace{10pt}\footnotesize{(b)}
\end{tabular}
\caption{\footnotesize{Robustness to a non-standard scenario: non-Gaussianities. (a) simulated original and glitchy noise for $o_{12}$; (b) PSDs of the simulated original and glitchy noise for $o_1$ and $o_{12}$. The level of data corruption is evident and glitches appear as high-frequency bumps around $\unit[0.2]{Hz}$.}}
\label{fig:sys_identification:robust2glitches}
\end{figure}

Glitchy noise is readily produced by coloring a white, zero-mean, unitary standard deviation input time-series, as in \sectref{sect:sys_identification:ifo_noise}, corrupted by random injections of glitches. \figref{fig:sys_identification:robust2glitches} shows how glitches appear in the interferometric differential readout and in the estimated PSDs, compared to the original noise stretches. The effect of glitches is that the PSD of the simulated noise scales linearly with the frequency, up to $\unit[4\e{-9}]{m\,Hz^{-\nicefrac{1}{2}}}$ and $\unit[6\e{-11}]{m\,Hz^{-\nicefrac{1}{2}}}$ around $\unit[0.2]{Hz}$ for the first and differential readout, respectively. This excess noise sums up to the original one and is shown as high-frequency components. Obviously, the noise statistic contains an excess at the tails. For example, $o_1$ has an excess kurtosis of $\oforder19$, compared to the original one of $-9\e{-3}$. No significant difference in skewness is detected since the statistic does not loose symmetry with the glitch injections.

Whitening filters are derived from the glitchy noise stretches with the same procedure described in \sectref{sect:sys_identification:whitening}. However, since the whitening process works assuming stationarity, glitches are not filtered out from the data.

\tabref{tab:sys_identification:robust2glitches} shows the results of three different parameter estimations with the definitions of the weighting functions in \eqref{eq:sys_identification:L_estimates}.
\begin{table}[!htbp]
\caption{\label{tab:sys_identification:robust2glitches}\footnotesize{Robustness to a non-standard scenario: non-Gaussianities. The comparison between three parameter estimations with the three definitions in \eqref{eq:sys_identification:L_estimates}. $\nu=79193$. The term in brackets is the error relative to the rightmost digit. In curly brackets the bias (absolute deviation from the real value in units of standard deviation) for each estimate.}}
\centering
\scalebox{.72}{
\begin{tabular}{l D{.}{.}{3.4} D{.}{.}{3.7} l D{.}{.}{3.7} l D{.}{.}{3.7} l D{.}{.}{3.4}}
\hline
\hline
\multicolumn{1}{l}{\multirow{3}{*}{Parameter}} & \multicolumn{1}{c}{\multirow{3}{*}{Real}} & \multicolumn{2}{c}{Best-fit} & \multicolumn{2}{c}{Best-fit} & \multicolumn{2}{c}{Best-fit} & \multicolumn{1}{c}{\multirow{3}{*}{Guess}} \\
& & \multicolumn{2}{c}{(mean sq.\,dev.)} & \multicolumn{2}{c}{(mean abs.\,dev.)} &  \multicolumn{2}{c}{(mean log.\,dev.)} \\
& & \multicolumn{2}{c}{$\chi^2=10$} & \multicolumn{2}{c}{$\chi^2=2.1$}     & \multicolumn{2}{c}{$\chi^2=0.95$} \\
\hline
$\omega_1^2\,[10^{-6}\,\text{s}^{-2}]$ & -1.32 & -1.320(1) & \{0.061\} & -1.3188(6) & \{2.0\} & -1.3192(4) & \{2.0\} & -1.3 \\
$\omega_{12}^2\,[10^{-6}\,\text{s}^{-2}]$ & -0.68 & -0.6798(7) & \{0.29\} & -0.68000(3) & \{0.011\} & -0.6804(2) & \{1.8\} & -0.7 \\
$S_{21}\,[10^{-4}]$ & 1.1 & 1.10(2) & \{0.074\} & 1.113(7) & \{1.8\} & 1.116(5) & \{3.4\} & 0 \\
$A_\mathrm{df}$ & 1.01 & 1.011(3) & \{0.29\} & 1.010(1) & \{0.23\} & 1.0109(8) & \{1.2\} & 1 \\
$A_\mathrm{sus}$ & 0.99 & 0.99000(5) & \{0.035\} & 0.98959(2) & \{20\} & 0.99001(1) & \{0.99\} & 1 \\
$\Delta t_1\,[\text{s}]$ & 0.1 & 0.100(3) & \{0.045\} & 0.090(1) & \{8.3\} & 0.1007(8) & \{0.90\} & 0 \\
$\Delta t_{12}\,[\text{s}]$ & 0.1 & 0.098(5) & \{0.36\} & -0.0290(2) & \{58\} & 0.098(2) & \{1.2\} & 0 \\
\hline
\hline
\end{tabular}
}
\end{table}
The most conservative least square estimator provides overestimated errors since they scale as $\oforder\sqrt{\chi^2}$. The absolute and logarithmic deviations provide better statistics and lower errors, but the first gives biased estimates of $A_\text{sus}$, $\Delta t_1$ and $\Delta t_{12}$ and the last one a slightly biased estimate of $S_{21}$. The analysis of residuals demonstrates that the three methods recover the noise shapes and are in agreement with each other, so the systematic errors are only in the estimated parameters. These estimators are also $30\%$ and $9\%$ faster than the Gaussian (mean squared deviation), as the outliers have less influence on the estimation chains.

By inspecting the results, it turns out that there is no absolute rule that can be applied when dealing with glitches. However, from the differences between the estimates it is possible to infer the sensitivity of each single parameter to glitches. For example, adopting the ratio between the biases as the a-posteriori criterion for comparing two methods, it tends to one if that parameter is not sensitive to glitches; otherwise, it tends to a very small or very large number. In view of this consideration, the comparison between the mean squared deviation and the mean logarithmic deviation gives that $S_{21}$ is the most sensitive parameter, whereas $\Delta t_{12}$ the least.

Starting from the fact that the three methods give the same results for purely Gaussian noise, a proposed recipe is the following:
\begin{enumerate}
\item apply the conservative approach (the ordinary mean squared deviation) directly to corrupted time series and try with different estimators (mean absolute deviation, mean logarithmic deviation, etc.);
\item start removing some outliers giving them negligible weight;
\item redo the analysis with all estimators;
\item check for convergence and agreement between the estimators.
\end{enumerate}
The overall process can be actually viewed as a reweighing analysis providing for robust uncertainties and, at the same time, the removal of outliers in a step-by-step smooth readjustment. Even though it would be possible in principle to clean up the data just before the estimation, in that case the results would likely be dependent on the statistical criterion used for such cleaning. Even though it is beyond the scope of this thesis to implement the idea, it is worth observing that the two main advantages of the preceding recipe are its robustness in definition and the fact that data polishing is smooth and model independent.

\section{Estimation of total equivalent acceleration noise} \label{sect:sys_identification:force_noise}

This section justifies the efforts in developing the techniques introduced so far with all tests and validation runs, showing the impact of system identification on the estimation of the total equivalent acceleration noise. As said throughout this thesis, the main objective of the LPF mission in view of a real GW astronomy with spaced-based detectors is the characterization of the Doppler link as the fundamental spacetime meter in terms of equivalent differential acceleration. Even if LPF is different in design with respect to LISA -- no faraway optical measurement between two SCs is actually implemented -- yet the principle and, most of all, the performances in sensitivity can be extrapolated and gather more confidence in the scientific scopes of any spaced-based GW detector.

Assessing the performance in sensitivity as equivalent input acceleration noise is a very effective way to put dynamics, sensing and control on the same footing as described in \sectref{sect:metrology:noise}. This can be achieved by means of the $\vect{\Delta}$ operator of \sectref{sect:dynamics:matrix_formalism}, connecting interferometric displacement readouts to total equivalent acceleration and at the same time compensating for TM couplings, SC jitter and sensing cross-talk.

Suppose that $\vect{S}_{\text{n},o}(\omega,\vect{p}_\text{true})$ is the measured interferometric noise PSD. Then, the estimated total equivalent acceleration noise PSD is given by
\begin{equation}
\vect{S}_{\text{n},f}(\omega,\vect{p}_\text{est}) = \vect{\Delta}(\omega,\vect{p}_\text{est})\,\vect{S}_{\text{n},o}(\omega,\vect{p}_\text{true})\,\ctranspose{\vect{\Delta}(\omega,\vect{p}_\text{est})}~,
\end{equation}
where $\vect{\Delta}(\omega,\vect{p}_\text{est})$ models the transfer from interferometric displacement readouts to total equivalent acceleration and $\vect{p}_\text{est}$ are the parameter estimates as obtained by system identification. It is worth noting that if $\vect{S}_{\text{n},o}$ was assumed constant to the parameter values in first approximation, the transfer to total equivalent acceleration would anyhow couple the output noise with the dynamics so that the estimated total equivalent acceleration noise becomes explicitly dependent on the parameter values. This shows that \textit{parameter estimation serves not only for system identification, but also for the actual identification of the total equivalent acceleration noise}.

Furthermore, suppose that $\vect{p}_\text{est}\simeq\vect{p}_\text{true}+\delta\vect{p}$, with $\delta\vect{p}$ the parameter biases being much larger than the statistical uncertainties on $\vect{p}_\text{est}$. It is easy to show that the parameter biases propagate to the differential operator $\vect{\Delta}_\text{est}\simeq\vect{\Delta}_\text{true}+\delta\vect{\Delta}$, where $\vect{\Delta}_\text{true}=\vect{\Delta}(\omega,\vect{p}_\text{true})$ and $\vect{\Delta}_\text{est}=\vect{\Delta}(\omega,\vect{p}_\text{est})$. Systematic errors found in the parameter values produce systematic errors in the recovered total equivalent acceleration noise
\begin{equation}
\delta\vect{S}_{\text{n},f} \simeq \delta\vect{\Delta}\,\vect{S}_{\text{n},o}\,\ctranspose{\vect{\Delta}}
+ \vect{\Delta}\,\vect{S}_{\text{n},o}\,\ctranspose{\delta\vect{\Delta}}~,
\end{equation}
where the subscript ``true'' is dropped out for clearness. As pointed out in \cite{ferraioli2011}, the statistical uncertainty on the parameter values are masked by the statistical uncertainty on the estimated spectrum. Despite this, systematic errors in the estimated parameters can fall well outside the confidence levels of the optimal spectrum and show themselves as not mere excess noise, but producing really different noise shapes. Hence, it is expected that \textit{the estimation of the total equivalent acceleration noise is biased if the parameter values are not correctly assessed from system identification}.

To demonstrate the impact of system identification on the estimation of the total equivalent acceleration noise, a very long noise run, $\oforder 6$ days, is simulated with the same procedures of \sectref{sect:sys_identification:ifo_noise}, i.e., by coloring a sequence of white Gaussian input time-series with cross-correlating noise shaping filters. The interferometric displacement noise model is derived in a non-standard configuration of LPF, as in \sectref{sect:sys_identification:initial_guess}, namely in the case of stronger-than-expected TM couplings, malfunctioning actuators and a higher sensing cross-talk. In this case, the estimation of the total equivalent acceleration noise with naively guessed parameter values will surely contain systematic errors.

The estimation of the total equivalent acceleration noise is readily performed on the multi-channel interferometric run with a scheme described in details in \cite{ferraioli2011,monsky2009}, by applying a time-domain version of the $\vect{\Delta}$ operator of \sectref{sect:dynamics:matrix_formalism}. The issues connected to numerical derivatives in LPF are extensively discussed and solved in \cite{ferraioli2009}. As said, system identification effectively helps in the calibration of the operator. In support of the statement, the numerical estimation of the total equivalent acceleration noise is performed assuming three different parameter sets that can be found in \tabref{tab:sys_identification:robust2guess}:
\begin{enumerate}
  \item the initial guess values, as it was \textit{without} a preliminary system identification: $\omega_1^2=\unit[-1.3\e{-6}]{s^{-2}}$, $\omega_{12}^2=\unit[-0.7\e{-6}]{s^{-2}}$, $S_{21}=0$, $A_\text{df}=1$, $A_\text{sus}=1$ (typically $\oforder10^4$ standard deviations away from the real values);
  \item the best-fit values, as it was \textit{with} a preliminary system identification, i.e., after having calibrated the differential operator: $\omega_1^2=\unit[-2.9998(2)\e{-6}]{s^{-2}}$, $\omega_{12}^2=\unit[-2.0000(1)\e{-6}]{s^{-2}}$, $S_{21}=-1.4998(1)\e{-3}$, $A_\text{df}=0.61994(8)$, $A_\text{sus}=0.599990(8)$;
  \item the true values, used for consistency checks: $\omega_1^2=\unit[-3\e{-6}]{s^{-2}}$, $\omega_{12}^2=\unit[-2\e{-6}]{s^{-2}}$, $S_{21}=-1.5\e{-3}$, $A_\text{df}=0.62$, $A_\text{sus}=0.6$.
\end{enumerate}
The result of the analysis is contained in \figref{fig:sys_identification:conv2acc}, showing the total equivalent differential acceleration noise, both numerically estimated and modeled, for the three different cases.

\figuremacroW{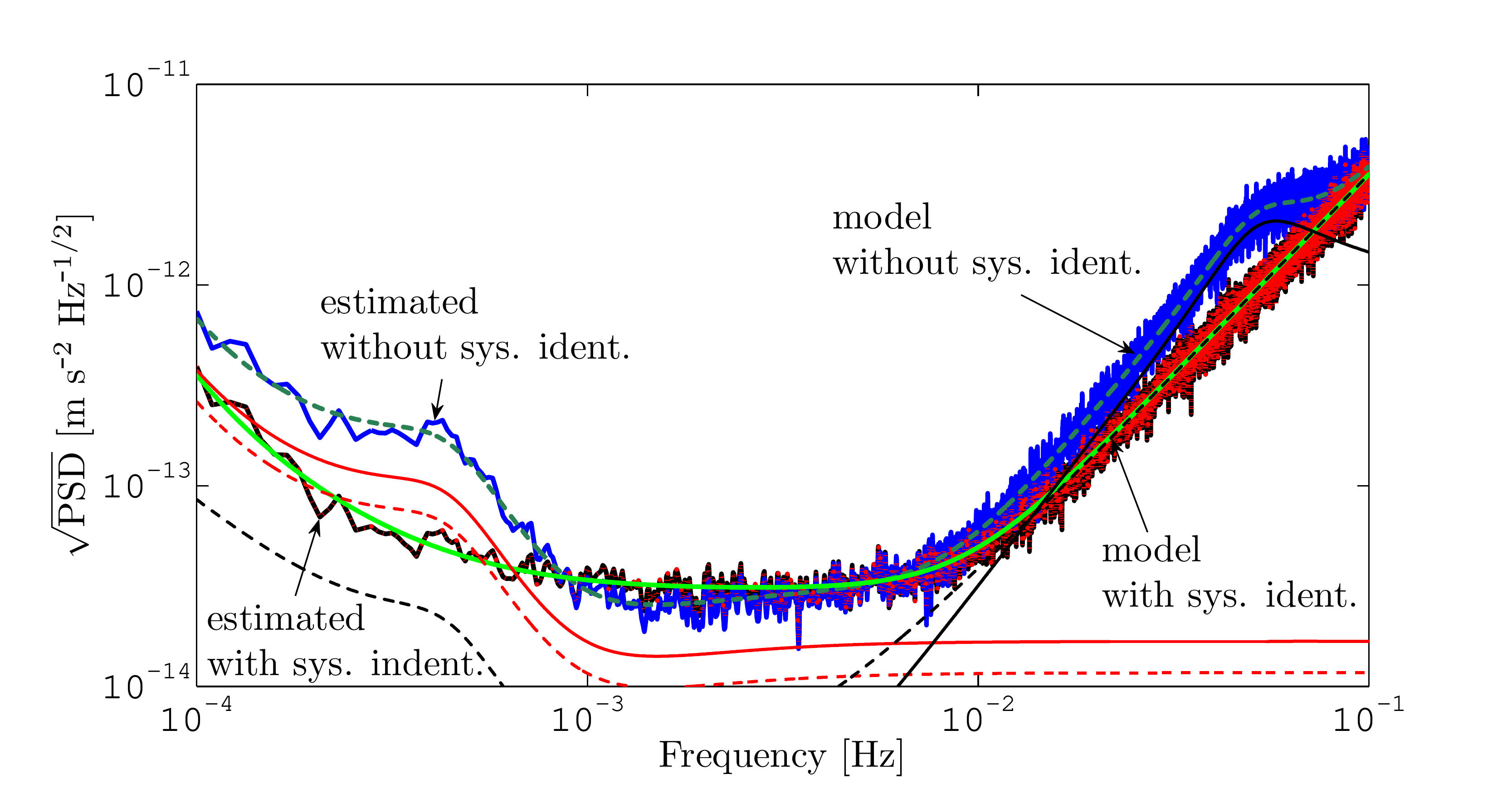}{Total equivalent differential acceleration noise numerically estimated on synthetic data and compared to theoretical noise models obtained by a full projection of fundamental noise sources. The estimation of the total out-of-loop equivalent acceleration can be performed either \textit{with} a preliminary system identification or \textit{without} it. The PSD estimated with a preliminary system identification completely overlaps the one of a hypothetical estimation assuming the knowledge of the true parameter values. The observed difference shows that a preliminary system identification is mandatory to avoid systematic errors in the reconstructed total equivalent acceleration noise. The solid thinner lines indicate the reasons of such a discrepancy. Around $\unit[50]{mHz}$ the bump is due to unsuppressed thruster noise exceeding the interferometric $o_{12}$ readout noise. At low frequency and around $\unit[0.4]{mHz}$, the two major contributions are the unsuppressed force couplings between the TMs and the SC and the capacitive actuation noise. Thanks to system identification, an improvement in performance of a factor 4 at low frequency is evident.
}{fig:sys_identification:conv2acc}{1}

First, the agreement between modeled and estimated total equivalent acceleration noise PSDs states that: (i) the generation of the interferometric noise is accurate to the assumed models at least to within the statistical uncertainty of the spectra; (ii) the numerical estimation of the total equivalent acceleration in time-domain is accurately explained by the frequency-domain transfer matrix from interferometric readouts to the total equivalent acceleration.

Second but more important, the total equivalent acceleration noise estimated with a preliminary system identification completely overlaps the one of a hypothetical estimation assuming the complete knowledge of the true values. Therefore it demonstrates that it is still possible to meet the sensitivity requirements during under-performing mission operations.

The observed systematic errors in the total equivalent acceleration noise estimated without identification show that system identification is strictly mandatory to avoid such problems and guarantee the scientific objectives. The systematic errors can be explained by the fact that the naive initial guess values are sensitively different from the true values. Since the operator is not calibrated on fiducial parameter values, it is not effective in compensating, in turn, the SC jitter due to the thruster actuation noise, the TM couplings and the capacitive actuation noise. In particular, around $\unit[50]{mHz}$ the bump is the unsuppressed thruster noise exceeding the interferometric $o_{12}$ readout noise: the effect is due to the uncalibrated drag-free gain $A_\text{df}$. At low frequency and around $\unit[0.4]{mHz}$, the major contributions are the coupling forces between the TMs and the SC (two contributions, accounting for $\unit[1.8\e{-13}]{m\,s^{-2}}$, almost the whole noise budget) and the capacitive actuation noise ($\unit[7\e{-14}]{m\,s^{-2}}$): the effect is due to the uncalibrated stiffness constants $\omega_1^2$ and $\omega_{12}^2$ and the suspension gain $A_\text{sus}$.

The final improvement in the estimated total equivalent acceleration noise with system identification is a factor $4$ around $\unit[0.4]{mHz}$ and a factor $2$ around $50\,\unit{mHz}$ in units of $\sqrt{\text{PSD}}$. The conclusion is that without a preliminary system identification -- robust to non-standard parameter values -- the performance of the mission and the characterization of the total equivalent acceleration noise would seriously be compromised.

\section{Suppressing transients in the total equivalent acceleration noise} \label{sect:sys_identification:transients}

This final section discusses on the suppression of system transients for realistic data produced by the OSE and provided by ESA. \sectref{sect:dynamics:transients} and in particular \eqref{eq:dynamics:transients_suppression} demonstrate that system transients can be suppressed to within the accuracy to which the differential operator $\vect{\Delta}$ has been calibrated on parameter values representative of the system. A supporting example is provided in what follows.

\figref{fig:sys_identification:transients} shows the first 3 hours of a typical noise run of the OSE. In complete realism, just after the TM release the system is firstly turned into accelerometer mode, then into science mode (around $\unit[1\e{4}]{s}$) \footnote{It is worth recalling that in accelerometer mode the TMs are both electrostatically suspended, whereas in the main science mode one of the two is in drag-free. The resulting noise is at least one order of magnitude lower in the second case, especially in the differential readout.}: transients appears as a direct consequence of the non-zero initial conditions. In fact, the initial positions are $\unit[0.24]{\mu m}$ ($o_1$) and $\unit[0.36]{\mu m}$ ($o_{12}$), whereas the estimated velocities \footnote{Since only an order of magnitude is needed, a two point forward difference is applied together with a low-pass filter with frequency cut at $\unit[100]{mHz}$.} are about $\unit[-500]{pm\,s^{-1}}$ ($o_1$) and $\unit[-4]{pm\,s^{-1}}$ ($o_{12}$). The transient in $o_1$ lasts for half a hour and in $o_{12}$ for about 2 hours -- the timescale of typical transients as predicted by \sectref{sect:dynamics:transients}. 

\figuremacroW{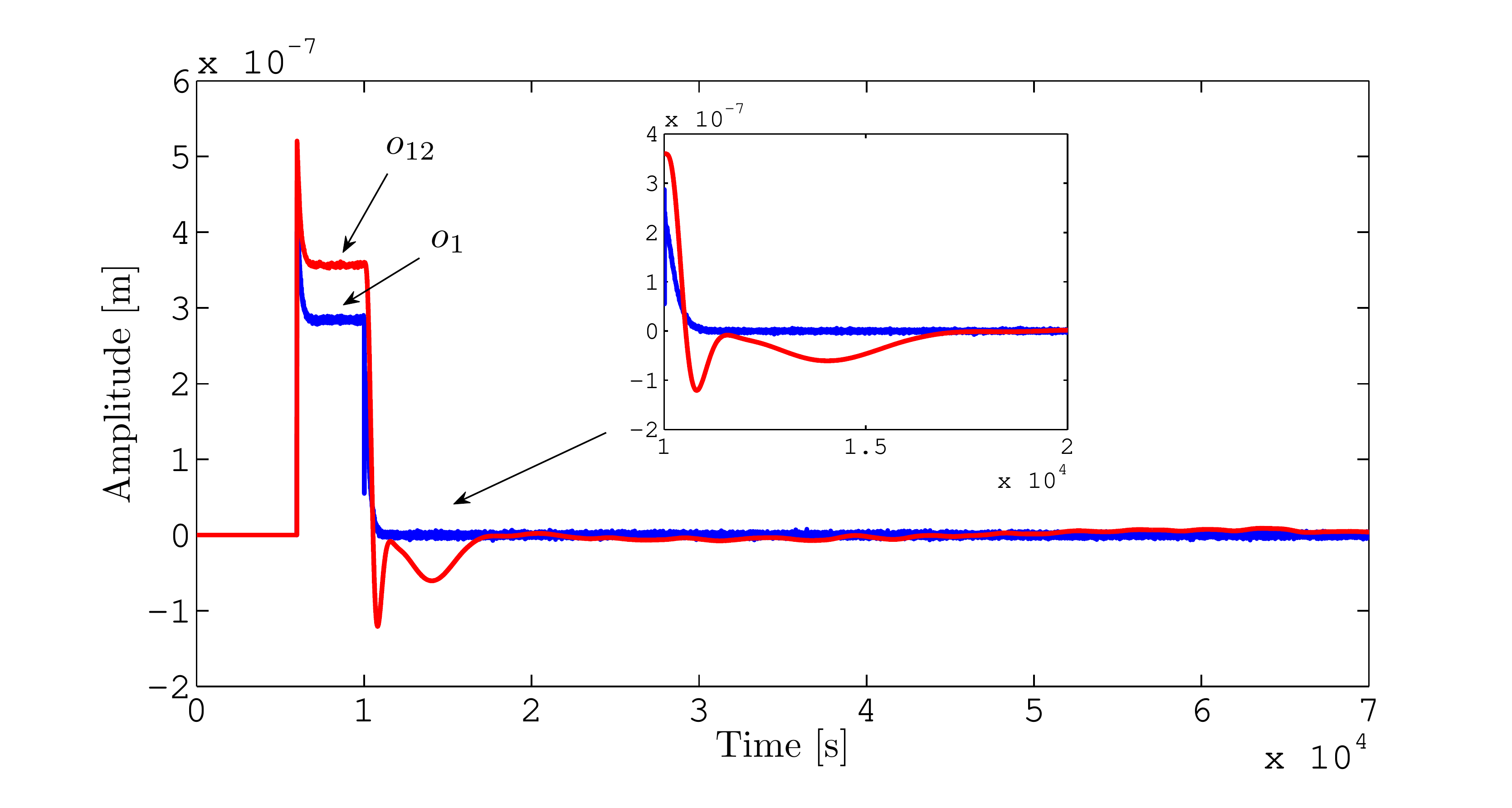}{The first 3 hours of a typical noise run of the OSE. After the TM release, the system is firstly turned into accelerometer mode, then into science mode around $\unit[1\e{4}]{s}$. The transient due to non-zero initial conditions lasts for half a hour in $o_1$ and about 2 hours in $o_{12}$.}{fig:sys_identification:transients}{0.8}

\begin{figure}[!htbp]
\centering
\begin{tabular}{*{2}{@{}c@{}}}
\includegraphics[width=0.5\columnwidth]{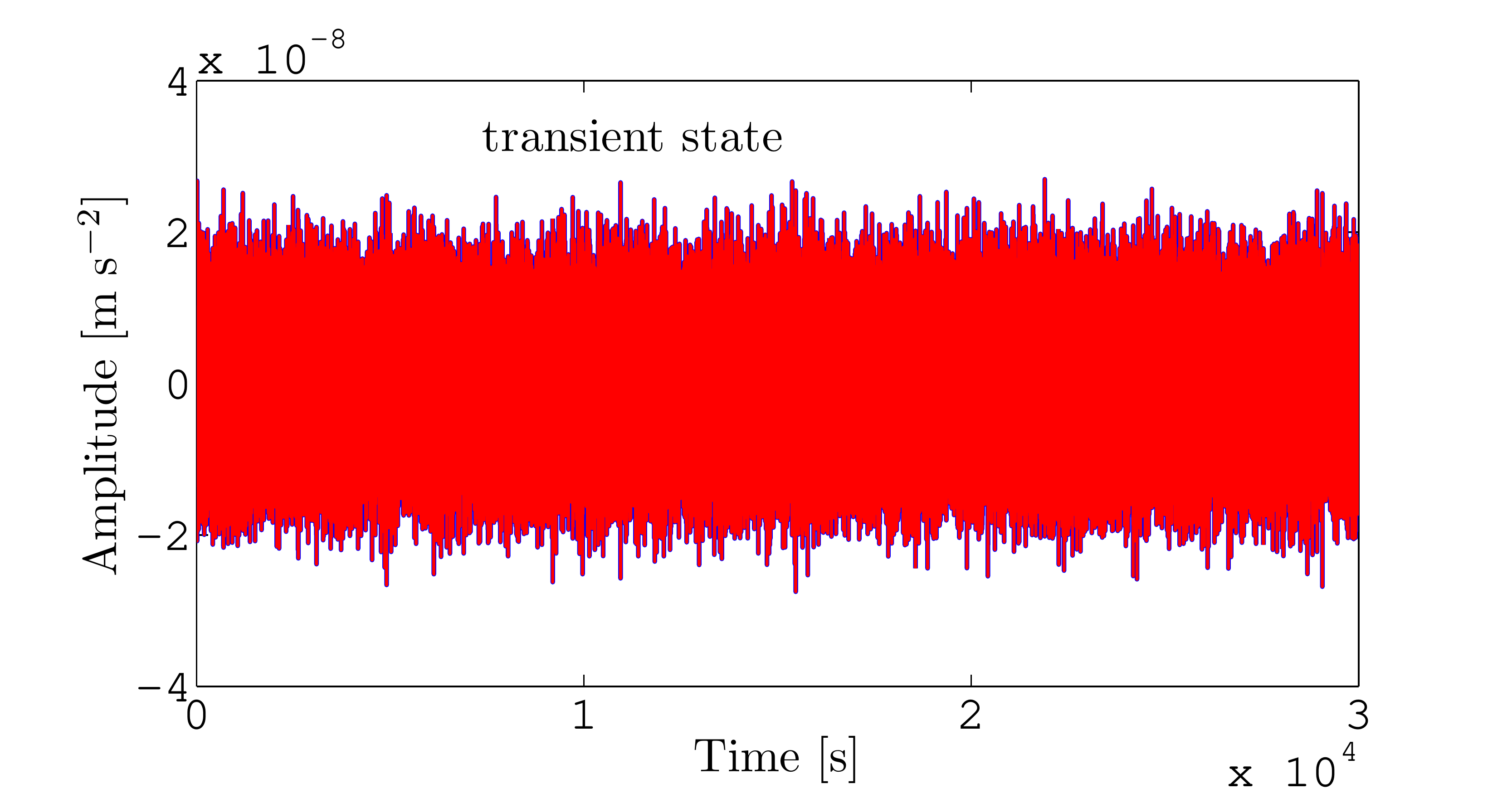} &
\includegraphics[width=0.5\columnwidth]{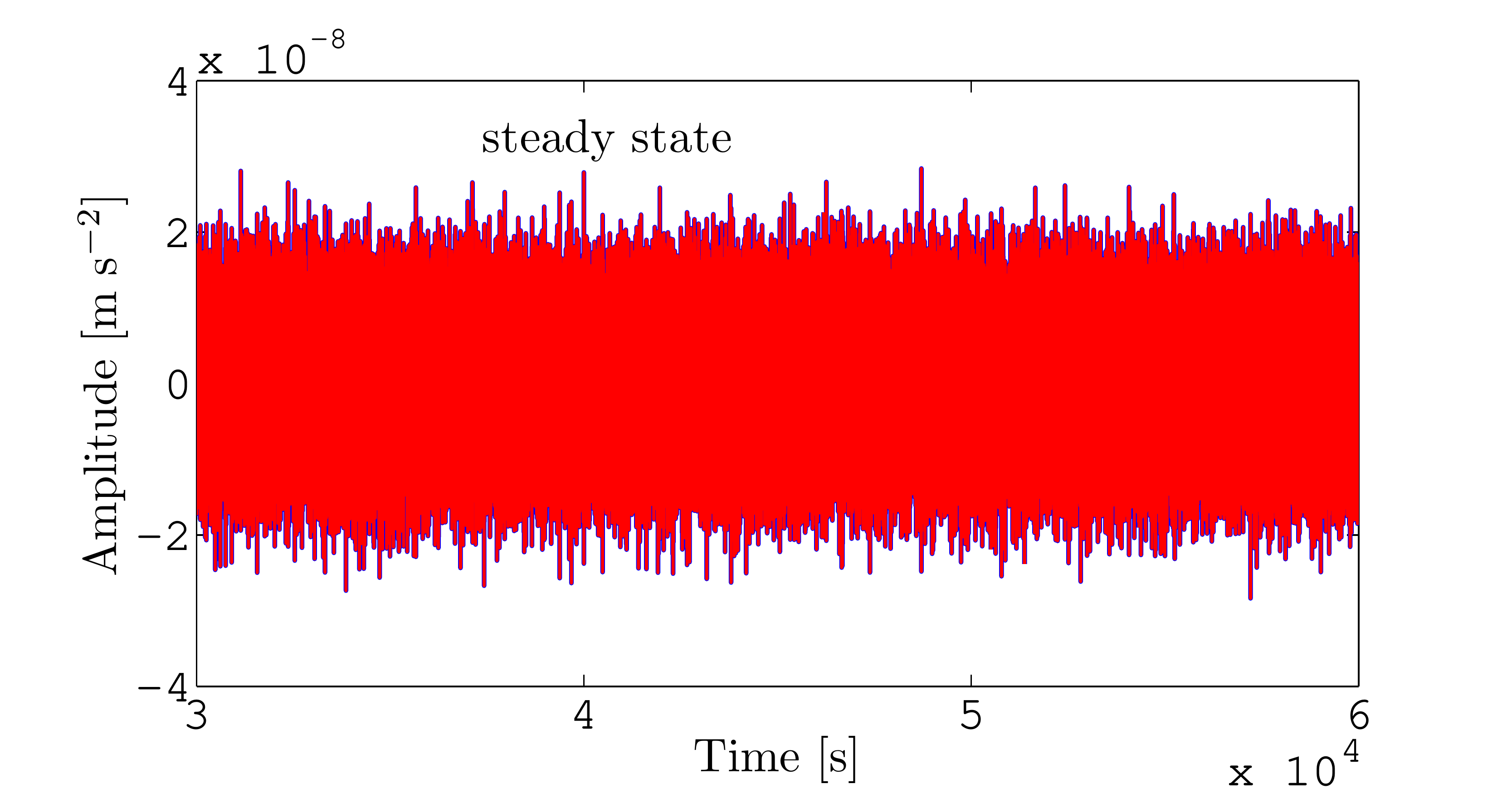} \\
\footnotesize{\hspace{10pt}(a)} &
\footnotesize{\hspace{10pt}(b)}
\end{tabular}
\caption{\footnotesize{Total equivalent differential acceleration noise numerically estimated on synthetic data produced by the OSE and shown in \figref{fig:sys_identification:transients}. The data are split into two segments: the first one just after the system is turned into science mode and containing the transient state; the second one follows it and is driven by the steady state. The estimation of the total equivalent out-of-loop acceleration can be performed either \textit{with} a preliminary system identification (fiducial parameter values) or \textit{without} it (approximate parameter values) (lines with different colors) and apparently there is no difference. (a) the transient is suppressed, compared to (b) where the system is dominated by the steady state.
}}
\label{fig:sys_identification:transients_acc}
\end{figure}

The estimation of the total equivalent differential acceleration noise is performed twice on the same data, including the initial transitory, assuming each time a different set of parameter values modeling the system. On one side, a fair approximation of those parameters -- the so-called initial guess -- reproduces the situation in which the estimation of the total equivalent acceleration is performed \textit{without} a preliminary system identification, as in the previous section. On the other side, a fiducial approximation of those parameters -- the so-called best-fit -- reproduces the situation in which the estimation of the total equivalent acceleration is performed \textit{with} a preliminary system identification. Moreover, the estimation is performed parallelly on two data segments lasting $\unit[3\e{4}]{s}$ each: the first one just after the system is turned into science mode and containing the transient state; the second one follows it and is driven by the steady state. \figref{fig:sys_identification:transients_acc} shows the estimated total equivalent differential acceleration noise for the two segments and for the two sets of parameter values. The comparison shows that the transient is suppressed, and there is no relevant difference between the two segments.

\figref{fig:sys_identification:transients_psd} reports the PSDs of the estimated total equivalent differential acceleration noise for the above time-series, i.e., assuming the two sets of parameter values for both segments. At low frequency, the noise level of the segment containing the transient is higher then the subsequent segment, but system identification helps in suppressing part of the noise around $\unit[1]{mHz}$. Below $\unit[0.7]{mHz}$ there is an evidence that there is an unsuppressed residual transitory in the data.


\figuremacroW{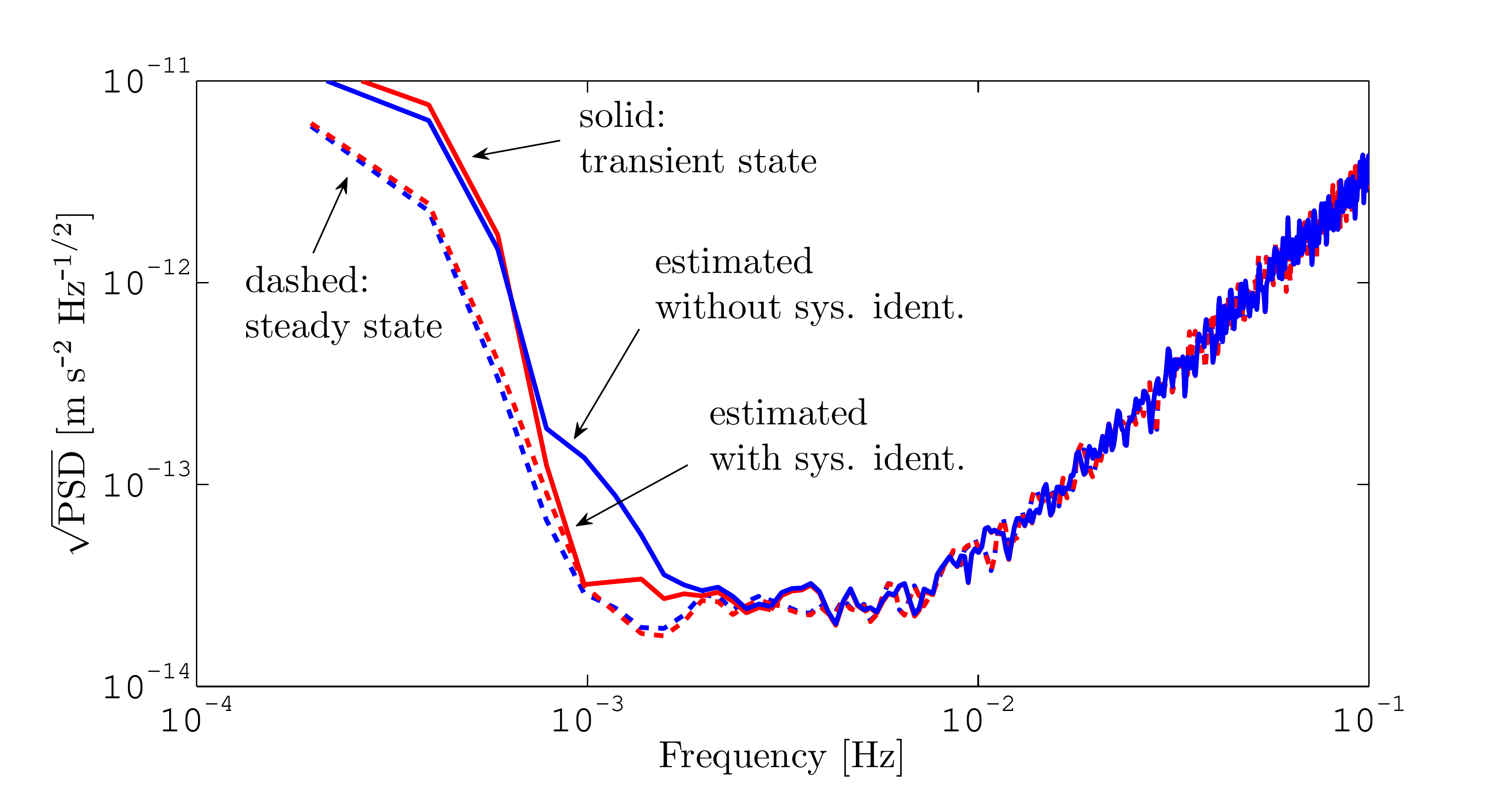}{Total equivalent differential acceleration noise numerically estimated on synthetic data produced by the OSE and shown in \figref{fig:sys_identification:transients_acc}. The estimation of the total equivalent out-of-loop acceleration can be performed either \textit{with} a preliminary system identification (fiducial parameter values) or \textit{without} it (approximate parameter values) on both segments: the first one dominated by the transient state and the second one dominated by the steady state. System identification helps in suppressing the transient around $\unit[1]{mHz}$.}{fig:sys_identification:transients_psd}{0.8}

The results of this section demonstrate how transients due to initial conditions can be suppressed with reasonably good approximation in the total equivalent acceleration time-series. The accuracy to which the suppression is effective depends on the accuracy to which the parameter values of the system are known. As shown, system identification helps in mitigating the effect due to transients in the data.

%





\ChangeFigFolder{5_optimal_design}


\chapter{Design of optimal experiments} \label{chap:optimal_design}


The previous chapter introduced system identification and its relevance for the unbiased estimation of the total equivalent acceleration noise of the LPF mission. A standard series of sine injections spanning the frequency band was utilized therein. Such bias injections make possible the estimation of the modeled system parameters the total equivalent acceleration noise depends on. Clearly, the goal of system identification is the parameter accuracy. A possible approach is also to search for optimized stimuli to assess the system parameters with better precision, with the final aim of a better estimate of the equivalent acceleration noise. 
The relevance is worth that this chapter addresses the question and provides for a solution.

\section{Review of the problem} \label{sect:optimal_design:problem}

Aiming at discriminating among different designs of the same system identification experiment, it is a rather natural consequence to enter into the field of the optimal design of experiments \cite{pukelsheim,melas}. This matter tries to answer to those physical problems characterized by a design matrix that shall be maximized in order to perform a targeted measurement with an optimized precision. This links to some very recent examples of practical applications of the optimal design theory, multidisciplinary and covering very different research fields: from dynamical systems \cite{brown2011}, to geophysics \cite{khodja2010}, quantum state tomography \cite{nunn2010} and even magnetic resonance in medical engineering \cite{choi2010}. A general review can be found in \cite{pronzato2008,sagnol2010}. In what follows the same philosophy is applied to the LPF mission with its peculiarities: the level of complexity is worth as the very example of a MIMO multi-degree-of-freedom dynamical system with coupled closed loops and subjected to various constraints.

As described in the previous chapter, system identification is targeted to measuring the system parameters $\vect{p}$ appearing within the transfer matrix $\vect{T}_{o_\text{i}\rightarrow o}(\omega,\vect{p})$ connecting applied controller biases to interferometric readouts, for the case of the investigation along the optical axis. If the inputs $\vect{o}_\text{i}$ are parameterized by a set of parameters $\vect{\theta}$, then the Fisher information matrix in \eqref{eq:sys_identification:info_matrix} becomes
\begin{equation}\label{eq:optimal_design:info_matrix}
\vect{\mathcal{I}}(\vect{\theta})=
\int
\ctranspose{\vect{o}_\text{i}(\omega,\vect{\theta})} \, \ctranspose{\nabla_{\vect{p}} \vect{T}_{o_\text{i}\rightarrow o}(\omega,\vect{p}_\text{est})} \,
\vect{S}_\text{n}(\omega)^{-1} \,
\nabla_{\vect{p}} \vect{T}_{o_\text{i}\rightarrow o}(\omega,\vect{p}_\text{est}) \, \vect{o}_\text{i}(\omega,\vect{\theta})\,\text{d}\omega~.
\end{equation}
Requiring that the estimates $\vect{p}_\text{est}$ should be given with the optimal precision implies that the preceding matrix must be optimized with respect to the design given by $\vect{\theta}$.

Theory provides for a solution of the problem. In fact, the optimal design is attained by building up a scalar estimator on the information matrix, $\phi[\vect{\mathcal{I}}]$, which is mathematically a functional over that matrix. With the parametrization introduced above, the functional simply becomes a scalar function of $\vect{\theta}$
\begin{equation}
\phi[\vect{\mathcal{I}}]=\phi(\vect{\theta})~,
\end{equation}
for given noise PSDs, interferometric readouts and estimated system parameters. Hereafter, three different choices of the functional $\phi$ are considered
\begin{equation}\label{eq:optimal_design:opt_criteria}
\phi(\vect{\theta})=
\begin{cases}
\det(\vect{\mathcal{I}}(\vect{\theta})) & \text{D optimality} \\
\min(\text{eig}(\vect{\mathcal{I}}(\vect{\theta}))) & \text{E optimality} \\
\text{tr}(\vect{\mathcal{I}}(\vect{\theta})) & \text{T optimality} \\
\end{cases}~,
\end{equation}
and the corresponding for the covariance matrix, obtained directly inverting \eqref{eq:optimal_design:info_matrix}, since maximum information is equivalent to minimum variance. The interpretation of each single criterion is readily discussed. The \textit{D optimality} is the determinant of the information matrix and averages the information along all terms, diagonal and off-diagonal. The \textit{E optimality} takes the minimum eigenvalue and tries to balance it with the others, hence regularizing the conditional number of the matrix \footnote{The conditional number is defined as the ratio between the minimum and maximum eigenvalues. It expresses the sensitivity of the matrix to numerical inversions. Round-off errors affect the operation when the conditional number is either very small or very large. A number of order 1 is considered stable to inversions.}. The \textit{T optimality} gives to the diagonal the highest weight and corresponds to the averaged information along all parameters. Even though quite different in definitions, the criteria share the same philosophy: maximizing/minimizing the information/covariance volume in the system parameter space around the minimum.

For LPF there is one more point adding much more complexity. The typical constraints that must be met during all operations and especially for the experiment design are:
\begin{enumerate}
  \item the general shape of the biases being injected;
  \item the dynamical range of the interferometer, $\oforder\unit[100]{\mu m}$;
  \item the force authority for thruster actuation, $\oforder\unit[100]{\mu N}$;
  \item the force authority for capacitive actuation, $\oforder\unit[2.5]{nN}$.
\end{enumerate}
Concerning the first one, the typical duration of an identification experiment shall not exceed $T\oforder3$ hours, mostly to ensure noise stationarity. The system can be stimulated with a series of sine-waves of constant duration each $\delta t\simeq\unit[1200]{s}$, as already described in \sectref{sect:sys_identification:experiments}. To simplify the problem, the duration is kept fixed during the optimization. Furthermore, the requirement of avoiding possible system transients at the beginning and the end of each cycle, suggests to set null Dirichlet boundary conditions (i.e., null initial and final values of the signals) and leave gaps of $\delta t_\text{gap}\simeq\unit[150]{s}$. The general expression of a guidance signal is a windowed series of sines
\begin{equation}\label{eq:optimal_design:input signal}
o_\text{i}(t) = \sum_{n=1}^{N_\text{inj}} a_n\,\sin(2\pi f_n\,t)\,\theta(t-t'_n)\,\theta(t''_n-t)~,
\end{equation}
where $\theta$ is the Heaviside unit-step, $f_n=n/\delta t$ is the injected frequency of the $n$-th cycle and $a_n$ the corresponding amplitude, through the maximum number of injections $N_\text{inj}=7$, and $t'_n=t_0+(n-1)(\delta t+\delta t_\text{gap})$ and $t''_n=t_0+n\,(\delta t+\delta t_\text{gap})-\delta t_\text{gap}$ the initial and final instants of the $n$-th injection cycle, with $t_0$ the starting instant of the experiment. Clearly, the frequencies $f_n$ are set by general requirements on the experiment duration, whereas the amplitudes $a_n$ by the other three requirements (dynamical range and force authority). It is rather obvious that the information matrix scales as the SNR of the signal, hence as $a_n$, so the amplitudes are chosen to be the maximum allowed, not exceeding namely $1\%$ of the operating range of the interferometer and $10\%$ of the maximum force authority.

The optimal design problem for LPF can now be stated as the following. The functional $\phi$ must be optimized for given noise PSDs and transfer matrix around the estimated system parameters, with respect to the design parameters $\vect{\theta}$ containing the frequencies of the injected biases. As the frequency changes, the amplitudes are updated accordingly while preserving the constraints elucidated above.

The dependence of the information matrix to the parameters of the injected bias signals is somewhat implicit and masked by the integral and the Fourier transform in \eqref{eq:optimal_design:info_matrix}. It should be also noticed that the criteria in \eqref{eq:optimal_design:opt_criteria} are de-facto producing a matrix that is as much diagonal as possible with respect to the choice of $\vect{\theta}$. The implicit parametric diagonalization of the information matrix is equivalent to the simultaneous diagonalization of noise and transfer matrices. In light of this, optimal design appears somehow related to an eigen-decomposition of the system with respect to the differential operator and noise at the same time.

\section{Optimizing the identification experiments} \label{sect:optimal_design:optimization}

Referring to \sectref{sect:dynamics:model_along_x} and \sectref{sect:sys_identification:dynamical_model} for the description of a LPF model along the optical axis, this section shows the improvement in the measured precision of the stiffness constants, $\omega_1^2$ and $\omega_{12}^2$, the sensing cross-talk, $S_{21}$, and the actuation gains, $A_\text{df}$ and $A_\text{sus}$. Possible delays in the application of guidance signals are left out from the analysis without lose of generality.

The two standard identification experiments described in \sectref{sect:sys_identification:experiments} can be optimized independently once an estimate of the parameter values is given by a preliminary system identification. The scheme proposed here -- to be followed during the mission -- is:
\begin{enumerate}
  \item estimate the parameter values with standard experiments as in the preceding chapter;
  \item optimize the experiments around the parameter estimates;
  \item estimate the parameter values with optimized experiments, as in this chapter, to get more confidence in the recovered total equivalent acceleration noise.
\end{enumerate}

As said, the design parameters on which the information matrix is optimized are the frequencies of the injected bias signals. Instead, the amplitudes are updated accordingly by meeting the requirements on the interferometer sensing range and force authorities. By means of the transfer matrices in \sectref{sect:dynamics:matrix_formalism}, the maximum amplitudes are conservatively computed by taking the minimum between the requirements in interferometer range and force authority
\begin{subequations}\label{eq:optimal_design:constraints}
\begin{align}
a_{o_{\text{i},1}} & = \min\{ T_{o_\text{i,1}\rightarrow o_1}\,o_{1,\text{max}}\, , \,
T_{o_\text{i,1}\rightarrow f_{\text{c},1}}\,f_{1,\text{max}}\}~, \\
a_{o_{\text{i},12}} & = \min\{ T_{o_\text{i,12}\rightarrow o_{12}}\,o_{12,\text{max}}\, , \,
T_{o_\text{i,12}\rightarrow f_{\text{c},12}}\,f_{12,\text{max}}\}~,
\end{align}
\end{subequations}
where $o_{1,\text{max}}=o_{12,\text{max}}=\unit[1]{\mu m}$ ($1\%$ of the interferometer range), $f_{1,\text{max}}=\unit[10]{\mu N}$ ($10\%$ of thruster authority) and $f_{12,\text{max}}=\unit[0.25]{nN}$ ($10\%$ of electrostatic suspension authority). For example, $T_{o_\text{i,1}\rightarrow o_1}$ represents the transfer from the guidance signal $o_\text{i,1}$ to the interferometric readout $o_1$; analogously, $T_{o_\text{i,1}\rightarrow f_{\text{c},1}}$ represents the transfer from the guidance signal $o_\text{i,1}$ to the commanded thruster force $f_{\text{c},1}$.

\figref{fig:optimal_design:oi1_max} shows how the amplitudes so far determined depend on the injection frequencies, for the first identification experiment. Analogously, \figref{fig:optimal_design:oi1_max} shows the same relationship for the second identification experiment. The interferometer range is the most stringent requirement, whereas force authority may limit at high frequency, especially for $o_{i,1}$. For this reason, only the first requirement is actually considered in the analysis, however limiting the maximum frequency to $\unit[50]{mHz}$ ($\nicefrac{1}{10}$ of the Nyquist frequency for data sampled at $\unit[1]{Hz}$).


\figuremacroW{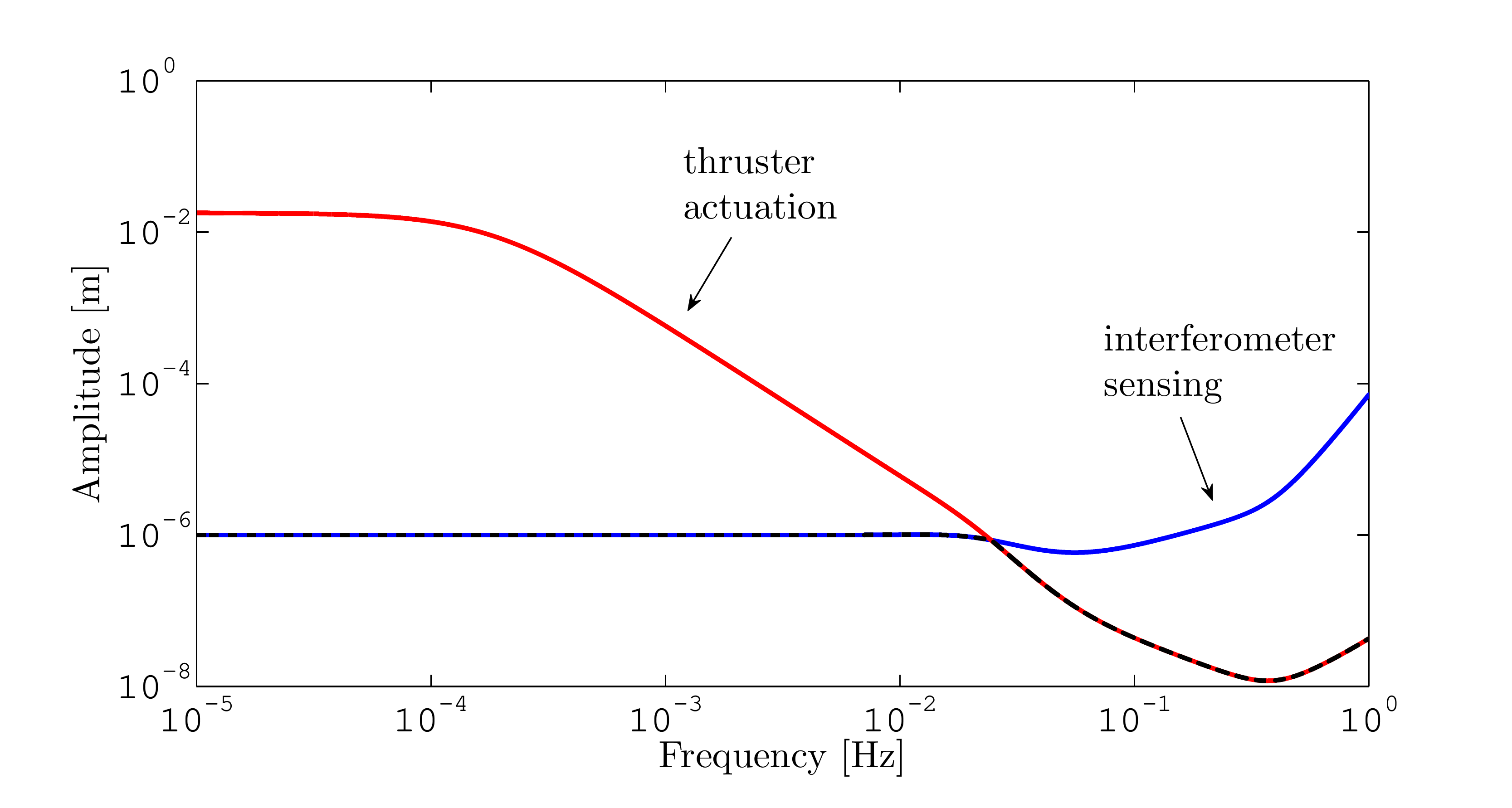}{Maximum allowed amplitude for bias injection $o_{i,1}$. The amplitude is limited by the interferometer operating range for almost the entire frequency band. Above $\unit[20]{mHz}$ it starts to be limited by thruster authority. Maximum amplitudes do not exceed $\unit[1]{\mu m}$ in interferometer range, $\unit[10]{\mu N}$ in thruster authority and $\unit[0.25]{nN}$ in electrostatic suspension authority. The combination of both requirements is shown as a dashed line.}{fig:optimal_design:oi1_max}{0.8}

\figuremacroW{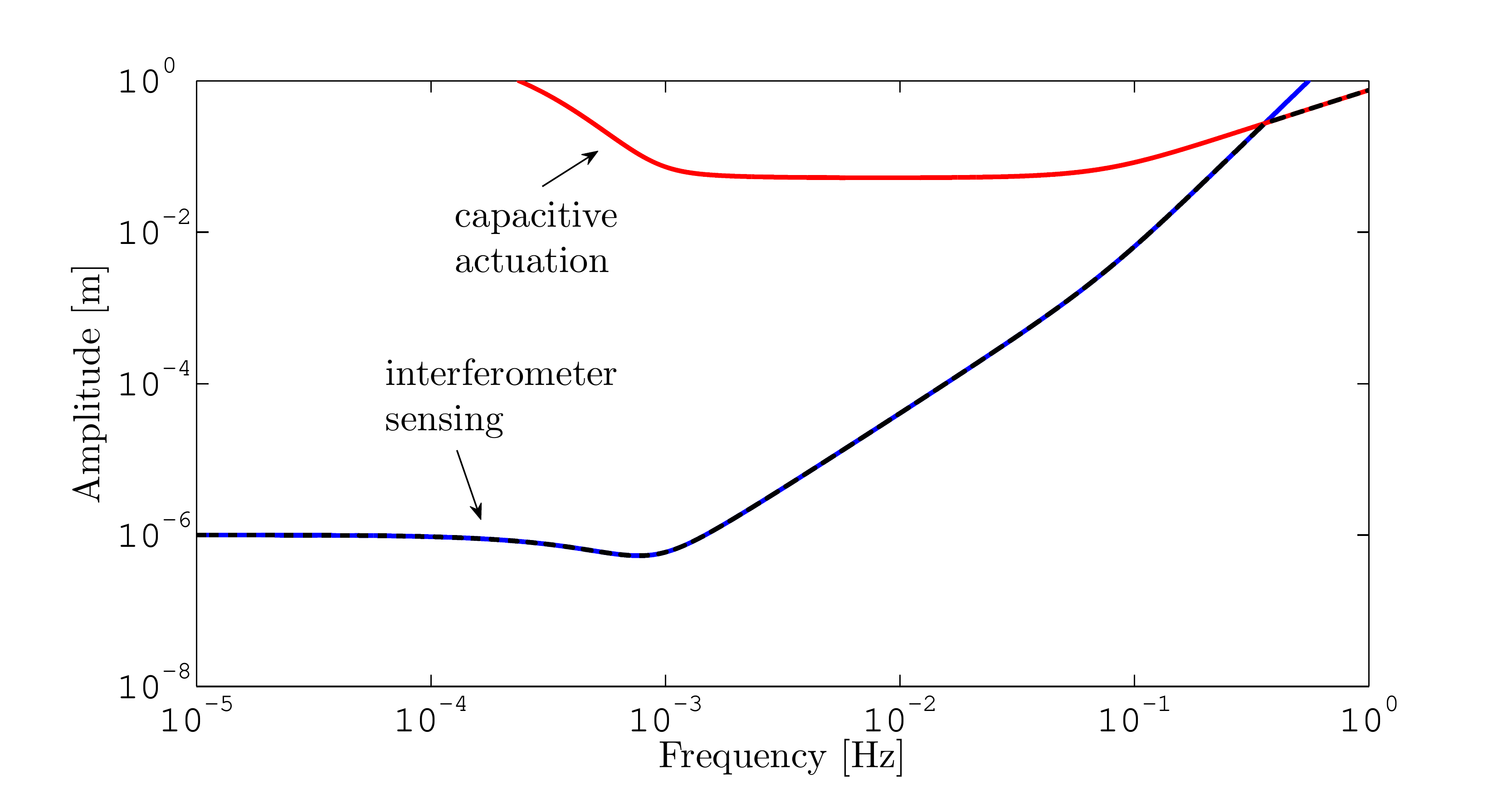}{Maximum allowed amplitude for bias injection $o_{i,12}$. The amplitude is limited by the interferometer operating range. Maximum amplitudes do not exceed $\unit[1]{\mu m}$ in interferometer range, $\unit[10]{\mu N}$ in thruster authority and $\unit[0.25]{nN}$ in electrostatic suspension authority. The combination of both requirements is shown as a dashed line.}{fig:optimal_design:oi12_max}{0.8}

The analysis of two experiments requires two independent optimizations on 7-dimensional discrete spaces, spanning the frequencies of each injection cycle. Despite the previous chapter where the optimization variables (the system parameters) were continuous, the injection frequency space must be discrete. In fact, each wave is required to start and stop at zero, so that transients can be avoided.

Discrete optimization is always more mathematically complicated than continuous optimization. The first can invoke refined mathematical techniques like graph theory. On the contrary, standard well-known numerical optimization algorithms frequently assume continuity and smoothness in the independent variable being optimized. Since investigating in sophisticated methods is out of the scope of this thesis, a trick is found here to overcome the problem of discrete numerical optimization. First of all, the choice naturally falls to \textit{direct methods}, like the simplex and pattern search \cite{press}. Those methods (i) do not make use of analytical derivatives, as such an evaluation for this problem introduces a very high level of complexity and (ii) are more robust to function discontinuities than other algorithms. The trick consists on overlapping a discrete grid to the continuous space, whose nodes are the pole of attraction for the independent continuous variables. The merit function consists of three main calculations:
\begin{enumerate}
  \item the information matrix $\vect{\mathcal{I}}$ for given noise, transfer matrix and parametric input signals, following \eqref{eq:optimal_design:info_matrix};
  \item the functional $\phi$ in \eqref{eq:optimal_design:opt_criteria};
  \item the rounding of the injection frequencies to the nearest grid node as the optimization carries on.
\end{enumerate}
Every time the merit function is called, the frequencies are forced to lay on the grid, but the side effect is that the surface becomes highly irregular. However, the optimization can be implemented with the standard direct search algorithms.

\section{Multi-experiment, single-input} \label{sect:optimal_design:optimized_exp1_exp2}

In view of comparing the performances of the 6 optimization criteria contained in \eqref{eq:optimal_design:opt_criteria} (both information and covariance matrices) for two identification experiments in a mission-like manner, here is the adopted analysis procedure:
\begin{enumerate}
  \item two standard identification experiments are simulated and the system parameters estimated according to the methods of the previous chapter;
  \item 6 independent optimizations around the best-fit values allow to find optimized experimental designs of the injection biases;
  \item 6 system identifications are performed along with those designs;
  \item optimal best-fit values and standard deviations are extracted from each fit.
\end{enumerate}

\tabref{tab:optimal_design:comparing_criteria} shows the results of the investigation, by comparing the standard experiment to the optimized ones. The standard deviations as estimated from the fit quantifies the precision of that design, whereas the estimate biases (deviation of the best-fit value from the real value in units of standard deviations) quantifies the accuracy. By inspecting the results, the T optimality criterion for the information matrix gives, in average, the best precision and accuracy. The estimate biases are within 1--2 standard deviations and the fit standard deviations are lower than the standard by a factor 2 for $\omega_1^2$ and $\omega_1^2$, 4 for $S_{21}$, 5 for $A_\text{sus}$ and 7 for $A_\text{df}$. Other criteria may worsen the measurement, especially for the covariance matrix: this is an indication that the numerical matrix inversion introduces an extra source of indetermination.
\begin{sidewaystable}[!htbp]
\caption{\label{tab:optimal_design:comparing_criteria}\footnotesize{Comparison of performances for different optimal designs. The fit standard deviations for all 5 parameters are reported for the 6 optimal design approaches, based on information and covariance matrices. In curly brackets the bias (absolute deviation from the real value in units of standard deviation) for each estimate. The T optimality criterion for the information matrix gives, in average, the best precision and accuracy.}}
\centering
\scalebox{.9}{
\begin{tabular}{l D{!}{\times}{0.10} D{!}{\times}{0.10} D{!}{\times}{0.10} D{!}{\times}{0.10} D{!}{\times}{0.10} D{!}{\times}{0.10} D{!}{\times}{0.10}}
\hline
\hline
\multicolumn{1}{l}{Parameter} & \multicolumn{1}{c}{\multirow{2}{*}{Standard}} & \multicolumn{3}{c}{Information} & \multicolumn{3}{c}{Covariance} \\
\multicolumn{1}{l}{st.\,dev.} & & \multicolumn{1}{c}{D} & \multicolumn{1}{c}{E} & \multicolumn{1}{c}{T} & \multicolumn{1}{c}{D} & \multicolumn{1}{c}{E} & \multicolumn{1}{c}{T} \\
\hline
$\sigma_{\omega_1^2}\,[\text{s}^{-2}]$ & 4 ! 10^{-10} \{1.4\} & 3 ! 10^{-10} \{0.48\} & 8 ! 10^{-9}  \{0.24\} & 2 ! 10^{-10} \{0.68\} & 9 ! 10^{-10} \{1.1\} & 3 ! 10^{-10} \{2.1\} & 2 ! 10^{-9} \{0.13\} \\
$\sigma_{\omega_{12}^2}\,[\text{s}^{-2}]$ & 2 ! 10^{-10} \{0.41\} & 2 ! 10^{-10} \{1.6\} & 8 ! 10^{-9}  \{0.23\} & 1 ! 10^{-10} \{2.0\} & 9 ! 10^{-10} \{0.97\} & 2 ! 10^{-10} \{2.7\} & 2 ! 10^{-9} \{0.16\} \\
$\sigma_{S_{21}}$ & 4 ! 10^{-7} \{0.086\} & 1 ! 10^{-7} \{0.55\} & 2 ! 10^{-7} \{0.77\} & 1 ! 10^{-7} \{1.1\} & 1 ! 10^{-7} \{0.047\} & 1 ! 10^{-7} \{0.056\} & 1 ! 10^{-7} \{0.58\} \\
$\sigma_{A_\text{df}}$ & 7 ! 10^{-4} \{1.6\} & 2 ! 10^{-4} \{0.61\} & 1 ! 10^{-4} \{1.0\} & 1 ! 10^{-4} \{0.50\} & 3 ! 10^{-4} \{0.36\} & 2 ! 10^{-4} \{2.1\} & 1 ! 10^{-4} \{1.8\} \\
$\sigma_{A_\text{sus}}$ & 1 ! 10^{-5} \{1.7\} & 1 ! 10^{-6} \{0.24\} & 1 ! 10^{-6}  \{0.27\} & 2 ! 10^{-6} \{0.28\} & 2 ! 10^{-6} \{0.38\} & 1 ! 10^{-6} \{0.95\} & 1 ! 10^{-6} \{1.5\} \\
\hline
\hline
\end{tabular}
}
\end{sidewaystable}

Choosing the T criterion for the information matrix as the reference for further comments, \tabref{tab:optimal_design:optim_exps} reports the optimal input frequencies and amplitudes compared to the standard ones for both experiments. Transparently, the system relaxes to only two relevant frequencies: the lowest, $\unit[0.83]{mHz}$, and the highest allowed, $\unit[49]{mHz}$. 
The result should not surprise since the previous chapter implicitly took to a similar conclusion: the two frequencies are indeed the two maxima of the transfer matrix in \figref{fig:sys_identification:ifo2ifo}. The transfer from $o_{i,1}$ and $o_{i,12}$ to $o_{12}$ are maximized at little less than $\unit[1]{mHz}$; the transfer from $o_{i,1}$ to $o_{1}$ is maximized at around $\unit[0.1]{Hz}$.
\begin{table}[!htbp]
\caption{\label{tab:optimal_design:optim_exps}\footnotesize{Comparison of input frequencies and amplitudes for the standard and optimal experiments. The injection cycles last $\unit[1200]{s}$ each and are separated by gaps of $\unit[150]{s}$. The system relaxes to only two relevant frequencies $\unit[0.83]{mHz}$ and $\unit[49]{mHz}$, namely the lowest and the highest allowed.}}
\centering
\begin{tabular}{D{.}{.}{3.3} D{.}{.}{3.3} D{.}{.}{3.3} D{.}{.}{3.3} D{.}{.}{3.3} D{.}{.}{3.3} D{.}{.}{3.3} D{.}{.}{3.3}}
\hline
\hline
\multicolumn{2}{c}{Standard Exp.\,1} & \multicolumn{2}{c}{Optimal Exp.\,1} & \multicolumn{2}{c}{Standard Exp.\,2} & \multicolumn{2}{c}{Optimal Exp.\,2} \\
\multicolumn{1}{c}{$\unit[f]{[mHz]}$} & \multicolumn{1}{c}{$\unit[a]{[\mu m]}$} & \multicolumn{1}{c}{$\unit[f]{[mHz]}$} & \multicolumn{1}{c}{$\unit[a]{[\mu m]}$} & \multicolumn{1}{c}{$\unit[f]{[mHz]}$} & \multicolumn{1}{c}{$\unit[a]{[\mu m]}$} & \multicolumn{1}{c}{$\unit[f]{[mHz]}$} & \multicolumn{1}{c}{$\unit[a]{[\mu m]}$} \\
\hline
0.83 & 1.0 & 0.83 & 1.0 & 0.83 & 0.80 & 0.83 & 0.55 \\
1.7 & 1.0 & 0.83 & 1.0 & 1.7 & 0.48 & 49 & 52 \\
3.3 & 1.0 & 49 & 0.55 & 3.3 & 0.19 & 49 & 52 \\
6.6 & 1.0 & 49 & 0.55 & 6.6 & 0.088 & 49 & 52 \\
13 & 0.59 & 0.83 & 1.0 & 13 & 0.096 & 49 & 52 \\
27 & 0.28 & 0.83 & 1.0 & 27 & 0.18 & 49 & 52 \\
53 & 0.14 & 49 & 0.55 & 53 & 0.46 & 49 & 52 \\
\hline
\hline
\end{tabular}
\end{table}

\begin{figure}[!htbp]
\centering
\begin{tabular}{*{2}{@{}c@{\hspace{-10pt}}}}
\includegraphics[width=0.5\columnwidth]{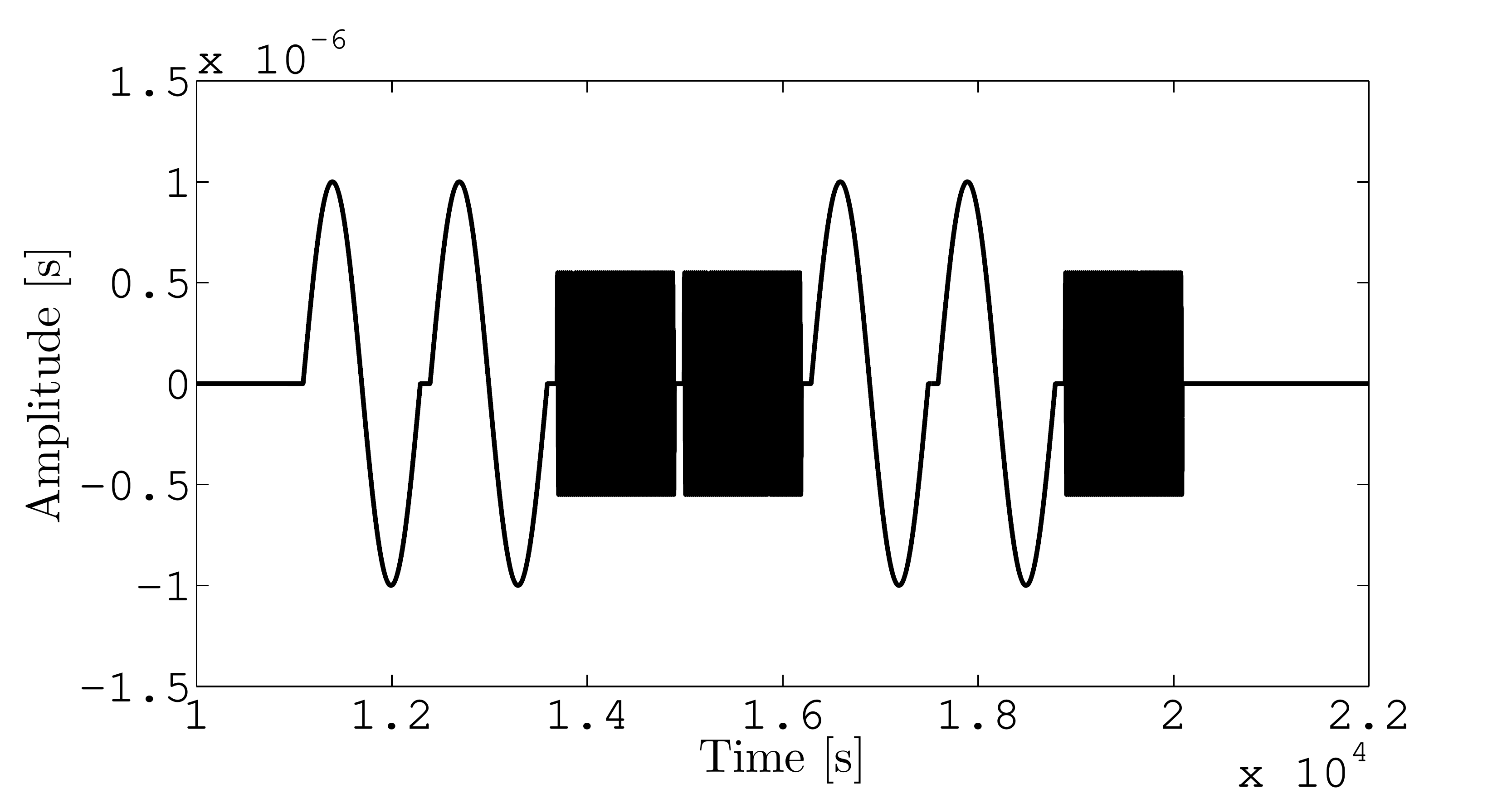} &
\includegraphics[width=0.5\columnwidth]{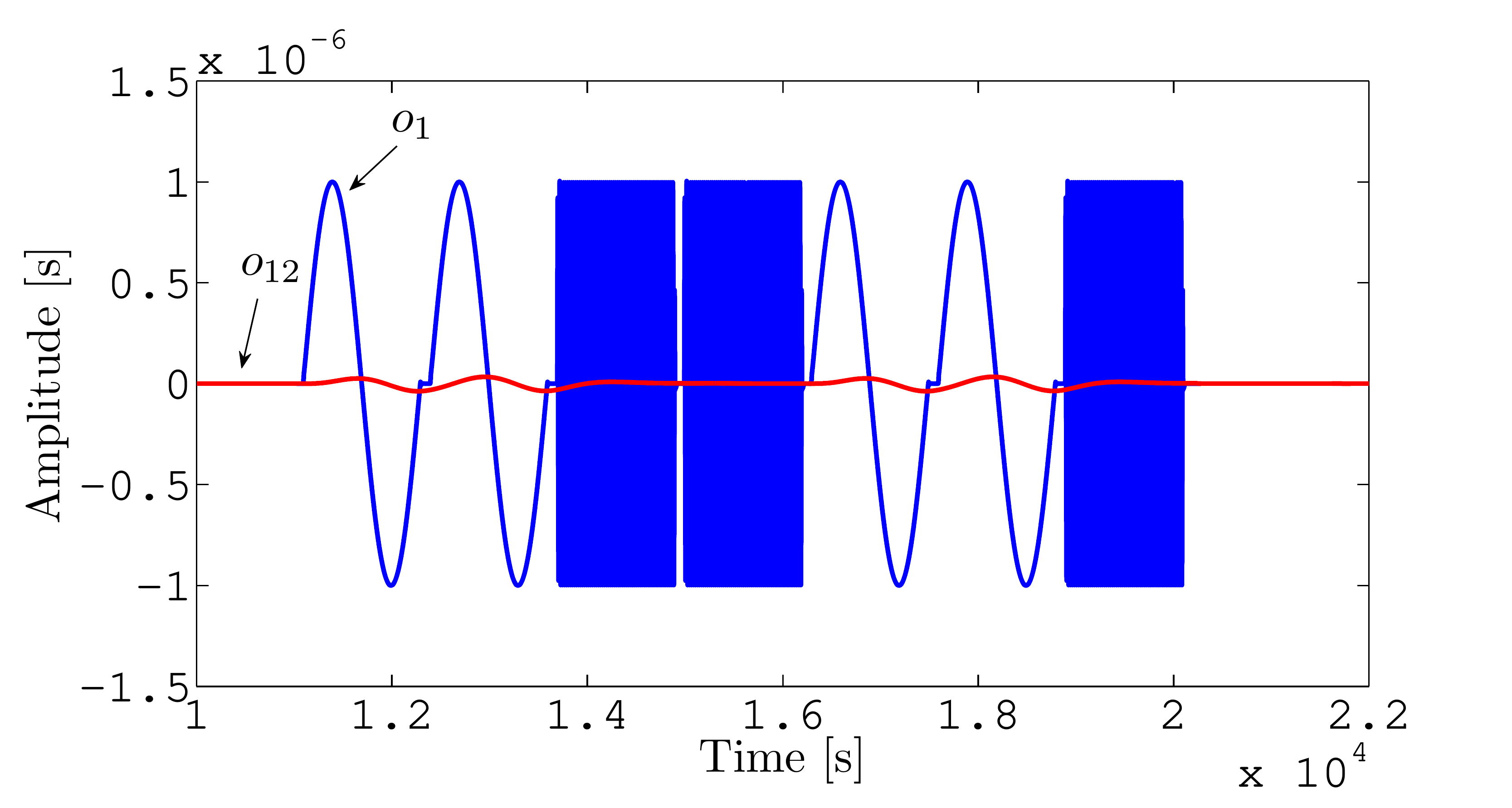} \\
\footnotesize{(a)} &
\footnotesize{(b)} \\
\includegraphics[width=0.5\columnwidth]{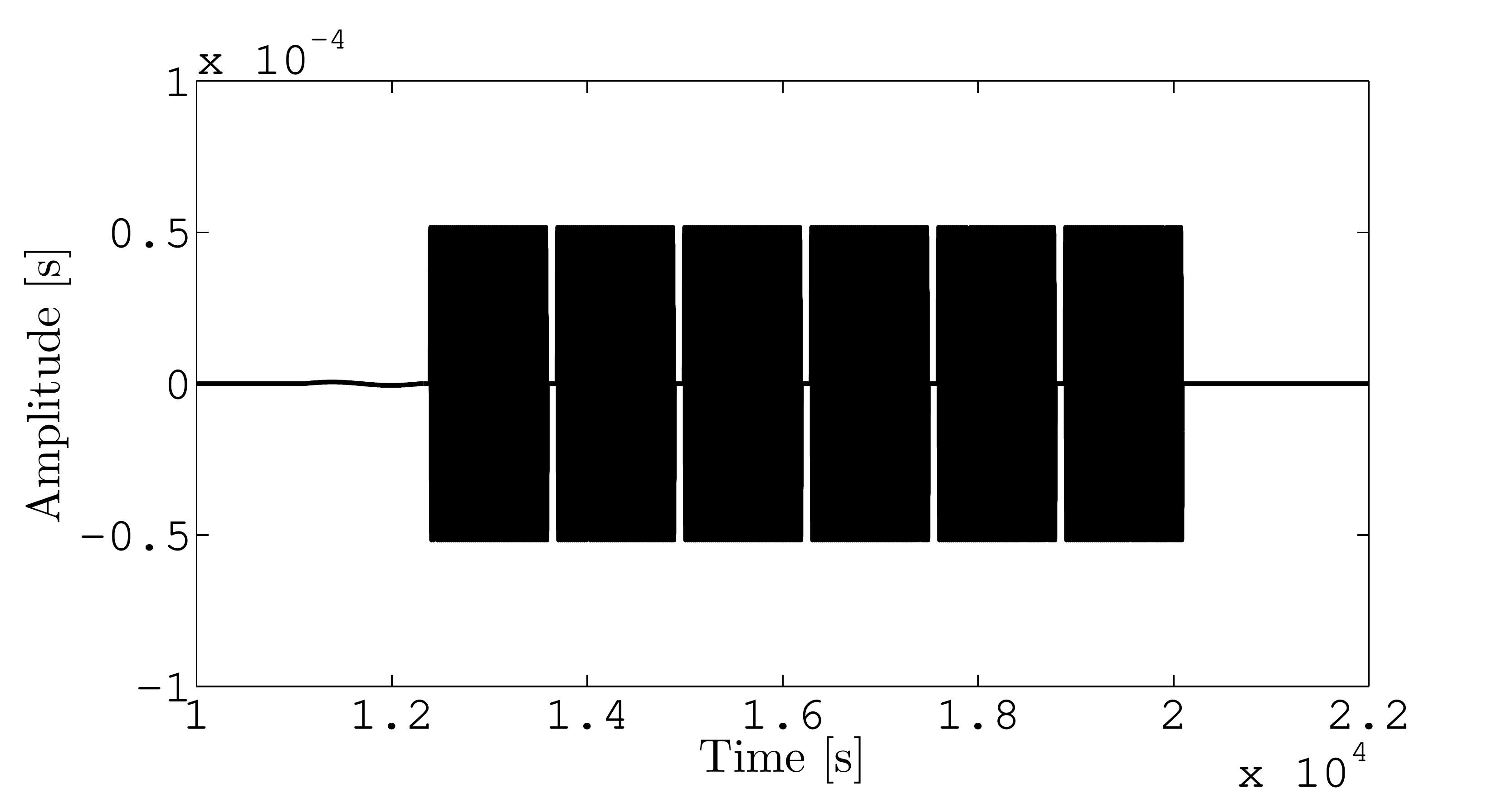} &
\includegraphics[width=0.5\columnwidth]{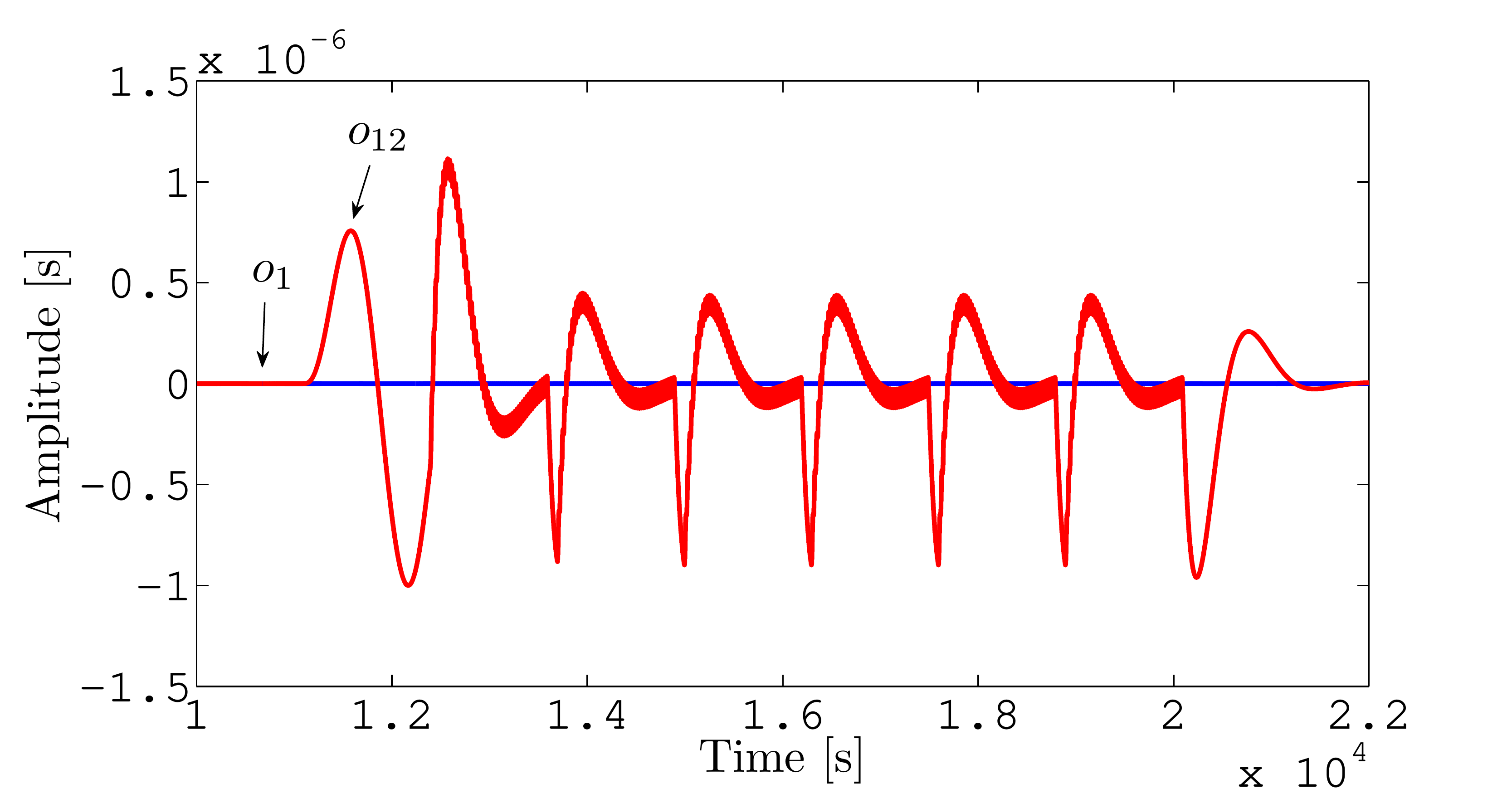} \\
\footnotesize{(c)} &
\footnotesize{(d)}
\end{tabular}
\caption{\footnotesize{Synthetic data for optimal design of Exp.\,1 (compare to \figref{fig:sys_identification:data_exp1}) and Exp.\,2 (compare to compare to \figref{fig:sys_identification:data_exp2}). The input signals, (a) and (c), and the interferometric readouts, (b) and (d), show that a better precision on the measurement of the system parameters can be attained by injecting only two relevant frequencies: the lowest and highest allowed.}}
\label{fig:optimal_design:signals}
\end{figure}

In \figref{fig:optimal_design:signals} the optimal bias signals are shown together with the system responses in both interferometric readouts for the two identification experiments. As usual, the bias signals are made of a series of sine-waves, whose frequencies and amplitudes are the ones described in \tabref{tab:optimal_design:optim_exps}. By inspecting the response in the second experiment (panel (d)) it naturally turns out that the big jumps are produced by the first derivative discontinuity of the Heaviside unit-step in \eqref{eq:optimal_design:input signal}. At that frequency, the discontinuity gives rise to a transient overlapping to the injected signal. However, the information on the system parameters is mostly carried by the injection frequency and not by the discontinuities. In fact, the simulation of another experiment with approximately the same duration and constituted by an injection of the same signal without the gaps proved that the same parameter precision can be attained.

A very interesting feature of the optimal design is its ability in improving the fit performances. It allows for the recovering of the best-fit values in fewer iterations than the standard design. It can be explained by the fact that the optimal design mitigates parameter correlations (the diagonalization of the information matrix implies lower correlation). \tabref{tab:optimal_design:corr} recaps some examples showing a clear improvement, 4 through 7 times better than the standard design, apart for $\text{Corr}[\omega_1^2,\omega_{12}^2]$ that remains unchanged.
\begin{table}[!htbp]
\caption{\label{tab:optimal_design:corr}\footnotesize{Different examples of parameter correlations. In some cases, the optimal design is capable in lowering the parameter correlation.}}
\centering
\begin{tabular}{l D{.}{.}{2.2} D{.}{.}{2.2}}
\hline
\hline
Correlation & \multicolumn{1}{c}{Standard} & \multicolumn{1}{c}{Optimal} \\
\hline
$\text{Corr}[S_{21},\omega_{12}^2]$ & -0.2 & -0.03 \\
$\text{Corr}[S_{21},\omega_1^2]$ & 0.09 & 0.02 \\
$\text{Corr}[A_\text{sus},\omega_1^2]$ & -0.7 & -0.2 \\
$\text{Corr}[\omega_1^2,\omega_{12}^2]$ & -0.5 & -0.5 \\
\hline
\hline
\end{tabular}
\end{table}

\section{Single-experiment, multi-input}

The preceding section has shown the optimization of the LPF experiments independently, by exploring the 7-dimensional input frequency space. The results are the improvement in precision, lower parameter correlation and the fact that only two input frequencies are actually needed. This section investigates on the possibility of defining a unique experiment in which bias signals are injected both at the same time.

Instead of two independent optimizations in 7-dimensional frequency spaces, for the experiment defined so far an optimization in a 14-dimensional frequency space is now needed. To actuate this program, the experiment, namely Exp.\,3, is defined for the simultaneous injection of $o_{\text{i},1}$ and $o_{\text{i},12}$. \tabref{tab:optimal_design:opt_design_exp3} reports the identification with such an optimized experiment, compared to the standard design, on one side, and the independently optimized designs, on the other side.
\begin{table}[!htbp]
\caption{\label{tab:optimal_design:opt_design_exp3}\footnotesize{Performances of optimal Exp.\,3 (simultaneous injection in both guidance signals), compared to the optimal Exp.\,1 \& Exp.\,2 of \sectref{sect:optimal_design:optimized_exp1_exp2} and the standard ones. The T optimality criterion is considered. The fit standard deviations for all 5 parameters are reported for the three cases. In curly brackets the bias (absolute deviation from the real value in units of standard deviation) for each estimate.}}
\centering
\begin{tabular}{l D{!}{\times}{2.4} l D{!}{\times}{2.4} l D{!}{\times}{2.4} l}
\hline
\hline
\multicolumn{1}{l}{Parameter} & \multicolumn{2}{c}{Standard} & \multicolumn{2}{c}{Optimal} & \multicolumn{2}{c}{Optimal} \\
\multicolumn{1}{l}{st.\,dev.} & \multicolumn{2}{c}{Exp.\,1 \& Exp.\,2} & \multicolumn{2}{c}{Exp.\,1 \& Exp.\,2} & \multicolumn{2}{c}{Exp.\,3} \\
\hline
$\sigma_{\omega_1^2}\,[\text{s}^{-2}]$ & 4 ! 10^{-10} & \{1.4\} & 2 ! 10^{-10} & \{0.68\} & 1 ! 10^{-10} & \{1.9\} \\
$\sigma_{\omega_{12}^2}\,[\text{s}^{-2}]$ & 2 ! 10^{-10} & \{0.41\} & 1 ! 10^{-10} & \{2.0\} & 8 ! 10^{-12} & \{0.42\} \\
$\sigma_{S_{21}}$ & 4 ! 10^{-7} & \{0.086\} & 1 ! 10^{-7} & \{1.1\} & 3 ! 10^{-8} & \{0.57\} \\
$\sigma_{A_\text{df}}$ & 7 ! 10^{-4} & \{1.6\} & 1 ! 10^{-4} & \{0.50\} & 8 ! 10^{-5} & \{0.73\} \\
$\sigma_{A_\text{sus}}$ & 1 ! 10^{-5} & \{1.7\} & 2 ! 10^{-6} & \{0.28\} & 1 ! 10^{-6} & \{0.16\} \\
\hline
\hline
\end{tabular}
\end{table}

The remarkable point to stress is the improvement in precision of an order of magnitude for almost all parameters with respect to the optimal experiments considered so far. Notice that the comparison should be taken with care. Since the information scale as the integration time $\mathcal{I}\propto T$ and the standard deviation scales as $\sigma\propto T^{-\nicefrac{1}{2}}$, then to compare the result of the third experiment to the other two experiments (that is approximately half long the total integration time of two independent experiments), its fit standard deviations must be divided by the factor $\oforder\sqrt{2}$. It should be also pointed out that parameter correlation does not improve, in fact a simultaneous injection may not be the best approach to disentangle degeneracies between the system parameters; a philosophy of the type \textit{the simpler the stimulus, the better the understanding of the system} should be adopted whenever possible.

\begin{figure}[!htbp]
\centering
\begin{tabular}{*{2}{@{}c@{\hspace{-10pt}}}}
\includegraphics[width=0.5\columnwidth]{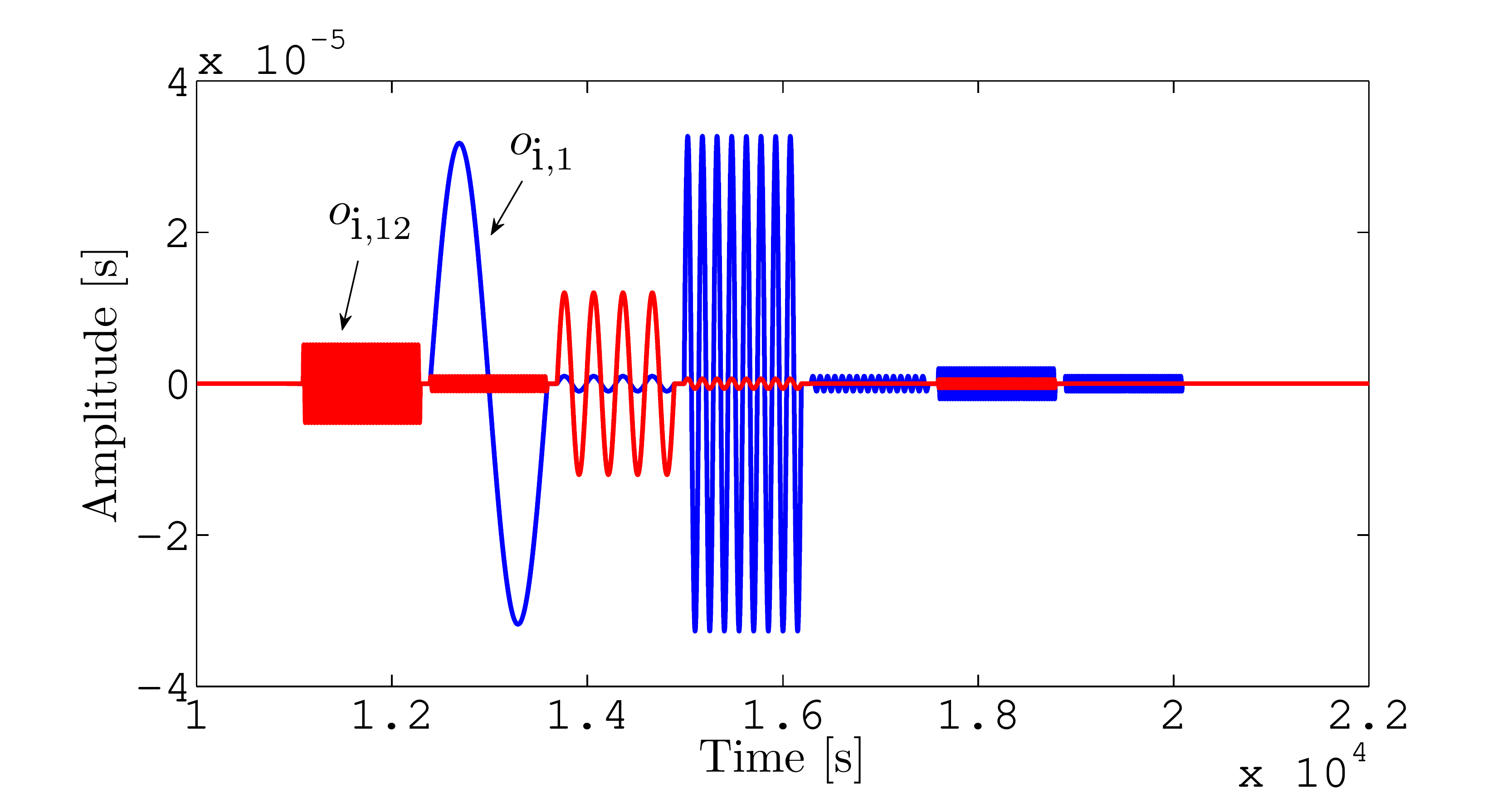} &
\includegraphics[width=0.5\columnwidth]{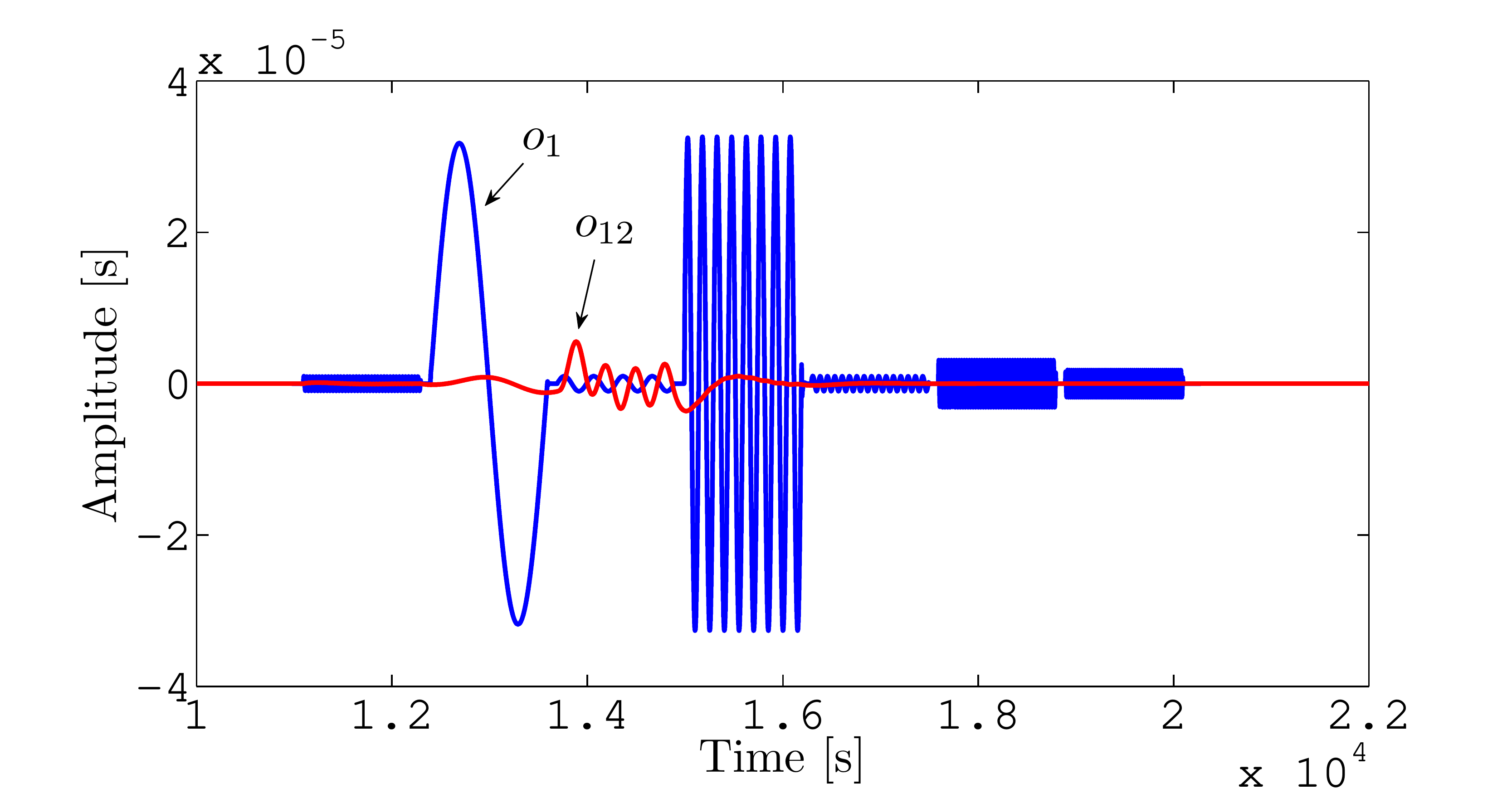} \\
\footnotesize{(a)} &
\footnotesize{(b)}
\end{tabular}
\caption{\footnotesize{Synthetic data for optimal design of Exp.\,3 (simultaneous injection in both guidance signals). (a) the input signals and (b) the interferometric readouts. The system does not appear relaxing to only two frequency as in the case of two independent injections.}}
\label{fig:optimal_design:signals_exp3}
\end{figure}

\figref{fig:optimal_design:signals_exp3} shows the signals being injected and the system responses in the interferometric readouts. As is clear, the level of numerical and conceptual complexity involved in the optimization of 14 input frequencies in the same experiments makes the interpretation very difficult. Contrary to the case of two independent injections, the optimization does not appear relaxing to two distinct frequencies. The reason could be conceptually matched to the simultaneous injection or due to intrinsical difficulties in the optimization. A mix of both causes is the most plausible explanation. Most important, the high amplitudes suggests that the constraints in interferometer range and force authority in \eqref{eq:optimal_design:constraints} should be rewritten in a more suitable form for promptly handling the problem.

The investigation on an experiment in which there is a simultaneous injection of both guidance signals shows a higher level of complexity in the optimization and the conceptual understanding of the system. As correlation could not be resolved in this experiment, i.e., the parameters could still remain correlated, such an experiment may not worth to be implemented during the LPF mission. Moreover, as the cross-talk from one degree of freedom to the optical axis is better identified with independent injections, this case seems much more controllable and easier to interpret than the case of simultaneous injections. The procedures developed in this chapter suggest that particular designs can even be found, at least for the simpler case of independent injections, providing in principle the measurement of the system parameters with optimized uncertainty/correlation.





\chapter{Conclusions and future perspectives} \label{chap:conclusion}


As a conclusion, it is worth to stress the key points presented in this thesis. This work can be ideally divided into the following parts:
\begin{enumerate}
  \item a theoretical contribution to the foundations of spacetime metrology that will be demonstrated with LPF, in which TMs are required to free fall with unprecedented pureness, to within $\unit[3\e{-14}]{m\,s^{-2}\,Hz^{-\nicefrac{1}{2}}}$ around $\unit[1]{mHz}$, and whose relative motion must be optically tracked with an accuracy better than $\unit[9\e{-12}]{m\,Hz^{-\nicefrac{1}{2}}}$ around $\unit[1]{mHz}$;
  \item a theoretical modeling of the dynamics of the LISA arm implemented in LPF;
  \item a description of the procedures developed for system identification, crucial for an accurate estimation of the total equivalent acceleration noise, and for the success of LPF.
\end{enumerate}

In particular, \chapref{chap:metrology} showed a derivation of the Doppler link response to GW signals, different from the well-known integration of null geodesics. The parallel transport of the emitter 4-velocity along the photon geodesic induces a time delay into the physical observable, the frequency shift. Hence, time delays track the effect of GWs on the Doppler link. The Doppler link is the measurement element of all space-based GW detectors, like LISA. The chapter also showed how curvature directly affects the frequency shift along the beam -- a measurement that is concurrent to both parasitic differential accelerations and non-inertial forces due to the particular choice of the reference frame. Moreover, there are many sources of non-idealities to be taken into account. The link is actually implemented with lower-measurements between four bodies in LISA and three bodies in LPF, so the body extension and misalignments in the optical elements couple with the main measurement axis, still affected by parasitic acceleration and non-inertial forces. It is useful for the discussion to treat all signals and noise sources as equivalent differential accelerations, input to the Doppler link reformulated as a time-delayed differential accelerometer.


LPF is the in-flight test of a down-scaled version of a single LISA arm. Most of the control philosophy, actuation and dynamics is inherited from the LISA design, with slight differences discussed in the text. As the control plays a crucial role in LPF for the compensation of the differential forces of the two TMs toward the SC, \chapref{chap:dynamics} described the sophisticated closed-loop dynamics of two TMs within a hosting SC, whose relative motion is sensed by an interferometer and capacitive sensors. The LPF dynamics can be modeled as vector equations in which operators describe dynamics, sensing and control -- three essential constituents of the system. In view of deriving a generalized equation of motion, a differential operator was identified. The operator has a twofold relevance: on one side, it allows for the conversion of the sensed relative motion into the total equivalent acceleration; on the other side, such an operation requires the calibration of the system through another operator exactly defined from the differential operator itself. The formalism effectively helps in the subtraction of the couplings, the control, the SC jitter and the system transitory. The chapter novelly showed that the accuracy to which transients can be suppressed depends on the accuracy to which the modeled system parameters have been estimated from targeted experiments. The chapter presented a dynamical model for LPF along the optical axis that was used in the analysis of this thesis and is planned to be employed during the mission. As the characterization of the dynamics along the optical axis -- the main measurement axis -- is the first target of the mission, the formalism was employed to derive the equations governing the cross-talk, with a supporting example, from other degrees of freedom to the nominal dynamics along the optical axis.

The estimation of the total equivalent acceleration noise requires the calibration of the differential operator converting the sensed motion into equivalent acceleration. The operator contains critical parameters modeling different non-ideality contributions like spring-like couplings between the TMs and the SC, sensing cross-talk coefficients, actuation gains and delays in the actual application of forces. The goal of \chapref{chap:sys_identification} was to describe the methods proposed, developed and tested to simulated experiments aimed at system identification, i.e., the identification of those parameters crucial for the estimation of the total equivalent acceleration noise, the substraction of couplings, control forces, cross-talk and system transients in the recovered acceleration time-series. The methods were applied to data simulated with the same model for validation purpose (Monte Carlo simulation), but also to data released by the OSE, the realistic simulator provided by ESA. In a mission-like approach, different non-standard scenarios were considered: under-performing actuators, under-estimated couplings and an example of non-Gaussianities. Since the estimated equivalent acceleration noise depends on the estimated system parameters, this chapter showed for the first time that system identification is mandatory for the estimation of the equivalent acceleration noise. Otherwise, systematic errors like the ones described in this chapter might compromise the scientific objectives of the mission. As said, system identification allows for mitigating transients in the data. In the end, the chapter showed an example of application -- completely in the transitory regime -- to data released by the OSE.

Since parameter estimation has fundamental importance for the achievement of the mission requirements, \chapref{chap:optimal_design} investigated on the design of optimal identification experiments. This allows for the estimation of the system parameters with better precision. Intuitively, better precision in the estimated parameters is equivalent to better confidence in the estimated equivalent acceleration noise. The standard theory of optimal design was applied by taking into account the peculiarities, constraints and complexity of LPF. The result was that the system can be stimulated with only two frequencies, obtaining a gain in precision, to within an order of magnitude in parameter standard deviation. It was also found that the two frequencies stimulate the system into two regimes: the high-frequency regime dominated by the sensing and the low-frequency regime dominated by the force couplings.

Evidently, this work covered only a restricted part of the experiments, the investigations, the measurements and the scientific returns of the LPF mission. First, more work could be done developing the theoretical description of the Doppler link as a differential accelerometer in \chapref{chap:metrology}. Second, the methods described in \chapref{chap:sys_identification} might be also employed, as they are, in an extensive investigation of the various possible cross-talk experiments and the calibration of LISA-like data. \chapref{chap:sys_identification} showed the robustness of the methods to a couple of non-standard scenarios that might happen during the mission. Additional investigation may be required for the possibility that the measured noise would contain non-Gaussian, non-stationary and transient components, even in the form of unmodeled transient signals. Finally, as a conclusion to the investigation of \chapref{chap:optimal_design}, the optimized designs should be also checked out with the OSE simulator.

This thesis showed the relevance of system identification for the correct assessment -- and the subtraction of various disturbances -- of the total equivalent differential acceleration. The total equivalent acceleration characterizes the performance in sensitivity of the LISA arm, viewed as a differential time-delayed accelerometer. Therefore, system identification is crucial for the success of LPF in demonstrating the principles of spacetime metrology needed for all future space-based missions.




\renewcommand{\appendixname}{}
\appendix



\ChangeFigFolder{9_backmatter}


\chapter{Appendix} \label{chap:appendix}


\section{A single galactic binary in LISA noise} \label{sect:appendix:introduction_binary}

\begin{figure}[!htbp]
\centering
\begin{tabular}{*{2}{@{}c@{}}} 
\multicolumn{2}{c}{\includegraphics[width=0.5\columnwidth]{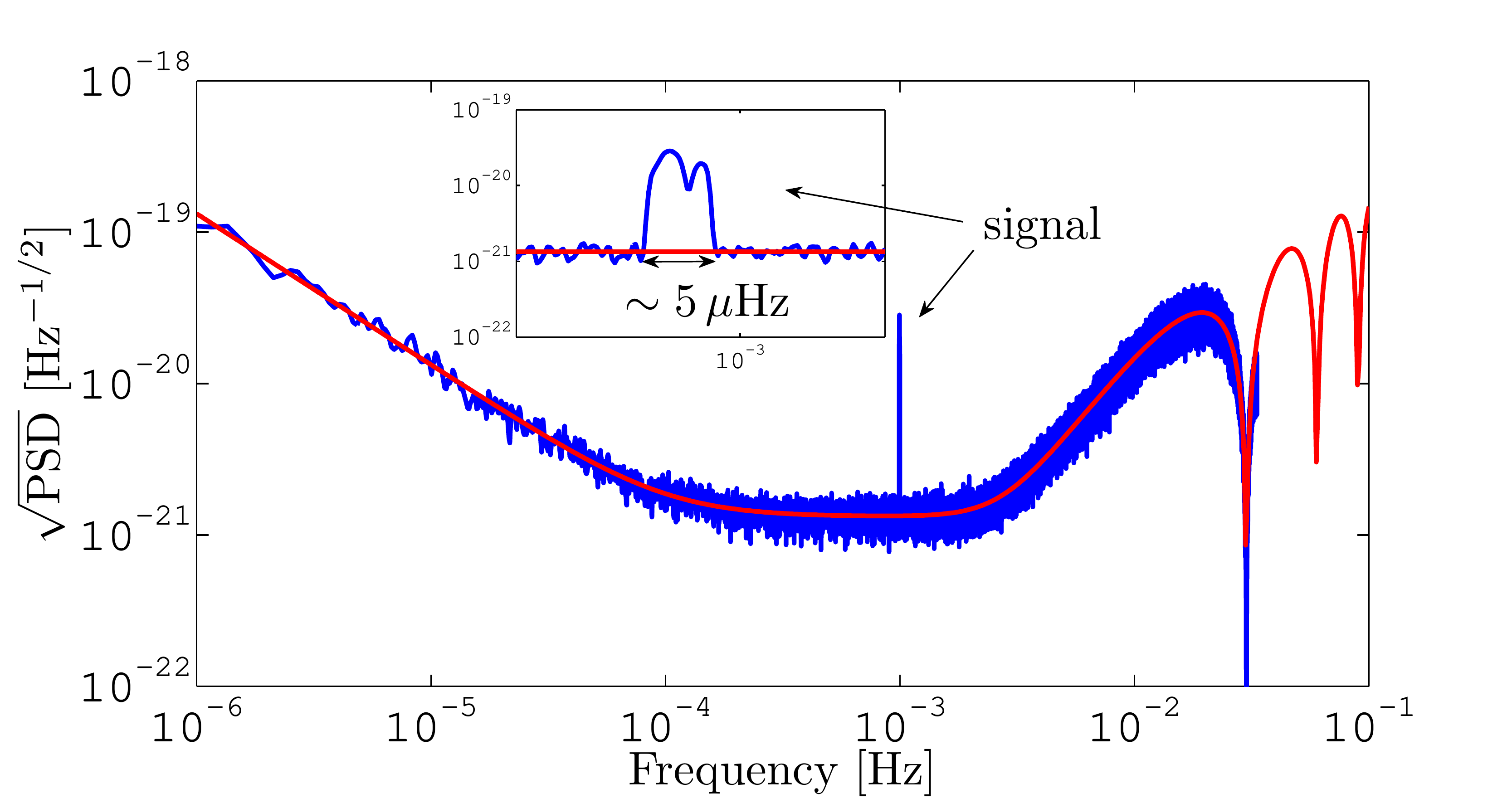}} \\
\multicolumn{2}{c}{\footnotesize{(a)}} \\
\includegraphics[width=0.5\columnwidth]{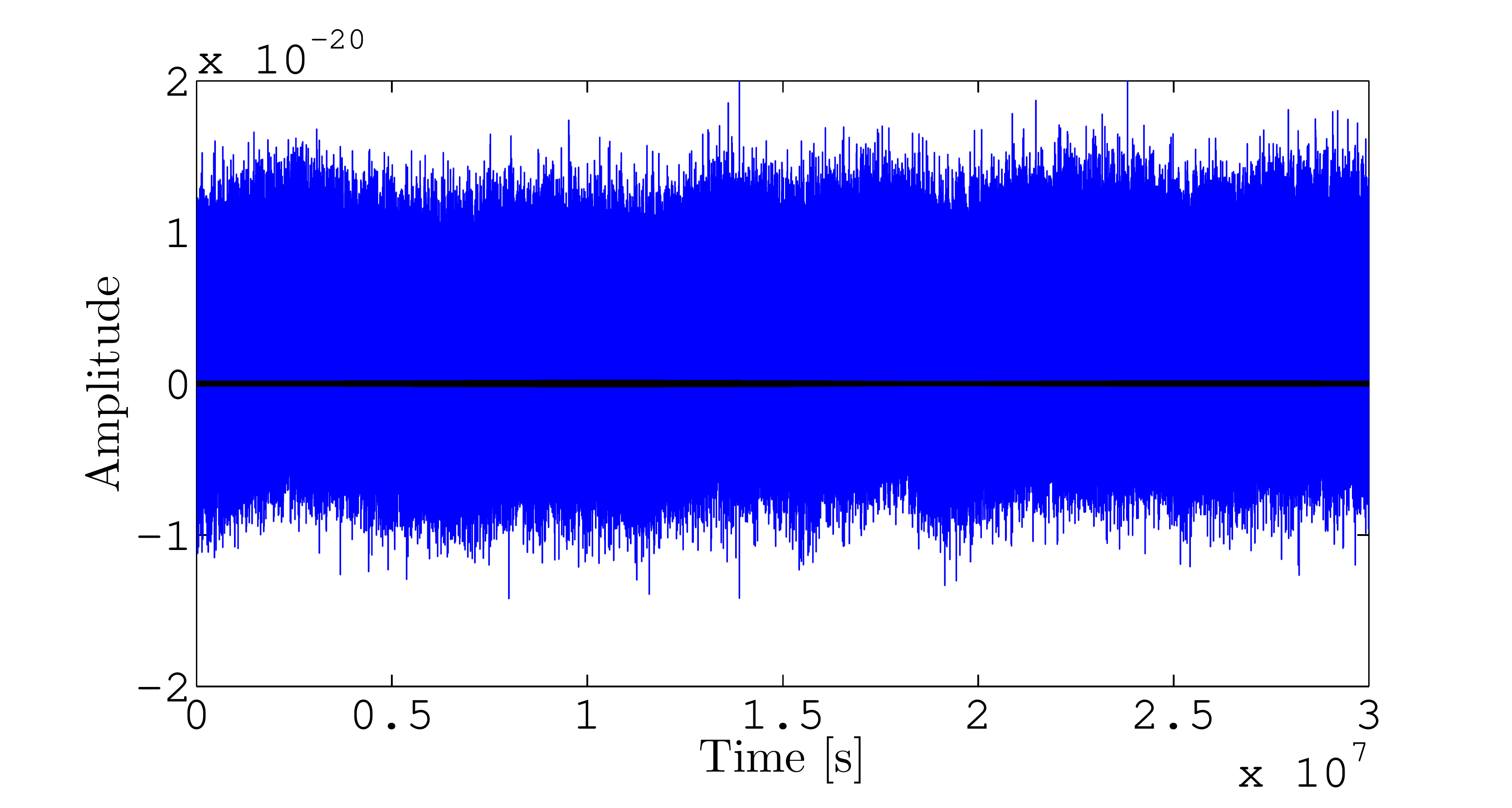} &
\includegraphics[width=0.5\columnwidth]{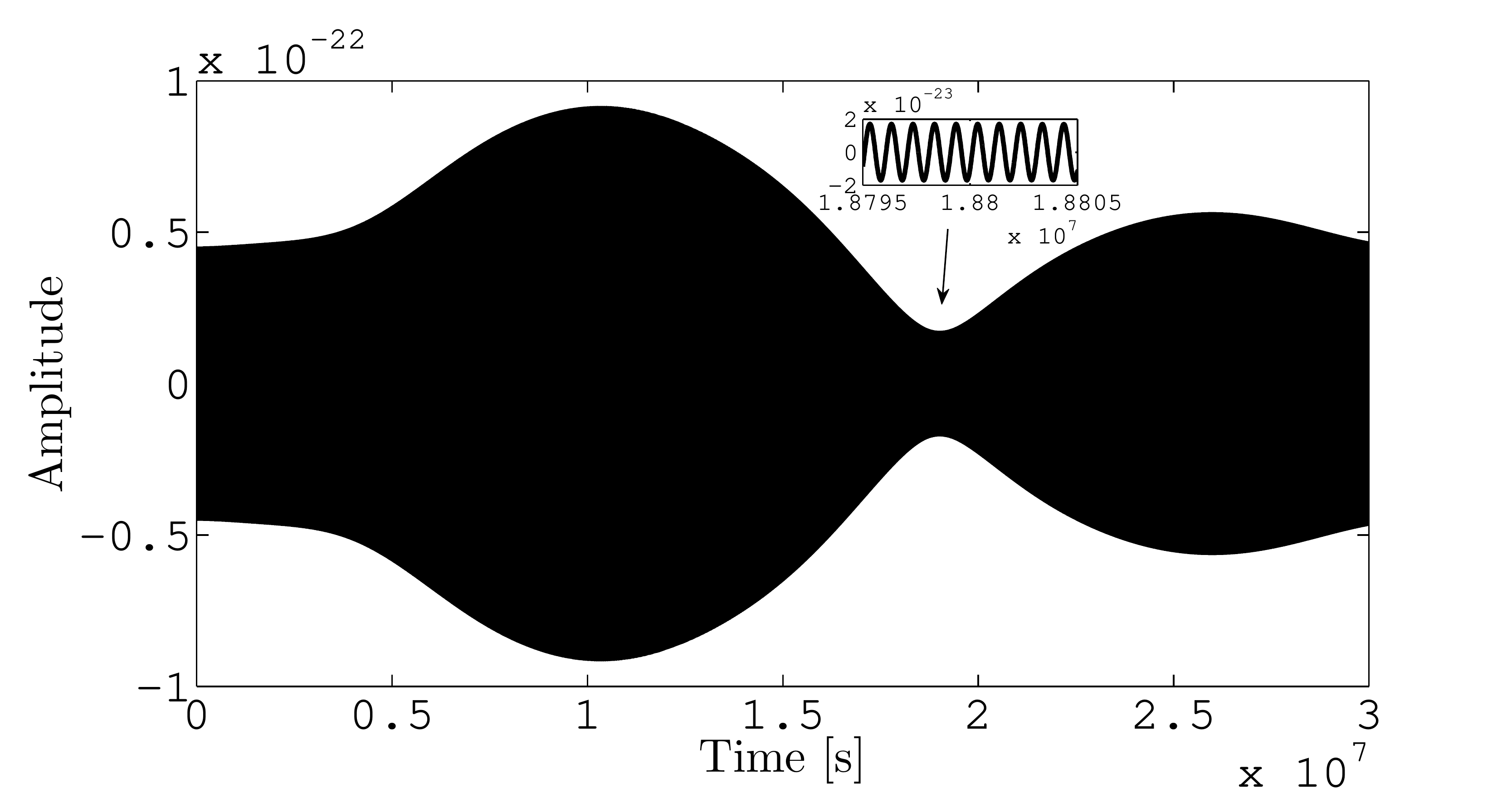} \\
\footnotesize{\hspace{10pt}(b)} & \footnotesize{\hspace{10pt}(c)}
\end{tabular}
\caption{\footnotesize{Simulation of a single galactic binary around $\unit[1]{mHz}$, weakly chirping at the rate of $\unit[10]{\mu Hz}$ over 2 years, in the detector noise as seen by the $X$ LISA interferometer. (a) the signal appears as a spike in the spectrum as large as the effect of the chirping is more prominent. (b) the relative time-domain signature containing the noisy time-series and the signal itself. (c) details of the source signal, where the annual modulation due to the revolution of the constellation around the Sun is evident. LISA will be able in detecting thousands of such sources, including signals from the merging of SMBHs (with overwhelmingly large SNR), the galactic binary foreground at low frequency and the EMRIs (with very low SNR).}}
\label{fig:appendix:binary}
\end{figure}

\figref{fig:appendix:binary} shows the response of the detector \cite{krolak2004} to the injection of a single galactic binary around $\unit[1]{mHz}$, weakly chirping at the rate of $\unit[10]{\mu Hz}$ over 2 years, in the $X$ (1st generation TDI) channel \cite{tinto2005a}. Noise is simulated according to the model described in \cite{estabrook2000}. The estimated PSD \footnote{Refer to footnote on pag.\;\pageref{foot:sys_identification:psd} for a brief description of the employed method for spectral estimation.} is also compared to the noise model showing the self-consistency of the data generation process \cite{ferraioli2010}. Notice the convolution of the signal with the detector inducing the annual modulation. The Doppler modulations are responsible of such complex LISA response, but allows for a very precise identification of parameters like the source position and polarization. LISA will be able in detecting thousands of such sources superimposed to the variety of signals as briefly described in the Introduction.

\section{Non-pure free fall and Fermi-Walker transport} \label{sect:appendix:metrology_fermi-walker}

The are two main differences between a realistic and an idealistic description of free-falling TMs making in practice a Doppler link:
\begin{enumerate}
\item the bodies are \textit{nearly} in free fall, i.e., accelerations are very small, but not zero;
\item the bodies have a finite extension coupling with the Doppler response and producing extra-acceleration.
\end{enumerate}
The \textit{Fermi-Walker transport} (FW) is the same underlying principle. In fact, the FW transport comes out every time the acceleration is different from zero and can even vary with time. In such situations, the best implementation of a local co-moving frame -- also defining the body reference frame -- is the one having gyroscopes attached to the three space axes. This construction prevents the space coordinates from rotating and forces them to be fixed as the time flows. In this reference frame, any 4-vector $x^\mu$ is differentiated with respect to the proper time of the geodesic following the rule \cite{misner}
\begin{equation}\label{eq:metrology:fermi_walker_transport}
\der{x^\mu}{\tau} = \Omega^{\mu\nu} x_\nu~.
\end{equation}
It is easy to recognize the ordinary cross product between the angular velocity and the vector itself in the non-relativistic regime, so the FW reference frame provides a generalization of the notion of angular velocity in GR. The antisymmetric tensor $\Omega^{\mu\nu}$ contains all Lorentz transformations (rotations and boosts), but no space rotations, and is given by
\begin{equation}\label{eq:metrology:fermi_walker_def}
\Omega^{\mu\nu} = \frac{1}{c^2}\left(v^\mu a^\nu - v^\nu a^\mu\right)~,
\end{equation}
where $v^\mu$ and $a^\mu$ are the body velocity and acceleration. When \eqref{eq:metrology:fermi_walker_def} holds true together with \eqref{eq:metrology:fermi_walker_transport}, then $x^\mu$ is said to be FW transported along the same geodesic. Hence, four orthogonal vectors being FW transported along the accelerated body geodesic define the local FW co-moving frame.

In LPF the philosophy is different: there no gyroscopes attached to the TMs, as the small linear and angular motion are indeed used to gather information on the acceleration noise affecting the TM geodesic. As pointed out in the thesis, the situation is more complicated since the TM-to-TM Doppler link is carried out with three independent lower-level measurements evidently introducing new couplings.

\section{Calculation in metrology without noise} \label{sect:appendix:metrology_calculation}

This section demonstrates the formula \eqref{eq:metrology:freqshift_nearby_h_term} with a detailed calculation in the TT gauge and in the instantaneous wave coordinate system. Indeed,
\begin{equation}
\begin{split}
h^\mu_{~\beta\,,\alpha} k_\mu k^\beta & = h_{\mu\beta\,,\alpha} k^\mu k^\beta \\
& = h_{1\beta\,,\alpha} k^1 k^\beta + h_{2\beta\,,\alpha} k^2 k^\beta \\
& = h_{11\,,\alpha} k^1 k^1 + h_{12\,,\alpha} k^1 k^2 + h_{21\,,\alpha} k^2 k^1 + h_{22\,,\alpha} k^2 k^2 \\
& = h_{+\,,\alpha} k_x^2 + h_{\times\,,\alpha} k_x k_y + h_{\times\,,\alpha} k_y k_x - h_{+\,,\alpha} k_y^2 \\
& = \left(k_x^2-k_y^2\right)h_{+\,,\alpha} + 2 k_x k_y\,h_{\times\,,\alpha}~,
\end{split}
\end{equation}
where the coefficients of $h_{+\,,\alpha}$ and $h_{\times\,,\alpha}$ are $K_+$ and $K_\times$.

\section{Linearized Einstein equations for Doppler link as differential accelerometer} \label{sect:appendix:metrology_einstein}

In this section the linearized Einstein equations are solved for the calculation of \sectref{sect:metrology:doppler_link_diff_acc}. The linearized Einstein equations \cite{misner} are given by
\begin{equation}
\begin{split}
h_{\mu\alpha,\nu}^{\quad~~\alpha} & + h_{\nu\alpha,\mu}^{\quad~~\alpha} - h_{\mu\nu,\alpha}^{\quad~~\alpha} - h^\alpha_{~\alpha,\mu\nu} \\
& - \eta_{\mu\nu}\left(h_{\alpha\beta}^{\quad,\alpha\beta} - h^{\alpha\quad\beta}_{~\alpha,\beta}\right) = 0~,
\end{split}
\end{equation}
where $\eta_{\mu\nu}$ is the flat Minkowski metric, $h_{\mu\nu}$ is the first-order perturbation and the gauge is arbitrary.

In the $(ct,x)$ coordinates
\begin{subequations}
\begin{align}
h^\alpha_{~\alpha} & = h_{00} - h_{11}~, \\
h_{\alpha\beta}^{\quad,\alpha\beta} & = h_{00,00} - 2h_{01,01} + h_{00,11}~, \\
\begin{split}
h^{\alpha\quad\beta}_{~\alpha,\beta} & = h^\alpha_{~\alpha,00} - h^\alpha_{~\alpha,11} \\
& = h_{00,00} - h_{11,00} - h_{00,11} + h_{11,11}~.
\end{split}
\end{align}
\end{subequations}
For $\mu=\nu=0$, the Einstein equations provide
\begin{equation}
\begin{split}
0 & = 2h_{0\alpha,0}^{\quad~~\alpha} - h_{00,\alpha}^{\quad~~\alpha} - h^\alpha_{~\alpha,00} \\
& \quad - h_{\alpha\beta}^{\quad,\alpha\beta} + h^{\alpha\quad\beta}_{~\alpha,\beta} \\
& = 2h_{00,00}-2h_{01,01}-h_{00,00}+h_{00,11}-h_{00,00}+h_{11,00} \\
& \quad - h_{00,00} + 2h_{01,01} - h_{00,11} + h_{00,00} - h_{11,00} - h_{00,11} + h_{11,11} \\
& = - h_{00,11} + h_{11,11}~.
\end{split}
\end{equation}
For $\mu=0$ and $\nu=1$,
\begin{equation}
\begin{split}
0 & = h_{0\alpha,1}^{\quad~~\alpha} + h_{1\alpha,0}^{\quad~~\alpha} - h_{01,\alpha}^{\quad~~\alpha} - h^\alpha_{~\alpha,01} \\
& = h_{00,01} - h_{01,11} + h_{01,00} - h_{11,01} - h_{01,00} + h_{01,11} - h_{00,01} + h_{11,01} \\
& = 0~.
\end{split}
\end{equation}
For $\mu=\nu=1$,
\begin{equation}
\begin{split}
0 & = 2h_{1\alpha,1}^{\quad~~\alpha} - h_{11,\alpha}^{\quad~~\alpha} - h^\alpha_{~\alpha,11} \\
& \quad + h_{\alpha\beta}^{\quad,\alpha\beta} - h^{\alpha\quad\beta}_{~\alpha,\beta} \\
& = 2h_{01,01} - 2h_{11,11} - h_{11,00} + h_{11,11} - h_{00,11} + h_{11,11} \\
& \quad + h_{00,00} - 2h_{01,01} + h_{00,11} - h_{00,00} + h_{11,00} + h_{00,11} - h_{11,11} \\
& = h_{00,11} - h_{11,11}~.
\end{split}
\end{equation}
Therefore, in the approximations of \sectref{sect:metrology:doppler_link_diff_acc}, the Einstein equations reduce to
\begin{equation}\label{eq:appendix:metrology_einstein}
h_{00,11} - h_{11,11} = 0~.
\end{equation}


\section{Demonstration of noise non-stationarity} \label{sect:appendix:non_stationary_noise}

This section demonstrates the validity of \eqref{eq:sys_identification_noise_variance}, i.e., that the fluctuation of any of the system parameter produces non-stationary noise. Expanding the noise around some nominal parameter value $p_0$, up to first order, and computing the variance of the interferometric noise, it reads
\begin{equation}\label{eq:appendix:non_stationary_noise}
\begin{split}
\var[o] & \simeq \var\left[o_0\right]+\var\left[o'\delta p\right]+2\cov\left[o_0,o'\delta p\right] \\
& = \var\left[o_0\right]+\var\left[o'\right]\delta p^2+2\cov\left[o_0,o'\right]\delta p~,
\end{split}
\end{equation}
where $\var\left[o'\right]$ and $\cov\left[o_0,o'\right]$ are the variance of the noise first derivative and the covariance between the zeroth order and the first derivative. So, for a zero-mean process with finite second moment, it holds
\begin{equation}
\begin{split}
\cov\left[o_0,o'\right] & = \text{E}\left[o_0 o'\right]-\text{E}\left[o_0\right]\text{E}\left[o'\right] \\
& = \text{E}\left[\frac{1}{2}\frac{\partial}{\partial p} n^2\right] \\
& = \frac{1}{2}\frac{\partial}{\partial p}\var[n]~.
\end{split}
\end{equation}
Substituting this result back into \eqref{eq:appendix:non_stationary_noise}, \eqref{eq:sys_identification_noise_variance} is finally demonstrated.

\section{Time-frequency analysis of non-stationary noise} \label{sect:sys_identification:non-stationary_noise_time_freq}

Noise stationarity is the most important assumption taken by the standard spectral estimation. There are cases where the estimated PSD or even more advanced techniques like the Kolmogorov-Smirnov test \cite{ferraioli2012} -- aimed at comparing the cumulative distribution function of the noise PSD compared to a reference (either another noise measurement or a model expectation) -- may fail in detecting noise excesses or small transient signals concentrated in very narrow time segments. To explain why, this section is devoted to showing an example where such a problem can be found and how to promptly deal with it by employing fast and efficient tools like the wavelet analysis \cite{mallat}.

Without loss of generality, in what follows it is considered that a short transient force (per unit mass) gradient is modeled as a Gaussian-shaped signal
\begin{equation}
f_{12,\text{tr}}(t) = a\,\exp\left[-\frac{(t-t_0)^2}{\tau^2}\right]~,
\end{equation}
of total duration $2\tau\oforder1$ hour ($\tau=\unit[1\e{3}]{s}$), is turned on after about 8 hours ($t_0=\unit[3\e{4}]{s}$) during a LPF noise run of about 12 hours and with amplitude $a=\unit[1.6\e{-13}]{m\,s^{-2}}$. The gradient might be either due to anomalous transient force couplings temporarily entering into the noise budget, or effectively unexpected signals.

An example is the prediction of a gradient surplus as a deviation from Newtonian gravity, described in \cite{magueijo2011,trenkel2010}, where a flyby of the saddle point of the Sun-Earth potential surface is proposed for the natural conclusion of the LPF mission toward the escaping trajectory to test for alternative theories of gravity. Even though there are different models claiming that a gradient with high SNR could be detected if the SC would cross the ``bubble'' around the saddle point with sufficient small impact parameter, in practice it is likely that the SC orbit will never reach such an accuracy. Hence, it is a good idea to start by looking for very small departures and putting thresholds to the observability of noise transients. So much far beyond the objectives of this section, the following gives a very first address of the problem.

\figref{fig:sys_identification:o12_excess} shows a simulation of a noise run of about 12 hours in the differential interferometer readout, together with the system response to the external signal (i.e., the gravity gradient excess as in the example above) of absolute peak $\oforder\unit[5]{nm}$. It is evident that: (i) the signal could be easily confused with the intrinsical noise fluctuation; (ii) PSD estimation \textit{can} warn of a change in the shape of the spectrum (a sign of non-stationarity), sometimes by a huge amount, sometimes by a negligible amount as in this example; but it does not say much about \textit{where} is changing and on what time scale, as the location is fundamentally important for the analysis of transients.

\figuremacroW{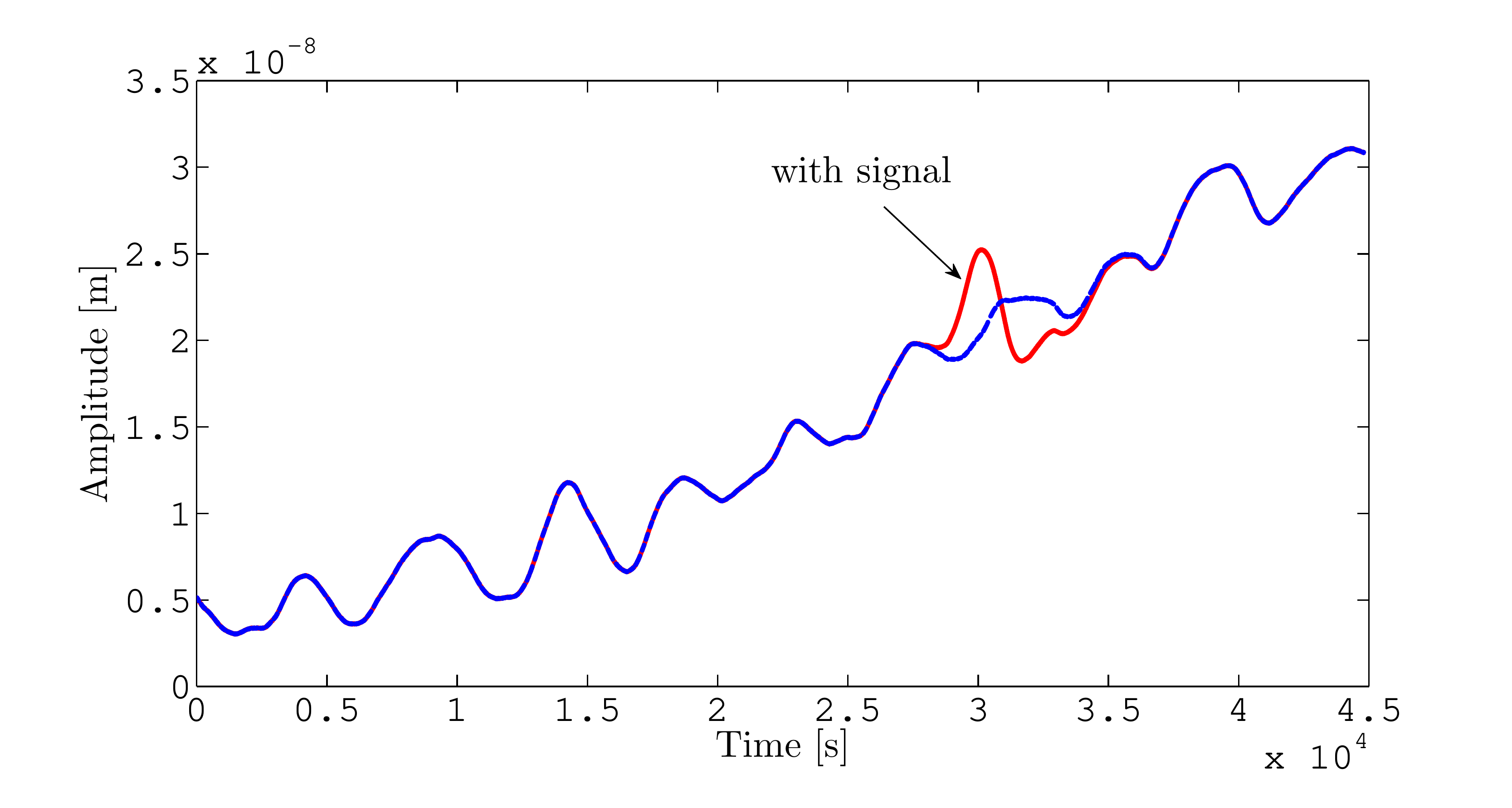}{A simulated noise run of the differential interferometer readout lasting for about 12 hours compared to the same with a gradient force injected into the system. The signal, of absolute peak $\oforder\unit[5]{nm}$, can be easily confused with the noise. PSD estimation is not capable to quantitatively assess the significance of the excess, both in term of frequency and position of the transient.}{fig:sys_identification:o12_excess}{0.8}

A solution is provided by the continuous wavelet transform that gives a full time-frequency representation of the data series. Without going through the mathematical details, the data stretch is decomposed into continuous waves, the wavelets, that are the equivalent to the Fourier sines. The Fourier transform is a function of frequency; the wavelet transform is function of both time and scale. The time dependency gives the energy content with respect to the wavelet location. The scale dependency gives the energy content with respect to the wavelet compression. Therefore, the scale is inversely proportionally to the frequency and, in fact, it is possible to associate an approximate frequency to the scale of a given wavelet. A detailed discussion can be found in \cite{mallat}.

\figref{fig:sys_identification:o12_spectrogram} reports the time-frequency representation, the spectrogram, of \figref{fig:sys_identification:o12_excess} for second-order Daubechies wavelets. The power is scaled to the total energy in the time-frequency bands, so that the spectrogram is normalized to one. The transient signal is visible as the narrow and darker line around the instant of injection. Notice that its power is more than two times the other peaks, so it can be easily identified in a quick-look search of unmodeled transient signals.

\figuremacroW{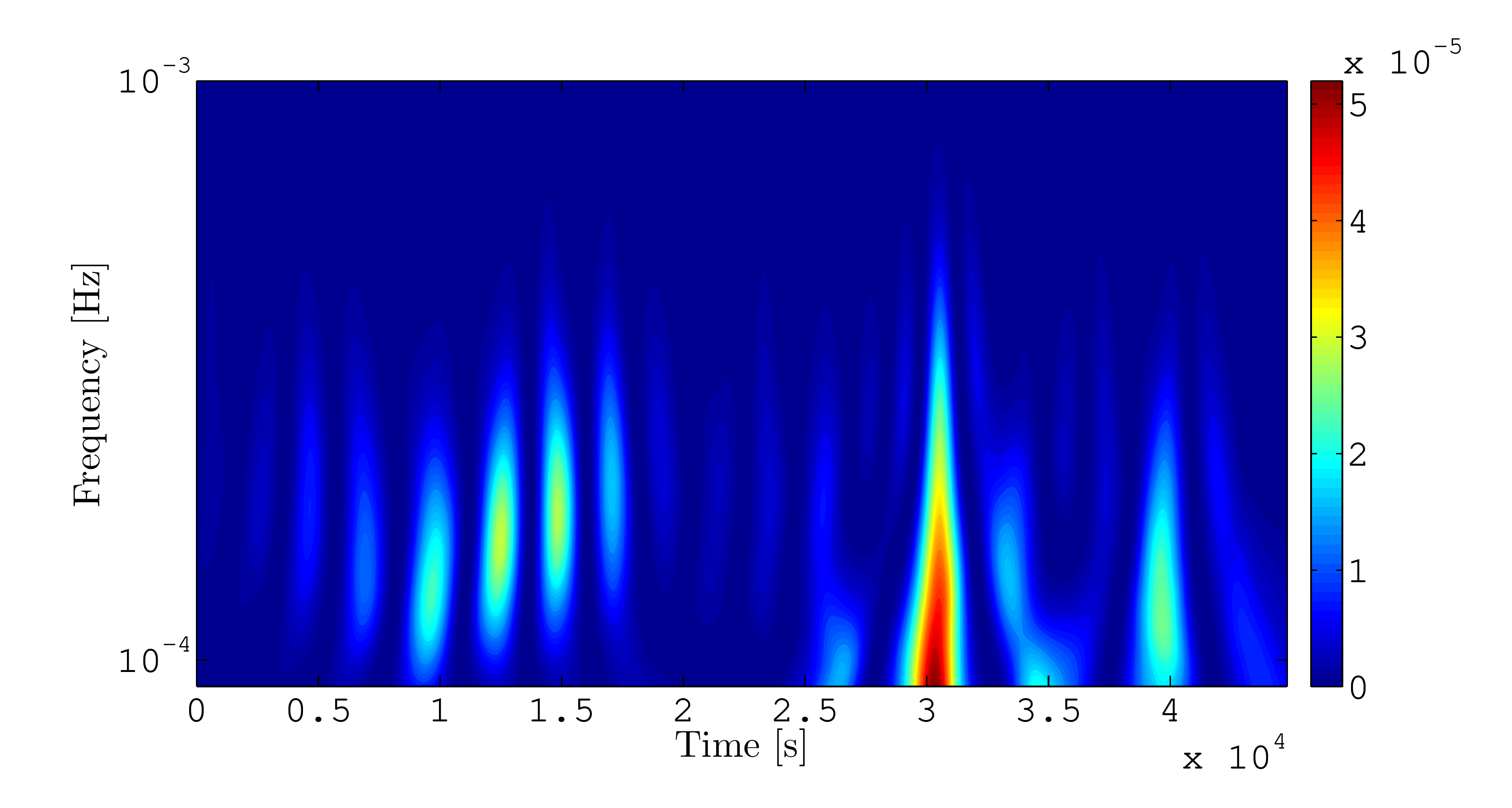}{Wavelet-based spectrogram of a simulated noise run of the differential interferometer readout lasting for about 12 hours in which a tiny force signal of peak amplitude $\unit[1.6\e{-13}]{m\,s^{-2}}$ is turned on after about 8 hours and producing the interferometric response showed in \figref{fig:sys_identification:o12_excess}. The transient signal is visible as the narrow and darker line at the instant of injection. The method allows for the identification of short unmodeled transient signals and excess noise.}{fig:sys_identification:o12_spectrogram}{0.8}

It is worth noting that an extensive investigation on this thematic -- and in particular on de-noising techniques with the discrete wavelet transform -- would surely improve the understanding of the non-stationary behavior of the LPF noise, in view of a fast identification of unmodeled transient signals.

\section{More on Monte Carlo validation} \label{sect:appendix:montecarlo}

This section investigates a little further on the Monte Carlo simulation of \sectref{sect:sys_identification:montecarlo}, that demonstrated that all parameters are unbiased and Gaussian distributed, as well as their variances.

\begin{figure}[htb]
\centering
\begin{tabular}{*{2}{@{\hspace{-10pt}}c@{\hspace{-10pt}}}}
\includegraphics[width=0.5\columnwidth]{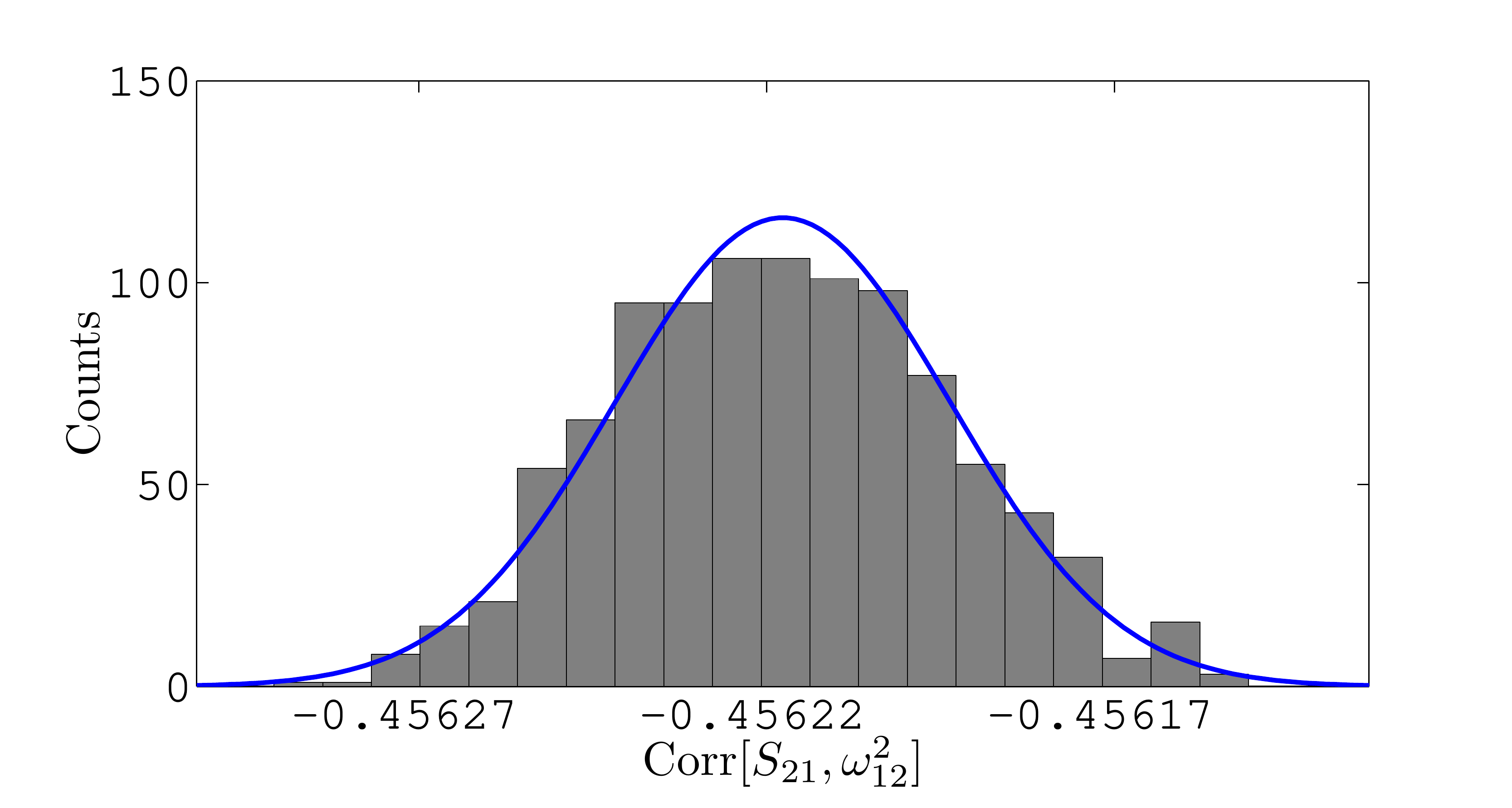} &
\includegraphics[width=0.5\columnwidth]{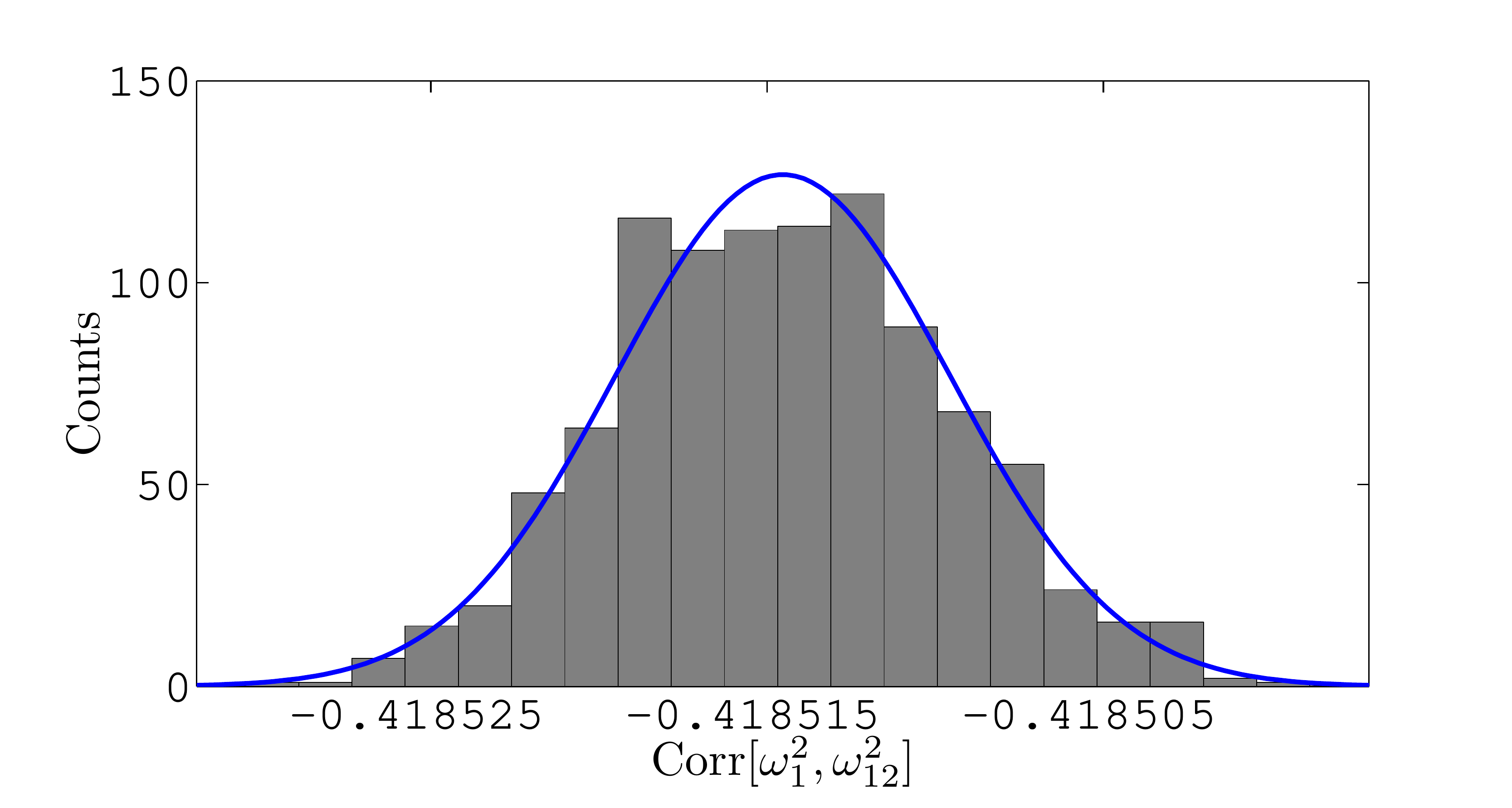} \\
\footnotesize{(a)} &
\footnotesize{(b)}
\end{tabular}
\caption{\label{fig:appendix:montecarlo_corr}\footnotesize{Monte Carlo validation of 1000 independent noise realizations on which parameter estimation is repeated identically at each step. The statistics is shown for two parameter correlations. The scaled Gaussian PDF is evaluated at the sample mean and standard deviation.}}
\end{figure}

Surprisingly, the correlations are also Gaussian distributed with good approximation. See \figref{fig:appendix:montecarlo_corr} for two examples.

\begin{figure}[htb]
\centering
\begin{tabular}{*{2}{@{\hspace{-5pt}}c@{\hspace{-5pt}}}}
\includegraphics[width=0.5\columnwidth]{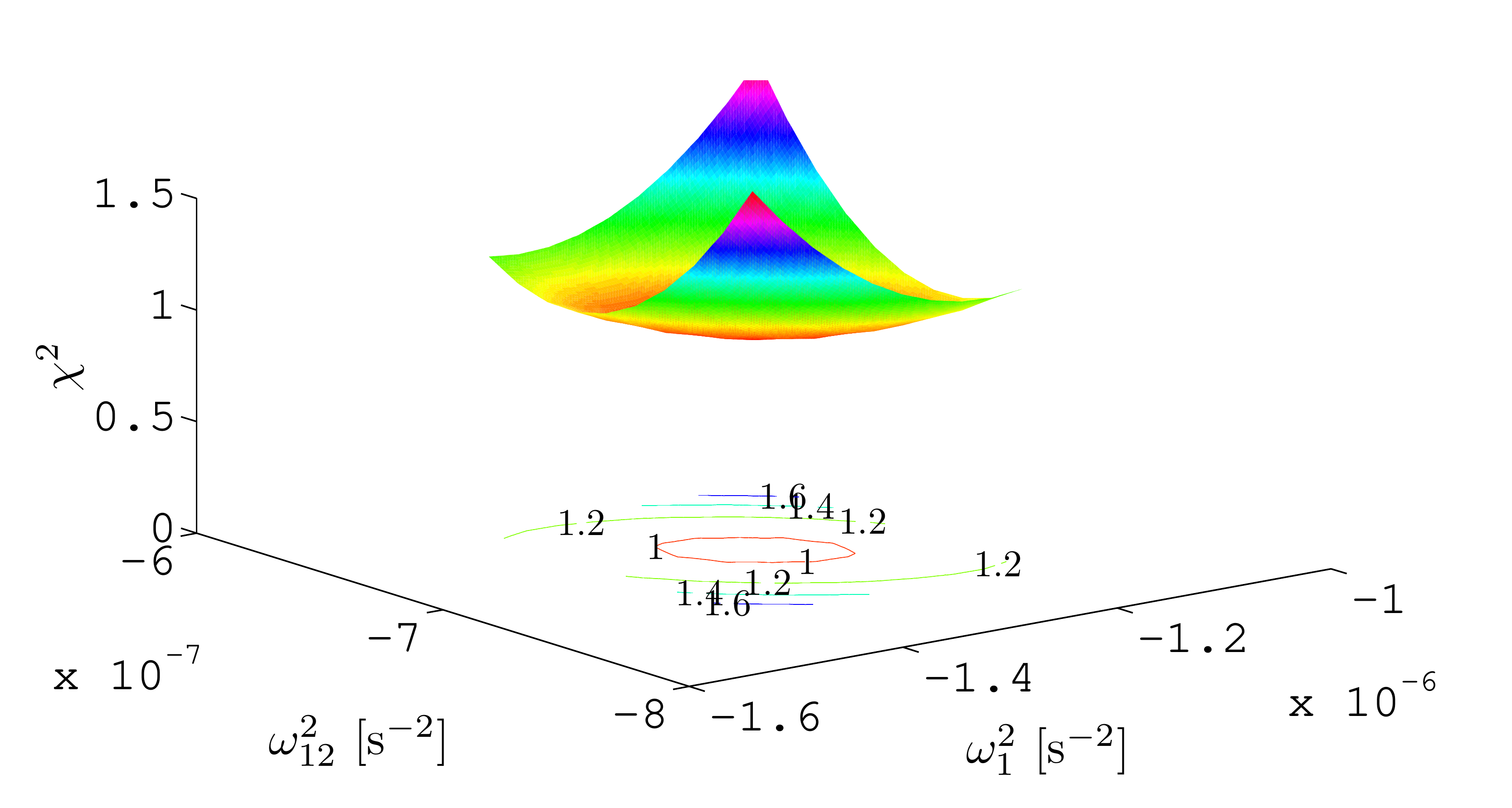} &
\includegraphics[width=0.5\columnwidth]{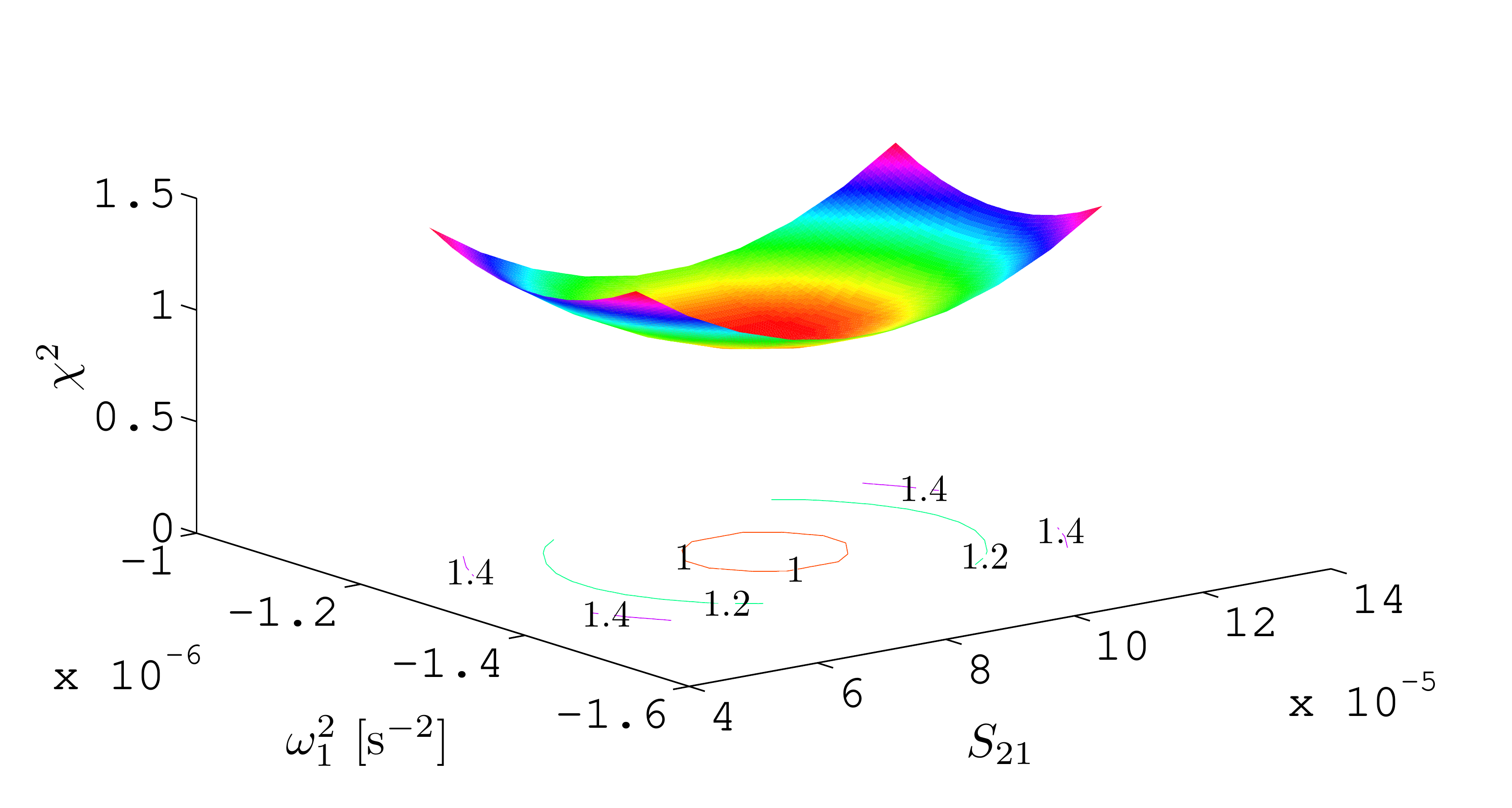} \\
\footnotesize{(a)} &
\footnotesize{(b)} \\
\includegraphics[width=0.5\columnwidth]{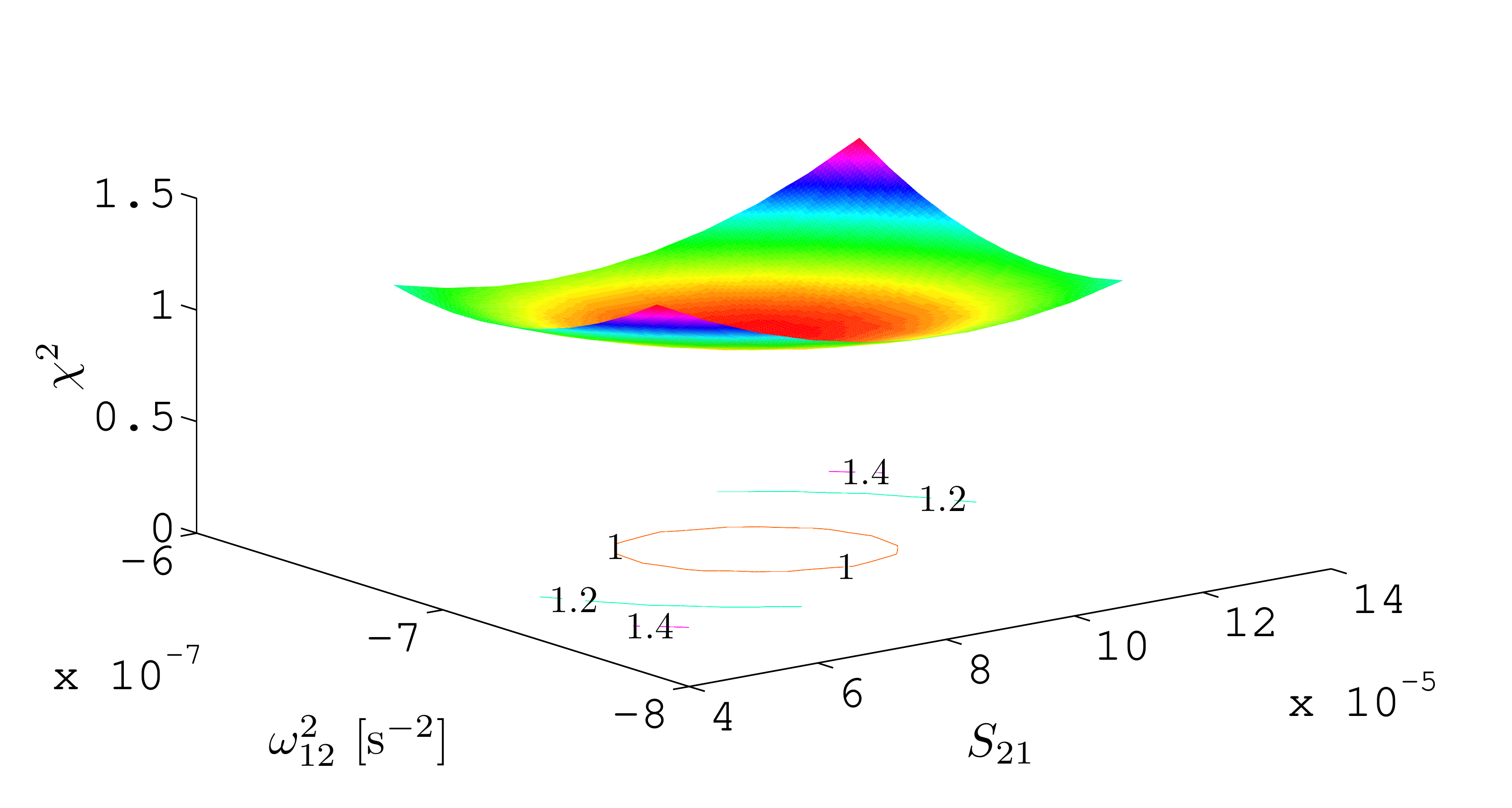} &
\includegraphics[width=0.5\columnwidth]{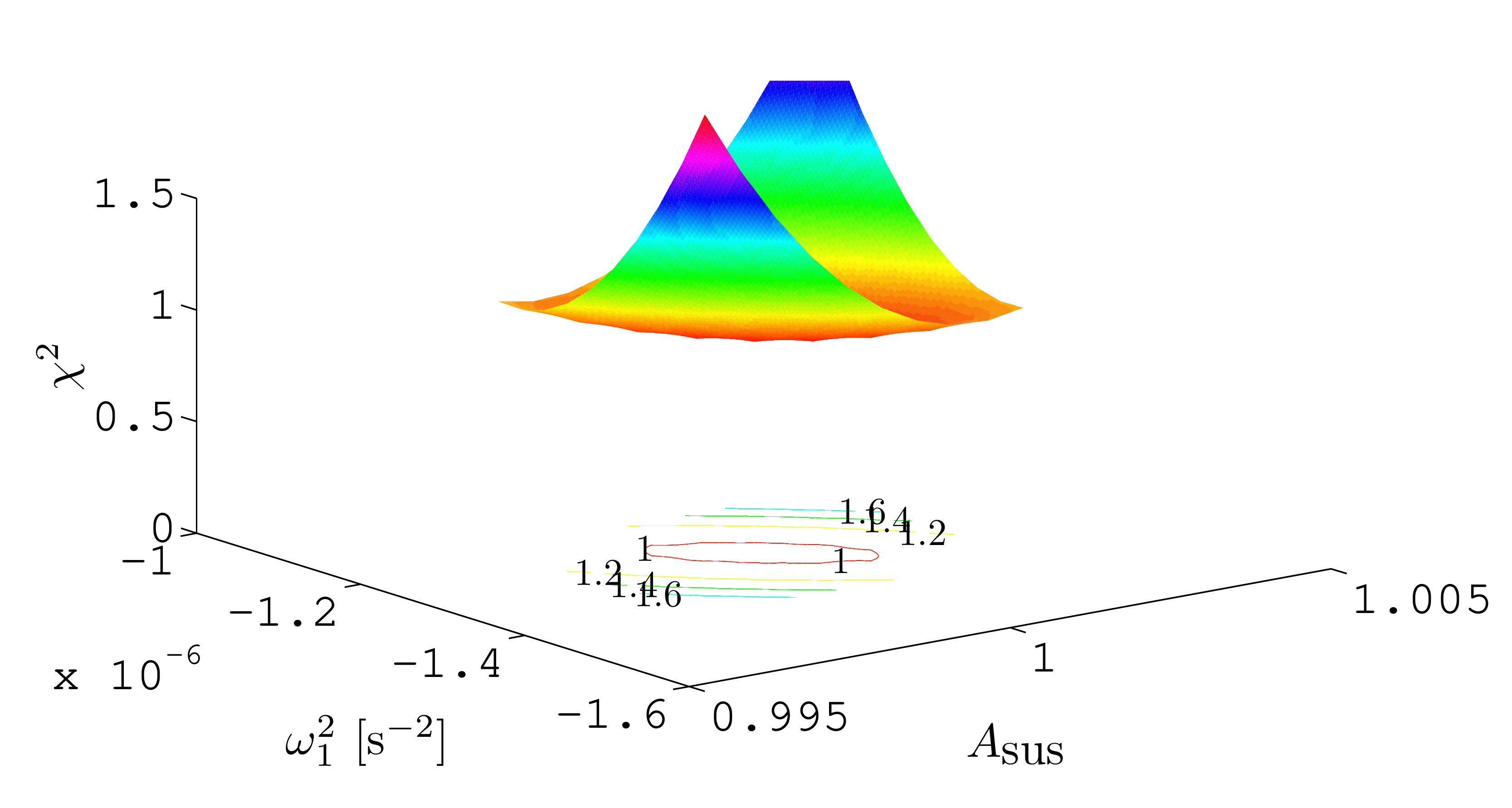} \\
\footnotesize{(c)} &
\footnotesize{(d)}
\end{tabular}
\caption{\footnotesize{$\chi^2$ log-likelihood curvature around the best-fit values. The 7-dimensional surface are projected onto two parameters at a time for some examples. Correlation is the reason why the surface can be rotated.}}
\label{fig:appendix:montecarlo_chi2_curv}
\end{figure}

The correlation between two parameters is somehow related to the rotation of the $\chi^2$ paraboloid principal axes around the minimum. To support this statement, \figref{fig:appendix:montecarlo_chi2_curv} shows few examples of projections of the $7$-dimensional surface onto two parameters at a time, around the best-fit values. Weakly correlated parameters, like $S_{21}$ and $\omega_1^2$ ($\oforder20\%$) (panel (b)), typically have the principal axes of the contour curves aligned with the $x$ and $y$ axis. Highly correlated parameters, like $A_\text{sus}$ and $\omega_1^2$ ($\oforder-70\%$) (panel (d)), have the principal axes that are significantly rotated.

\figref{fig:appendix:montecarlo_chi2_chains} shows a record history of all Monte Carlo estimation chains. The scatter of the chains is due to the noise fluctuation along the Monte Carlo iterations. There are clearly some chains that are far away from the accumulation zone: this behavior is quite unexpected as one would think the noise to have little impact on the chain locations. Despite the big scatter, the asymptotic distribution is Gaussian, as elucidated in \figref{fig:sys_identification:montecarlo_chi2}.

\figuremacroW{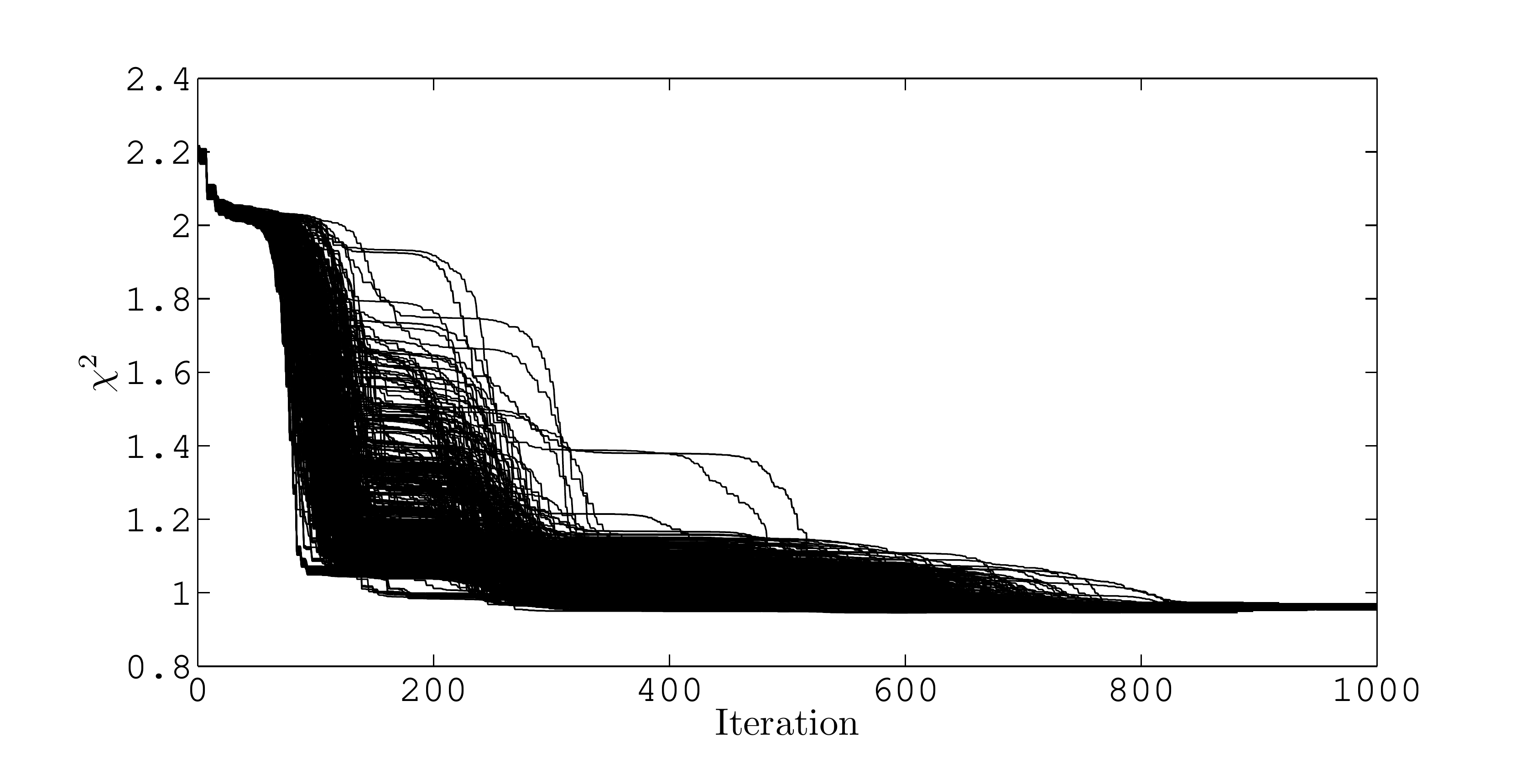}{Monte Carlo fit $\chi^2$ chains. The processes typically last for $\oforder1000$ iterations and stop when either the function or the variable tolerance is below $1\e{-4}$.}{fig:appendix:montecarlo_chi2_chains}{0.8}



\begin{multicols}{2} 
\begin{scriptsize} 

\bibliographystyle{Latex/Classes/PhDbiblio-url2} 
\renewcommand{\bibname}{References} 

\bibliography{9_backmatter/references} 

\end{scriptsize}
\end{multicols}








\end{document}